\documentclass[11pt, twoside, openright, a4paper]{report}

\usepackage{etoolbox, ifthen}
\newbool{rectoVerso}

\setbool{rectoVerso}{true} 
\numdef{\entreeBiblio}{1} 

\newcommand{\titreArticle}{Cosmology with weak-lensing peak counts}
\newcommand{\nomBiblio}{Bibliography} 

\usepackage[paper=a4paper, left=27mm, right=27mm, top=32mm, headheight=5mm, headsep=5mm, bottom=32mm, footskip=10mm]{geometry} 
\usepackage{chngpage}

\usepackage[english]{babel} 
\usepackage[super]{nth}
\usepackage{lmodern}

\usepackage[utf8]{inputenc}
\usepackage[T1]{fontenc}

\usepackage{fancyhdr}
\ifbool{rectoVerso}{
    \fancyhead[LE]{\sf \nouppercase{\leftmark}}
    \fancyhead[RO]{\sf \nouppercase{\rightmark}}
    \fancyfoot[RE]{\sf PhD thesis of Chieh-An Lin}
    \fancyfoot[LO]{\sf \titreArticle}
    \fancyfoot[LE,RO]{\sf \thepage}
}{
    \fancyhead[RE,RO]{\sf \nouppercase{\rightmark}}
    \fancyfoot[LE,LO]{\sf \titreArticle}
    \fancyfoot[RE,RO]{\sf \thepage}
}
\fancypagestyle{plain}{
	\fancyhf{}

	\fancyfoot[C]{\sf \thepage}
}

\usepackage{titletoc}
\usepackage{tocbibind}
\titlecontents{chapter}[0em]{\vskip2ex \bfseries\sffamily \large}{\chaptername\ \thecontentslabel\ ---\ }{}{\hfill\contentspage}
\titlecontents{section}[2em]{}{\sf \thecontentslabel\hspace{1em}}{}{\titlerule*[0.8em]{.}\sf\contentspage}
\titlecontents{subsection}[4em]{}{\sf \thecontentslabel\hspace{1em}}{}{\titlerule*[0.8em]{.}\sf\contentspage}
\usepackage{titlesec}
\titleformat{\chapter}[display]{\bfseries\sffamily \huge \raggedleft}{\Huge{\chaptername\ \Roman{chapter}}}{1em}{}[\titlerule]
\titleformat{\section}[block]{\bfseries\sffamily \Large}{\arabic{chapter}.\arabic{section}}{1em}{}
\titleformat{\subsection}[block]{\bfseries\sffamily \large}{\arabic{chapter}.\arabic{section}.\arabic{subsection}}{1em}{}
\titleformat{\subsubsection}[block]{\bfseries\sffamily}{}{1em}{}
\titleformat{\paragraph}[runin]{\bfseries\sffamily}{}{1em}{}

\usepackage{soul}

\usepackage{natbib} 
\bibpunct{(}{)}{;}{a}{}{,} 
\usepackage{imakeidx}
\makeindex[intoc]

\usepackage[font=footnotesize, labelfont=sc]{caption}
\usepackage{graphicx, subfig}
\usepackage{longtable, array, amsmath}
\graphicspath{{Figures/}} 

\usepackage{amsmath, amssymb, amsfonts, amsthm}
\usepackage[squaren, Gray, cdot]{SIunits}
\usepackage{bbm}
\numberwithin{equation}{chapter} 

\usepackage{Paper_Linc}
\usepackage{Math_Linc}

\usepackage[encapsulated]{CJK}
\usepackage{ucs} 
\newcommand{\zh}[1]{\begin{CJK}{UTF8}{bsmi}#1\end{CJK}}

\defcitealias{Lin_Kilbinger_2015}{Paper I}
\defcitealias{Lin_Kilbinger_2015a}{Paper II}
\defcitealias{Lin_etal_2016a}{Paper III}
\defcitealias{Liu_etal_2014}{Liu X. et al.}
\defcitealias{Liu_etal_2014a}{Liu J. et al.}
\defcitealias{Liu_etal_2015}{Liu J. et al.}
\defcitealias{Liu_etal_2015a}{Liu X. et al.}
\defcitealias{Liu_etal_2016}{Liu X. et al.}
\newcommand{\PaperI}{\citetalias{Lin_Kilbinger_2015}}
\newcommand{\PaperII}{\citetalias{Lin_Kilbinger_2015a}}
\newcommand{\PaperIII}{\citetalias{Lin_etal_2016a}}

\newcommand{\nOne}{\widetilde{N}}
\newcommand{\nTwo}{\widetilde{N}_{,1}}
\newcommand{\nThree}{\widetilde{N}_{,2}}
\newcommand{\nFour}{\widetilde{N}_{,11}}
\newcommand{\nFive}{\widetilde{N}_{,22}}
\newcommand{\nSix}{\widetilde{N}_{,12}}
\newcommand{\kOne}{\widetilde{K}}
\newcommand{\kTwo}{\widetilde{K}_{,1}}
\newcommand{\kThree}{\widetilde{K}_{,2}}
\newcommand{\kFour}{\widetilde{K}_{,11}}
\newcommand{\kFive}{\widetilde{K}_{,22}}
\newcommand{\kSix}{\widetilde{K}_{,12}}
\newcommand{\knTwo}{\widetilde{K}_{N,1}}
\newcommand{\knThree}{\widetilde{K}_{N,2}}
\newcommand{\kSum}{\widetilde{K}_\mathrm{sum}}
\newcommand{\kDiff}{\widetilde{K}_\mathrm{diff}}

\newcommand{\abd}{\mathrm{abd}}
\newcommand{\pct}{\mathrm{pct}}
\newcommand{\cut}{\mathrm{cut}}

\newcommand{\MRLens}{\acro{\textsc{MRLens}}}

\newbool{couvertureEnFrancais}
\setbool{couvertureEnFrancais}{false}

\begin{document}



\newcommand{\lastname}[1]{\textsc{#1}}
\newcommand{\highlight}[1]{\textsf{#1}}

\ifbool{couvertureEnFrancais}{
\begin{titlepage}
	\thispagestyle{empty}
    \sf \small \raggedright \noindent
	Service d'Astrophysique, CEA Saclay\\[0.5ex]
	École doctorale Astronomie et Astrophysique d'Île-de-France\\[0.5ex]
	Université Paris Diderot 7 - Sorbonne Paris Cité
	\vspace*{\stretch{1}}
	
	\sf \LARGE \raggedleft 
	Thèse en astrophysique
	\vspace*{\stretch{0.25}}
	
	\bfseries \sffamily \Huge
	\rule{\textwidth}{0.5pt}\\[1.5ex]
	Statistiques d'ordres\\
	supérieurs appliquées à\\
	l'effet de lentille gravitationnelle\\[0.5ex]
	\rule{\textwidth}{0.5pt}
	\vspace*{\stretch{0.25}}
	
	\sf \LARGE
	Chieh-An \lastname{Lin}
	\vspace*{\stretch{2}}
	
	\rm \normalsize \raggedright
	\begin{tabular}{r@{\hspace{1em}}l}
		Le 28 septembre 2016              & \highlight{Date de soutenance}\\[2ex]
		\highlight{Directeur de thèse}    & Martin \lastname{Kilbinger}\\
		\highlight{Co-directeur de thèse} & Marc \lastname{Sauvage}\\[2ex]
		Simona \lastname{Mei}             & \highlight{Présidente du jury}\\
		David \lastname{Bacon}            & \highlight{Rapporteur}\\
		Cristiano \lastname{Porciani}     & \highlight{Rapporteur}\\
		Matteo \lastname{Maturi}          & \highlight{Examinateur}\\
		Sophie \lastname{Maurogordato}    & \highlight{Examinatrice}
	\end{tabular}
\end{titlepage}
}{}

\clearpage
\thispagestyle{empty}
\cleardoublepage

\begin{titlepage}
	\thispagestyle{empty}
    \sf \small \raggedright \noindent
	Service d'Astrophysique, CEA Saclay\\[0.5ex]
	École doctorale Astronomie et Astrophysique d'Île-de-France\\[0.5ex]
	Université Paris Diderot 7 - Sorbonne Paris Cité
	\vspace*{\stretch{1}}
	
	\sf \LARGE \raggedleft 
	PhD Thesis in Astrophysics
	\vspace*{\stretch{0.25}}
	
	\bfseries \sffamily \Huge
	\rule{\textwidth}{0.5pt}\\[1.5ex]
	Cosmology with\\
	weak-lensing peak counts\\[0.5ex]
	\rule{\textwidth}{0.5pt}
	\vspace*{\stretch{0.25}}
	
	\sf \LARGE
	Chieh-An \lastname{Lin}
	\vspace*{\stretch{2}}
	
	\rm \normalsize \raggedright
	\begin{tabular}{r@{\hspace{1em}}l}
		September \nth{28}, 2016       & \highlight{Defense date}\\[2ex]
		\highlight{Advisor}            & Martin \lastname{Kilbinger}\\
		\highlight{Co-advisor}         & Marc \lastname{Sauvage}\\[2ex]
		Simona \lastname{Mei}          & \highlight{Jury president}\\
		David \lastname{Bacon}         & \highlight{Referee}\\
		Cristiano \lastname{Porciani}  & \highlight{Referee}\\
		Matteo \lastname{Maturi}       & \highlight{Examinator}\\
		Sophie \lastname{Maurogordato} & \highlight{Examinator}
	\end{tabular}
\end{titlepage}

\clearpage
\thispagestyle{empty}
\cleardoublepage

\thispagestyle{empty}
\mbox{}\vskip6cm
\mbox{}\hfill
\begin{minipage}{10cm}
	\raggedleft \it
	To those who struggle to find a balance between\\
	life, career, and curiosity
\end{minipage}

\clearpage
\thispagestyle{empty}
\cleardoublepage


\chapter*{Acknowledgements}
\addcontentsline{toc}{chapter}{Acknowledgements} 
\fancyhead[LE]{\sf \nouppercase{Acknowledgements}}
\fancyhead[RO]{}

Happy theses are all alike; every unhappy thesis is unhappy in its own way \citep{Tolstoy_1877}. However, whether happy or not, a thesis cannot be accomplished without various helps that its author appreciates sincerely from his/her deepest mind.

My first acknowledgement goes to \textit{Le Livre de poche} collection of novels and fictions for keeping my company during my entire commute life. As far as commute is concerned, what falls to my lot is not the burden, but the unbearable shortness of reading \citep{Kundera_1984}. I also very appreciate arXiv's contribution in this part. 

I want to thank R\'egion d'\^Ile-de-France for its generous funding for my thesis under the grant \acro{DIM-ACAV}. If on a winter's night a traveler with a poster tube is heading for somewhere outside of Paris, it is due to the financial supports from \acro{DIM-ACAV} and \acro{PNCG} \citep{Calvino_1979}.

The jury members, especially the referees, are thanked for their commitment. The thickness of my manuscript is the evidence of my guilt. Take care, Madam, Sir: those over there are not windmills but just a two-hundred-page book \citep{Cervantes_1605}.

I would like to express my full gratitude to my family and stepfamily, who always prioritize my journey back to Taiwan. I do not know how to reward back but only regret not to be present during all these years. To them, I am sorry for missing the Sunday-night reunion with grandma (and political discussions). I am also sorry for those who have been waiting for me at the Mahjong table during the Chinese new year. Godot apologizes \citep{Beckett_1953}.

Despite spending nine years in Paris, I have been fortunate to be able to stay in touch with many of my friends back in Taiwan. I want to thank to \zh{嘉維} for being the permanent organizer of dinners between our childhood friends; to \zh{昱光}, \zh{信億}, \zh{玠涵}, \zh{蛇蛇}, \zh{美如}, \zh{慧詩}, \zh{俊智}, \zh{老趙}, \zh{阿呆}, \zh{承學}, \zh{名偉}, \zh{曜哥}, \zh{美良}, and others, thank you for supporting me with food and wine. Most of you were my classmates, some up to 11 years. Knowing that time is making our friendship precious, I regret that this life of 0.09 hundred years of different latitude from you guys seems to have a second opportunity to continue on postdoc \citep{Marquez_1967}.

My decision to dedicate my life to a career as a scientist would not have been possible without the mentorship of my teachers. It was the best of choices; well, probably also the worst of choices \citep{Dickens_1859}. I would like to thank \zh{老蔡} for initiating my scientific interests, \zh{阿乾} for giving me a solid foundation, Roland Lehoucq for inciting my passion in physics, David Langlois who introduced General Relativity, \zh{蔡炳坤} (principal of my high school) for convincing me to study in France, and Ang\'elique for teaching me French.

It is confusing to wake up, finding myself living in Paris. While it is certainly not as terrifying as finding oneself transformed into an enormous insect, still, the life of an expatriate is an experience impossible to be expressed to anyone, sometimes even to myself \citep{Kafka_1915}. To \zh{宜學}, Mu, \zh{邁克}, \zh{哲宇}, \zh{霽庭}, \zh{紘翎}, \zh{學庸}, \zh{迪凱}, \zh{嘉麗}, Lucho, Daniel, Anna, Pony, \zh{鈺霖}, \zh{杜力}, \zh{兔子}, \zh{小癸}, \zh{莉雯}, \zh{驊庭}, \zh{敏慧}, \zh{怪獸}, and others who share the same foreign experiences with me. I particularly thank Chloe, expatriate in the USA, for correcting this text which was in my awful English.

One is not born French-speaking; one becomes Fre..., euh, \textit{francophone} \citep{Beauvoir_1949}. I am grateful to my French friends for helping me mingle into their circles during \textit{la pr\'epa} and \textit{l'X}. It is wonderful to know my superb ex-roommates Paul and JM, as well as Florent, Tim, PEG, Clotilde, Ludovic, Clara, Elsa, H\'el\`ene, M\'elanie, Julien, and many others with whom I regret not being able to meet more frequently.

I will not forget \zh{志煌}, \zh{有蓉}, and \zh{杰森}, with whom I co-founded a mini-magazine, \textit{Aurore Formosane}, that introduces Taiwanese culture to the French. A magazine, well, it is nothing but living in our respective shelter-like studio and recording people's life under the thread of an unfriendly neighbor nation \citep{Frank_1947}. I thank Lisa, \zh{威賦}, \zh{涵雁}, Ouliann, and \zh{蔡曜} for providing regular helps.

Few of my friends know that I used to play bridge. For those who are too modern to understand, it is rather an out-of-fashion card game that I started playing in high school. To bid or not to bid; that is the question \citep{Shakespeare_1603}. Thanks to Jeanne, S\'ebastien, and J\'er\'emy, I continued to enjoy this game in Paris. I am sorry that you will need to find a fourth. Make sure that he/she also counts the prime numbers.

My laboratory life at CEA would not have been as wonderful without Lucie, S\'ebastien, Tuc, Ma\"elle, M\'elanie, Sarah, R\'emy, Olivier, Orianne, Victor, Mathieu, J\'er\'emy, \zh{悅寧} (Yueh-Ning), Marc, \zh{雨晏} (Yu-Yen), Amandine, Remco, Paula, Sofia, Kyle, Koryo, and others. At CEA, every Tuesday is a new day and it is better to be lucky; but I would rather be exact, then when free breakfast comes you are ready \citep{Hemingway_1952}. I would also like to thank J\'er\^ome and Pascale for taking care of the PhD students, Marie for her patiently answering all my technical problems, and Dominique for being a fantastic secretary. 

Attending conferences is another source of pleasure in my research life. Without my conference wing(wo)men: Florent, Santi, Vivien, Nico, Alexandre, R\'emy, Olga, Thibault, Markus, Malin, Jenny, Jia, Emmanuel, Carolina, \zh{彥吉} (Yen-Chi), and \zh{之藩} (Geoff), the coffee breaks would have been boring, and my ``attacks'' on famous academics would have been even more awkward. Thank you, and let us keep the motto in mind: ``one for all, and all except for one.'' \citep{Dumas_1844}

Now, I want to thank the whole group of CosmoStat for being a strong supports to my thesis with both work-wise and life-wise aspects. On the one hand, it is such a great pleasure to have a nice working atmosphere with talents: Jean-Luc, J\'er\^ome, Florent, Sandrine, St\'ephane, Bertrand, Adrienne, Paniez, Julien, Austin, Sam, Joana, Daniel, J\'er\'emy, Simon, Fred, C\'ecile (rou$\sim$), \zh{姜明} (Ming), and Morgan. On the other hands, as \citet{Descartes_1637} said: I work, therefore I drink. I am glad to have participated so many drinks always full of happiness.

All people are equally thankful but some people are more equally thankful than others \citep{Orwell_1945}. I thank Michael Vespe for introducing \acro{ABC} to me (which has an important role in my thesis) during the Cosmo21 conference in Lisbon in May 2014.

Next, I need to mention \zh{品廷} (Pinting) who, everytime when he hurries out from a metro station, always reminds me of a rabbit in waistcoat with a pocket watch \citep{Carroll_1865}. You are always the first person I think of when I need to ask for a favor. I guess the fact that we learnt French and came to France together bonded our friendship tightly, even though our personalities differ a lot. This makes me believe that the opposite characters do not stop people from becoming close friends.

My special thanks goes to Fran\c{c}ois for two main reasons. You were my officemate for two years. When you work, just like me, you are incredibly noisy. I have no idea how Ming and Paniez could ignore our crappy chats and silly ideas. I am also impressed by your sense of humor, geekness, diligence, and, most notably your perpetual willingness to help others with their problems. Whether it's about cosmology, signal processing, coding, or even \LaTeX, you never say no and would always have come up with some brilliant solutions. I know if I ask you to draw me a sheep immediately without asking why \citep{Saint-Exupery_1943}. Your helps concerning various subjects are considerable.

My deepest thanks goes to my advisor Martin. It is a truth universally acknowledged, that a naive student in possession of no ideas must be in want of a good advisor, and I am extremely fortunate to have a great one \citep{Austen_1813}. You let me interrupt you with my up-to-hour discussions on all kinds of topics. You helped me learn to bring initiatives to projects and build collaborations. You encouraged me to travel as much as I want. I will never forget the investigation of our common academic ancestors in Heidelberg and the PhD Comics you posted on my office door. You are friendly and considerate; you think with a scientific mind and a sense of humor. No one can ask for more.

\clearpage
\thispagestyle{empty}
\cleardoublepage


\chapter*{Abstract}
\addcontentsline{toc}{chapter}{Abstract} 
\fancyhead[LE]{\sf \nouppercase{Abstract}}
\fancyhead[RO]{}

\subsubsection{English version}

Weak gravitational lensing (WL) causes distortions of galaxy images and probes massive structures on large scales, allowing us to understand the late-time evolution of the Universe. One way to extract the cosmological information from WL is to use peak statistics. Peaks are tracers of massive halos and therefore probe the mass function. They retain non-Gaussian information and have already been shown as a promising tool to constrain cosmology. In this work, we develop a new model to predict WL peak counts. The model generates fast simulations based on halo sampling and selects peaks from the derived lensing maps. This approach has three main advantages. First, the model is very fast: only several seconds are required to perform a realization. Second, including realistic conditions is straightforward. Third, the model provides the full distribution information because of its stochasticity. We show that our model agrees well with N-body simulations. Then, we study the impacts of the cosmology-dependent covariance on constraints and explore different parameter inference methods. A special focus is put on approximate Bayesian computation (ABC), an accept-reject sampler without the need to estimate the likelihood. We show that ABC is able to yield robust constraints with much reduced time costs. Several filtering techniques are studied to improve the extraction of multiscale information. Finally, the new model is applied to the CFHTLenS, KiDS DR1/2, and DES SV data sets. Our preliminary results agree with the Planck constraints assuming the Lambda-CDM model. Overall, the thesis forges an innovative tool for future WL surveys.

\subsubsection{French version}

L'effet de lentille gravitionnelle faible (WL) déforme les images des galaxies observées. Il porte des informations sur les grandes structures et nous apprend l'évolution de l'Univers. Une façon d'extraire des informations de WL est d'utiliser les statistiques d'ordres supérieurs à deux, en particulier le comptage de pics. Les pics indiquent la présence des halos et mesurent la fonction de masse. Ils sont considérés comme un bon outil pour contraindre la cosmologie. Dans cette thèse, on développe un nouveau modèle de prédiction sur le nombre des pics. Celui-ci génère des simulations des halos et sélectionne les pics à partir de la carte de WL résultante. Cette approche jouit de trois avantages. D'abord, le modèle est rapide: une réalisation s'obtient en quelques secondes. Ensuite, il est aisé d'y inclure les effets observationnels. Enfin, on a accès à la distribution complète des observables grâce à son caractère stochastique. On montre que notre modèle est en accord avec les simulations à N-corps. Puis, on examine des méthodes de contraintes variées. Un accent est mis sur la computation bayesienne approximative (ABC) qui est un processus d'acceptation/rejet ne nécessitant pas d'évaluer la fonction de vraisemblance. On montre que ABC fournit des contraintes cohérentes en un temps plus faible que la méthode classique. Des méthodes de filtrage, pour améliorer l'extraction des informations multi-échelles, sont étudiées. Enfin, le nouveau modèle est appliqué sur les données des relevés. Nos résultats préliminaires sont en accord avec ceux de Planck, en admettant le modèle Lambda-CDM. Dans l'ensemble, cette thèse bâtit un chemin pionier pour les futurs relevés de WL.

\clearpage
\thispagestyle{empty}
\cleardoublepage

\fancyhead[LE]{\sf \nouppercase{Content}}
\fancyhead[RO]{}
\tableofcontents
\clearpage
\thispagestyle{empty}
\cleardoublepage




\chapter{Introduction}
\fancyhead[LE]{\sf \nouppercase{\leftmark}}
\fancyhead[RO]{\sf \nouppercase{\rightmark}}

\hfill\textit{\large Measure what is measurable, and make measurable what is not so.}

\hfill--- Galileo Galilei
\vskip4ex

Since existence, human beings have never stopped their curiosity about the most outer frontier of the world. From bare-eye observations to telescopes, cosmology --- the science of the Universe --- has been developed little by little. What makes up the Universe? What is the history of it? These might be the two ultimate questions that all cosmologists try to answer. So far, we estimate that the age of the Universe is roughly 13.8 billion years, and that up to 95\% is actually composed of ``dark components'' \citep{PlanckCollaboration_etal_2015}: with \textit{dark matter}, we manage to explain the luminous mass deficit from observations; and \textit{dark energy} is responsible for the accelerating cosmic expansion. Only about 5\% of the Universe is ordinary matter: baryons in the form of ions, atoms, and molecules.

During the last twenty years, the probes of the cosmic microwave background confirm that our Universe can actually be described by a simple model with only six parameters \citep{Spergel_etal_2003}, namely ``\acro{$\LCDM$}''. This success, both for the probes and for the model, defines an important benchmark in cosmology. Until now, almost all observations are consistent with the \acro{$\LCDM$} model. For this reason, cosmologists aim more and more for improving the uncertainty of the measurements. We have been passing through the era of ``precision cosmology'' for a while. However, several recent studies \citep{MacCrann_etal_2015, PlanckCollaboration_etal_2015a} reveal the existence of tensions in the \acro{$\LCDM$} model, which opens a door for possible new physics in the future.

To verify if new physics exists, one of the viable methods is gravitational lensing, or lensing in short. The origin of lensing is the light deflection by gravity: when light travels in the Universe, the presence of massive structures perturb the local space-time curvature, so that the straight light path is deflected. This phenomenon has been well predicted and modelled by the famous theory of general relativity \citep{Einstein_1915}. Due to this deflection, distant sources yield distorted images when the emitted light reaches observers. This distortion encodes cumulatively the gravitational information along the line of sight. Therefore, by measuring properly the image distortion, cosmologists can measure the distribution of mass in the Universe.

When cosmologists observe multiple images of the same source, or ring- or arc-like patterns, since such objects could not exist in the Universe, they know that images have been strongly lensed. In this case, we call it ``strong lensing''. On the contrary, the lensing effect is so weak for most parts of the sky such that distinguishing lensed and unlensed images by eye is almost impossible. In this case, we call it ``weak lensing''. Despite its low signal level, weak lensing still contains very rich cosmological information.

To extract cosmological information from weak lensing, a very common way is to use second-order statistics. However, this only retains the Gaussian information \citep{Gauss_1809}, and it misses the rich nonlinear part of structure evolution encoded on small scales. To compensate for this drawback, several non-Gaussian statistics have been proposed, for example higher order moments, the three-point correlation function, Minkowski functionals, or peak statistics.

The non-Gaussian observable that is studied in this thesis is peak counts. Weak-lensing peaks are defined as the local maxima on a convergence map. Since the weak-lensing convergence represents the projected mass, peaks are tracers of massive regions in the Universe, which are characterized by the mass function. For this reason, we probe the mass function and constrain the cosmology simply by counting peaks.

However, not all the peaks are true mass clusters. The galaxy shape noise as well as real-world observing conditions create selection effects, such as false positives, false negatives, or changes of peak height. In the literature, we see that early studies tended to use only very high signal-to-noise ratio ($\gtrsim 5$) peaks. In this way, cosmologists can guarantee that all peaks are true clusters and proceed with the physical analysis with identifications. However, recent studies show that including selection effects directly into the model allows us to explore medium and low peaks, which increases cosmological information extraction.

So far, there exist three approaches which predict peak counts by taking selection effects into account: analytical modelling, $N$-body simulations, and a fast stochastic model (this work). Analytical models are based on Gaussian random field theory and peak theory. This approach is not very flexible. It suffers from real-world effects and can be biased. On the other hand, $N$-body simulations can include observational conditions into forward computations. However, they are very expensive to run and evaluate. That is how the motivation for a new model emerges: we want to benefit from the advantages of $N$-body modelling, find a shortcut to avoid complex $N$-body processes, and yield a correct peak-count prediction.

Nowadays, statistical advances are changing the perspectives of cosmology. The applications of sparse methods, Bayesian analysis, machine learning, etc. to cosmology has been prosper. As the volume of data and the requirement of the precision on results grow, using sophisticated statistical tools becomes indispensable for cosmological studies. 

Recently, approximate Bayesian computation (\acro{ABC}) has been gathering a strong growth of attention from the astrophysical community. \acro{ABC} is a likelihood-free parameter inference method. Usually, it is used when the likelihood is expensive to evaluate or impossible to define. However, even if the likelihood can be computed in a straightforward way, \acro{ABC} is still powerful and it has shown a great potential to provide cosmological constraints.

The objective of this thesis is to address the following questions:
\begin{itemize}
	\item How to model weak-lensing peak counts properly in realistic conditions?
	\item Which parameter constraint method is preferred?
	\item How to compare different filtering techniques?
	\item What cosmological information can we extract from peaks?
\end{itemize}

The content of the thesis is structured as follows. In Chaps. \ref{sect:cosmology}, \ref{sect:structure}, and \ref{sect:lensing}, some concepts of cosmology, structure formation, and weak gravitational lensing that are used in this work will be successively presented. A review of the weak-lensing-peak studies in the literature will be given at the end of \chap{sect:lensing}. In \chap{sect:modelling}, I will introduce the problem of peak-count modelling and propose a solution to overcome it. A comparison to $N$-body simulation and to the model from \citet{Fan_etal_2010} will be provided. In \chap{sect:constraint}, I will discuss the results from different constraint strategies. A particular attention will be put on the cosmology-dependent-covariance effect. In \chap{sect:ABC}, approximate Bayesian computation will be introduced and the resulting cosmological constraints will be shown. In \chap{sect:filtering}, I will compare different filtering techniques, including linear and nonlinear ones. In \chap{sect:data}, the new peak-count model and \acro{ABC} will be applied to data retrieved from three different lensing surveys. Finally, I will summarize the discoveries and the results of this thesis in \chap{sect:conclusion}.

\clearpage
\thispagestyle{empty}
\cleardoublepage


\chapter{Modern cosmology}
\label{sect:cosmology}
\fancyhead[LE]{\sf \nouppercase{\leftmark}}
\fancyhead[RO]{\sf \nouppercase{\rightmark}}

\subsubsection{Overview}

This chapter will introduce the cosmological basis for the thesis, particularly Friedmann's equations, cosmological distances, and a brief description of the history of the Universe.

\section{Cosmic expansion history}

\subsection{$\vect{\Lambda}$CDM model}

From the current observations, the Universe is very well described by the \acro{$\LCDM$} model, where $\Lambda$ stands for the cosmological constant, the simplest parametrization of ``dark energy'' and \acro{CDM} stands for cold ``dark matter''. 

The concept of \textit{dark matter} was first proposed by \citet{Zwicky_1933} to describe the mass deficit from luminosity compared to the gravitational mass necessary for galaxy random movement in cluster virial theorem. Even by considering non-luminous baryons, the deficit can still not be fully explained because the baryonic fraction is well constrained by Big-Bang nucleosynthesis and baryonic acoustic oscillations. Some possible candidates of dark matter are elementary particles, for example those which are motivated by supersymmetry theory. Alternatively, this mass deficit problem can also be solved by modifying the gravity laws. However so far, no direct detection of dark matter particles has been claimed, and no modified gravity theory can match current observation results in a satisfactory way. 

On the other hand, \textit{dark energy} is a key to explain the acceleration of the expansion of the Universe, which was confirmed by \citet{Riess_etal_1998}. The simplest dark energy model is the cosmological constant, which conducts a constant pressure per volume which pushes the Universe out.

The \acro{$\LCDM$} model contains in total six parameters. Six is the minimal number of parameters to fit the observations, so the \acro{$\LCDM$} model can be considered as the simplest model. Apart from the physical fractions of baryon, cold dark matter, and dark energy ($\OmegaB h^2$, $\Omega_\rmc h^2$, $\OmegaL$), three other parameters are the amplitude and spectral index of the scalar field power spectrum ($\Delta_\mathcal{R}^2$, $n_\rms$) and the reionization optical depth $\tau$.  According to recent studies \citep{PlanckCollaboration_etal_2015}, ordinary matter only accounts for 5\% of the total energy of the Universe; 26\% are dark matter; and 69\%, the majority, are dark energy, the nature of which is totally unknown.

\subsection{Einstein field equation}

The current understanding of the Universe is founded on the theory of General Relativity \citep{Einstein_1915}. In the theory, time and space are related together by the \textit{metric} $\rmd s^2\equiv g_{\mu\nu}\rmd x^\mu\rmd x^\nu$ and form one single space called space-time. Then, gravity, the dominant force at cosmic scales, is described as the distortion of the space-time curvature, which explains the deviation from straight lines of the trajectories of travelling particles. For example, light (or photons) will always travel on the zero-geodesics.

Let us consider an homogeneous and isotropic universe. This ideal scenario is described by the \textit{Friedmann-Lemaître-Robertson-Walker} (\acro{FLRW}) \textit{metric}\index{Friedmann-Lemaître-Robertson-Walker (\acro{FLRW}) metric} as
\begin{align}
	\rmd s^2 = -\rmc^2\rmd t^2 + a^2(t) &\left[\frac{\rmd R^2}{1 - KR^2} + R^2 \left(\rmd\theta^2 + \sin^2\theta\ \rmd\phi^2 \right)\right] \notag\\
	= -\rmc^2\rmd t^2 + a^2(t) &\left[\rmd w^2 + f_K^2(w) \left(\rmd\theta^2 + \sin^2\theta\ \rmd\phi^2 \right)\right] \label{for:cosmology:FLRW_metric}
\end{align}
where $\rmc$ is the light speed, $t$ is the cosmic time, $a(t)$ is the scale factor, $R$, $\theta$, and $\phi$ are spherical coordinates in comoving space, $K$ is the flatness of the universe which characterizes the relation between space and time, and $w$ and $f_K(w)$ are respectively comoving radial and transverse distances (see \sect{sect:cosmology:distances:distances}). A flat universe corresponds to $K = 0$, whereas a closed (or elliptic) universe has $K > 0$ and an open (or hyperbolic) universe has $K < 0$.

The relation between the curvature distortion and cosmic components (matter, radiation, etc.) is described by the Einstein field equation. It links the space-time curvature to the energy-momentum. The version with the cosmological constant $\Lambda$ is
\begin{align}
	G_{\mu\nu} = \frac{8\pi \rmG}{\rmc^4} T_{\mu\nu} - \Lambda g_{\mu\nu},
\end{align}
where $\rmG$ is the gravitational constant, $g_{\mu\nu}$ is the metric, and $G_{\mu\nu}$ is the Einstein tensor which can be written with the Ricci tensor $R_{\mu\nu}$ and the Ricci scalar $R$, as
\begin{align}
	G_{\mu\nu} \equiv R_{\mu\nu} - \frac{1}{2} Rg_{\mu\nu},
\end{align}
and $T_{\mu\nu}$ is the \textit{stress-energy tensor}\index{Stress-energy tensor}.

\subsection{Friedmann's equations}

For an isotropic homogeneous universe, one can set the $(1,1)-$order tensor $T^\mu\mbox{}_\nu$ to
\begin{align}
	T^\mu\mbox{}_\nu =
	\begin{pmatrix}
		-\rho(t) & 0	    & 0    & 0\\
		0 	 	 & p(t) & 0    & 0\\
		0	   	 & 0    & p(t) & 0\\
		0		 & 0    & 0    & p(t)
	\end{pmatrix},
\end{align}
where $\rho(t)$ denotes the mass density and $p(t)$ denotes the pressure. This allows one to simplify the Einstein equation into a simpler form. The 00-component and the trace respectively lead to
\begin{align}
	&\frac{\dot{a}^2}{a^2} + \frac{\rmc^2K}{a^2} = \frac{8\pi \rmG}{3}\rho + \frac{\rmc^2\Lambda}{3}, \label{for:cosmology:Friedmann_equations_1}\\
	&\frac{\ddot{a}}{a} = -\frac{4\pi \rmG}{3}\left(\rho + \frac{3p}{\rmc^2}\right) + \frac{\rmc^2\Lambda}{3}, \label{for:cosmology:Friedmann_equations_2}
\end{align}
where the notation $\dot{\mbox{}\mbox{}}$ is the derivative with regard to the cosmic time $t$. Equations \eqref{for:cosmology:Friedmann_equations_1} and \eqref{for:cosmology:Friedmann_equations_2}, called \textit{Friedmann's equations}\index{Friedmann equations}, allow one to derive the evolutions of each cosmic component and of the whole Universe. Taking the derivative of \for{for:cosmology:Friedmann_equations_1} and using \for{for:cosmology:Friedmann_equations_2}, one has
\begin{align}
	\frac{8\pi \rmG}{3}\dot{\rho} &= 2\left(\frac{\dot{a}}{a}\right) \left(\frac{a\ddot{a}-\dot{a}^2}{a^2}\right)-\frac{2\rmc^2K\dot{a}}{a^3}\notag\\
	&= 2\left(\frac{\dot{a}}{a}\right) \left(\frac{\ddot{a}}{a}-\frac{\dot{a}^2}{a^2} - \frac{\rmc^2K}{a^2}\right)\notag\\
	&= 2\left(\frac{\dot{a}}{a}\right) \left[\left(-\frac{4\pi \rmG}{3}\left(\rho + \frac{3p}{\rmc^2}\right) + \frac{\rmc^2\Lambda}{3}\right) - \left(\frac{8\pi \rmG}{3}\rho + \frac{\rmc^2\Lambda}{3}\right)\right]\notag\\
	&= \frac{8\pi \rmG}{3} \left(\frac{\dot{a}}{a}\right) \left( -\rho-\frac{3p}{\rmc^2}-2\rho \right),
\end{align}
so
\begin{align}
	\dot{\rho} = -3\left(\rho + \frac{p}{\rmc^2}\right)\frac{\dot{a}}{a}.
\end{align}
It is then useful to introduce the notion of the \textit{equation of state}\index{Equation of state}, which is the link between the density and the pressure. For each component $\alpha$, we may write
\begin{align}
	\frac{p_\alpha}{\rmc^2} = w_\alpha \rho_\alpha,
\end{align}
so that the general form of the component evolution is
\begin{align}\label{for:cosmology:rho_evolution}
	\frac{\rho_\alpha}{\rho_{\alpha,0}} = \left(\frac{a}{a_0}\right)^{-3(1+w_\alpha)},
\end{align}
where the subscript $_0$ means that the quantity is evaluated at the present time $t=t_0$. We set naturally $a_0=1$.

For non-relativistic components, such as baryon and cold dark matter, $w=0$; for ultra-relativistic components such as photon and neutrinos, $w = 1/3$. We usually simplify these as two families: matter and radiation. For dark energy, one can imagine a scenario where $w=-1$. This actually corresponds to the case of the cosmological constant $\Lambda$. To see this, one only needs to consider a ``density of cosmological constant'' $\rho_\Lambda$ such that $\Lambda = 8\pi\rmG\rho_\Lambda/\rmc^2$. By doing this, the contribution from dark energy is included in the $\rho$ and $p/\rmc^2$ prescriptions and $\Lambda$ can be dropped. From \for{for:cosmology:rho_evolution}, one can see that $\rho_\Lambda$ is independent from the evolution of the Universe since $w=-1$, which justifies the name of ``cosmological constant''. However, a more complex model could be possible. For example, 
\begin{align}
	w_\mathrm{de}(a) = \wZero + (1-a) w_a^\mathrm{de}
\end{align}
is a commonly considered equation of state for dark energy. The case of $w_\mathrm{de}<-1/3$ will be compatible with the expansion of the Universe.

Consider now an universe composed with the matter (labelled \texttt{m}), the radiation (labelled \texttt{r}), and the cosmological constant. We introduce the \textit{Hubble parameter}\index{Hubble parameter} to simplify the notation:
\begin{align}
	H(t) \equiv \frac{\dot{a}(t)}{a(t)},
\end{align}
which is also the relative expansion rate of the Universe. The first Friedmann's equation (Eq. \ref{for:cosmology:Friedmann_equations_1}) becomes 
\begin{align}\label{for:cosmology:Friedmann_equations_4}
	H^2 + \frac{\rmc^2K}{a^2} = \frac{8\pi \rmG}{3}(\rho_\rmm + \rho_\rmr) + \frac{\rmc^2\Lambda}{3}.
\end{align}
Defining now the dimensionless fraction of different components \textbf{at the present time} as
\begin{align}\label{for:cosmology:fraction}
	\Omega_K \equiv \frac{-\rmc^2K}{a_0^2H_0^2},\ \ \ \OmegaL \equiv \frac{\rmc^2\Lambda}{3H_0^2},\ \ \ \Omega_\rmm \equiv \frac{8\pi \rmG}{3H_0^2}\rho_\rmm,\ \ \ \Omega_\rmr \equiv \frac{8\pi \rmG}{3H_0^2}\rho_\rmr,
\end{align}
we obtain, by evaluating \for{for:cosmology:Friedmann_equations_4} at $t=t_0$, an energy balance:
\begin{align}
	\Omega_K + \OmegaL + \OmegaM + \Omega_\rmr = 1.
\end{align}
Here, both cosmological constant and flatness have been treated as a sort of energy. Following this reasoning, the total density of the Universe at the present time, called the \textit{critical density}\index{Critical density}, is
\begin{align}
	\rho_\crit \equiv \frac{3H_0^2}{8\pi \rmG} = 2.775\dixd{11}\ \Msol h^2 / \Mpc^3.
\end{align}
This quantity has an unity of mass volume density. In this thesis, the mass unity will always be $\Msol/h$, the solar mass over $h$; and the distance will always be $\Mpc/h$, where $h$ is the dimensionless Hubble parameter\index{Hubble parameter, dimensionless}, defined as
\begin{align}
	H_0 \equiv 100\cdot h\ \text{km/s/Mpc} = 10^5\cdot h\ \text{m/s/Mpc}.
\end{align}

\begin{figure}[tb]
	\centering
	\includegraphics[scale=0.65]{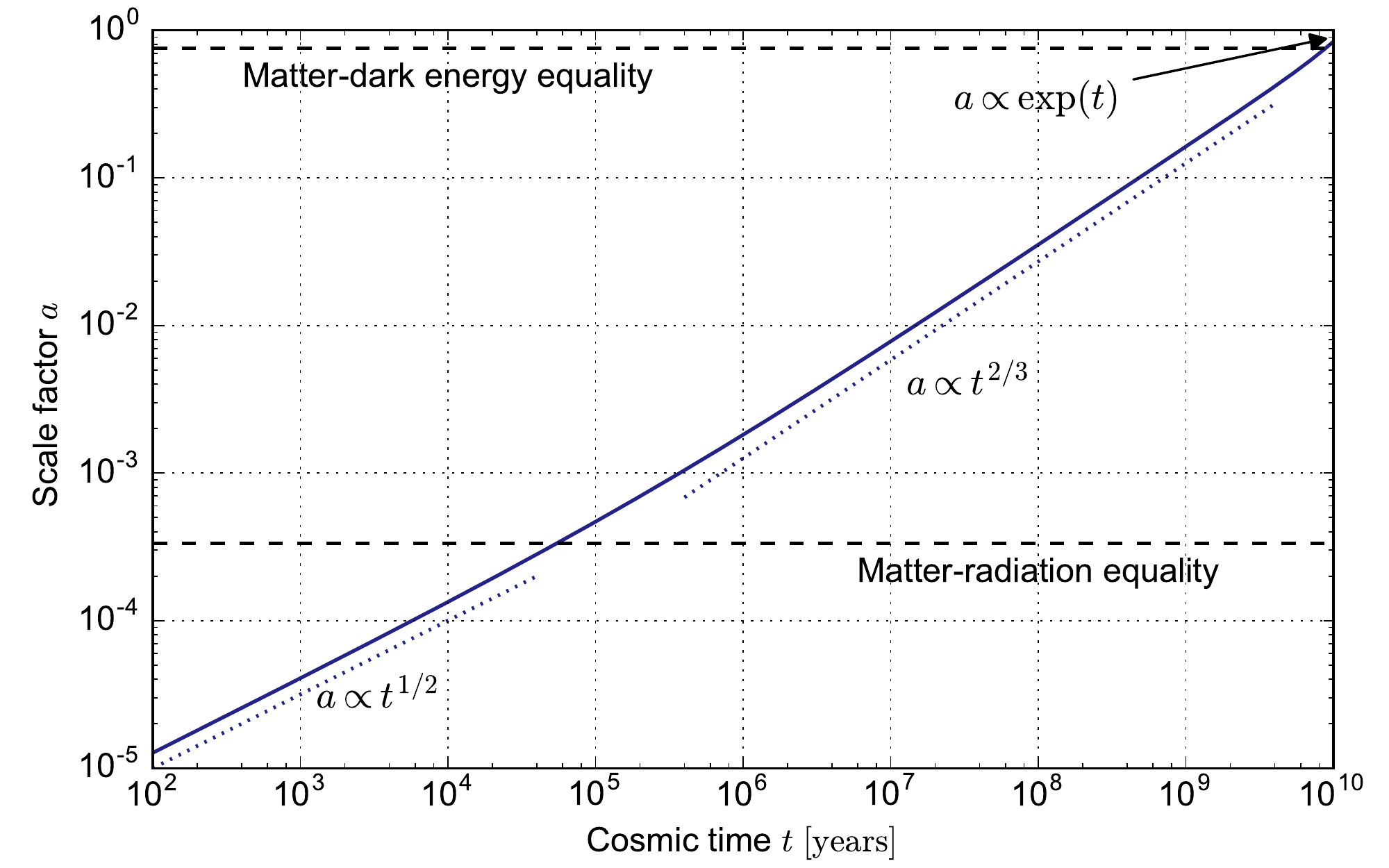}
	\caption{Evolution of the scale factor $a(t)$ with regard to the cosmic time under a \acro{$\LCDM$} cosmology.}
	\label{fig:cosmology:Scale_factor}
\end{figure}

Given the current decomposition of the Universe, what is its evolution history? This can be derived from \for{for:cosmology:Friedmann_equations_4}. Using Eqs. \eqref{for:cosmology:rho_evolution} and \eqref{for:cosmology:fraction}, We actually have
\begin{align}\label{for:cosmology:Friedmann_equations_5}
	\frac{H^2}{H_0^2} = \Omega_K\left(\frac{a}{a_0}\right)^{-2} + \OmegaL + \OmegaM\left(\frac{a}{a_0}\right)^{-3} + \Omega_\rmr\left(\frac{a}{a_0}\right)^{-4}.
\end{align}
Since $a_0 = 1$, we usually express \for{for:cosmology:Friedmann_equations_5} in redshift $z\equiv a\inv-1$ (see \sect{sect:cosmology:distances:redshift} for the definition of the redshift), so that
\begin{align}\label{for:cosmology:E}
	\frac{H^2(z)}{H_0^2}=E^2(z),\ \ \ \text{with}\ \ \ E(z) = \sqrt{\OmegaL + \Omega_K(1+z)^2 + \OmegaM(1+z)^3 + \Omega_\rmr(1+z)^4}.
\end{align}
This means that, to reconstruct the evolution, we need to measure the fraction of each component and the Hubble rate, all at the present time. Taking these quantities as external parameters, \for{for:cosmology:E} allows one to integrate over the lookback time and trace the evolution of $a$ (\fig{fig:cosmology:Scale_factor}). For example, we have $a\propto t^{1/2}$ in the radiation-dominated era and $a\propto t^{2/3}$ in the matter dominated era. 

In the \acro{$\LCDM$} model, $\OmegaM$ is composed of cold dark matter and baryons, while $\Omega_K, \Omega_\rmr\approx0$.

\section{Distances in the Universe}

The notion of ``distance'' is very subtle in the cosmological context. The distances and scales that cosmologists refer to are often so large that even light takes millions of years to travel. This evokes the concept of horizon and causality. Implicitly for an observer, a distanced object is also an ``old'' object, since information that the observer receives was emitted long time ago. As a consequence, the information from different distances can be considered as different stages on a time sequence. Although these emissions come from different physical positions, cosmologist can still take advantage to study space-invariant properties, such as correlation functions.

\subsection{Redshift}
\label{sect:cosmology:distances:redshift}

The time-distance duality above is formally characterized by \textit{redshift}\index{Redshift}. The redshift is an observational phenomenon indicating that the spectrum of sources is shifted toward red sequences. This can be identified by atomic emission and absorption wavelengths. The cause of the redshift can be the Doppler effect, but also the expansion of the Universe. Actually, from the metric (Eq. \ref{for:cosmology:FLRW_metric}), we have
\begin{align}
	\frac{\rmc\rmd t}{a(t)} = \rmd w
\end{align}
since $\rmd^2 s = 0$ for light. If we focus on a light ray emitted at $t_\rme$ (with wavelength $\lambda_\rme$) and received at $t_\rmr$ (with $\lambda_\rmr$), then
\begin{align}
	\int_{t_\rme}^{t_\rmr} \frac{\rmc\rmd t}{a(t)} = \int_0^w \rmd w = \int_{t_\rme+\lambda_\rme/\rmc}^{t_\rmr+\lambda_\rmr/\rmc}\frac{\rmc\rmd t}{a(t)},
\end{align}
which leads to
\begin{align}
	\int_{t_\rme}^{t_\rme+\lambda_\rme/\rmc} \frac{\rmc\rmd t}{a(t)} = \int_{t_\rmr}^{t_\rmr+\lambda_\rmr/\rmc}\frac{\rmc\rmd t}{a(t)},
\end{align}
Therefore, we can relate the wavelengths at emission and reception to their respective scale factors $a_\rme=a(t_\rme)$ and $a_\rmr=a(t_\rmr)$ by
\begin{align}
	\frac{\lambda_\rmr}{\lambda_\rme} = \frac{a_\rmr}{a_\rme},
\end{align}
and the shift is defined as
\begin{align}
	z  \equiv \frac{\lambda_\rmr - \lambda_\rme}{\lambda_\rme}.
\end{align}
In this case, we call it cosmological redshift. This term is dominant in most cosmological contexts. In the absence of expansion, $a_\rmr = a_\rme$ and the cosmological redshift disappears. Therefore, if we take $a_\rmr=a_0=1$ and $a_\rme=a$, the redshift becomes simply
\begin{align}
	a  = \frac{1}{1+z}.
\end{align}
As we can see from \for{for:cosmology:Friedmann_equations_5}, $a$ is a monotonic function of $t$ if all $\Omega_i>0$. This is the case of the current Universe. Thus, $z$ is implicitly a one-to-one function of $t$. We can then parametrize the time by the redshift. At the present time, we have $z(t_0)=0$.

\subsection{Cosmological distances}
\label{sect:cosmology:distances:distances}

To convert redshift to distance, we should be aware of the fact that the definition of cosmological distance is not unique. Here, we will always consider a light ray emitted at redshift $z_\rme$ and received by an observer at redshift $z_\rmr < z_\rme$.

\begin{figure}[tb]
	\centering
	\includegraphics[scale=0.65]{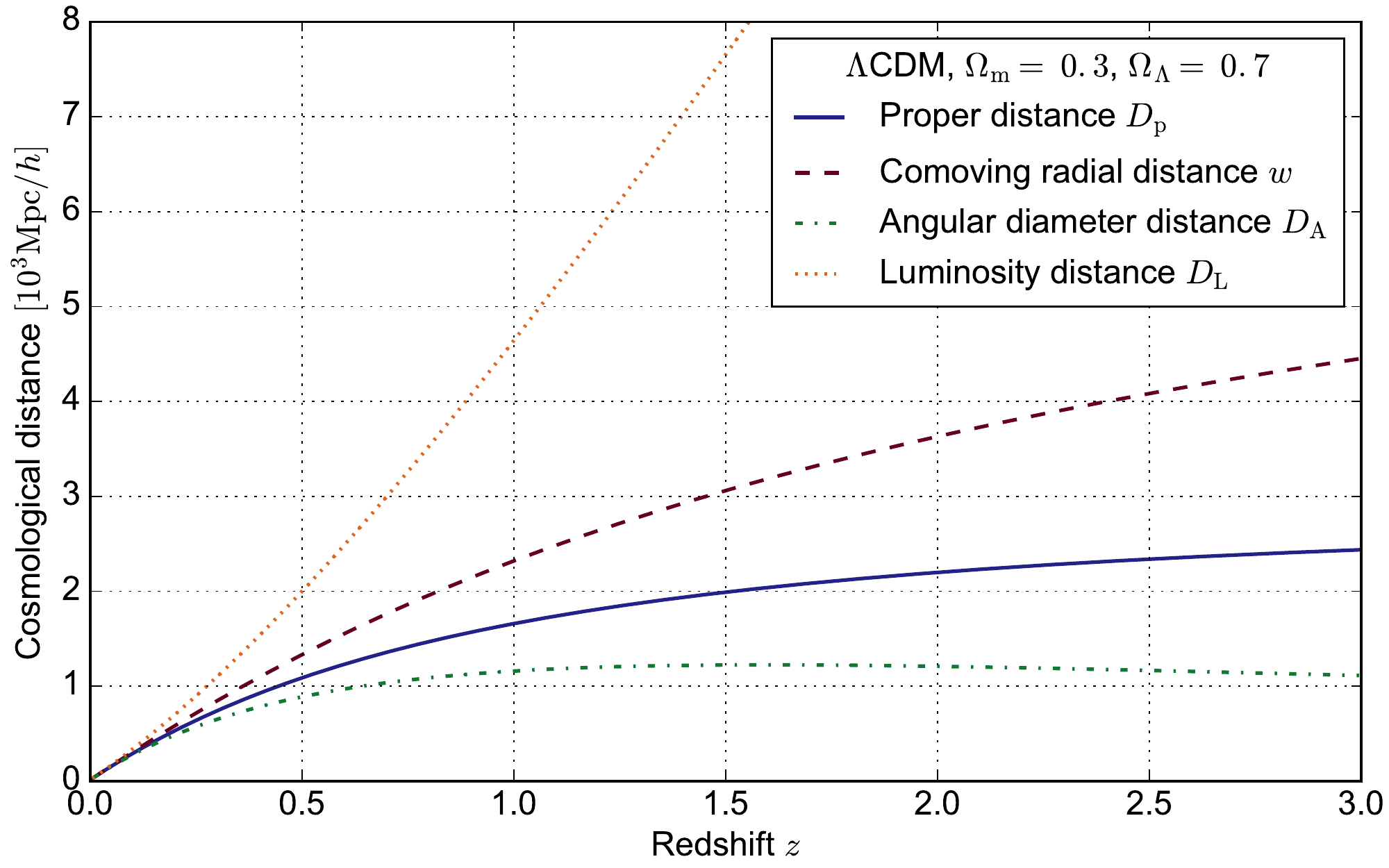}
	\caption{Different distances under a \acro{$\LCDM$} cosmology.}
	\label{fig:cosmology:Cosmological_distances}
\end{figure}

\paragraph{Proper distance} The proper distance\index{Distance, proper} $D_\rmp$ is the distance that a photon ``sees''. It refers to the collection of infinitesimal light paths despite the expansion of the Universe. Mathematically, it is defined as $\rmd D_\rmp = \rmc\rmd t$. Since the light speed is a constant, the proper distance is nothing but 
\begin{align}
	D_\rmp(z_\rmr, z_\rme) \equiv \rmc\big(t(z_\rmr) - t(z_\rme)\big).
\end{align}
\vspace*{-2.5ex}

\paragraph{Comoving radial distance} The comoving radial distance (or simply comoving distance)\index{Distance, comoving (radial)} $w$ is defined as $\rmd w = \rmc a\inv\rmd t$. From the point of view of the metric, it takes the infinitesimal proper distance and rescales it by $a\inv$ due to the cosmic flow. From a more global aspect, this is just the physical distance between the emission point and the reception point measured at $z=0$. The word ``comoving'' always refer to a scaling to $z=0$. Mathematically, one has $\rmd w = \rmc a\inv\rmd t = \rmc(a\dot{a})\inv\rmd a = \rmc(a^2H)\inv\rmd a$, hence
\begin{align}
	w(z_\rmr, z_\rme) \equiv D_\rmH\int_{z_\rmr}^{z_\rme} \frac{\rmd z}{E(z)},
\end{align}
where
\begin{align}
	D_\rmH \equiv \frac{\rmc}{H_0} = 2997.92458\ \Mpc/h
\end{align}
is the Hubble distance, and $E(z)$ is given by \for{for:cosmology:E}.
\vspace*{-2.5ex}

\paragraph{Comoving transverse distance} The comoving transverse distance\index{Distance, comoving transverse} $f_K$ is the ratio of the comoving separation between two points at $w$ to their separation angle. Once again, we scale the distance from the emission epoch $z_\rme$ to $z=0$. So, a better interpretation is to stop the cosmic flow at $z=0$, and to measure the (physical) distance between these two distanced points. This is purely geometric, so $f_K$ only depends on $w$ and is related to the curvature $K$ defined in the metric (Eq. \ref{for:cosmology:FLRW_metric}). We have:
\begin{align}\label{for:cosmology:comoving_transverse_dist}
	f_K(w) \equiv \left\{
	\begin{matrix}
		\displaystyle\frac{1}{\sqrt{-K}} \sinh\left(\sqrt{-K}w\right) & \text{if}\ K<0,\\[2ex]
		w\hfill\mbox{} & \text{if}\ K=0,\\[1ex]
		\displaystyle\frac{1}{\sqrt{K}} \sin\left(\sqrt{K}w\right)\hfill\mbox{} & \text{if}\ K>0.
	\end{matrix}\right.
\end{align}
\vspace*{-2.5ex}

\paragraph{Angular diameter distance} The angular diameter distance\index{Distance, angular diameter} $D_\rmA$ is the ratio of a physical separation at the emission epoch (with comoving radial distance $w$) to its angular separation from an observer. This is used for calculating the size of bound objects, such as galaxies and halos, because we usually require their original size, not rescaled to $z=0$. To recover this, we only need to scale the comoving quantity back to the emission epoch $z_\rme$. Hence, the angular diameter distance is simply
\begin{align}
	D_\rmA(z_\rmr, z_\rme) \equiv \frac{f_K(w(z_\rmr, z_\rme))}{1+z_\rme}.
\end{align}
Particularly, if the Universe is flat and the observer is at $z_\rmr=0$, the angular diameter distance becomes
\begin{align}
	D_\rmA(0, z_\rme) = \frac{D_\rmH}{1+z_\rme} \int_0^{z_\rme} \frac{\rmd z'}{E(z')}.
\end{align}
\vspace*{-2.5ex}

\paragraph{Luminosity distance} The luminosity distance\index{Distance, luminosity} $D_\rmL$ is defined to relate the luminosity $L$ and the brightness $B$, such that 
\begin{align}
	D_\rmL \equiv \sqrt{\frac{L}{4\pi B}}.
\end{align}
The dependency on redshifts is omitted. It acts only on $B$ as the flux emitted at $z_\rme$ and received at $z_\rmr$. From Etherington's reciprocity theorem \citep{Etherington_1933, Ellis_2007}, we obtain a further relation:
\begin{align}
	D_\rmA = \frac{f_K}{1+z} = \frac{D_\rmL}{(1+z)^2}
\end{align}
for all observers at redshift 0 and sources at redshift $z$.

\paragraph{Comoving volume} The comoving volume\index{Comoving volume} is simply the volume defined in comoving space. On small scales, where one can locally neglect the variation of the cosmic flow, it is the physical volume at the present time which corresponds to the physical volume of the same Lagrangian points at the considered epoch. However, if we consider a long time span, e.g. a lightcone constructed from $z=1$ to the present time, this interpretation is not valid anymore. This difficulty reveals the fact that the comoving volume is an ill-defined concept. The problem can only be resolved if we cut the total volume into small pieces and rescale them to $z=0$ with their respective scale factors. For these reason, in this work, except for specific precision, all volumes are physical volumes measured at the considered epoch.

\section{Modern cosmological paradigm}

\subsubsection{Fine-tuning problems}

The standard model of cosmology provided by Friedmann's equations is subject to problems evoked by two observational facts. The first is the \textit{flatness problem}. The flatness of the Universe that we observe today is very close to zero. If we look back to the birth time, it means that the initial density of different cosmic components should be set to a specific value at extremely high precision. The second one is the \textit{horizon problem}. It turns out that our Universe is very homogeneous on large scales. However, these scales are so large that the interactions, at most propagated at the speed of light, have not yet occurred (scales have not ``entered the horizon''). One of the solutions to these two problems is the \textit{cosmological inflation} \citep{Guth_1981}. During the inflation, the flatness would be forced very close to zero and the anisotropies would be suppressed, so that the Universe becomes what we observe now, flat and homogeneous.

\subsubsection{A brief history of the Universe}

\begin{figure}[tb]
	\centering
	\includegraphics[width=0.9\textwidth]{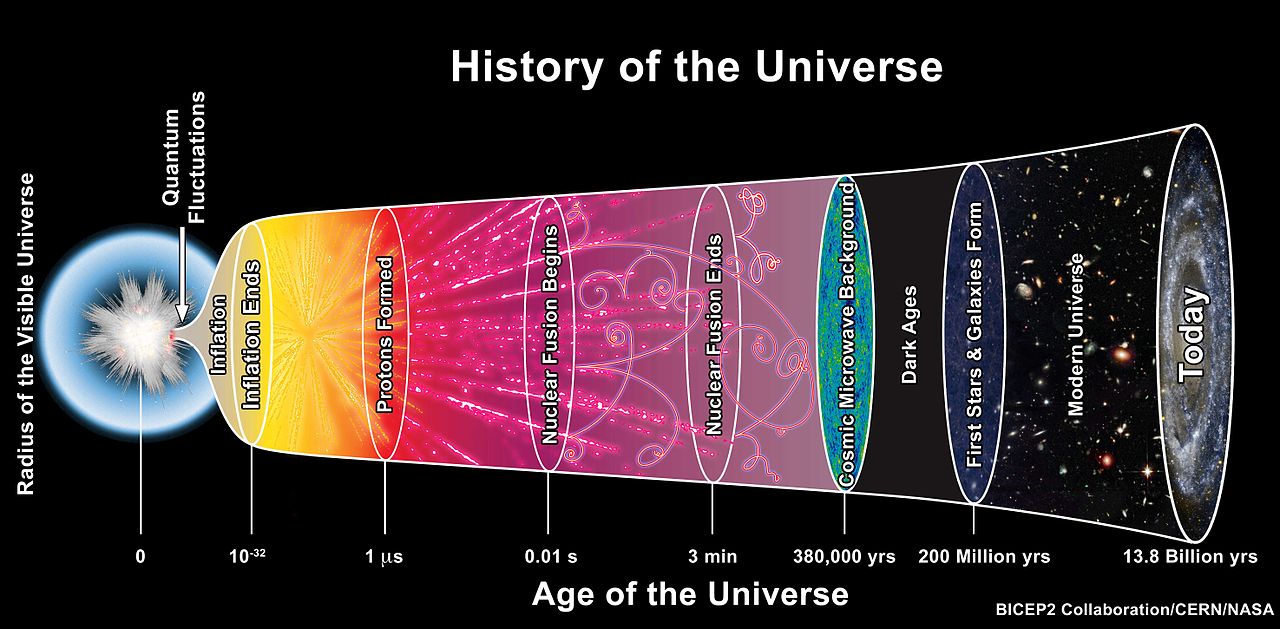}
	\caption{Illustration of the history of the Universe. (Source: BICEP2/CERN/NASA)}
	\label{fig:cosmology:History_of_the_Universe}
\end{figure}

Today, the most commonly accepted theory for the beginning of the Universe is the \textit{Big Bang} theory, which admits a singularity in the space-time. Very soon after, cosmological inflation occurs and the primordial (quantum) perturbation signatures get dragged and remain in the Universe. The temperature drops considerably and symmetries between fundamental forces break. Successively, antibaryons annihilate, neutrinos decouple, and the \textit{Big Bang nucleosynthesis} occurs, which forms atomic nuclei. All these processes are estimated to happen during the first few minutes of the history of the Universe. 

As time goes on, the Universe continues to expand and the temperature continues to decrease. The matter-radiation equality is achieved roughly at $z\approx3600$. Then, at $z\approx1100$, which is about 380,000 years after the Big Bang, the temperature of the Universe falls below 4000~K. The plasma of photons, electrons, and protons is not at thermal equilibrium anymore. Electrons and protons are bound together forming hydrogen atoms, which is called \textit{recombination}. Meanwhile, photons are decoupled from the matter and start to travel freely. This turns the Universe transparent and these photons become the cosmic microwave background (\acro{CMB}) that we observe today.

After the recombination, the Universe enters into the dark age where no radiation other than \acro{CMB} is emitted. During this epoch, matter collapses and structures form due to gravity. At $z\approx20$, highly dense regions start to form stars and galaxies. These astrophysical objects emit new light, which is energetic enough to rip electrons from electroneutral atoms. This is called \textit{reionization}. While the most distanced galaxy that humans observe so far is recorded at $z=11.1$ \citep{Oesch_etal_2016}, the primary form of galaxy clusters only appear after $z\approx 6$. At $z\approx0.4$, the Universe passes through the matter-dark energy equality era. Today, it is believed that the dominant component in the Universe is dark energy.

\subsubsection{Summary}

In this small chapter, we have reviewed briefly Friedmann's equations which describe the evolution of the scale factor. We have also seen that the word ``distance'' could have various definitions depending on the context.

The \acro{$\LCDM$} model is the current standard model for cosmology. With the \acro{$\LCDM$} model, the flatness and the horizon problems are explained with the cosmological inflation.

In next chapter, we will focus on the formation and the evolution of cosmological structures.

\clearpage
\thispagestyle{empty}
\cleardoublepage


\chapter{Structure formation}
\label{sect:structure}
\fancyhead[LE]{\sf \nouppercase{\leftmark}}
\fancyhead[RO]{\sf \nouppercase{\rightmark}}

\subsubsection{Overview}

The goal of this chapter is to introduce two important aspects of structures in the Universe for weak-lensing peak counts: the mass function and the universal halo density profile. To have a better understanding, I will first review some key concepts such as Fourier analysis, cosmological perturbation theory, and the spherical collapse model to define matter fluctuations, the growth factor, and the linear contrast threshold. After that, different models of the mass function and the density profile will be presented.

\section{Fourier analysis}

For numerous reasons, analysis in Fourier space is often preferred than in direct space for cosmological studies. For example, the idea of horizon crossing --- distant perturbations start to ``communicate'' with each other due to the expanding Universe --- can be straightforwardly summarized as wave modes entering into the horizon. Also, randomness has been included in cosmological models to describe the observed universe (e.g. \acro{CMB} temperature and large-scale structures), and an efficient way to extract cosmological information from it is to study the statistical quantities measured in the Fourier space, e.g. the power spectrum. In this section, the power spectrum, the quantity $\sigma_8$, and the formalism of the Fourier analysis will be introduced.

\subsection{Two-point-correlation function and power spectrum}

Let $f(\vect{r})$ be a \acro{3D} field defined at the real space position $\vect{r}$. The Fourier transform of $f(\vect{r})$ is defined as 
\begin{align}
	\tilde{f}(\vect{k}) \equiv \int_{\mathbb{R}^3}\rmd^3\vect{r}\ \rme^{-\rmi \vect{k}\vect{r}} f(\vect{r}),
\end{align}
where $\vect{k}$ is a wave vector in the Fourier space, and the inverse transform as 
\begin{align}
	f(\vect{r}) \equiv \int_{\mathbb{R}^3}\frac{\rmd^3\vect{k}}{(2\pi)^3}\ \rme^{\rmi \vect{k}\vect{r}} \tilde{f}(\vect{k}).
\end{align}
Consider now two fields $f_1$, $f_2$ and their convolution product:
\begin{align}
	(f_1\ast f_2)(\vect{r}) \equiv \int\rmd^3\vect{r}'\ f_1(\vect{r}-\vect{r}')f_2(\vect{r}').
\end{align}
This is identical to the product of $f_1$ and $f_2$ in the Fourier space since
\begin{align}
	(f_1\ast f_2)(\vect{r}) &= \int\rmd^3\vect{r}'\frac{\rmd^3\vect{k}}{(2\pi)^3}\frac{\rmd^3\vect{k}'}{(2\pi)^3}\ \rme^{\rmi \vect{k}(\vect{r}-\vect{r}')}\rme^{\rmi \vect{k}'\vect{r}'} \tilde{f}_1(\vect{k})\tilde{f}_2(\vect{k}') \notag\\
	&= \int\frac{\rmd^3\vect{k}}{(2\pi)^3}\frac{\rmd^3\vect{k}'}{(2\pi)^3}\ \left(\int\rmd^3\vect{r}'\ \rme^{-\rmi (\vect{k}-\vect{k}')\vect{r}'}\right)\rme^{\rmi\vect{k}\vect{r}}  \tilde{f}_1(\vect{k})\tilde{f}_2(\vect{k}') \notag\\
	&= \int\frac{\rmd^3\vect{k}}{(2\pi)^3}\frac{\rmd^3\vect{k}'}{(2\pi)^3}\ (2\pi)^3 \delta^{(3)}(\vect{k}-\vect{k}')\rme^{\rmi\vect{k}\vect{r}}  \tilde{f}_1(\vect{k})\tilde{f}_2(\vect{k}') \notag\\
	&= \int\frac{\rmd^3\vect{k}}{(2\pi)^3}\ \rme^{\rmi\vect{k}\vect{r}}  \Big(\tilde{f}_1\cdot\tilde{f}_2\Big)(\vect{k}), 
\end{align}
knowing that the Fourier transform of 1 is a Dirac function. In other words,
\begin{align}\label{for:structure:convolution_3}
	\widetilde{f_1\ast f_2}=\tilde{f}_1\cdot\tilde{f}_2
\end{align}

For any \acro{3D} field $f$, the \textit{two-point-correlation function}\index{Two-point-correlation function} (\acro{2PCF}) $C(\vect{r})$ of $f$ is defined as
\begin{align}
	C(\vect{r}) = \left\langle f(\vect{r}')f(\vect{r}'+\vect{r}) \right\rangle_{\vect{r}'} \equiv\int_{\mathbb{R}^3}\rmd^3\vect{r}'\ f(\vect{r}')f(\vect{r}'+\vect{r}), 
\end{align}
where the subscript $_{\vect{r}'}$ stands for averaging over $\vect{r}'$. This is interpreted as the expectation value of the product of two field points which are separated by the vector $\vect{r}$. Note that this is not exactly a convolution product for the reason of sign. Actually, we have
\begin{align}\label{for:structure:2PCF_2}
	C(\vect{r}) = \int\frac{\rmd^3\vect{k}}{(2\pi)^3}\ \rme^{\rmi\vect{k}\vect{r}} \tilde{f}(\vect{k})\tilde{f}(-\vect{k}) = \int\frac{\rmd^3\vect{k}}{(2\pi)^3}\ \rme^{\rmi\vect{k}\vect{r}} \left|\tilde{f}(\vect{k})\right|^2,
\end{align}
since $\tilde{f}(-\vect{k}) = \left.\tilde{f}\right.^*(\vect{k})$.

Let us now focus on the the contrast field and its variance. The contrast $\delta$ of a field $f$ is given by  
\begin{align}
	\delta(\vect{r}) \equiv \frac{f(\vect{r})-\bar{f}}{\bar{f}}\ \ \ \text{with}\ \ \ \bar{f}\equiv \lim_{V\rightarrow +\infty}\frac{1}{V}\int_V\rmd^3\vect{r}\ f(\vect{r}).
\end{align}
Since the average of the contrast is zero by construction, the variance $\sigma^2$ is just the integral of $\delta^2$ over the whole space, which leads to
\begin{align}\label{for:structure:variance_1}
	\sigma^2 \equiv \left\langle \delta^2(\vect{r}) \right\rangle_{\vect{r}} = \int_{\mathbb{R}^3}\rmd^3\vect{r}\ \delta^2(\vect{r}).
\end{align}
Note that, depending on the field, this quantity can diverge. We can see that \for{for:structure:variance_1} is equivalent to \for{for:structure:2PCF_2} by substituting $f$ with $\delta$. Keeping the same symbol for the \acro{2PCF}, $\sigma^2$ now becomes
\begin{align}\label{for:structure:variance_2}
	\sigma^2 = C(0) = \int\frac{\rmd^3\vect{k}}{(2\pi)^3}\ \left|\tilde{\delta}(\vect{k})\right|^2 = \int\frac{k^2\rmd k\ \rmd^2\Omega}{(2\pi)^3}\ \left|\tilde{\delta}(\vect{k})\right|^2.
\end{align}
At this stage, we can define the \textit{power spectrum}\index{Power spectrum} as the average of $|\tilde{\delta}(\vect{k})|^2$ over all directions:
\begin{align}\label{for:structure:power_spectrum_1}
	P(k) \equiv \frac{1}{4\pi}\int\rmd^2\Omega\ \left|\tilde{\delta}(\vect{k})\right|^2.
\end{align}
If we consider only the case in which $\delta$ is isotropic, then the power spectrum can be written directly as $P(k) = |\tilde{\delta}(\vect{k})|^2$. Sometimes, an alternative definition can also be found in the literature:
\begin{align}
	\left\langle \tilde{\delta}(\vect{k})\tilde{\delta}^*(\vect{k}') \right\rangle_\Omega \equiv (2\pi)^3 P(k) \delta^{(3)}(\vect{k}-\vect{k}'),
\end{align}
where $\delta$ at the right-hand side is the Dirac function. This is strictly equivalent to \for{for:structure:power_spectrum_1}. 

Meanwhile, by inserting \for{for:structure:power_spectrum_1} into \for{for:structure:variance_2}, we obtain
\begin{align}\label{for:structure:variance_3}
	\sigma^2 = \int\frac{4\pi k^2\rmd k}{(2\pi)^3}\ P(k) = \int\frac{\rmd k}{k}\ \Delta^2(k),
\end{align}
where
\begin{align}
	\Delta^2(k) \equiv \frac{k^3}{2\pi^2}P(k)
\end{align}
is called the \textit{dimensionless power spectrum}. In most cases, $\Delta^2(k)$ is modelled by power laws. For example, the power spectrum $\Delta_\mathcal{R}^2(k)$ of the primordial scalar field $\mathcal{R}$ usually takes the following form:
\begin{align}
	\Delta_\mathcal{R}^2(k) = A\left(\frac{k}{k_0}\right)^{n_\rms-1},
\end{align}
where $A$ is the amplitude, $n_\rms$ is the scalar spectral index, and $k_0$ is a pivot scale arbitrarily chosen. In early years, cosmologists assume that $n_\rms=1$, so that the primordial perturbation is scale-independent. The power is identical for all modes. However, recent studies show that $n_\rms$ is slightly smaller than 1. The case $n_\rms=1$ has been excluded at 3.6-$\sigma$ \citep{Hinshaw_etal_2013}.

A third definition of the power spectrum can be provided with \for{for:structure:variance_3}. As mentioned above, the variance $\sigma^2$ can actually diverge as \for{for:structure:variance_3} extends over $\mathbb{R}$. However, the larger $|k|$ is, the smaller the scale, and the infinitely small scales can not be reached in observations. Therefore, it is convenient to define an upper cutoff for the integral of $\sigma^2$. In the end, if we define the cutoff variance $\tilde{\sigma}^2$ as
\begin{align}
	\tilde{\sigma}^2(k) \equiv \int_0^k \frac{\rmd k'}{k'}\ \Delta^2(k'), 
\end{align}
the dimensionless power spectrum can be considered as the variation of the cutoff variance per logarithmic scale, shown by
\begin{align}
	\Delta^2(k) = \frac{\rmd\tilde{\sigma}^2(k)}{\rmd\ln k}.
\end{align}

\subsection{Matter fluctuation}

Let us focus on the case of the matter field. Let $\delta$ be the matter density contrast of the Universe. Instead of setting a upper cutoff, the integral \eqref{for:structure:variance_1} can also be regularized by filtering the field. Denote $W_R(\vect{r})$ as a filter function whose size is characterized by a physical size $R$. Then, the new quantity
\begin{align}
	\sigma^2(R) \equiv \left\langle (W_R\ast\delta)^2(\vect{r}) \right\rangle_{\vect{r}}
\end{align}
is finite if $W_R$ has a finite support. Using Eqs. \eqref{for:structure:convolution_3}, \eqref{for:structure:variance_2} and \eqref{for:structure:power_spectrum_1}, we find
\begin{align}
	\sigma^2(R) &= \int\frac{k^2\rmd k\ \rmd^2\Omega}{(2\pi)^3}\ \left|\left(\widetilde{W_R\ast\delta}\right)(\vect{k})\right|^2 \notag\\
	&= \int\frac{k^2\rmd k\ \rmd^2\Omega}{(2\pi)^3}\ \left|\left(\widetilde{W}_R\cdot\tilde{\delta}\right)(\vect{k})\right|^2 \notag\\
	&= \int\frac{k^2\rmd k}{2\pi^2}\ P(k)\left|\widetilde{W}_R(\vect{k})\right|^2,
\end{align}
where $P(k)$ is the matter power spectrum. A common choice for $W_R$ is the \acro{3D} top-hat filter with radius $R$: $W_R(\vect{r})=(3/4\pi R^3)\cdot\Theta(R-r)$, where $\Theta$ is the Heaviside step function. In this case,
\begin{align}
	\widetilde{W}_R(k) = \frac{3}{k^3R^3} \Big[\sin(kR)-kR\cos(kR)\Big].
\end{align} 
So, the power spectrum is passed through a window function such that the integration does not diverge anymore. Particularly, 
\begin{align}
	\sigEig^2 \equiv \sigma^2(R = 8~\Mpc/h)
\end{align}
represents the variance of the matter density contrast smoothed with a spherical top-hat filter with physical radius $8~\Mpc/h$. It is a parameter often considered in lensing analyses. Sometimes, the size of the window function is determined by a given ``mass scale''. That means the radius of the top-hat sphere is set so that the mass of enclosed matter is $M$. Denote $\bar{\rho}_0$ as the background mass density at the present time. The definition of the mass-based matter fluctuation leads to
\begin{align}\label{for:structure:sigma_M}
	\sigma^2(M) \equiv \sigma^2\left(R=\left(\frac{3M}{4\pi\bar{\rho}_0}\right)^{1/3}\right).
\end{align}

\section{Linear perturbation theory}

The Universe on large enough scales can be considered as homogeneous. However, on smaller scales, primordial fluctuations leave inhomogeneous footprints and get transformed into matter and radiation. When these large scales enter the horizon, local inhomogeneity starts to evolve. In comoving coordinates, the evolution is the combination of two effects: attraction by gravity and repulsion by the fluid pressure. For modes which cross the horizon relatively late, the evolution is still in its early stage and can be well described by linear perturbation theory. This linear regime will be the focus of this section.

\subsubsection{Fluid dynamic equations}

Consider a universe in the epoch after the matter-radiation equality, so that we can assume a matter- or dark-energy-dominated universe. Following \citet{Peebles_1993}, we will model the matter as an ideal pressureless fluid whose state is characterized by the density $\rho(t,\vect{r})$ and the velocity $\vect{u}(t,\vect{r})$, both defined at cosmic time $t$ in physical coordinates $\vect{r}$. The dynamic of Newtonian physics can de described by
\begin{align}
	\text{Continuity equation:}\hspace*{1em} & \frac{\rmd\rho}{\rmd t}(t, \vect{r}) + \nabla\cdot \big(\rho(t, \vect{r})\vect{u}(t, \vect{r})\big) = 0, \label{for:structure:continuity_eq_1}\\
	\text{Euler equation:}\hspace*{1em} & \frac{\rmd\vect{u}}{\rmd t}(t, \vect{r}) + (\vect{u}\cdot\nabla) \vect{u}(t, \vect{r}) = -\nabla\Phi(t, \vect{r}), \label{for:structure:Euler_eq_1}\\
	\text{Poisson equation:}\hspace*{1em} & \Delta\Phi(t, \vect{r}) = 4\pi\rmG \sum_\alpha (1+3w_\alpha)\rho_\alpha(t, \vect{r}), \label{for:structure:Poisson_eq_1}
\end{align}
where $\Phi$ is the Newtonian potential, $\alpha$ runs over different cosmological components (matter, photons, neutrinos, dark energy, etc.), and $\rho_\alpha$ and $w_\alpha$ are corresponding density and equation of state of each component. 

Equations \eqref{for:structure:continuity_eq_1} \eqref{for:structure:Euler_eq_1}, and \eqref{for:structure:Poisson_eq_1} represent respectively mass conservation, momentum conservation, and gravity. The right-hand side of \for{for:structure:Euler_eq_1} has already been reduced to a single gravity term because of considering a pressureless fluid. If we want to establish the dynamic for the radiation-dominated epoch, then the pressure term needs to be accounted for. The unusual form of the Poisson equation is the general expression when all cosmological components are accounted for. Actually, the Poisson equation can be seen as a classical limit of the first Friedmann equation (Eq. \ref{for:cosmology:Friedmann_equations_1}), which is derived from the 00 component of the Einstein equation. If we take $\alpha = \rmm$ (matter), then $w_\alpha=0$ and the usual Poisson equation is recovered. If in addition, the cosmological constant is considered, then the right-hand side of \for{for:structure:Poisson_eq_1} becomes $4\pi\rmG\rho_\rmm -\rmc^2\Lambda$, since $w_\de = -1$ and $\rho_\de = \rmc^2\Lambda/8\pi\rmG$.

To avoid the possible confusion in the follows, I will use the total derivative with regard to $t$ instead of the partial one. These two notations are equivalent in the Eulerian spatial representation, since a \acro{3D} Eulerian point is independent from time. 

\subsubsection{Change of variables}

Let us now write these equations in comoving space, by first introducing some notations. Let $\vect{R}\equiv \vect{r}/a$ be the comoving coordinates associated to $\vect{r}$. Since $\vect{R}$ actually depends on $t$ via $a$, the total derivative $\rmd/\rmd t$ is not equal to $\partial/\partial t$ anymore. For a scalar field $f$, denote $f_{\vect{r}}$ and $f_{\vect{R}}$ as its mathematical forms in $\vect{r}$ and $\vect{R}$ spaces respectively. Then, the change of variables $\vect{R}(t) =\vect{r}/a(t)$ leads to
\begin{align}
	f_{\vect{r}}(t, \vect{r}) = f_{\vect{r}}(t, a(t)\vect{R}(t)) = f_{\vect{R}}(t, \vect{R}(t)),
\end{align}
and
\begin{align}
	\frac{\rmd f_{\vect{r}}}{\rmd t}(t, \vect{r}) &=\frac{\rmd f_{\vect{R}}}{\rmd t}(t, \vect{R}(t)) \notag\\
	&= \frac{\partial f_{\vect{R}}}{\partial t} + \frac{\partial\vect{R}}{\partial t}\cdot(\nabla f_{\vect{R}}) \notag\\
	&= \frac{\partial f_{\vect{R}}}{\partial t} + \frac{\dot{a}}{a^2}\vect{r}\cdot(\nabla f_{\vect{R}}) \notag\\
	&= \frac{\partial f_{\vect{R}}}{\partial t} + H(t)\vect{R}(t)\cdot(\nabla f_{\vect{R}}), \label{for:structure:change_of_variables}
\end{align}
where $\dot{a} = \rmd a/\rmd t$ and $H(t)=\dot{a}/a$ is the Hubble parameter. If $\vect{f}$ is a vector field, then $\vect{R}(t)\cdot(\nabla f_{\vect{R}})$ from \for{for:structure:change_of_variables} should be replaced by $(D\vect{f}_{\vect{R}})(\vect{R})$ as the differential of $\vect{f}_{\vect{R}}$ applied at $\vect{R}$. One verifies at ease that
\begin{align}\label{for:structure:differential}
	(D\vect{f}_{\vect{R}})(\vect{R}) = (\vect{R}\cdot\nabla)\vect{f}_{\vect{R}}
\end{align}
by expanding component by component.

On the other hand, the velocity $\vect{u}$ describes the total velocity with regard to an Eulerian reference point, which takes also the expansion of the Universe into account. If one extracts the cosmic flow from it, the remaining part is called \textit{peculiar velocity}\index{Peculiar velocity}, denoted here as $\vect{v}$. Precisely, one can write
\begin{align}
	\vect{u}(t, \vect{r}) = \vect{v}(t, \vect{r}) + H(t) \cdot \vect{r},
\end{align}
or in comoving coordinates
\begin{align}\label{for:structure:peculiar_velocity_2}
	\vect{u}(t, \vect{R}) = \vect{v}(t, \vect{R}) + \dot{a}(t) \cdot \vect{R}.
\end{align}
Here, as I will do for the rest of the section, the subscripts $_{\vect{r}}$ and $_{\vect{R}}$ have been omitted. Since in equations we will be dealing with the physical quantities which should be the same in comoving and physical spaces, the confusion does not exist anymore. Particularly, $\vect{u}(t, \vect{R})$ and $\vect{v}(t, \vect{R})$ should be interpreted as physical velocities parametrized by comoving coordinates, not comoving velocities.

Finally, the nabla operator here is the gradient derivative which depends implicitly on the space in which the field is parametrized. For example, omitting the subscripts $_{\vect{r}}$ and $_{\vect{R}}$, the change of variables leads to
\begin{align}
	\nabla f(t, \vect{r}) = \frac{1}{a(t)} \nabla f(t, \vect{R}).
\end{align}

\subsubsection{Dynamic in comoving coordinates}

Let us start with writing the continuity equation in comoving space. With Eqs. \eqref{for:structure:change_of_variables} and \eqref{for:structure:peculiar_velocity_2}, \for{for:structure:continuity_eq_1} becomes
\begin{align}
	\left(\frac{\partial\rho}{\rmd t}(t, \vect{R}) - H\vect{R}\cdot\nabla\rho(t, \vect{R})\right) + \frac{1}{a}\nabla\cdot\Big(\rho(t, \vect{R})\big(\vect{v}(t, \vect{R}) + \dot{a}\vect{R}\big)\Big) = 0.
\end{align}
After simplification, we obtain
\begin{align}\label{for:structure:continuity_eq_3}
	\frac{\partial\rho}{\partial t} - 3H\rho + \frac{1}{a}\nabla\cdot (\rho\vect{v}) = 0,
\end{align}
where $\nabla\cdot\vect{R} = 3$ is needed.

Then, we derive an Euler equation for $\vect{v}(t, \vect{R})$. Applying Eqs. \eqref{for:structure:change_of_variables} and \eqref{for:structure:peculiar_velocity_2}, we obtain
\begin{align}
	\frac{\partial\vect{u}}{\partial t}(t, \vect{R}) - \frac{\dot{a}}{a}(D\vect{u})\cdot\vect{R} + \frac{1}{a}\Big((\vect{v}+\dot{a}\vect{R})\cdot\nabla\Big) \vect{u}(t, \vect{R}) = -\frac{1}{a}\nabla\Phi(t, \vect{R}), 
\end{align}
and the left-hand side can be simplified by \for{for:structure:differential}, leading to
\begin{align}\label{for:structure:Euler_eq_3}
	\frac{\partial\vect{u}}{\partial t}(t, \vect{R}) + \frac{1}{a}(\vect{v}\cdot\nabla) \vect{u}(t, \vect{R}) = -\frac{1}{a}\nabla\Phi(t, \vect{R}). 
\end{align}
The first term from \for{for:structure:Euler_eq_3} involves $\partial(\dot{a}\vect{R})/\partial t$. Although $\vect{R}$ depends on $t$, its dependence should be neglected in the partial derivative. Therefore, $\partial(\dot{a}\vect{R})/\partial t = \ddot{a}\vect{R}$. Using \for{for:structure:peculiar_velocity_2} and $(\vect{v}\cdot\nabla)\vect{R}=\vect{v}$, we get
\begin{align}
	\frac{\partial\vect{v}}{\partial t} + \ddot{a}\vect{R} + H\vect{v} + \frac{1}{a}(\vect{v}\cdot\nabla) \vect{v} = -\frac{1}{a}\nabla\Phi. 
\end{align}

For the Poisson equation, it is useful to decompose the potential into homogeneous and inhomogeneous parts, such that
\begin{align}\label{for:structure:Poisson_eq_2}
	\Delta\Phi(t, \vect{r}) = \Delta\phi(t, \vect{r}) + 4\pi\rmG \sum_\alpha(1+3w_\alpha) \bar{\rho}_\alpha(t)
\end{align}
with the \textit{reduced Newtonian potential}\index{Newtonian potential, reduced} $\phi$:
\begin{align}\label{for:structure:Poisson_eq_3}
	\Delta\phi(t, \vect{r}) \equiv 4\pi\rmG (\rho(t, \vect{r}) - \bar{\rho}(t)),
\end{align}
where $\rho=\rho_\rmm$ for simplicity. Moreover, $\rho-\bar{\rho} = \bar{\rho}\delta = \bar{\rho}_0 a\invCb\delta$ and $\OmegaM=8\pi\rmG\bar{\rho}_0/3H_0^2$, so $4\pi\rmG(\rho-\bar{\rho}) = 4\pi\rmG\bar{\rho}_0\delta a^{-3} = 3H_0^2\OmegaM\delta/2a^3$. Thus, \for{for:structure:Poisson_eq_3} becomes
\begin{align}\label{for:structure:Poisson_eq_4}
	\Delta\phi(\vect{r}) = \frac{3H_0^2\OmegaM}{2}\frac{\delta}{a^3}.
\end{align}
The second term at the right-hand side of \for{for:structure:Poisson_eq_2} is independent from $\vect{r}$ since all components except for matter are assumed to be distributed uniformly. This homogeneity can be described by the Friedmann equation (\ref{for:cosmology:Friedmann_equations_2}). In comoving coordinates, we end up with
\begin{align}
	\frac{1}{a^2}\Delta\Phi(t, \vect{R}) = \frac{3H_0^2\OmegaM}{2a^3}\delta(t, \vect{R}) - \frac{3\ddot{a}}{a}.
\end{align}

It is convenient to replace $\rho(t, \vect{R})$ by $\bar{\rho}(t)(1+\delta(t, \vect{R}))$ and to reason in the density contrast. This can simplify \for{for:structure:continuity_eq_3} knowing that $\bar{\rho}(t) = \bar{\rho}_0 a^{-3}(t)$. 
At the end of the day, the dynamic equations for $\delta$ and $\vect{v}$ parametrized by comoving coordinates are
\begin{align}
	&\frac{\partial\delta}{\partial t} + \frac{1}{a}\nabla\cdot \big((1+\delta)\vect{v}\big) = 0, \label{for:structure:continuity_eq_4}\\
	&\frac{\partial\vect{v}}{\partial t} + \ddot{a}\vect{R} + H\vect{v} + \frac{1}{a}(\vect{v}\cdot\nabla) \vect{v} = -\frac{1}{a}\nabla\Phi, \label{for:structure:Euler_eq_5}\\
	&\frac{1}{a^2}\Delta\Phi = \frac{3H_0^2\OmegaM}{2a^3}\delta - \frac{3\ddot{a}}{a}. \label{for:structure:Poisson_eq_6}
\end{align}

\subsubsection{Linearization}

The system of equations above describes the evolution of the density contrast with time. However, it is nonlinear and the analytical solution is hard to construct. Using perturbation theory, we can work on the linearized system. The obtained solution will be a good approximation when $\delta$ and $\vect{v}$ stay small. This corresponds to the early stage of the evolution, therefore to large scales which enter the horizon late. In other words, if we cut the Universe into some large enough pieces and associate a density contrast to each of them, then the interaction between these regions will be triggered late, and the contrast field will be well described by the linearized equations showed as follows.

Dropping terms with $\delta\vect{v}$ and $\vect{v}^2$, Eqs. \eqref{for:structure:continuity_eq_4}, \eqref{for:structure:Euler_eq_5}, and \eqref{for:structure:Poisson_eq_6} become
\begin{align}
	&\frac{\partial\delta}{\partial t} + \frac{1}{a}\nabla\cdot\vect{v} = 0,\\ 
	&\frac{\partial\vect{v}}{\partial t} + \ddot{a}\vect{R} + H\vect{v} = -\frac{1}{a}\nabla\Phi,\\ 
	&\frac{1}{a^2}\Delta\Phi = \frac{3H_0^2\OmegaM}{2a^3}\delta - \frac{3\ddot{a}}{a}. 
\end{align}
One can deduce easily
\begin{align}
	\frac{\partial}{\partial t}\left(a\frac{\partial\delta}{\partial t}\right) &= -\nabla\cdot\frac{\partial\vect{v}}{\partial t} \notag\\
	&= \ddot{a}\nabla\cdot\vect{R} + H\nabla\cdot\vect{v} + \frac{1}{a}\Delta\Phi \notag\\
	&= 3\ddot{a} - \dot{a}\frac{\partial\delta}{\partial t} + \frac{3H_0^2\OmegaM}{2a^2}\delta - 3\ddot{a}, \label{for:structure:delta_evolution_1}
\end{align}
which results in
\begin{align}
	\frac{\partial^2\delta}{\partial t^2} + 2H\frac{\partial\delta}{\partial t} - \frac{3H_0^2\OmegaM}{2a^3}\delta = 0.
\end{align}
This is the linearized equation for the mass density contrast, an ordinary differential equation which admits the general solution of the form \citep{Peebles_1993, Peacock_1999, Dodelson_2003}
\begin{align}
	\delta(t, \vect{R}) = D_+(t)\delta_+(\vect{R}) + D_-(t)\delta_-(\vect{R}),
\end{align}
where $\delta_+$ and $\delta_-$ are two independent particular solutions, which should be interpreted as proportional to the contrast field of the considered scale at horizon crossing. The coefficients of \for{for:structure:delta_evolution_1} indicates that it exists necessarily a \textit{growing mode}, labelled with \texttt{+}, and a \textit{decaying mode}, labelled with \texttt{-}. 

\begin{figure}[tb]
	\centering
	\includegraphics[scale=0.65]{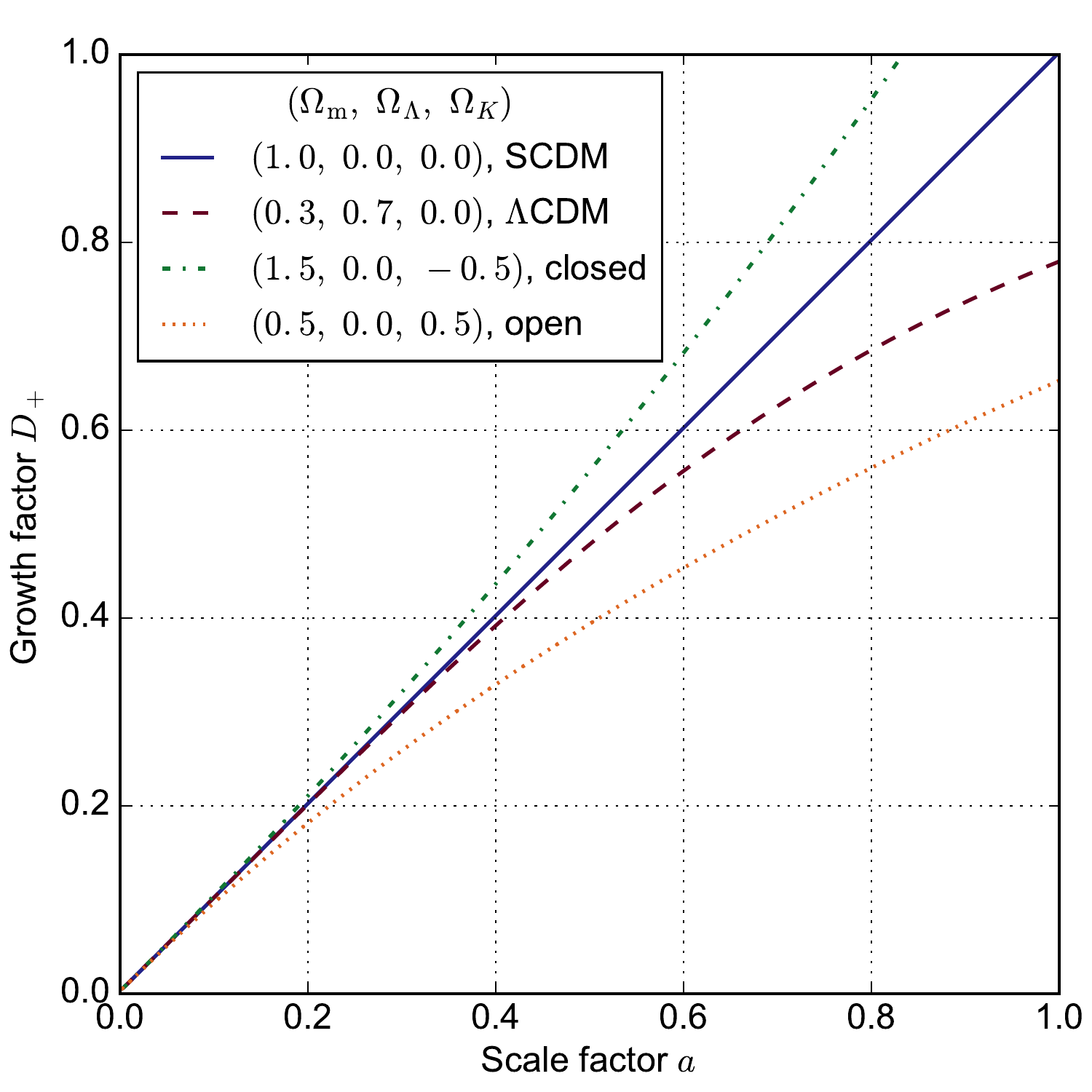}
	\caption{Evolution of the growth factor under different cosmologies. In the standard cold dark matter (\acro{SCDM}) model, we have $D_+(a)=a$.}
	\label{fig:lensing:Growth_factor}
\end{figure}

In the linear regime, the density contrast will keep track on the initial condition, only evolving proportionally with the time-dependent factors $D_+$ and $D_-$. Supposing that the decaying mode vanishes fast enough for relevant $t$, the density contrast can then be simply expressed by the growing mode whose general form is
\begin{align}
	D_+(t) - D_+(t_\mathrm{eq}) \propto H(t) \int_{t_\mathrm{eq}}^t \frac{\rmd t}{\dot{a}^2(t)},
\end{align}
where $t_\mathrm{eq}$ is the time of the matter-radiation equality. $D_+$ is called the \textit{growth factor}\index{Growth factor}, and is conventionally normalized so that for a flat universe, $D_+(a)=a$ during the matter-dominated era. In this case, we can write, for $a>a_\mathrm{eq}$, 
\begin{align}\label{for:structure:D_plus_a}
	D_+(a) = \frac{5\OmegaM}{2}\frac{H(a)}{H_0}\int_0^a \frac{\rmd a'}{(a'H(a')/H_0)^3}.
\end{align}
Several examples of $D_+$ evolution are shown in \fig{fig:lensing:Growth_factor}. It is also useful to normalize the growth factor with regard to its value at the current time. Define
\begin{align}\label{for:structure:D}
	D(z) = \frac{D_+(a=(1+z)\inv)}{D_+(a=1)},
\end{align}
then $D(z=0)=1$ and the linear density contrast at redshift $z$ can be simply written as $\delta(z, \vect{R}) = D(z)\delta(\vect{R})$, where $\delta(\vect{R})$ represents the fluctuation at the present time.

\section{Spherical collapse model}

Concerning the nonlinear evolution of the Universe, cosmologists observe from $N$-body simulations \citep[e.g.][]{Hahn_etal_2007, Chen_etal_2015} that initial perturbations form sheet-like structures by self-gravitation. Then, sheets fold into filaments, and  filaments collapse to form bound objects at the end. In $\LCDM$-cosmology, these bound objects are composed principally of dark matter, called \textit{dark-matter halos}. The halos contain also baryons which leads to forming galaxies. Since dark matter is not visible, what we observe is clusters of galaxies. For this reason, clusters (of matter) and halos are often interchangeable in the literature.

\begin{figure}[tb]
	\centering
	\includegraphics[scale=0.65]{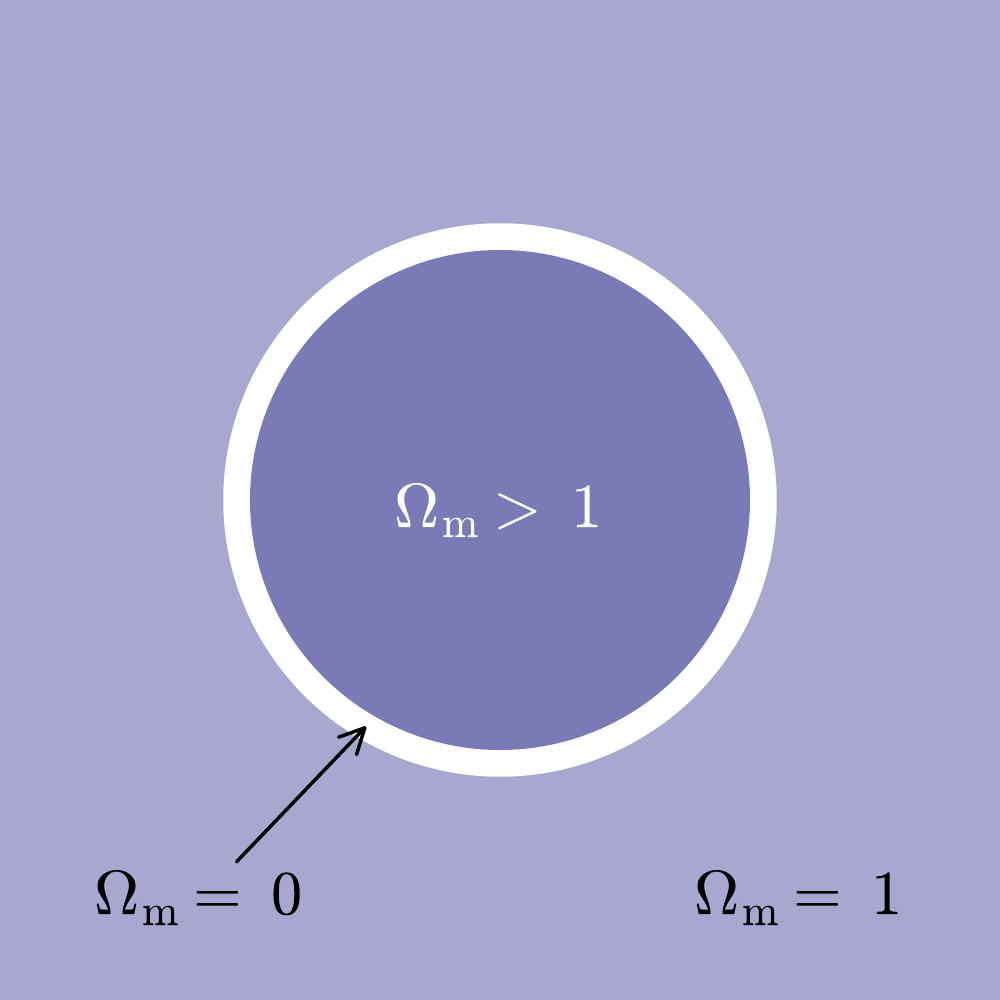}
	\caption{Illustration of an idealized scenario for the spherical collapse. In an homogeneous Einstein-de Sitter universe ($\OmegaM=1$), we empty the mass on a thin shell and put it inside, so that $\OmegaM>1$ for the inner region.}
	\label{fig:lensing:Empty_shell}
\end{figure}

The formation of the halos in $N$-body simulations is complex. Here, we take a simplified approach. We will skip sheets and filaments, and use the spherical collapse model\index{Spherical collapse model} to explain the cluster formation. Following \citet{Peebles_1980}, we consider a globally homogeneous and isotropic universe and a spherical region inside. We take out the mass contained within a thin shell of the sphere, and add it to the inner region, so that the inner mass density is slightly higher than the background. We further assume that the background universe is flat and matter-dominated, so that $\OmegaM\upp{\rmb} = 1$ and $\Omega\upp{\rmb}_K = 0$. The spherical region is overdense and can be modelled as a closed universe with $0 < -\Omega_K = \OmegaM -1 \ll 1$ (\fig{fig:lensing:Empty_shell}).

We are going to solve the Friedmann equation to obtain the evolution of the scale factor. Contrary to the prescription from \chap{sect:cosmology} where all densities have been scaled and expressed with the fraction at the present time, here we will scale to the initial time, since $\OmegaM$, $\Omega_K$, $\OmegaM\upp{\rmb}$, and $\Omega_K\upp{\rmb}$ are defined at the moment where the ``split'' takes place. Thus, in this section, $t_0$ is refered to the initial time, with $a(t_0)=1$ and $H_0 = \dot{a}(t_0)/a(t_0)$ to simplify the calculation.

\subsubsection{Collapse: evolution of a closed universe}

For the overdense region, Friedmann's equation (Eq. \ref{for:cosmology:E}) leads to
\begin{align}
	\left(\frac{\rmd a}{\rmd t}\right)^2 = H_0^2 \left(\frac{\OmegaM}{a} + (1 - \OmegaM) \right).
\end{align}
Let $\theta = H_0\tau\sqrt{\OmegaM-1}$ where $\tau$ is the conformal time in the overdense region. The derivative of $a$ with regard to $\theta$ is 
\begin{align}
	\left(\frac{\rmd a}{\rmd \theta}\right)^2
	&= \left(\frac{\rmd t}{\rmd \theta}\frac{\rmd a}{\rmd t}\right)^2
	= \frac{a^2}{H_0^2 (\OmegaM -1)}\cdot H_0^2 \left( \frac{\OmegaM}{a} + (1 - \OmegaM) \right)\notag\\
	&= \frac{\OmegaM a}{\OmegaM - 1} - a^2. \label{for:structure:da_over_dtheta}
\end{align}
To obtain the evolution of $a$ with regard to $\theta$, we need to rearrange \for{for:structure:da_over_dtheta} and integrate. Denote $A = \OmegaM/2(\OmegaM-1)$ and $b = a/A - 1$, the change of variables leads to
\begin{align}
	\rmd\theta = \frac{\rmd a}{\sqrt{-(a - A)^2 + A^2}} = \frac{1}{A}\frac{\rmd a}{\sqrt{1-(a/A - 1)^2}} = \frac{\rmd b}{\sqrt{1-b^2}} = -\rmd(\arccos b).
\end{align}
The primitive function of $1/\sqrt{1-b^2}$ can be $\arcsin b$ or $\arccos b$. The latter has been chosen because we wish $\theta$ to be positive. Therefore, the scale factor of the overdense region parametrized by $\theta$ is
\begin{align}\label{for:structure:a_region}
	a(\theta) = \frac{\OmegaM}{2(\OmegaM -1)}(1 - \cos\theta)
\end{align}
for $\theta_0 \leq\theta < 2\pi$, where $\theta_0$ satisfies $a(\theta_0) = 1$. Then, from the relation between $\theta$ and $\tau$, we can derive $a\rmd\theta = H_0\sqrt{\OmegaM -1}\ \rmd t$, so
\begin{align}
	t(\theta) = \frac{\OmegaM}{2H_0(\OmegaM -1)^{3/2}}(\theta - \sin\theta)
\end{align}
for $\theta_0 \leq\theta < 2\pi$. Equation \eqref{for:structure:a_region} shows that the size of the overdense region increases first, reaches a maximum at $\theta = \theta_\maxx = \pi$, and decreases to 0 at the end. This means that the closed universe will expand until a turn-around point, then collapse. At the turn-around point, the corresponding corresponding scale factor and time are
\begin{align}
	a_\maxx &= \frac{\OmegaM}{\OmegaM-1},\\ 
	t_\maxx &= \frac{\pi}{2H_0}\frac{\OmegaM}{(\OmegaM - 1)^{3/2}}. 
\end{align}
We will use these two quantities to normalize $a$ and $t$, which yield the normalized scale factor $\tilde{a}$ and the dimensionless time $\tilde{t}$:
\begin{align}
	\tilde{a}(\theta) &\equiv \frac{a(\theta)}{a_\maxx} = \frac{1-\cos\theta}{2}, \label{for:structure:a_region_normalized}\\
	\tilde{t}(\theta) &\equiv \frac{t(\theta)}{t_\maxx} = \frac{\theta - \sin\theta}{\pi}. 
\end{align}

Meanwhile, the background evolves as a matter-dominated flat universe. The solution for the Friedmann equation is
\begin{align}
	a\upp{\rmb}(t) = \left(\frac{3H_0t}{2}\right)^{2/3},
\end{align}
where $t\geq t_0$. After normalization, for $\theta>\theta_0$, the background scale factor is 
\begin{align}
	\tilde{a}\upp{\rmb}(\theta) &\equiv \frac{a\upp{\rmb}(\theta)}{a_\maxx} = \frac{\OmegaM -1}{\OmegaM}\left(\frac{3H_0t(\theta)}{2}\right)^{2/3} \notag\\
	&= \frac{\OmegaM -1}{\OmegaM}\left(\frac{3H_0}{2} \frac{t(\theta)}{t_\maxx} \cdot\frac{\pi}{2H_0}\frac{\OmegaM}{(\OmegaM - 1)^{3/2}}\right)^{2/3} \notag\\
	&= \frac{1}{\OmegaM^{1/3}} \left(\frac{3\pi}{4}\tilde{t}(\theta)\right)^{2/3}
	\approx \left(\frac{3}{4}(\theta - \sin\theta)\right)^{2/3}, \label{for:structure:a_background_normalized}
\end{align}
since $\OmegaM \approx 1$. 

\subsubsection{Mass overdensity}

Now, we would like to know the evolution of the \textit{mass overdensity}\index{Mass overdensity} $\Delta$ and the density contrast $\delta$ described by the spherical collapse model. These two quantities are defined as 
\begin{align}
	\Delta \equiv 1+\delta = \frac{\rho(a)}{\rho\upp{\rmb}\big(a\upp{\rmb}\big)} = \frac{\rho_0a^{-3}}{\rho_0\upp{\rmb}\left.a\upp{\rmb}\right.^{-3}}.
\end{align}
The ratio of mass density at the initial time $t_0$ is nothing but $\OmegaM/\OmegaM\upp{\rmb}$, which is close to 1. As a result, with Eqs. \eqref{for:structure:a_region_normalized} and \eqref{for:structure:a_background_normalized},
\begin{align}\label{for:structure:contrast_in_theta}
	\Delta = 1+\delta \approx \left(\frac{\tilde{a}\upp{\rmb}}{\tilde{a}}\right)^3 = \frac{9(\theta - \sin\theta)^2}{2(1-\cos\theta)^3}. 
\end{align}
What is the density contrast at $t_\maxx$? If we take $\theta = \theta_\maxx = \pi$, then
\begin{align}
	1 + \delta_\maxx = \frac{9\pi^2}{16} \approx 5.55,
\end{align}
which means that the collapse starts when the density contrast reaches about 5.55.

We can observe that in this model, if $a \rightarrow 0$, then $\delta\rightarrow \infty$. However, in the real universe, the collapse will stop before an infinite density is reached. One of the ways to regularize this problem is to assume the final state predicted by the virial theorem for the system. This means that the kinetic energy $E_\rmk$ and the potential energy $E_\rmp$ will satisfy
\begin{align}\label{for:structure:virial_theorem}
	E_\rmk = -\frac{1}{2}E_\rmp.
\end{align}
It is logical to suppose that $E_\rmk=0$ at the turn-around point because when collapse starts, the ``speed'' $\dot{a}$ is zero. Let $r_\maxx$ be the physical radius of the region at $t_\maxx$. The potential energy, which is also the total energy, leads to 
\begin{align}
	E = E_\rmp = -\frac{3\rmG M^2}{5r_\maxx}.
\end{align}
It is not difficult to find that at $r_\vir = r_\maxx/2$, $E - E_\rmp$ and $E_\rmp$ verifies \for{for:structure:virial_theorem}. This condition imposes that the collapse stops at the half of its maximal size. Let us set $a_\vir$ to $a_\maxx/2$. The corresponding $\theta_\vir$ turns out to be $3\pi/2$. Thus, \for{for:structure:contrast_in_theta} yields
\begin{align}
	1 + \delta_\vir = \frac{9}{2} \left(\frac{3\pi}{2}+1\right)^2 \approx 147.
\end{align}

However, for historical reasons, cosmologists derive the virial overdensity differently. Let $\theta_\mathrm{end}=2\pi$ be the value of the time parameter at the end of the evolution. By taking $\theta=\theta_\mathrm{end}$ in the numerator of \for{for:structure:contrast_in_theta} and $\theta=\theta_\vir$ in the denominator, this strange choice leads to
\begin{align}\label{for:structure:contrast_end}
	1 + \delta_\mathrm{end} \equiv \frac{9(\theta_\mathrm{end} - \sin\theta_\mathrm{end})^2}{2(1-\cos\theta_\vir)^3} = 18\pi^2 \approx 178, 
\end{align}
which is the result more commonly found in the literature. Whatever the numerical result is, the conclusion is the following: perturbations form gravitationally bound structures when they become about 147--178 times denser than the background. This simple model is consistent with $N$-body simulations.

\subsubsection{Collapse in linear theory}

At the end of this section, we will analyze the collapse in linear theory. What would be the linear part of the evolution when the spherical collapse is about to start? To answer to this question, let us develop $a$ and $t$ in power series of $\theta$. The results are
\begin{align}
	\tilde{a}_\lin &\equiv \frac{a_\lin}{a_\maxx} = \frac{1}{4}\theta^2 - \frac{1}{48}\theta^4, \label{for:structure:a_lin_normalized}\\
	\tilde{t}_\lin &\equiv \frac{t_\lin}{t_\maxx} = \frac{1}{6\pi}\theta^3 - \frac{1}{120\pi}\theta^5. \label{for:structure:t_lin_normalized}
\end{align}
Let us try to express $\tilde{a}_\lin$ in terms of $\tilde{t}_\lin$. Denote $x = 6\pi \tilde{t}_\lin$. Applying a linear approximation to \for{for:structure:t_lin_normalized}:
\begin{align}
	\theta^3 = x + \frac{1}{20}\theta^5 &\approx x + \frac{1}{20} x^{5/3} = x\left(1 + \frac{1}{20} x^{2/3}\right), 
\end{align}
and inserting it into \for{for:structure:a_lin_normalized}, we find
\begin{align}
	\tilde{a}_\lin &= \frac{1}{4}\theta^2 \left(1 - \frac{1}{12}\theta^2\right) \notag\\
	&\approx \frac{1}{4} x^{2/3} \left(1 + \frac{1}{20} x^{2/3}\right)^{2/3} \left(1 - \frac{1}{12}x^{2/3}\left(1 + \frac{1}{20} x^{2/3}\right)^{2/3}\right) \notag\\
	&\approx \frac{1}{4} x^{2/3} \left(1 + \frac{1}{30} x^{2/3}\right)\left(1 - \frac{1}{12}x^{2/3}\right) \notag\\
	&\approx \frac{1}{4} x^{2/3} \left(1 - \frac{1}{20} x^{2/3}\right). 
\end{align}
Therefore, the density contrast in linear theory is
\begin{align}
	1 + \delta_\lin = \left( \frac{\tilde{a}\upp{\rmb}}{\tilde{a}_\lin} \right)^3
	&= \left(\frac{1}{4} x^{2/3}\right)^3 \left(\frac{1}{4} x^{2/3} \left(1 - \frac{1}{20} x^{2/3}\right)\right)^{-3} \notag\\
	&\approx 1 + \frac{3}{20}x^{2/3} \notag\\
	&= 1 + \frac{3}{20} \Big(6\pi\tilde{t}(\theta) \Big)^{2/3}. 
\end{align}
Taking $\theta=\theta_\maxx=\pi$, we see that the linear density contrast at the turn-around point is $1+\delta_{\maxx,\lin} \approx 2.06$. Taking $\theta=\theta_\mathrm{end}=2\pi$, the linear density contrast at the end of the collapse is $1+\delta_{\mathrm{end},\lin} = 1 + (3/20)(12\pi)^{2/3} \approx 2.686$. This value $\delta_\rmc \equiv 1.686$ is usually considered as a \textit{density contrast threshold} above which a sub-region of the universe is assumed collapsed, with the actual overdensity $\Delta = 178$.

\section{Mass function}
\label{sect:structure:massFct}

\subsection{The Press-Schechter formalism}

The abundance of the halos is characterized by the \textit{mass function}\index{Mass function}, which depends on mass and redshift. It is an important halo statistic for cosmology. A simple way to model the mass function is proposed by \citet[][see also \citealt{Peacock_1999} for interpretation]{Press_Schechter_1974}. Consider the density contrast threshold $\delta_\rmc$ in the linear theory and a contrast field $\delta$ smoothed with a top-hat filter of size $r$. The reasoning of \citet{Press_Schechter_1974} is that a (Lagrangian) point is supposed to fall into a collapsed region if the smoothed contrast is larger than $\delta_\rmc/\sigma(r)$, and the characteristic size of each of these regions is larger than $r$. More precisely, given a mass threshold $M$, and the fluctuation level $\sigma(M)$ defined by this mass (Eq. \ref{for:structure:sigma_M}), the total mass represented by a smoothed contrast larger than $\delta_\rmc/\sigma(M)$ would be organized as a cluster with mass larger than $M$. 

However, there exists a small problem with this approach: the underdense regions are not taken into account. This can be easily shown by taking $M\rightarrow 0$. When $M$ is arbitrarily small, the total mass above $M$ is just the mass of the Universe, but since $\sigma(M)$ becomes arbitrarily large, the mass represented by Lagrangian points with smoothed contrast larger than $\delta_\rmc/\sigma(M)$ only accounts for a part of the mass in the Universe. To fix this discrepancy, \citet{Press_Schechter_1974} simply propose to add a factor of 2, since they model the distribution of $\delta$ of Gaussian random field such that the positive and the negative part are equal. As a result, by denoting $n(z, M)\rmd M$ as the halo number density for mass within $[M,M+\rmd M[$, we have
\begin{align}\label{for:structure:PS_integral}
	\int_M^{+\infty}\rmd M\ \frac{Mn(z,M)}{\bar{\rho}(z)} = 2\int_\nu^{+\infty} \frac{\rmd u}{\sqrt{2\pi}}\ \exp\left(-\frac{u^2}{2}\right),
\end{align}
where $\nu\equiv\delta_\rmc/D(z)\sigma(M)$ which takes into account the linear redshift dependence, $\bar{\rho}(z)$ is the mean density at $z$, and $u$ should be understood as the smoothed density contrast \footnote{The interpretations of both sides of \for{for:structure:PS_integral} are not physically consistent. While the left-hand side stands for a mass fraction, the right-hand side is a fraction of volume. A more rigorous approach can be obtained by adding a factor of $1+\sigma(M)\cdot u$ to the integrand of the right-hand side}. Deriving both sides with regard to $\ln\nu$, we obtain
\begin{align}
	\frac{Mn(z,M)}{\bar{\rho}(z)}\frac{\rmd M}{\rmd\ln\nu} = \sqrt{\frac{2}{\pi}}\ \nu\exp\left(-\frac{\nu^2}{2}\right).
\end{align}

In the literature, there exist different conventions for noting a mass function. In this thesis, I will adopt the one used by \citet{Jenkins_etal_2001}. First, let $n(z, <M)$ be the number of halos with mass smaller than $M$. Immediately, we have $\rmd n(z, <M) = n(z, M)\rmd M$. Now, let us define the \textit{mass function}\index{Mass function} as 
\begin{align}\label{for:structure:massFct}
	f(\nu) \equiv \frac{M}{\bar{\rho}(z)}\frac{\rmd n(z, <M)}{\rmd\ln\nu}.
\end{align}
The interpretation of $f$ is straightforward with \for{for:structure:PS_integral}: it characterizes the correspondence between virialized objects of mass $M$ and linear contrasts in log scale, and the issue of negative contrasts leads to impose the condition $\int\rmd(\ln\nu)\ f(\nu) = 1$ for $f$. Sometimes $f$ is called the \textit{multiplicity function}. With the prescription of \for{for:structure:massFct}, the Press-Schechter mass function is
\begin{align}
	f_\mathrm{PS}(\nu) = \sqrt{\frac{2}{\pi}}\ \nu\exp\left(-\frac{\nu^2}{2}\right).
\end{align}
In the end, the halo number density depends on cosmology via the Friedmann equation and the power spectrum. The former governs the evolution of $\bar{\rho}(z)$, and the latter is essential for determining the level of $\sigma(M)$.

\subsection{Other models}

Since then, better mass function models, fitted by $N$-body simulations, have been proposed. \citet{Sheth_Tormen_1999, Sheth_Tormen_2002} gave the following expression which stands for the ellipsoidal collapse model:
\begin{align}
	f_\mathrm{ST}(\nu) = A\ \sqrt{\frac{a\nu^2}{2\pi}}\cdot \left(1+\left(a\nu^2\right)^{-p}\right) \exp\left(-\frac{a\nu^2}{2}\right),
\end{align}
where $p = 0.3$, $a = 0.75$ and $A$ is the normalization constant, which is about 0.32218 in this case. If we set $p=0$ and $a=1$, $A$ turns out to be 2, and the case of \citet{Press_Schechter_1974} is recovered.

\citet{Jenkins_etal_2001} provided a fitted model by using the Virgo Consortium simulations, leading to
\begin{align}\label{for:structure:massFct_J01}
	f_\mathrm{J01}(\nu=\delta_\rmc/x) = 0.315 \cdot \exp\left(-|\ln x\inv + 0.61|^{3.8}\right),
\end{align}
where $x = \delta_\rmc/\nu = D(z)\sigma(M)$.

\citet{Warren_etal_2006} derived their model from the following fitting form:
\begin{align}
	f_\mathrm{W06}(\nu) = 0.7234\cdot \left(x^{-1.625}+0.2538\right) \exp\left(-\frac{1.1982}{x^2}\right).
\end{align}

\citet{Tinker_etal_2008a} found that the universality of the mass function is broken, providing a redshift-dependent model:
\begin{align}
	f_\mathrm{T08}(\nu) = A(z)\left(\left(\frac{x}{b(z)}\right)^{-a(z)}+1\right) \exp\left(-\frac{c}{x^2}\right),
\end{align}
where $A(z)=0.186/(1+z)^{0.14}$, $a(z)=1.47/(1+z)^{0.06}$, $b(z)=2.57\cdot(1+z)^{\alpha(200)}$, $\log\alpha(\Delta)=-(0.75/\log(\Delta/75))^{1.2}$, and $c=1.19$.

\citet{Bhattacharya_etal_2011} generalized the model from \citet{Sheth_Tormen_1999} with redshift dependency:
\begin{align}
	f_\mathrm{B11}(\nu) = A(z)\sqrt{\frac{2}{\pi}} \left(a(z)\nu^2\right)^{q/2} \left(1+\left(a(z)\nu^2\right)^{-p}\right) \exp\left(-\frac{a(z)\nu^2}{2}\right),
\end{align}
with $A(z) = 0.333/(1+z)^{0.11}$, $a(z) = 0.788/(1+z)^{0.01}$, $p = 0.807$, and $q = 1.795$.

\begin{figure}[tb]
	\centering
	\includegraphics[scale=0.65]{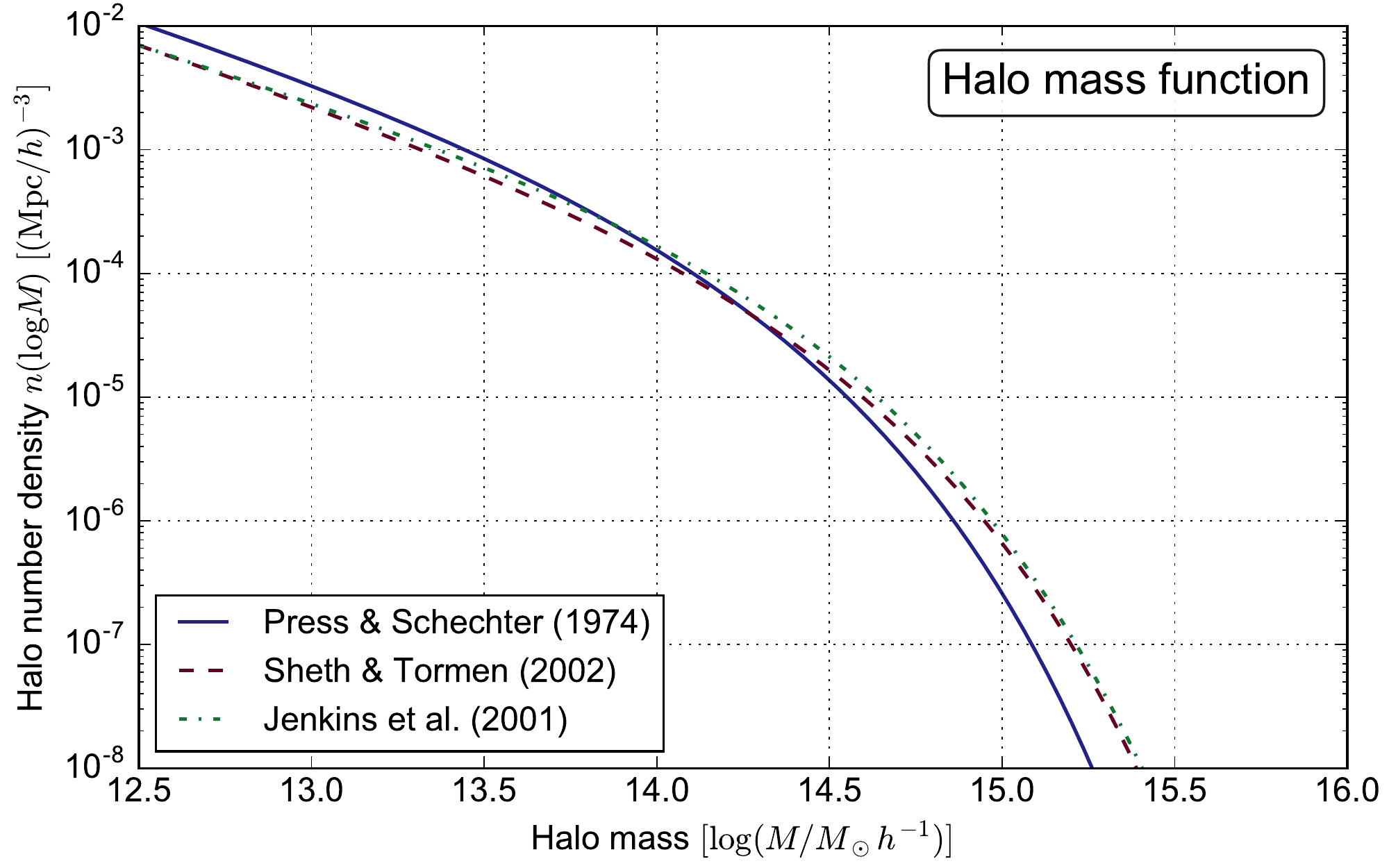}
	\caption{Various mass function models.}
	\label{fig:lensing:Mass_function_model}
\end{figure}

\figFull{fig:lensing:Mass_function_model} shows some mass function models. I would like to highlight that the $y$-axis, which is the halo number density at redshift $z$, is determined as
\begin{align}\label{for:structure:dn_over_dlogM}
	n(z, \log M) \equiv\frac{\rmd n(z, \lessM)}{\rmd\log M} = \ln 10\cdot \frac{\bar{\rho}(z)}{M}\frac{\rmd\ln\nu}{\rmd\ln M}f(\nu).
\end{align}
This is the volume density within a log mass width $\rmd\log M$. Thus, the unit of $n(z, \log M)$ is $(\Mpc/h)^{-3}$ as physical length because of $\bar{\rho}(z)$, while the unit of $n(z, M)$ is actually $h^4/(\Msol\Mpc^3)$. The other subtlety is the factor $\ln 10\approx 2.3$, which is indispensable if the $x$-axis is displayed in decimal logarithmic scale.

\section{Halo density profile}

\subsection{The NFW profiles}
\label{sect:structure:profile:NFW}

\begin{figure}[tb]
	\centering
	\includegraphics[scale=0.65]{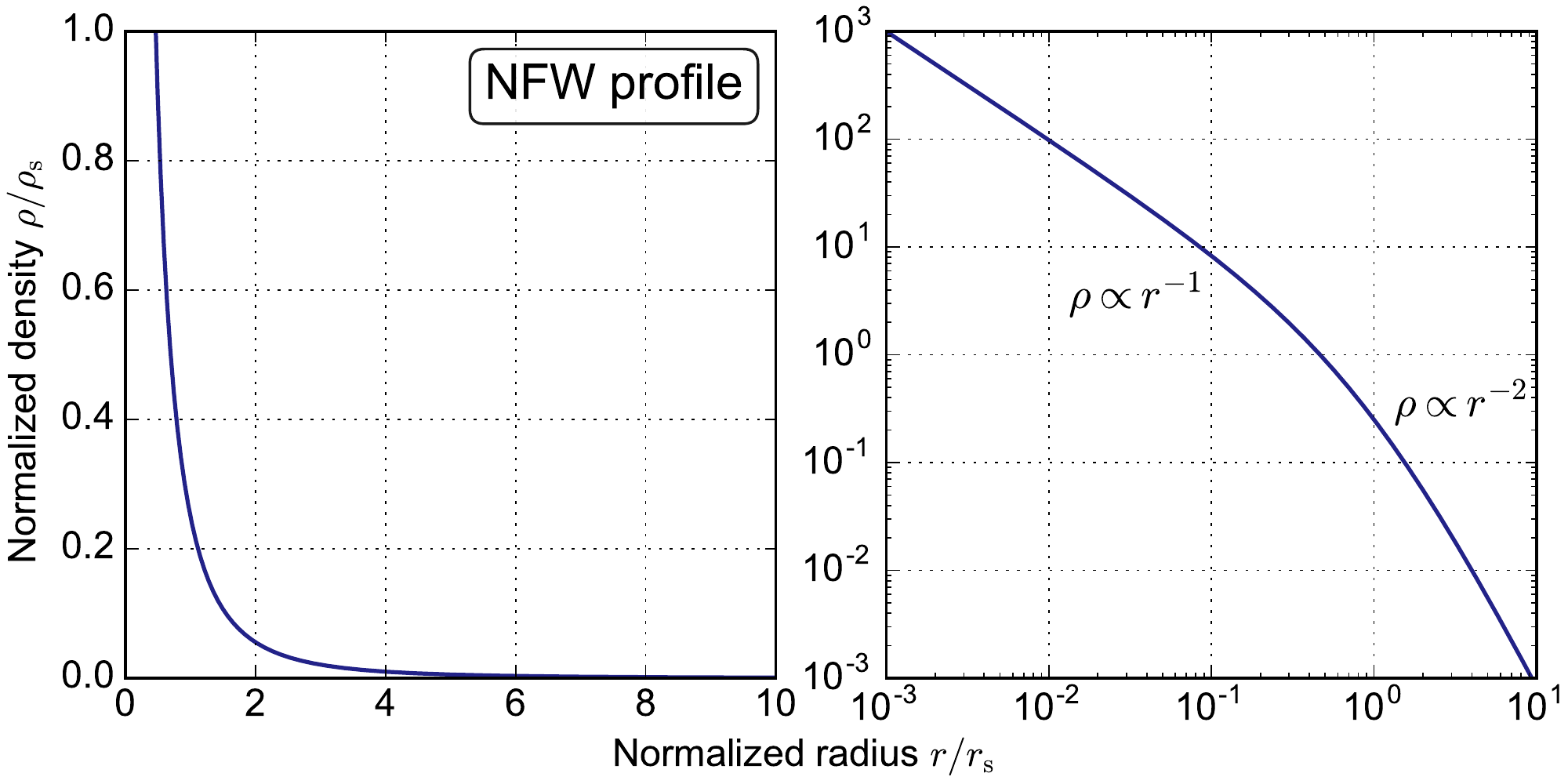}
	\caption{\acro{NFW} profile in normalized coordinates. The concentration is set to 10 as an example. The left panel shows the profile in real space, and the right panel represents the same plot in log space.}
	\label{fig:lensing:NFW}
\end{figure}

Let us address the inner structure of a gravitationally bound object. When structures collapse, the distribution of the matter can depend on the position with regard to the center, thus yields a density profile. Using $N$-body simulations, \citet{Navarro_etal_1995, Navarro_etal_1996, Navarro_etal_1997} provided a halo density profile which is given as follows,\index{Navarro-Frenk-White (\acro{NFW}) profile}
\begin{align}\label{for:structure:NFW_profile}
	\rho_\mathrm{NFW}(\vect{r}) = \frac{\rho_\rms}{(r/r_\rms)(1+r/r_\rms)^2}.
\end{align}
The Navarro-Frenk-White (\acro{NFW}) profile is parametrized by two numbers: the central mass density $\rho_\rms$ and the scalar radius $r_\rms$. Depending on the convention, these two
quantities can have different definitions. A universal way to express them is as follows:
\begin{align}
	\rho_\rms &\equiv \rho_\mathrm{ref}\Delta\cdot \frac{fc^3}{3}, \label{for:structure:NFW_rho_s}\\
	r_\rms &\equiv \frac{r_\Delta}{c}. 
\end{align}
The quantity $\rho_\mathrm{ref}\Delta$ is the mass density inside the halo region, which is $\Delta$ times larger than the reference density $\rho_\mathrm{ref}$. The corresponding radius for the considered $\Delta$ is denoted as $r_\Delta$. The $\rho_\rms$ and $r_\rms$ are related by the concentration parameter $c$ and a derived quantity $f$ given by
\begin{align}\label{for:structure:NFW_f}
	f \equiv \frac{1}{\ln(1+c)-c/(1+c)}.
\end{align}

\begin{figure}[tb]
	\centering
	\includegraphics[width=0.8\textwidth]{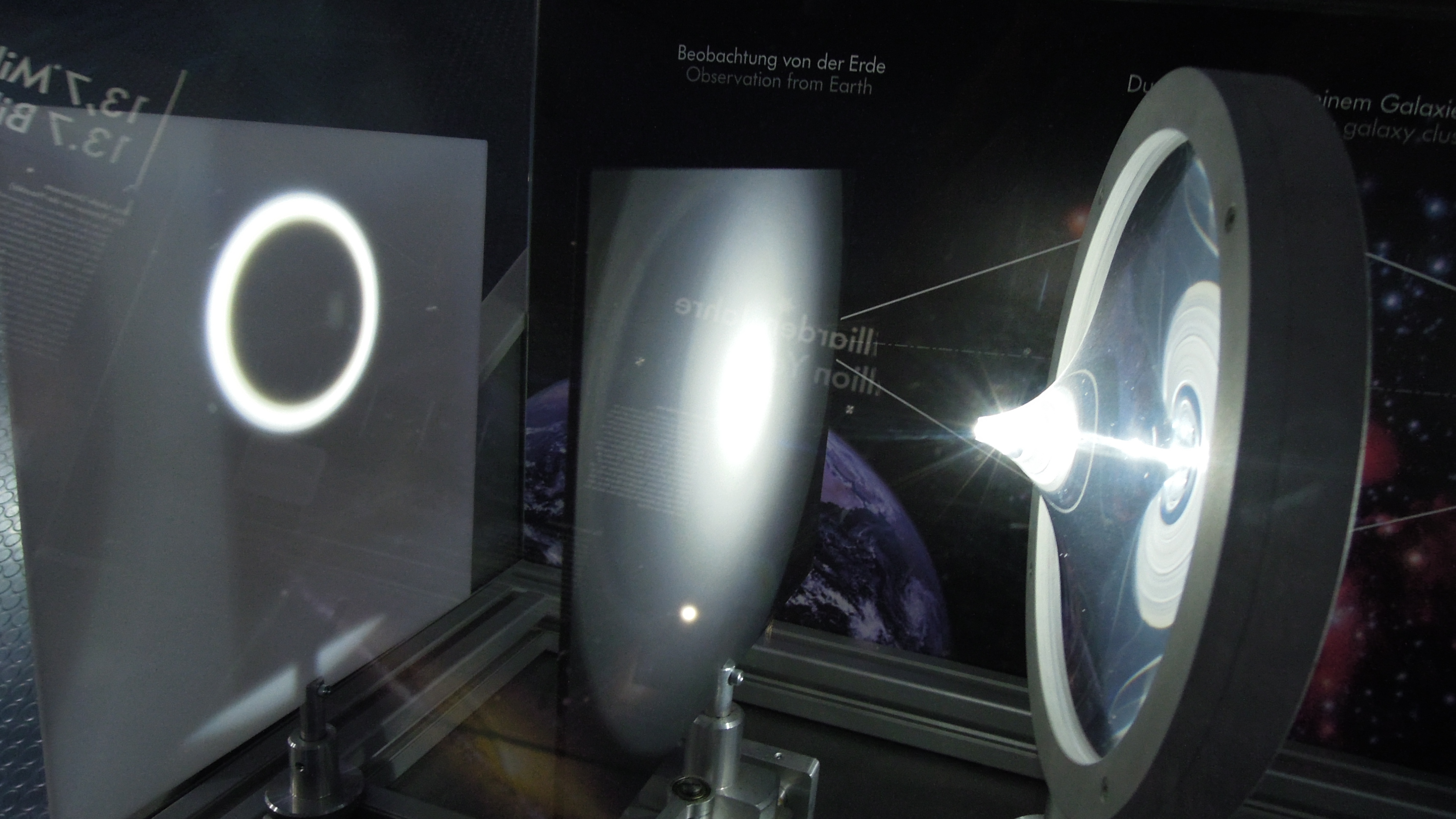}
	\caption{Creation of an Einstein ring by an optical NFW-like profile! Photo taken at the \textit{Deutsches Museum} in Munich.}
	\label{fig:lensing:Optical_NFW}
\end{figure}

\subsubsection{Choice of the convention}

Depending on choices, $\rho_\mathrm{ref}$ may be $\rho_{\crit,0}$, $\rho_\crit(z)$, $\bar{\rho}_{\rmm,0}$, or $\bar{\rho}_\rmm(z)$ (critical and matter densities), and $\Delta$ may be a redshift-dependent relation $\Delta_\vir(z)$ or a constant such as 200 or 500. There is no agreement or consensus on this topic. For example, studies on the mass function tend to use $\bar{\rho}_\rmm(z)$, because dark energy, which becomes dominant in the late-time Universe, is not involved in gravitational collapse. However, studies on the mass-concentration relation \citep{Duffy_etal_2008, DeBoni_etal_2013, Bhattacharya_etal_2013, Meneghetti_etal_2014, Umetsu_etal_2014, Merten_etal_2015, Diemer_Kravtsov_2014, Diemer_Kravtsov_2015} prefer to use $\rho_\crit(z)$ as the reference density. One reason for this less physical choice is historical, since during long time $\Omega_\de$ had been considered close to 0. The other reason is that $\rho_\crit$ is higher than $\bar{\rho}_\rmm$, so the associated radius is smaller (e.g. $r_{200\rmc}<r_{200\rmm}$) and smaller radii lead to a better definition of halo mass (less unbound particles at the outskirt) and scale radius. Particularly, \citet{Diemer_Kravtsov_2014} pointed out that the universality of the \acro{NFW} profiles is better when the density and the radius are scaling respectively by $\rho_\crit$ and $r_{200\rmc}$.

Concerning the choice of $\Delta$, a common one is 200 as an approximation of $18\pi^2$ obtained from \for{for:structure:contrast_end}. Many X-ray studies usually pick up $\Delta=500$, for the same reason as having $\rho_\crit$: enhance the threshold and reduce the radius to focus on the inner region. However, we can also adopt a cosmology-dependent virial overdensity \citep[see e.g.][]{Bryan_Norman_1998, Henry_2000, Weinberg_Kamionkowski_2003}. This is usually fitted from calculations of collapses under different cosmologies.

In this thesis, I adopt the following choice for the \acro{NFW} profiles:
\begin{align}
	r_\rms = \frac{r_\vir}{c},\ \ \ \rho_\rms = \bar{\rho}_\rmm(z)\Delta_\vir(z)\cdot\frac{fc^3}{3}\ \ \ \text{and}\ \ \ 
	\Delta_\vir(z) \equiv \frac{\rho_\vir(z)}{\bar{\rho}_\rmm(z)}.
\end{align}
By the definition of the virial density, the mass of a halo is nothing but
\begin{align}
	M = \bar{\rho}_\rmm(z)\Delta_\vir(z)\cdot\frac{4}{3}\pi r_\vir^3 = 4\pi\int_0^{r_\vir}r^2\rmd r\ \rho_\mathrm{NFW}(r).
\end{align}
The second equality also explains the factor of $fc^3/3$ from \for{for:structure:NFW_rho_s}. Note that the mass is actually not defined if the integral extends to infinity. For this reason, it is useful to define the truncated \acro{NFW} profiles (labelled as \texttt{TJ} for \citealt{Takada_Jain_2003a}):
\begin{align}\label{for:structure:TJ_profile}
	\rho_\mathrm{TJ}(\vect{r}) = \frac{\rho_\rms}{(r/r_\rms)(1+r/r_\rms)^2} \Theta(r_\vir-r),
\end{align}
where $\Theta$ is the Heaviside step function.

\subsection{Mass-concentration relation}

The concentration parameter $c$ from the \acro{NFW} profiles is the link between $\rho_\rms$ and $r_\rms$. That means we can reduce of number of profile parameters to two: mass $M$ and concentration $c$. Studies show that $c$ is not a constant but varies with $M$. Therefore, the \acro{NFW} prescription is always accompanied with a mass-concentration relation\index{Mass-concentration relation} (\acro{$M$-$c$} relation). Inspired by \citet{Bullock_etal_2001}, \citet{Takada_Jain_2002} proposed a \acro{$M$-$c$} relation under a simple form:
\begin{align}\label{for:structure:M_c_relation}
	c(z,M) = \frac{c_0}{1+z}\left(\frac{M}{M_\star}\right)^{-\beta},
\end{align}
where $M_\star$ is a pivot mass. In this thesis, I will follow the choice of \citet{Takada_Jain_2002} where $M_\star$ satisfies $\delta_\rmc = D(z=0)\sigma(M_\star)$, but it is also common to take $M_\star$ as a constant.

\subsection{Other profiles}

A physically motivated model is the singular isothermal sphere (\acro{SIS}) profile:
\begin{align}
	\rho_\mathrm{SIS}(\vect{r}) = \frac{\sigma_v^2}{2\pi\rmG r^2},
\end{align}
where $\sigma_v$ is the \acro{1D} velocity dispersion. This profile varies as $r\invSq$ and fails to describe the inner part of the halos.

Recently, a focus has been put again on the Einasto profile \citep{Einasto_1965}. This profile is governed by the exponential function:
\begin{align}
	\rho_\mathrm{E65}(\vect{r}) = \rho_{-2}\exp\left(-\frac{2}{\alpha}\Big( (r/r_{-2})^\alpha - 1 \Big)\right)
\end{align}
with a shape parameter $\alpha$. The subscript $_{-2}$ stands for the position where the density varies as $1/r^2$.

From $N$-simulations, cosmologists observe that halos are rather triaxial than spherical. On the necessity to model this phenomenon, \citet{Jing_Suto_2002} proposed to parametrize $r$ in \for{for:structure:NFW_profile} by 
\begin{align}
	r^2 = c^2\left( \frac{x^2}{a^2} + \frac{y^2}{b^2} + \frac{z^2}{c^2}  \right),
\end{align}
where $x$, $y$, and $z$ are chosen to be aligned with the ellipsoid axes with $a\leq b\leq c$. \citet{Jing_Suto_2002} also derived from $N$-body simulations a numerical fit of the distribution of $a/c$ and $b/c$.

Finally, \citet{Baltz_etal_2009} provided a version of the \acro{NFW} profiles with a smoothed truncation, which takes form:
\begin{align}
	\rho_\mathrm{BMO}(\vect{r}) = \frac{\rho_\rms}{(r/r_\rms)(1+r/r_\rms)^2(1+r/r_\rmt)^n},
\end{align}
where $r_\rmt$ is the truncation radius and $n$ is the sharpness parameter. A great advantage of this profile is that it models more correctly the regime between the one- and two-halo terms. \citet{Baltz_etal_2009} also derived the projected mass from this profile.

\subsubsection{Summary}

Cosmological structures are the source of weak gravitational lensing. We have reviewed some important mechanisms of structure formation in this chapter. 

On large scales, matter structures can be well described by linear perturbation theory, summarized as the evolution of the growth factor. 

On small scales, the linear theory is still served as the baseline to understand the structure. Via the spherical collapse model, a connection between the linear and the nonlinear evolution of a virialized object is established, and this connection is used for more detailed studies such as for the mass function.

Two topics have been particularly addressed: the mass function and the halo density profile. Various models have been introduced. These two subjects are essential in the framework of this thesis. 

After understanding structure formation, I will focus on developing the theory of gravitational lensing in the next chapter.

\clearpage
\thispagestyle{empty}
\cleardoublepage


\chapter{Weak gravitational lensing}
\label{sect:lensing}
\fancyhead[LE]{\sf \nouppercase{\leftmark}}
\fancyhead[RO]{\sf \nouppercase{\rightmark}}

\subsubsection{Overview}

Weak gravitational lensing will be discussed in detail in this chapter. The presentation will start with deriving lensing equations from light deflection. Then, after introducing the convergence and the shear, their expected values from clusters will be introduced. After that, readers will see how the lensing signal can be reconstructed and how cosmological parameters can be extracted from different observables, Gaussian or non-Gaussian. Finally, I will review existing work on weak-lensing peak counts in the literature.

\section{Light deflection}

\subsection{Different regimes of gravitational lensing}

Gravitational lensing is one of the phenomena predicted by general relativity proposed by \citet{Einstein_1915}. According to this theory, massive objects perturb the spacetime geometry, so that light which always travels on geodesics appears to be bent in a \acro{3D} space. For observers who inspect distant objects, images appear at different positions from the original ones, since the direction of the deflected light path converging on the observer is no longer identical to the direct line of sight direction to the source. Therefore, images are subject to transformations such as offset, magnification, and distortion.

This light deflection can also be modelled in classical physics. However, given a point mass, the deflection angle predicted by the classical model is smaller than that in general relativity by a factor 2. This factor was confirmed when the first gravitational lens was observed. In 1919, \citet{Dyson_etal_1920} measured the positions of stars when they passed behind the Sun. The observation was feasible thanks to a total solar eclipse, and the results were consistent with Einstein's theory.

\begin{figure}[tb]
	\centering
	\includegraphics[width=0.7\textwidth]{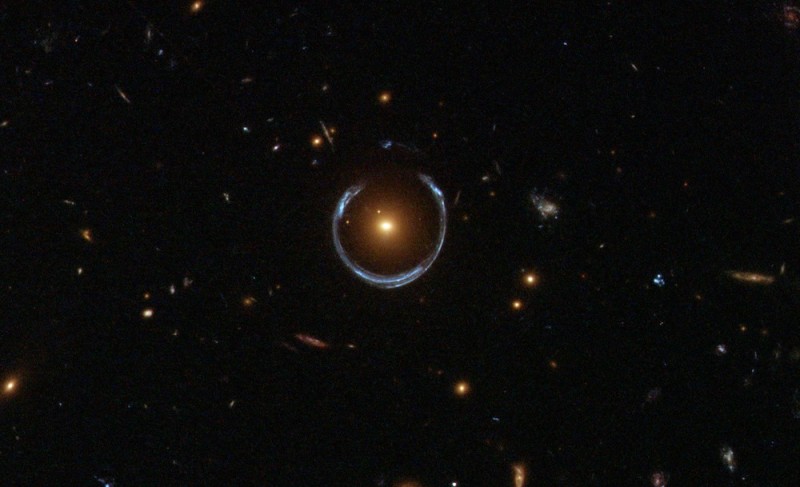}
	\caption{Image taken by the Hubble Space Telescope. This is a strong lens named LRG 3-757. The blue background galaxy is lensed and is split into multiple images. This ring-like pattern is called Einstein ring. (Source: ESA-Hubble/NASA)}
	\label{fig:lensing:Einstein_ring}
\end{figure}

Depending on the detection, gravitational lensing can be separated into three regimes. First, when light rays may take more than one path from the source to the observer, the deflection usually results in multiple images or extreme shapes such as arcs and rings. Lensed objects can be easily identified. In this case, the phenomenon is recognized as \textit{strong lensing}\index{Lensing, strong}. An example of strong lenses is shown in \fig{fig:lensing:Einstein_ring}. Readers can see that the blue background galaxy has been lensed into at least two arc-like images, forming a near-perfect ring.

\begin{figure}[tb]
	\centering
	\includegraphics[scale=0.65]{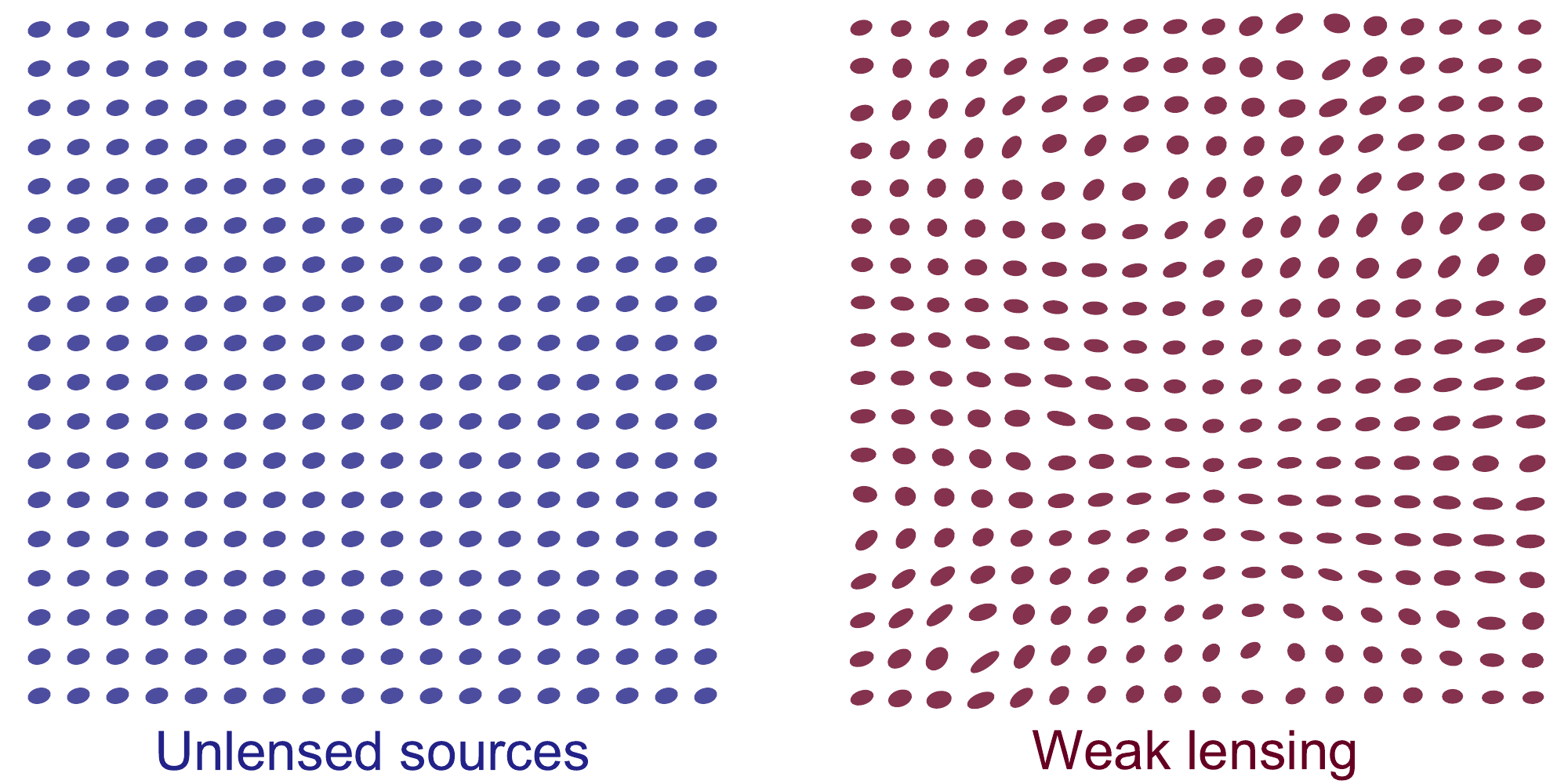}
	\caption{Illustration of the weak-lensing effect.}
	\label{fig:lensing:Weak_lensing}
\end{figure}

The second regime is \textit{weak lensing}\index{Lensing, weak} (\acro{WL}). \figFull{fig:lensing:Weak_lensing} shows an exaggerated illustration. In this regime, one can not tell whether images are distorted or not on individual objects. However, by measuring precisely image shapes and by using statistics, signals can still be identified. Cosmologists have started to report significant \acro{WL} measurements since the 90s: first by cluster lensing \citep[e.g.][]{Bonnet_etal_1994, Luppino_Kaiser_1997}, then by galaxy-galaxy lensing \citep[e.g.][]{Brainerd_etal_1996, Fischer_etal_2000}, and finally by lensing from large-scale structures \citep[cosmic shear,][]{Bacon_etal_2000, VanWaerbeke_etal_2000, Wittman_etal_2001}. Since then, \acro{WL} has been considered as a tool for probing cosmology, especially for measuring $\OmegaM$, $\sigEig$, and $\wZero$. Recently, the Canada-France-Hawaii Telescope Lensing Survey (\acro{CFHTLenS}) has set a milestone by providing interesting \acro{WL} cosmological constraints. Many important probes are still ongoing, e.g. the Kilo-Degree Survey (\acro{KiDS}), the Dark Energy Survey (\acro{DES}), the Subaru Hyper Suprime-Cam (\acro{HSC}) survey, and the Javalambre Physics of the Accelerating Universe Astrophysical Survey (\acro{J-PAS}). Furthermore, we look forward in the future to very large lensing data sets from the Euclid mission, the Large Synoptic Survey Telescope (\acro{LSST}), and the Wide Field Infrared Survey Telescope (\acro{WFIRST}). To enhance statistical power, these surveys require very precise shape measurements. The objective is to reach the precision of $2\dixx{-4}$ on the average ellipticity \citep{Laureijs_etal_2011}, which is an order of magnitude smaller than the deviation of the Earth from a perfect sphere.

Last, if the lens mass is as low as a planet or a star, we enter the \textit{microlensing}\index{Microlensing} regime. In this case, distortion are not measurable, but the difference on apparent magnitude is. Recently, this phenomenon is widely used for searching for exoplanets.

\subsection{Geodesic deviation equation}

\begin{figure}[tb]
	\centering
	\includegraphics[scale=0.65]{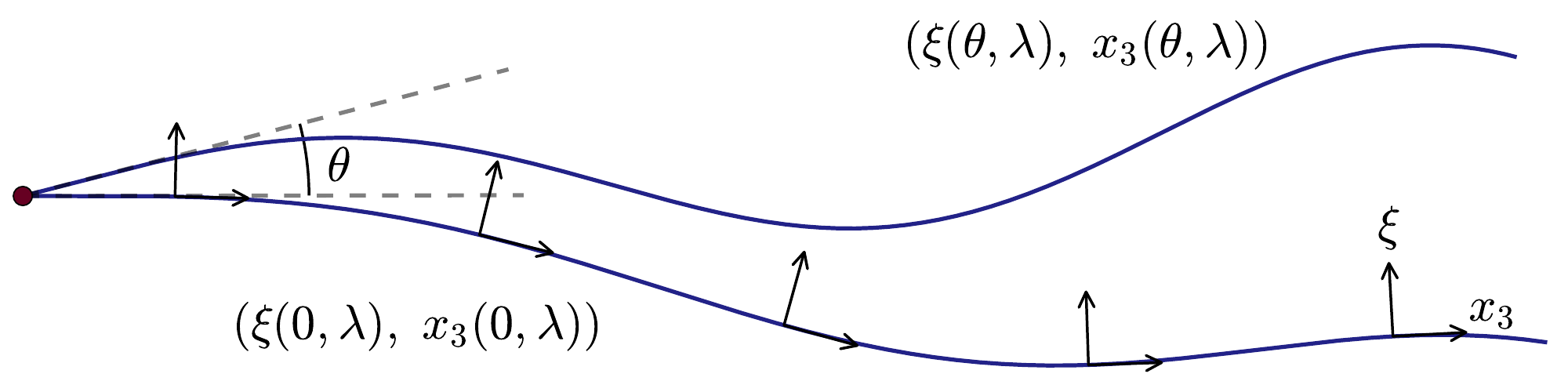}
	\caption{Illustration of the geodesic deviation. The bottom light ray is chosen as the reference; the top light ray is traced with an exaggerated separation angle $\theta$. Both are parametrized by the affine parameter $\lambda$. The space coordinates are decomposed in to a normal part $\bxi$ and a tangential part $x_3$. On the reference line, we have naturally $\bxi(0, \lambda) = (0,0)$.}
	\label{fig:lensing:Geodesic_illustration}
\end{figure}

In general relativity, the trajectory of light is defined by null geodesics. Therefore, given any metric, light deflection from gravitational lensing can be derived from the geodesic equation. Following \citet{Seitz_etal_1994} and \citet{Bartelmann_2010}, let us select a right ray as the reference geodesic, parametrized by an affine parameter $\lambda$. Consider a nearby light ray from a direction characterized by a Cartesian vector $\btheta=(\theta_1, \theta_2)$ defining a position on the sky, we define $\bxi(\btheta, \lambda)=(x_1, x_2)$ as the physical separation vector between two geodesics, i.e. the transverse components which are perpendicular to the tangential direction $x_3$ of the reference ray in the physical \acro{3D} space (\fig{fig:lensing:Geodesic_illustration}). Then, the geodesic deviation equation\index{Geodesic deviation equation}, which characterizes the evolution of $\bxi$, leads to
\begin{align}\label{for:lensing:geodesic}
	\frac{\rmd^2\bxi(\btheta, \lambda)}{\rmd\lambda^2} = \mathcal{T}(x_1, x_2, x_3)\bxi(\btheta, \lambda),
\end{align} 
where $\mathcal{T}$ is the \textit{optical tidal matrix}\index{Optical tidal matrix}. This matrix can be decomposed into two terms:
\begin{align}\label{for:lensing:tidal_matrix}
	\mathcal{T} = \mathcal{T}_\mathrm{bg} + \mathcal{T}_\mathrm{cl},
\end{align}
where the background term $\mathcal{T}_\mathrm{bg}$ is given by the homogeneous and isotropic Universe, and the clump term $\mathcal{T}_\mathrm{cl}$ characterizes local inhomogeneities. Both terms can be expressed by the Riemann, the Ricci, and the Weyl tensors \citep[see e.g.][]{Seitz_etal_1994}. Consider now a \acro{FLRW} metric perturbed by a weak Newtonian gravitational potential $\Phi$. Assuming that $\Phi$ is quasi-static and $\Phi\ll\rmc^2$, the metric is given by
\begin{align}
	\rmd s^2 = a^2(\tau) \left[-\left(1+\frac{2\Phi}{\rmc^2}\right) \rmc^2\rmd\tau^2 + \left(1-\frac{2\Phi}{\rmc^2}\right) \left(\rmd w^2 + f_K^2(w)\rmd\Omega^2\right)\right],
\end{align}
where $a$ is the scale factor, $\tau$ the conformal time, $\Omega$ the solid angle, $\rmc$ the light speed, and $w$ and $f_K(w)$ are respectively the comoving radial and transverse distances. In this case, \citet{Seitz_etal_1994} showed that 
\begin{align}
	\left(\mathcal{T}_\mathrm{bg}\right)_{ij} &= -\frac{4\pi\rmG}{\rmc^2}\frac{\bar{\rho}_0}{a^5}\ \delta_{ij}, \label{for:lensing:T_bg}\\
	\left(\mathcal{T}_\mathrm{cl}\right)_{ij} &= -\frac{1}{\rmc^2a^2}\left(2\frac{\partial^2}{\partial x_i\partial x_j} + \delta_{ij}\frac{\partial^2}{\partial x_3^2}\right) \phi, \label{for:lensing:T_cl}
\end{align}
where $i,j\in\{1,2\}$, $\rmG$ is the gravitational constant, $\bar{\rho}_0$ the background density at the current time, $\delta_{ij}$ the Kronecker delta, and $\phi$ the reduced Newtonian potential (defined by Eq. \ref{for:structure:Poisson_eq_3}). Here, the reduced potential is needed since $\mathcal{T}_\mathrm{bg}$ has already extracted the background part of the potential. Note that $x_3$ is the line-of-sight direction and $x_1$ and $x_2$ are transverse components.

Now, we consider that light is propagated in the \textit{thin-lens approximation}\index{Thin-lens approximation}, which means that inhomogeneities are geometrically thin. This allows us to consider \acro{2D} clump mass distributions in isolation. In this case, the term $\partial^2\phi/\partial x_3^2$ vanishes due to an integration over $x_3$ \citep{Seitz_etal_1994}. Hereafter, we will always stay in this approximation and consider $\partial^2\phi/\partial x_3^2$ to drop out from \for{for:lensing:T_cl}.

Let us rewrite \for{for:lensing:geodesic} in the comoving space. Due to the redshift effect, the affine parameter satisfies $\rmd\lambda=a(\rmc\rmd t)=a^2(\rmc\rmd\tau)=a^2\rmd w$. Let $(X_1, X_2, w)=(x_1/a, x_2/a, x_3/a)$ be the comoving coordinates. Define $\bXi\equiv\bxi/a=(X_1, X_2)$ as the separation vector in the comoving space. Denoting $a'=\rmd a/\rmd\lambda$, $\bxi'=\rmd \bxi/\rmd\lambda$, we have
\begin{align}\label{for:lensing:comoving_geodesic_1}
	\frac{\rmd^2\bXi}{\rmd w^2} = a^2\frac{\rmd}{\rmd\lambda}\left( a\bxi'-\bxi a'\right) = a^3 \bxi'' - a^2a''\bxi.
\end{align} 
On the one hand, \for{for:lensing:geodesic} leads to $\bxi'' = (\mathcal{T}_\mathrm{bg} + \mathcal{T}_\mathrm{cl})\bxi$. On the other hand, 
\begin{align}
	a'' = \frac{\rmd^2a}{\rmd\lambda^2} = \frac{1}{\rmc^2 a}\frac{\rmd}{\rmd t}\left(\frac{\dot{a}}{a}\right) = \frac{\dot{a}}{\rmc^2a} \frac{\rmd}{\rmd a}\left(\frac{\dot{a}}{a}\right) = \frac{1}{2\rmc^2}\frac{\rmd}{\rmd a}\left(\frac{\dot{a}}{a}\right)^2,
\end{align} 
where $(\dot{a}/a)^2$ can be given by the Friedmann equation, \for{for:cosmology:Friedmann_equations_2}, which results in
\begin{align}\label{for:lensing:a_prime_prime_2}
	a'' = -\frac{4\pi\rmG}{\rmc^2}\frac{\bar{\rho}_0}{a^4} + \frac{K}{a^3},
\end{align}
where $K$ is the spacetime curvature. Using Eqs. \eqref{for:lensing:T_bg}, \eqref{for:lensing:T_cl}, and \eqref{for:lensing:a_prime_prime_2}, \for{for:lensing:comoving_geodesic_1} leads to
\begin{align}
	\frac{\rmd^2 X_i}{\rmd w^2} = \sum_{j=1,2} \left(-\frac{K}{a}\delta_{ij} - \frac{2a}{\rmc^2}\frac{\partial^2 \phi}{\partial x_i \partial x_j}\right)x_j = -KX_i - \frac{2}{\rmc^2}\sum_{j=1,2} \left(\frac{\partial^2 \phi}{\partial X_i \partial X_j}\right)X_j.
\end{align}
Since for any \acro{2D} differentiable function $f$, for small $X_1$ and $X_2$, we have
\begin{align}\label{for:lensing:differentiable}
	f(X_1, X_2) - f(0) = \frac{\partial f}{\partial X_1}X_1 + \frac{\partial f}{\partial X_2}X_2,
\end{align}
applying \for{for:lensing:differentiable} to $(\partial \phi/\partial X_i)(\cdot, w)$ yields the comoving geodesic deviation equation under the thin-lens approximation:
\begin{align}\label{for:lensing:comoving_geodesic_3}
	\frac{\rmd^2 \bXi}{\rmd w^2} + K\bXi = -\frac{2}{\rmc^2}\Big[\nabla_\perp \phi(\bXi, w) - \nabla_\perp \Phi(0, w) \Big],
\end{align}
where $\nabla_\perp=(\partial/\partial X_1, \partial/\partial X_2)$. The dependency of $\bXi$ on $w$ is implicit.

\subsection{Born approximation and distortion matrix}
\label{sect:lensing:deflection:matrix}

To solve \for{for:lensing:comoving_geodesic_3}, the technique of the Green's function\index{Green's function} is used. Given any linear differential operator $L$ and any source function $S$, the system $L(y(x)) = S(x)$ can be solved by finding a Green's function $G(x, x')$ such that
\begin{align}
	L(G(x, x')) = \delta(x-x'),
\end{align}
where $\delta$ is the Dirac distribution. If $G$ is provided, the particular solution $y_\rmp$ can be found by 
\begin{align}
	y_\rmp(x) = \int \rmd x'\ G(x-x')S(x'),
\end{align}
with homogeneous boundary conditions (i.e. under the form of $y\upp{i}(x_0)=0$), and the general solution is 
\begin{align}
	y(x) = y_\rmp(x) + y_\rmc(x),
\end{align}
where $y_\rmc$ is the solution of the homogeneous equation $Ly = 0$ with the boundary conditions of the original problem.

Applying the above paragraph to \for{for:lensing:comoving_geodesic_3}, readers easily find
\begin{align}
	L(\bXi) &= \frac{\rmd^2 \bXi}{\rmd w^2} + K\bXi,\\ 
	S(w) &= -\frac{2}{\rmc^2}\Big[\nabla_\perp \phi(\bXi, w) - \nabla_\perp \phi(0, w) \Big]\Theta(w), 
\end{align}
where the Heaviside step function $\Theta$ is added since $S(w)$ is not defined for $w<0$, and the boundary conditions are $\bXi(0)=0$ and $(\rmd\bXi/\rmd w)|_{w=0}=\btheta$. For $w\neq w'$, $\delta(w-w')=0$, so the Green's function possesses the following form:
\begin{align}
	G(w, w') = \left\{\begin{array}{ll}
		A\cdot f_K(w-w') + B\cdot g_K(w-w') & \text{if $w>w'$,}\\
		C\cdot f_K(w-w') + D\cdot g_K(w-w') & \text{if $w<w'$,}
	\end{array}\right.
\end{align}
where $A$, $B$, $C$, and $D$ are constants to be determined and
\begin{align}
	f_K(w) = \left\{\begin{array}{ll}
		\displaystyle\frac{1}{\sqrt{K}}\sin(\sqrt{K}w) & \text{if $K>0$,}\\[2ex]
		w & \text{if $K=0$,}\\[1ex]
		\displaystyle\frac{1}{\sqrt{K}}\sinh(\sqrt{K}w) & \text{if $K<0$,}
	\end{array}\right.
	\hspace{2em}
	g_K(w) = \left\{\begin{array}{ll}
		\displaystyle\frac{1}{\sqrt{K}}\cos(\sqrt{K}w) & \text{if $K>0$,}\\[2ex]
		1 & \text{if $K=0$,}\\[1ex]
		\displaystyle\frac{1}{\sqrt{K}}\cosh(\sqrt{K}w) & \text{if $K<0$.}
	\end{array}\right.
\end{align}
We can see that the function $f_K$ here is the same as \for{for:cosmology:comoving_transverse_dist}. The Green function satisfies the homogeneous boundary conditions, i.e. $G(0,w')=(\partial G/\partial w)|_{w=0}=0$. For a fixed point $w'>0$, this enforces $C=D=0$. Then, the order of $L$ is two. This guarantees the continuity at $w=w'$ at zero order and $(\partial G/\partial w)|_{w=w'_+}=(\partial G/\partial w)|_{w=w'_-}+1$, which implies
\begin{align}
	\left\{\begin{matrix}
		A\cdot f_K(0) + B\cdot g_K(0) = 0\\
		A\cdot f_K'(0) + B\cdot g_K'(0) = 1
	\end{matrix}\right.,
\end{align}
which results in $A=1$ and $B=0$. Therefore, the Green function and the particular solution are
\begin{align}
	&G(w, w') = \Theta(w-w')f_K(w-w'),\\ 
	&\bXi_\rmp(w) = -\frac{2}{\rmc^2}\int_0^w \rmd w'\ f_K(w-w')\Big[\nabla_\perp \phi(\bXi, w') - \nabla_\perp \phi(0, w') \Big]. 
\end{align}
On the other hand, the solution to the homogeneous equation is also in the form of $A\cdot f_K(w) + B\cdot g_K(w)$. With the boundary condition, we obtain at ease $A=\btheta$ and $B=0$. At the end of the day, the solution to \for{for:lensing:comoving_geodesic_3} is 
\begin{align}\label{for:lensing:Xi}
	\bXi(\btheta, w) = f_K(w)\btheta - \frac{2}{\rmc^2}\int_0^w \rmd w'\ f_K(w-w')\Big[\nabla_\perp \phi(\bXi, w') - \nabla_\perp \phi(0, w') \Big].
\end{align}

Now we come up with an expression for the separation vector in spite of its recursive definition. This can be resolved by inserting iteratively \for{for:lensing:Xi} into itself. The term $\nabla_\perp \phi$ can be developed with regard to $\bXi$ and if the series is truncated at the first term, this is called \textit{Born approximation}\index{Born approximation}, originally used in quantum mechanics for scattering theory. Applying the Born approximation to \for{for:lensing:Xi} is simply replacing $\bXi$ at the right-hand side with $f_K(w)\btheta$. Physically, it assumes that the potential evaluated on the perturbed light path does not differ substantially from that evaluated on the unperturbed line of sight. 

Consider now that $\bXi$ at the right-hand side of \for{for:lensing:Xi} is replaced with $f_K(w)\btheta$ by the Born approximation. If a light ray is received from the direction $\btheta$, the set of $\bXi(\btheta, w\geq 0)$ constructs its trajectory, and any point on this path can be a possible position of the source. Say that the source is located at $w$. Then, the apparent angular position is obviously $\btheta$. However, if the inhomogeneity is not present, the unlensed angular position is known. This is $\bbeta(\btheta, w)\equiv \bXi(\btheta, w)/f_K(w)$ since $\bXi$ is defined in comoving space. As a result, \for{for:lensing:Xi} yields a mapping from the lensed space (observed position $\btheta$) to the unlensed space (true position $\bbeta$). We can characterize this transformation using the 2 $\times$ 2 \textit{distortion matrix}\index{Distortion matrix} $\mathcal{A}$ defined as
\begin{align}\label{for:lensing:distortion_matrix_1}
	\mathcal{A}(\btheta, w) \equiv \frac{\partial\bbeta}{\partial\btheta}=\frac{1}{f_K(w)}\frac{\partial\bXi}{\partial\btheta}.
\end{align}
By inserting \for{for:lensing:Xi} into \for{for:lensing:distortion_matrix_1}, this yields
\begin{align}\label{for:lensing:distortion_matrix_2}
	\mathcal{A}_{ij}(\btheta, w) = \delta_{ij} - \frac{2}{\rmc^2}\int_0^w \rmd w'\ \frac{f_K(w-w')f_K(w')}{f_K(w)}\partial_i\partial_j \phi(f_K(w')\btheta, w'),
\end{align}
where $\partial_i\equiv\partial/\partial X_i$ and $i,j\in\{1,2\}$. It is useful to introduce the \textit{lensing potential}\index{Lensing potential} $\psi$, defined as
\begin{align}
	\psi(\btheta, w) \equiv \frac{2}{\rmc^2}\int_0^w \rmd w'\ \frac{f_K(w-w')}{f_K(w)f_K(w')}\ \phi(f_K(w')\btheta, w'),
\end{align}
which can be considered as a distance-weighted Newtonian potential. This simplifies the components of the distortion matrix into
\begin{align}\label{for:lensing:distortion_matrix_3}
	\mathcal{A}_{ij}(\btheta, w) = \delta_{ij} - \partial_i\partial_j\psi(\btheta, w),
\end{align}
where $\partial_i=\partial/\partial \theta_i$.

\subsection{Convergence and shear}

\index{Lensing, weak}From the previous section, we derive from general relativity the distortion of images from sources due to gravitational lensing. In the linear regime, the distortion is characterized by \for{for:lensing:distortion_matrix_3}. To interpret the transformation, we parametrize the distortion matrix as
\begin{align}
	\mathcal{A}(\btheta, w) = \begin{pmatrix}
		1-\kappa - \gamma_1 &          - \gamma_2\\
		         - \gamma_2 & 1-\kappa + \gamma_1
	\end{pmatrix}
	= (1-\kappa)\begin{pmatrix}
		1 - g_1 &   - g_2\\
		  - g_2 & 1 + g_1
	\end{pmatrix},
\end{align}
where $\ g_i(\btheta, w) \equiv \gamma_i(\btheta, w)/(1 - \kappa(\btheta, w))$. Cosmologists usually use the complex notation to denote $\gamma_i$ and $g_i$:
\begin{align}\label{for:lensing:shear}
	\gamma \equiv \gamma_1 + \rmi\gamma_2 = |\gamma|\cdot \rme^{2\rmi\varphi}, \ \ g \equiv g_1 + \rmi g_2 = |g|\cdot \rme^{2\rmi\varphi},
\end{align}
where $\varphi$ is the ``rotation angle''\index{Rotation angle} that will be discussed later. In this parametrization, the scalar $\kappa$ is called the \textit{convergence}\index{Convergence}; the complex number $\gamma$ is called the \textit{shear}\index{Shear} and $g$ the \textit{reduced shear}\index{Shear, reduced}.  It is straightforward to relate $\kappa$ and $\gamma$ to $\psi$ by
\begin{align}\label{for:lensing:psi_relation_1}
	\kappa = \frac{\partial_1^2+\partial_2^2}{2}\psi,\ \ \gamma_1 = \frac{\partial_1^2-\partial_2^2}{2}\psi, \ \ \gamma_2 = \partial_1\partial_2\psi.
\end{align}
Therefore, as we can see, $\kappa$ and $\gamma$ are actually related to one another. In the \acro{WL} regime, we have $|\kappa|<1$ and $|g|<1$, and in most cases we often assume $|\kappa|\ll 1$ and $|g|\ll 1$.

\begin{figure}[tb]
	\centering
	\includegraphics[scale=0.65]{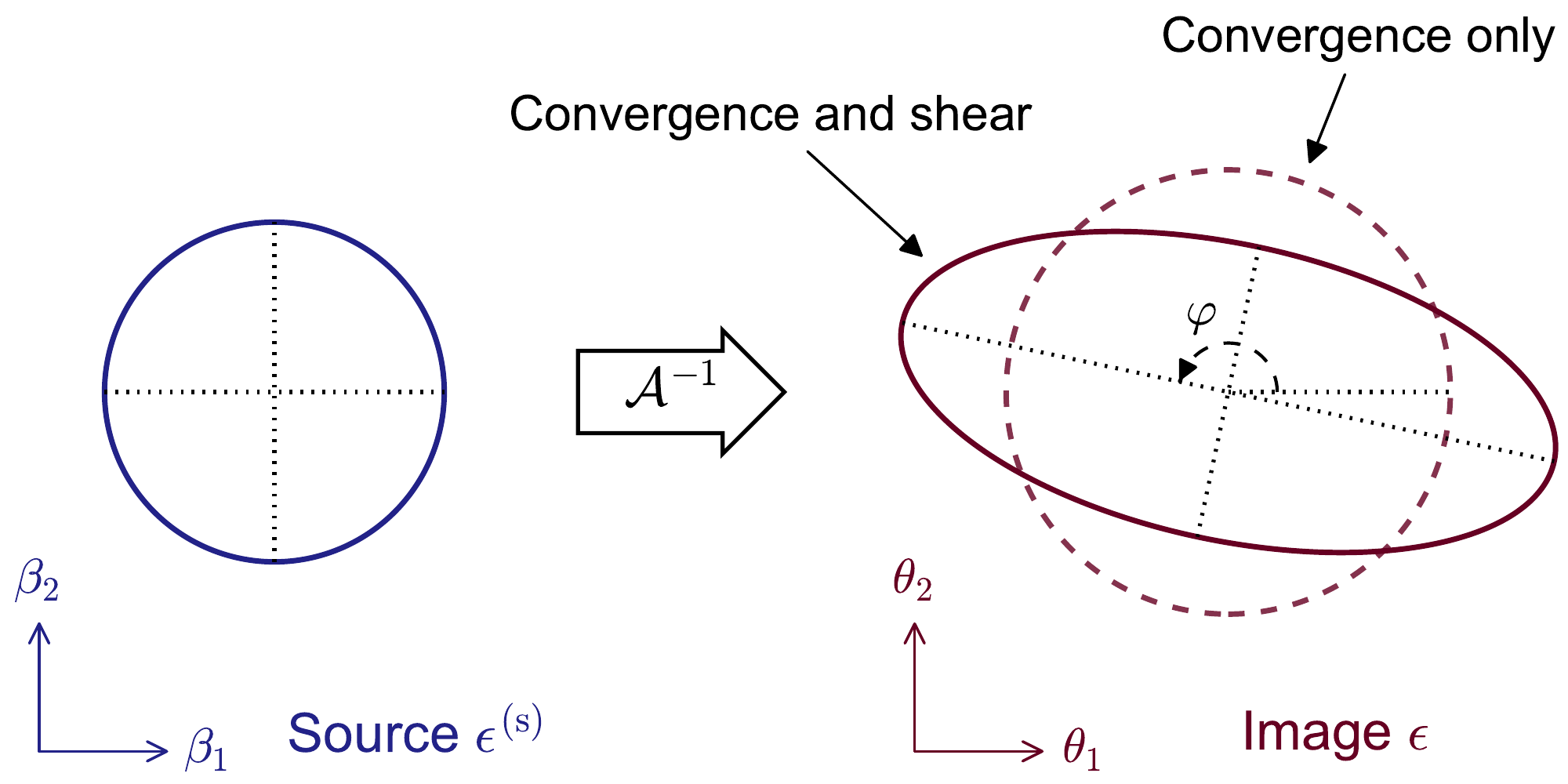}
	\caption{Illustration of the transformation generated by the distortion matrix.}
	\label{fig:lensing:Distortion_matrix}
\end{figure}

\subsubsection{Interpretation}

What is the interpretation of these quantities? A summary is provided by \fig{fig:lensing:Distortion_matrix}. Consider a circular source $S$ with radius $\beta$. The border of this source can be parametrized as $S = (\beta\cos t, \beta\sin t)$, $t\in \mathbb{R}$. Distorted by lensing, the equation of the image $I$ is
\begin{align}\label{for:lensing:circular_example_1}
	I = \mathcal{A}^{-1}S = \frac{\beta}{(1-\kappa)(1-|g|^2)}\begin{pmatrix}
		\cos t + |g| \cos(2\varphi - t)\\
		\sin t + |g| \sin(2\varphi - t)
	\end{pmatrix}.
\end{align}
If we rotate the image by $-\varphi$, the rotated image $I_\mathrm{rot}$ is 
\begin{align}
	I_\mathrm{rot} &= \begin{pmatrix}
		  \cos(-\varphi) & \sin(-\varphi)\\
		- \sin(-\varphi) & \cos(-\varphi)
	\end{pmatrix}I \notag\\
	&=\frac{\beta}{(1-\kappa)(1-|g|^2)}\begin{pmatrix}
		(1+|g|)\cos(t+\varphi)\\
		(1-|g|)\sin(t+\varphi)
	\end{pmatrix}
	=\begin{pmatrix}
		\cos(t+\varphi)\cdot \beta/(1-\kappa-|\gamma|)\\
		\sin(t+\varphi)\cdot \beta/(1-\kappa+|\gamma|)
	\end{pmatrix}. 
\end{align}
This is the equation of a parametrized ellipse whose semi-axes are respectively $\beta/(1-\kappa\mp |\gamma|)$. If the shear vanishes, the image turns out to be a circle with radius $\beta/(1-\kappa)$. Thus, $\kappa$ indicates the isotropic part of the transformation. On the other hand, $\gamma$ contains the distortion information. The norm $|\gamma|$ characterizes the flatness of the ellipse and the rotation angle $\varphi$ indicates the distortion direction. 

\begin{figure}[tb]
	\centering
	\includegraphics[scale=0.65]{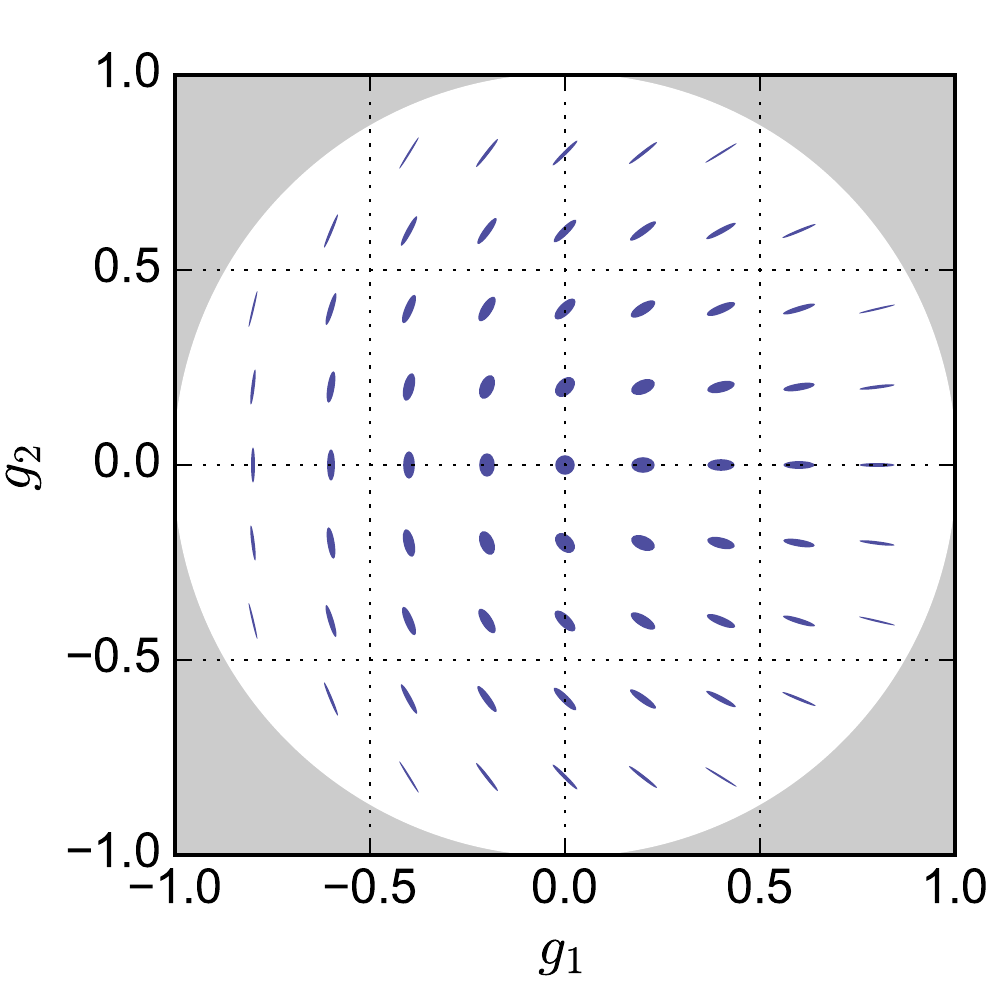}
	\caption{Illustration of different values of $g$. Each spot is the image of the same circular source for a given lensing shear. The grey zone corresponds to the strong lensing regime. Readers will also see in \sect{sect:lensing:extraction:shape} that this also excludes ellipticities without physical interpretation.}
	\label{fig:lensing:Shear_visualization}
\end{figure}

\figFull{fig:lensing:Shear_visualization} illustrates the distorted image for different values of the shear, for the same circular source. The ellipses of the same morphological shape are located on the same circle centered at $(g_1, g_2)=(0, 0)$. Note that the rotation angle of the ellipse is not the phase of the shear. Their relation can be obtained easily by reversing \for{for:lensing:shear}, which yields
\begin{align}
	\varphi = \frac{1}{2}\arctan\left(\frac{\gamma_2}{\gamma_1}\right).
\end{align}

\subsubsection{Another interpretation}

Another interpretation exists for the convergence. Let us consider the Gauss theorem\index{Gauss theorem} applied to gravity: 
\begin{align}\label{for:lensing:Gauss_theorem_1}
	\int_V\rmd V\ \nabla \vect{g} = \oint_S \rmd S\ \vect{g}\cdot\vect{n},
\end{align}
where $\vect{g}=-(\rmG\rho/r^2)\cdot\vect{n}\rmd V$ is the Newtonian gravitational field and $\vect{n}$ is the elementary vector normal to an elementary surface $\rmd S$. The right-hand side of \for{for:lensing:Gauss_theorem_1} represents the total gravitational flux passing through a Gauss surface; the left-hand side is the total mass wrapped inside. By taking the Gauss surface as an infinitesimal sphere, \for{for:lensing:Gauss_theorem_1} becomes $\nabla\vect{g} = -4\pi\rmG\rho$. The field is the opposite of the gradient of the potential $\vect{g}=-\nabla\Phi$, which yields the Poisson equation: $\Delta\Phi = 4\pi\rmG\rho$.

However, this can not be applied directly in the \acro{WL} context. If the Gauss surface is submerged in a homogeneous Universe, then the flux vanishes everywhere. In order to account for this effect properly, we need to cancel the density by its background value $\bar{\rho}$. By doing this, we recover \for{for:structure:Poisson_eq_3}, which is a Poisson equation for the reduced Newtonian potential $\phi$: $\Delta\phi = 4\pi\rmG(\rho-\bar{\rho})$. Equation \eqref{for:structure:Poisson_eq_4} gives a simplified expression for $\Delta\phi$ in physical coordinates. In comoving space, the \acro{3D} Poisson equation\index{Poisson equation} is
\begin{align}\label{for:lensing:Poisson_equation}
	\Delta\phi = \frac{3H_0^2\OmegaM}{2}\frac{\delta}{a}.
\end{align}

Since $\kappa=1-(\mathcal{A}_{11} + \mathcal{A}_{22})/2$, using \for{for:lensing:distortion_matrix_2}, we find
\begin{align}
	\kappa(\btheta, w) = \frac{1}{\rmc^2}\int_0^w \rmd w'\ \frac{f_K(w-w')f_K(w')}{f_K(w)}(\Delta - \partial_3^2) \phi(f_K(w')\btheta, w').
\end{align}
By recalling the thin-lens approximation, $\partial_3^2 \Phi$ can be eliminated. Thus, with help of \for{for:lensing:Poisson_equation}, finally we get
\begin{align}\label{for:lensing:convergence_2}
	\kappa(\btheta, w) = \frac{3H^2_0 \OmegaM}{2\rmc^2} \int_0^w \rmd w'\ \frac{f_K(w-w')f_K(w')}{f_K(w)} \frac{\delta\left( f_K(w')\btheta, w' \right)}{a(w')}.
\end{align}
This formula is an integration of the density contrast over the line of sight, weighted by comoving transverse distances and the scale factor. Therefore, \textit{the convergence $\kappa$ can also be considered as the projected overdensity, or simply the projected mass}. If the source redshift distribution is known as $p(w)$, then it is useful to write \for{for:lensing:convergence_2} into another form which yields the expectation value of $\kappa$:
\begin{align}
	\kappa(\btheta) = \frac{3H^2_0 \OmegaM}{2\rmc^2} \int_0^{w_\maxx} \rmd w\ g(w)f_K(w) \frac{\delta\left( f_K(w)\btheta, w \right)}{a(w)},
\end{align}
where
\begin{align}\label{for:lensing:lens_efficiency}
	g(w) \equiv \int_w^{w_\maxx}\rmd w'\ p(w')\frac{f_K(w'-w)}{f_K(w')}
\end{align}
is the lens efficiency for $p(w)$, and $w_\maxx\equiv w(a=0)$ is the comoving horizon distance.

\section{Weak lensing by a massive cluster}

As we have seen earlier, lensing collects the information of mass distributions along the line of sight. Usually, the signal is insignificant unless a bound massive object, typically a dark matter halo or a cluster of galaxies, has been crossed by the light ray. In this case, we can often neglect the additional lensing by unbound structures, and focus only on the cluster contribution.

In the following, I am going to derive the \acro{WL} signal in the framework of cluster lensing. This is essential for this thesis work since weak-lensing peak counts depend tightly on cluster population, as we will discuss in \chap{sect:modelling}.

\subsection{Mass projection}
\label{sect:lensing:cluster:projection}

\index{Projected mass}Consider an isotropic and homogeneous Universe with background density $\bar{\rho}(z)=\rho_\crit\OmegaM(1+z)^3$. We put a dark matter halo at $w_\ell$ (with the corresponding scale factor $a_\ell$) and neglect the enhancement of the background density. For a source located at $w_\rms$ (with the corresponding scale factor $a_\rms$), the convergence $\kappa$ is provided by \for{for:lensing:convergence_2} as
\begin{align}\label{for:lensing:convergence_3}
	\kappa(\btheta, w_\rms) = \frac{3H^2_0 \OmegaM}{2\rmc^2} \int_0^{w_\rms} \rmd w'\ \frac{f_K(w_\rms-w')f_K(w')}{f_K(w)} \frac{\rho\left( f_K(w')\btheta, w' \right)-\bar{\rho}(z(w'))}{\bar{\rho}(z(w'))a(w')}.
\end{align}
Outside the halo region, the density is equal to the background $\bar{\rho}(z)$, so the integral only extends over the halo region. We neglect here the variation of the scale factor and the comoving distances inside the halo since the halo radius is small compared to $w_\ell$ in general. In this case, \for{for:lensing:convergence_3} yields
\begin{align}
	\kappa(\btheta, w_\ell, w_\rms) = \kappa_\halo(\btheta, w_\ell, w_\rms) - \Delta\kappa(w_\ell, w_\rms),
\end{align}
where 
\begin{align}
	\kappa_\halo(\btheta, w_\ell, w_\rms) &=\frac{3H^2_0 \OmegaM}{2\rmc^2} \frac{f_K(w_\rms - w_\ell)f_K(w_\ell)}{f_K(w_\rms) a_\ell} \int_\halo \rmd w'\ \frac{\rho (f_K(w')\btheta, w')}{\bar{\rho}(a_\ell)}, \label{for:lensing:kappa_halo_1}\\
	\Delta\kappa(w_\ell, w_\rms) &=\frac{3H^2_0 \OmegaM}{2\rmc^2} \frac{f_K(w_\rms - w_\ell)f_K(w_\ell)}{f_K(w_\rms) a_\ell} \int_\halo \rmd w'. 
\end{align}
By analyzing the order of magnitude, we can neglect $\Delta\kappa$: the integral is roughly $\sim (3H_0^2/2\rmc^2)$ $\cdot R_\vir\cdot f(w_\ell)$, where $R_\vir$ is the comoving quantity which corresponds to the physical virial radius; the factor $3H_0^2/2\rmc^2$ is $\sim\dix{-8}\ h^2 / \Mpc^2$; taking $R_\vir\sim 10\ \Mpc/h$ and $f(w_\ell)\sim 10^3\ \Mpc/h$, an optimistic estimation yields $\Delta\kappa\sim\dix{-4}$ which justifies that it is negligible.

Let us now concentrate on $\kappa_\halo$, the convergence computed as the projected mass instead of the density contrast. Define $(x_1, x_2, x_3)$ as the physical coordinates associated to $(f_K(w')\btheta, w')$ where $x_3$ is aligned with the line of sight. We aim to rewrite \for{for:lensing:kappa_halo_1} in physical coordinates. First, by the definition of $\OmegaM$, 
\begin{align}
	\frac{3H^2_0 \OmegaM}{2\rmc^2} = \frac{4\pi\rmG}{\rmc^2}\bar{\rho}_0,\ \ \ \text{since}\ \OmegaM = \frac{8\pi \rmG}{3H_0^2}\bar{\rho}_0.
\end{align}
Second, by replacing $f_K$ with $D_\rmA$, readers will find
\begin{align}
	\frac{f_K(w_\rms-w_\ell)f_K(w_\ell)}{f_K(w_\rms)a_\ell} = \frac{(D_\rmA(w_\ell, w_\rms)/a_\rms) \cdot (D_\rmA(0, w_\ell)/a_\ell)}{(D_\rmA(0, w_\rms)/a_\rms)a_\ell} = \frac{\DL\DLs}{\DS a_\ell^2},
\end{align}
where $\DL$ is the angular diameter distance between the lens and the observer, $\DS$ between the source and the observer, and $\DLs$ between the lens and the source. Third, $\rmd w'=\rmd x_3/a_\ell$. Finally, the matter density varies as the inverse cube of the scale factor, $\bar{\rho}(a_\ell) = \bar{\rho}_0 a_\ell^{-3}$. Putting all together, we get
\begin{align}\label{for:lensing:kappa_halo_2}
	\kappa_\halo(\btheta, w_\ell, w_\rms) = \frac{\Sigma_\halo(\btheta)}{\Sigma_\crit(w_\ell, w_\rms)},
\end{align}
where $\Sigma_\halo$ is the projected surface density of the halo:
\begin{align}\label{for:lensing:projected_surface_density}
	\Sigma_\halo(\btheta)\equiv \int \rmd x_3\ \rho_\halo(\DL\btheta, x_3),
\end{align}
and $\Sigma_\crit$ is the \textit{critical surface density}\index{Critical surface density}:
\begin{align}\label{for:lensing:critical_surface_density}
	\Sigma_\crit(w_\ell, w_\rms)\equiv\frac{\rmc^2}{4\pi\rmG}\frac{\DS}{\DL\DLs}.
\end{align}
 
\index{Projected mass, correction for the}In the literature, one will often find that \for{for:lensing:kappa_halo_2} is used rather than \for{for:lensing:convergence_2} for deriving the convergence from a cluster. However, one can not forget the condition in which \for{for:lensing:kappa_halo_2} is valid: \textit{the density outside the halo is not zero but equal to the background value}. Thus, if one considers a lightcone populated with halos, the space which is not occupied by halos is actually \textbf{not} empty! If one uses \for{for:lensing:kappa_halo_2} to compute $\kappa$ in this case, it needs to be corrected, since the real background mass has been enhanced and this enhancement may not be negligible.

\subsection{Tangential shear profile}

In gravitational lensing, sources situated near a point mass will be distorted tangentially to the impact vector. A direct consequence are arc-like features surrounding strong gravitational lenses. In \acro{WL}, this tangential stretch can still be observed and can be used for cluster identification.

\begin{figure}[tb]
	\centering
	\includegraphics[scale=0.65]{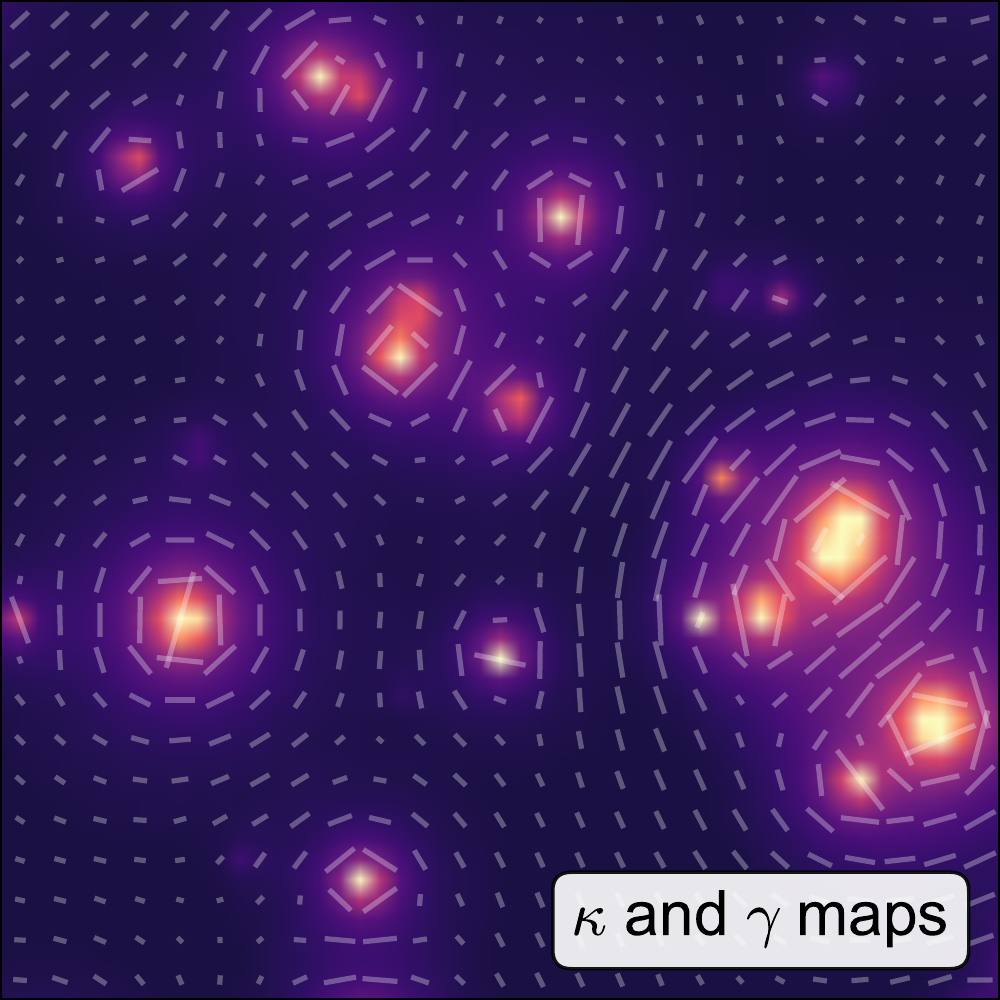}
	\caption{Illustration of the tangential shear. The colored map represents the convergence, while the ``whisker plot'' stand for the shear. The length and the orientation of the ``whiskers'' are decided respectively by the norm and the rotation angle of the tangential shear.}
	\label{fig:lensing:kappa_gamma_maps_3}
\end{figure}

In an ideal scenario where the cluster is perfectly spherical, the rotation angle of the shear will align with the tangential direction and the norm varies depending on the angular distance between the lens and the source. To extract the stretch information, we only have to choose the tangential direction as the reference axis. This depends on the relative position between the cluster and the galaxy. Let $\phi$ be the phase angle of the position vector of the source pointed from the lens. The \textit{tangential shear}\index{Shear, tangential} $\gamma_+$ and the \textit{cross shear}\index{Shear, cross} $\gamma_\times$ components are
\begin{align}
	\gamma_+ \equiv -\Re\left[\gamma\rme^{-2\rmi\phi}\right],\ \ \ \gamma_\times \equiv -\Im\left[\gamma\rme^{-2\rmi\phi}\right].
\end{align}
\figFull{fig:lensing:kappa_gamma_maps_3} shows a convergence map overlapping with a shear map, represented by a ``whisker'' plot. Readers can easily figure out that the minus sign and $\exp(-2\rmi\phi)$ account for a rotation by $\pi/2-\phi$. Similar definitions also exist for the ellipticity $\epsilon_+$ and $\epsilon_\times$ that we will see later.

In more general cases, azimuthally averaging over the tangential shear helps recover the mass enclosed inside the corresponding circular region. This introduces the \textit{tangential shear profile} as $\gamma_+(\theta)$. Given any reference point, the tangential shear profile can be determined from the convergence inside the disk of radius $\theta$ without the need of circular symmetry. To establish this relation, let us start with applying the \acro{2D} Gauss theorem to $\nabla\psi$ on a disk of radius $\theta$:
\begin{align}
	\theta\int_0^{2\pi} \rmd\phi\ \nabla\psi\cdot\vect{n} = \int_{\theta'=0}^\theta \int_{\phi=0}^{2\pi} \theta'\rmd\theta'\rmd\phi\ \nabla\cdot\nabla\psi.
\end{align}
From \for{for:lensing:psi_relation_1}, $\nabla\cdot\nabla\psi = \Delta\psi = 2\kappa$. Also, $\nabla\psi\cdot\vect{n}=\partial\psi/\partial\theta$, so
\begin{align}\label{for:lensing:dpsi_dtheta_1}
	\theta\cdot \left\langle\frac{\partial\psi}{\partial\theta}\right\rangle_\rmc(\theta) = 2\int_0^\theta \theta'\rmd\theta'\ \langle\kappa\rangle_\rmc(\theta'),
\end{align}
where 
\begin{align}
	\langle f\rangle_\rmc(\theta)\equiv \frac{1}{2\pi}\int_0^{2\pi}\rmd\phi\ f(\theta, \phi), 
\end{align}
is the circular average of any quantity $f$ defined in a \acro{2D} space. By deriving \for{for:lensing:dpsi_dtheta_1} with regard to $\theta$, we obtain
\begin{align}\label{for:lensing:dpsi_dtheta_2}
	\left\langle\frac{\partial\psi}{\partial\theta}\right\rangle_\rmc(\theta) + \theta\cdot \left\langle\frac{\partial^2\psi}{\partial\theta^2}\right\rangle_\rmc(\theta) = 2\theta\cdot \langle\kappa\rangle_\rmc(\theta).
\end{align}
The first term at the left-hand side of \for{for:lensing:dpsi_dtheta_2} can be replaced with \for{for:lensing:dpsi_dtheta_1}. For the second term, $\partial^2\psi/\partial\theta^2$ is actually identical to $\kappa-\gamma_+$. This can be justified easily when $\phi=0$ since in this case, $\gamma_+=-\gamma_1$, and the general cases can be derived by rotation. The resulting formula is
\begin{align}
	\frac{2}{\theta} \int_0^\theta \theta'\rmd\theta'\ \langle\kappa\rangle_\rmc(\theta') - \theta\cdot \langle\gamma_+\rangle_\rmc(\theta) = \theta\cdot \langle\kappa\rangle_\rmc(\theta).
\end{align}
Defining $\overline{f}(<\theta)$ as the mean of any quantity $f$ over a disk of radius $\theta$:
\begin{align}
	\pi\theta^2\cdot \overline{f}(<\theta) \equiv \int_\mathrm{disk} \rmd\btheta'\ f(\btheta') = 2\pi\int_0^\theta \theta'\rmd\theta'\ \langle f\rangle_\rmc(\theta'), 
\end{align}
we obtain finally
\begin{align}\label{for:lensing:tangential_shear_profile_1}
	\langle\gamma_+\rangle_\rmc(\theta) = \overline{\kappa}(<\theta) - \langle\kappa\rangle_\rmc(\theta).
\end{align}
If the reference point is the center of a dark matter halo with the projected mass $\Sigma_\halo$, from \for{for:lensing:kappa_halo_2}, the derived tangential shear profile is
\begin{align}\label{for:lensing:tangential_shear_profile_2}
	\langle\gamma_{+,\halo}\rangle_\rmc(\theta) = \frac{\overline{\Sigma}_\halo(<\theta) - \langle\Sigma_\halo\rangle_\rmc(\theta)}{\Sigma_\crit}.
\end{align}

\subsection{Example: NFW profiles}

In this section, we show the convergence and the tangential shear profile for the \acro{NFW} profiles \citep{Navarro_etal_1996, Navarro_etal_1997}, for both truncated and non-truncated cases. If the profile is spherically symmetric, \for{for:lensing:projected_surface_density} can be rewritten into 
\begin{align}
	\Sigma_\halo\left(\theta=\frac{x_1}{\DL}\right) = \int\rmd x_3\ \rho_\halo\left(\sqrt{x_1^2+x_3^2}\right).
\end{align}
Let $u=x_1/r_\rms$ and $v=x_3/r_\rms$ be the coordinates normalized by the scale radius $r_\rms$ of the \acro{NFW} profiles. The projected mass becomes
\begin{align}\label{for:lensing:projected_surface_density_NFW}
	\Sigma_\halo\left(\theta=\frac{r_\rms}{\DL}u\right) &= \int_{-r_\rms v_\maxx}^{r_\rms v_\maxx}\rmd(r_\rms v)\ \frac{\rho_\rms}{\sqrt{u^2+v^2}\cdot(1+\sqrt{u^2+v^2})^2} \notag\\
	&= 2\rho_\rms r_\rms \int_0^{v_\maxx} \frac{\rmd v}{\sqrt{u^2+v^2}\cdot(1+\sqrt{u^2+v^2})^2},
\end{align}
with
\begin{align}
	v_\maxx=\left\{\begin{array}{ll}
		+\infty              & \text{for the exact \acro{NFW} profiles,}\\
		\sqrt{c^2 - u^2} & \text{for the profiles truncated at $r_\vir$,}
	\end{array}\right.
\end{align}
where $\rho_\rms$ is the \acro{NFW} characteristic density, $r_\vir$ the virial radius, and $c$ the concentration parameter. Thus, the problem is reduced to the calculation of the dimensionless projected mass $G_{\kappa}(u)$:
\begin{align}\label{for:lensing:G_kappa}
	G_{\kappa}(u) \equiv \int_0^{v_\maxx} \frac{\rmd v}{\sqrt{u^2+v^2}\cdot(1+\sqrt{u^2+v^2})^2}.
\end{align}

\begin{figure}[tb]
	\centering
	\includegraphics[scale=0.65]{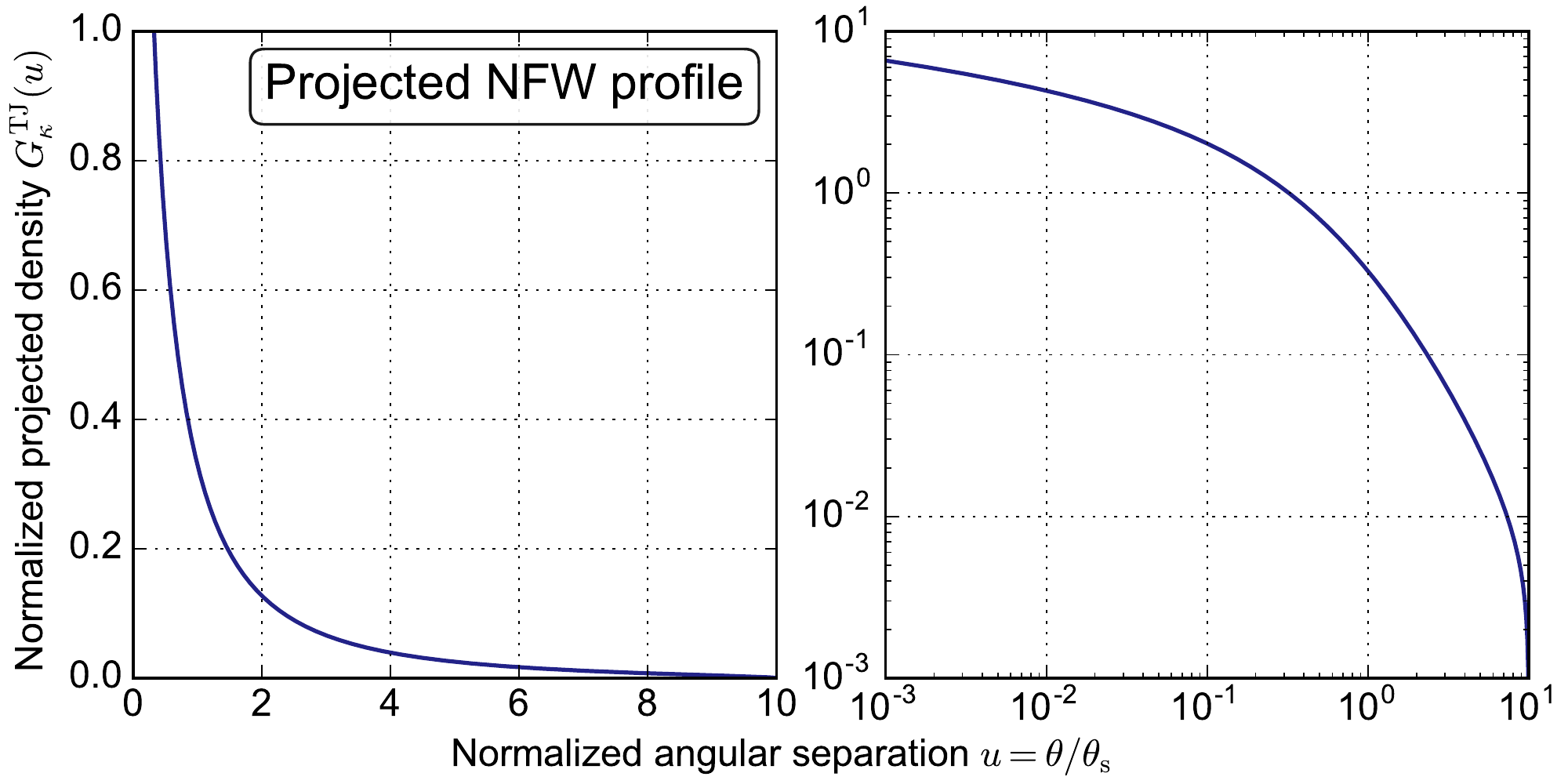}
	\caption{$G_\kappa^\mathrm{TJ}$ as a function of $u$. This is the projection of the truncated \acro{NFW} profile shown in normalized coordinates. The concentration is set to 10 as an example. The left panel shows the profile in linear space, and the right panel represents the same plot in log space.}
	\label{fig:lensing:Projected_NFW}
\end{figure}

Following \citet{Wright_Brainerd_2000} and \citet{Takada_Jain_2003a}, the projected masses for the original \acro{NFW} profiles (Eq. \ref{for:structure:NFW_profile}) and the truncated ones (labelled \texttt{TJ}, Eq. \ref{for:structure:TJ_profile}) are 
\begin{align}
	G_\kappa^\mathrm{NFW}(u) = \small\left\{
	\begin{array}{ll}
		\displaystyle \text{not defined} & \text{if $u=0$,}\\[3ex]
		\displaystyle -\frac{1}{1-u^2} + \frac{1}{(1-u^2)^{3/2}}\arcosh \left(\frac{1}{u}\right) & \text{if $0<u<1$,}\\[3ex]
		\displaystyle \frac{1}{3} & \text{if $u=1$,}\\[3ex]
		\displaystyle \frac{1}{u^2-1} - \frac{1}{(u^2-1)^{3/2}} \arccos \left(\frac{1}{u}\right) & \text{if $u>1$,}
	\end{array}\right. \normalsize
\end{align}
and 
\begin{align}\label{for:lensing:G_kappa_TJ}
	G_\kappa^\mathrm{TJ}(u) = \small\left\{
	\begin{array}{ll}
		\displaystyle \text{not defined} & \text{if $u=0$,}\\[3ex]
		\displaystyle -\frac{1}{1-u^2} \frac{\sqrt{c^2-u^2}}{c+1} + \frac{1}{(1-u^2)^{3/2}}\arcosh \left[\frac{u^2+c}{u(c+1)}\right] & \text{if $0<u<1$,}\\[3ex]
		\displaystyle \frac{\sqrt{c^2-1}}{c+1} \cdot \frac{c+2}{3(c+1)} & \text{if $u=1$,}\\[3ex]
		\displaystyle \frac{1}{u^2-1} \frac{\sqrt{c^2-u^2}}{c+1} - \frac{1}{(u^2-1)^{3/2}} \arccos \left[\frac{u^2+c}{u(c+1)}\right] & \text{if $1<u\leq c$,}\\[3ex]
		\displaystyle 0 & \text{if $u>c$.}
	\end{array}\right.\normalsize
\end{align}
These functions are not defined at $u=0$ since the \acro{NFW} profiles diverge to infinity when $r\rightarrow 0$. Note that $\arcosh (1/u)= 2\artanh \sqrt{\frac{1-u}{1+u}}$ for $0<u<1$ and $\arccos(1/u)= 2\arctan \sqrt{\frac{u-1}{u+1}}$ for $u>1$. Therefore, gathering Eqs. \eqref{for:lensing:kappa_halo_2}, \eqref{for:lensing:critical_surface_density}, \eqref{for:lensing:projected_surface_density_NFW}, and \eqref{for:lensing:G_kappa}, the convergence projected from the \acro{NFW} profiles is
\begin{align}\label{for:lensing:kappa_halo_NFW}
	\kappa_\halo(\btheta) = \frac{2\rho_\rms r_\rms}{\Sigma_\crit}\cdot G_\kappa \left(\frac{\theta}{\theta_\rms}\right) = \frac{4\pi\rmG}{\rmc^2} \frac{\DL\DLs}{\DS} \cdot \frac{Mfc^2}{2\pi r_\vir^2} \cdot G_\kappa\left(\frac{c\theta}{\theta_\vir}\right),
\end{align}
where $\theta_\rms\equiv r_\rms/D_\ell$, $\theta_\vir\equiv r_\vir/D_\ell$, $M$ is the halo mass, and $f$ is given by \for{for:structure:NFW_f}. The equality $2\rho_\rms r_\rms=(Mfc^2)/(2\pi r_\vir^2)$ is useful for implementation.

From \for{for:lensing:tangential_shear_profile_2}, we can derive the tangential shear profile. The profiles are spherical, so we simplify the notation $\langle\Sigma_\halo\rangle_\rmc(\theta)$ to $\Sigma_\halo(\theta)$ and $\langle\gamma_{+,\halo}\rangle_\rmc(\theta)$ to $\gamma_{+,\halo}(\theta)$ . Inserting \for{for:lensing:projected_surface_density_NFW} into \for{for:lensing:tangential_shear_profile_2}, we obtain 
\begin{align}\label{for:lensing:gamma_halo_NFW}
	\gamma_{+,\halo}(\theta) = \frac{2\rho_\rms r_\rms}{\Sigma_\crit}\cdot G_\gamma \left(\frac{\theta}{\theta_\rms}\right),
\end{align}
with
\begin{align}
	G_\gamma(u) = \overline{G}_\kappa(<u) - G_\kappa(u) = \frac{2}{u^2}\int_0^u u'\rmd u'\ G_\kappa(u') - G_\kappa(u).
\end{align}
The $G_\gamma$ for the original \acro{NFW} profiles and the truncated ones (labelled \texttt{TJ}) are given by \citet{Wright_Brainerd_2000} and \citet{Takada_Jain_2003b} as 
\begin{align}
	G_\gamma^\mathrm{NFW}(u) = \small\left\{
	\begin{array}{ll}
		\displaystyle \text{not defined} & \text{if $u=0$,}\\[2ex]
		\displaystyle \frac{1}{1-u^2} + \frac{2}{u^2}\ln\frac{u}{2} +\frac{1}{\sqrt{1-u^2}}\left(\frac{-1}{1-u^2} + \frac{2}{u^2}\right)\arcosh\left(\frac{1}{u}\right) & \text{if $0<u<1$,}\\[3ex]
		\displaystyle \frac{5}{3} + 2\ln\frac{1}{2} & \text{if $u=1$,}\\[2ex]
		\displaystyle \frac{-1}{u^2-1} + \frac{2}{u^2}\ln\frac{u}{2} + \frac{1}{\sqrt{u^2-1}}\left(\frac{1}{u^2-1} + \frac{2}{u^2}\right) \arccos\left(\frac{1}{u}\right) & \text{if $u>1$,}
	\end{array}\right.\normalsize
\end{align}
and
\begin{align}
	G_\gamma^\mathrm{TJ}(u) = \small\left\{
	\begin{array}{ll}
		\displaystyle \text{not defined} & \text{if $u=0$,}\\[3ex]
		\displaystyle \frac{1}{u^2(c+1)} \left[\frac{2-u^2}{1-u^2}\sqrt{c^2-u^2} - 2c\right] +\frac{2}{u^2}\ln\left[\frac{u(c+1)}{c+\sqrt{c^2-u^2}}\right] & \\[3ex]
		\hspace*{4em}\displaystyle +\frac{2-3u^2}{u^2(1-u^2)^{3/2}}\arcosh\left[\frac{u^2+c}{u(c+1)}\right] & \text{if $0<u<1$,}\\[5ex]
		\displaystyle \frac{1}{3(c+1)} \left[\frac{11c+10}{c+1}\sqrt{c^2-1} - 6c\right] + 2\ln\left[\frac{c+1}{c+\sqrt{c^2-1}}\right] & \text{if $u=1$,}\\[6ex]
		\displaystyle \frac{1}{u^2(c+1)} \left[\frac{2-u^2}{1-u^2}\sqrt{c^2-u^2} - 2c\right] +\frac{2}{u^2}\ln\left[\frac{u(c+1)}{c+\sqrt{c^2-u^2}}\right]\\[3ex]
		\hspace*{4em}\displaystyle -\frac{2-3u^2}{u^2(u^2-1)^{3/2}}\arccos\left[\frac{u^2+c}{u(c+1)}\right] & \text{if $1<u\leq c$,}\\[3ex]
		\displaystyle \frac{2}{fu^2} & \text{if $u>c$.}
	\end{array}\right.\normalsize
\end{align}

\section{Extraction of cosmological information}
\label{sect:lensing:extraction}

\subsection{From shape to shear}
\label{sect:lensing:extraction:shape}

Earlier, we have used the convergence $\kappa$ and the shear $\gamma$ to characterize the \acro{WL} effect. To estimate these lensing quantities, we need to measure the difference between galaxies' lensed and unlensed shapes. Unfortunately, we can not measure the unlensed form of galaxies. However, if cosmologists assume the isotropy of the galaxy intrinsic orientation, then the average shape over some galaxies should be circular. If now we detect a significant elliptical average, then we know that images have been lensed. From this ``elliptical average'' we should also be able to measure the lensing signal. To do this, we first need to define what the ellipticity is.

\subsubsection{Ellipticity definition}

The definition with moments is adopted for ellipticity in this work. Let $\btheta$ be \acro{2D} Cartesian coordinates for angular position, and $I(\btheta)$ be the surface brightness (brightness density) of an image at $\btheta$, and we assume that the image is not contaminated by other images. We define the center of the image as
\begin{align}
	\bar{\btheta} \equiv \frac{\int\rmd^2\btheta\ \omega\big(I(\btheta)\big)I(\btheta)\cdot\btheta}{\int\rmd^2\btheta\ \omega\big(I(\btheta)\big)I(\btheta)},
\end{align}
where $\omega(I)$ is a weight function of suitable choice, which can be used to define the image border. Then, the second-order moments of the image are
\begin{align}
	Q_{ij} \equiv \frac{\int\rmd^2\btheta\ \omega\big(I(\btheta)\big)I(\btheta)\cdot(\theta_i-\bar{\theta}_i)(\theta_j-\bar{\theta}_j)}{\int\rmd^2\btheta\ \omega\big(I(\btheta)\big)I(\btheta)}.
\end{align}
With this definition, the trace of $Q$ describes the image size, whereas the traceless part contains the distortion. From $Q_{ij}$, we define the \textit{ellipticity}\index{Ellipticity} $\epsilon$ as
\begin{align}
	\epsilon \equiv\epsilon_1+\rmi\epsilon_2,\ \ \text{and}\ \ \ &\epsilon_1\equiv\frac{Q_{11}-Q_{22}}{Q_{11}+Q_{22}+2\sqrt{Q_{11}Q_{22}-Q_{12}^2}}, \notag\\[1ex]
	&\epsilon_2\equiv\frac{2Q_{12}}{Q_{11}+Q_{22}+2\sqrt{Q_{11}Q_{22}-Q_{12}^2}}. \label{for:lensing:ellipticity}
\end{align}
Note that similar to the shear, the ellipticity is described by a complex number whose norm is always smaller than one, since 
\begin{align}
	|\epsilon|^2 = \frac{Q_{11}+Q_{22}-2\sqrt{\Delta}}{Q_{11}+Q_{22}+2\sqrt{\Delta}},\ \ \Delta\equiv Q_{11}Q_{22}-Q_{12}^2.
\end{align}

In analogy, we can define the intrinsic ellipticity in the unlensed space (or the source space). Following the same notation as \sect{sect:lensing:deflection:matrix}, we denote $\bbeta$ the coordinates of angular position in the source space. The corresponding second-order moments lead to
\begin{align}
	Q\src_{ij} \equiv \frac{\int\rmd^2\bbeta\ \omega\big(I\src(\bbeta)\big)I\src(\bbeta)\cdot(\beta_i-\bar{\beta}_i)(\beta_j-\bar{\beta}_j)}{\int\rmd^2\bbeta\ \omega\big(I\src(\bbeta)\big)I\src(\bbeta)},
\end{align}
where $I\src(\bbeta)$ is the surface brightness distribution in the source space, which satisfies the equality $I\src(\bbeta)=I(\btheta)$ (see \sect{sect:lensing:extraction:muBias} for details). The intrinsic ellipticity\index{Ellipticity, intrinsic} $\epsilon\src$ is defined as 
\begin{align}
	\epsilon\src \equiv\frac{Q\src_{11}-Q\src_{22}+2\rmi Q\src_{12}}{Q\src_{11}+Q\src_{22}+2\sqrt{Q\src_{11}Q\src_{22}-Q_{12}^{(\rms)2}}},
\end{align}

By the definition of $\mathcal{A}$, $\beta_i-\bar{\beta}_i = \sum_{j=1,2}\mathcal{A}_{ij}\cdot(\theta_j-\bar{\theta}_j)$, so
\begin{align}
	(\beta_i-\bar{\beta}_i)(\beta_j-\bar{\beta}_j) = \sum_{k=1,2}\sum_{\ell=1,2}\mathcal{A}_{ik}(\theta_k-\bar{\theta}_k)(\theta_\ell-\bar{\theta}_\ell)\mathcal{A}^T_{\ell j}.
\end{align}
This results in $Q\src=\mathcal{A}Q\mathcal{A}^T=\mathcal{A}Q\mathcal{A}$. After some simple algebra \citep{Seitz_Schneider_1997}, the intrinsic ellipticity is related to the observed one by the reduced shear $g$ as
\begin{align}\label{for:lensing:intrinsic_observed_relation_1}
	\epsilon\src = \left\{\begin{array}{ll}
		\displaystyle\frac{\epsilon - g}{1 - g^*\epsilon}   & \text{if $|g|\leq 1$,}\\[3ex]
		\displaystyle\frac{1-g\epsilon^*}{\epsilon^* - g^*} & \text{if $|g|\geq 1$.}
	\end{array}\right.
\end{align}

\begin{figure}[tb]
	\centering
	\includegraphics[scale=0.65]{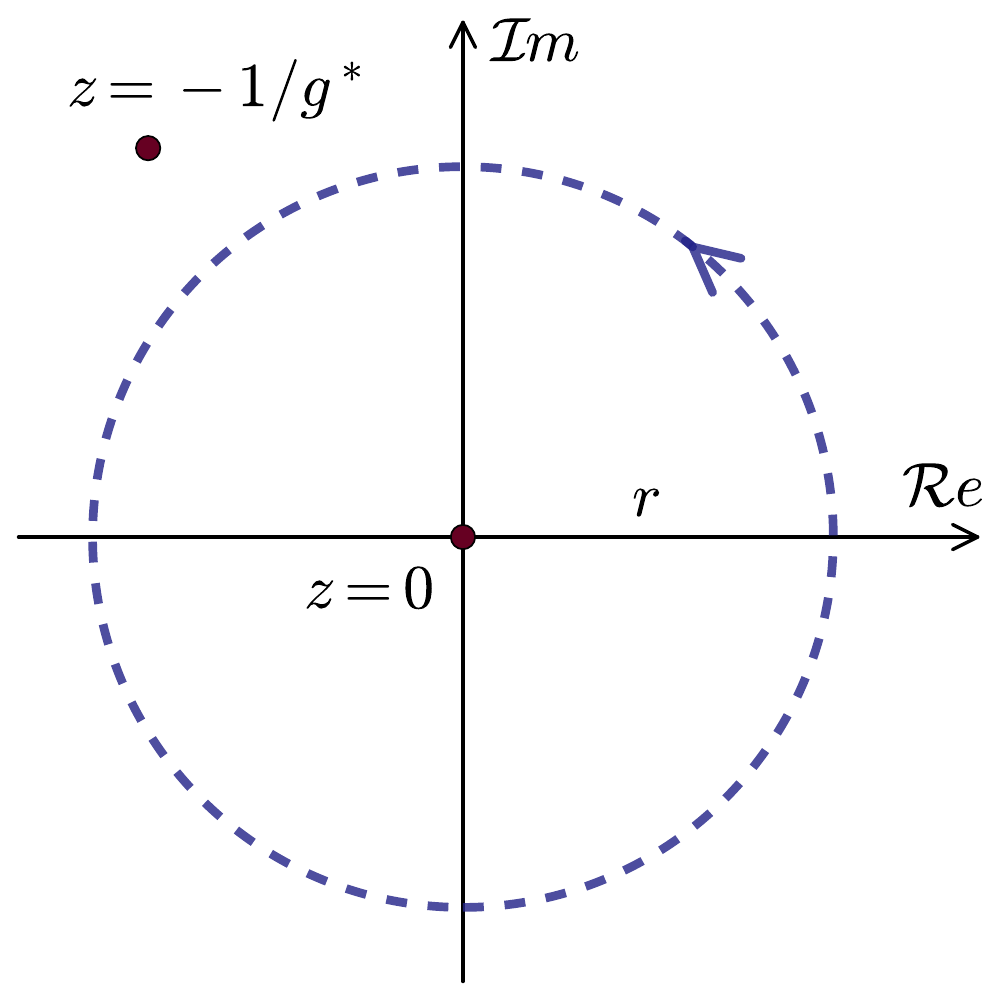}
	\caption{Illustration of the residue theorem. For the case of $|g|\leq 1$, only one pole out of two stays inside the integration contour.}
	\label{fig:lensing:Integration_poles}
\end{figure}

\subsubsection{Shear estimator}

Now, the question becomes: how to estimate $g$? More precisely, if we average $\epsilon$ locally, what would be the expected value if images are lensed and if $\epsilon\src$ is statistically isotropic? To figure this out, let us inverse \for{for:lensing:intrinsic_observed_relation_1} into
\begin{align}\label{for:lensing:intrinsic_observed_relation_2}
	\epsilon = \left\{\begin{array}{ll}
		\displaystyle\frac{\epsilon\src+g}{1+g^*\epsilon\src} & \text{if $ |g|\leq 1$,}\\[3ex]
		\displaystyle\frac{1+g\epsilon^{(\rms)*}}{\epsilon^{(\rms)*}+g^*} & \text{if $|g|\geq 1$.}
	\end{array}\right.
\end{align}
Let $\epsilon\src=r\rme^{2i\varphi}$. For all function $f(\epsilon\src)$, the weighted average of $f$ with regard to the intrinsic ellipticity distribution is defined as
\begin{align}\label{for:lensing:average_ellipticity}
	\langle f\rangle\equiv \frac{\int r\rmd r\rmd\varphi\ P(r)\cdot f}{\int r\rmd r\rmd\varphi\ P(r)},
\end{align}
where $P(r)$ is the probability distribution of $\epsilon\src$. The dependency of $P$ on $\varphi$ has been omitted because of isotropy. If $\epsilon$ is lensing-free, then the isotropic statement will yield $\langle\epsilon\rangle=\langle\epsilon\src\rangle=0$. In general cases, if $|g|\leq 1$, the numerator of \for{for:lensing:average_ellipticity} is
\begin{align}\label{for:lensing:numerator}
	\int r\rmd r\rmd\varphi\ P(r)\cdot \epsilon = \int_0^1r\rmd r\ P(r)\cdot I(r), 
\end{align}
with 
\begin{align}
	I(r) = \int_0^{\pi}\rmd\varphi\ \frac{r\rme^{2\rmi\varphi}+g}{1+g^*r\rme^{2\rmi\varphi}}.
\end{align}
This expression can be integrated in the complex plane using the residue theorem\index{Residue theorem}. Fix $r$ and let $z = r\rme^{2\rmi\varphi}$, we obtain $\rmd z = 2\rmi r\rme^{2\rmi\varphi} \rmd\varphi = 2\rmi z\rmd\varphi$, so that
\begin{align}
	I(r) = \frac{1}{2\rmi}\oint\rmd z\ \frac{z+g}{z(1+g^*z)} = \frac{1}{2\rmi}\oint\rmd z\ \left(\frac{g}{z} + \frac{1-|g|^2}{1+g^*z}\right).
\end{align}
Since $0\leq r\leq 1$ and $|g|\leq 1$, we have $r\leq 1 \leq |1/g^*|$. There is only one pole $z=0$ inside the contour (\fig{fig:lensing:Integration_poles}). One obtains
\begin{align}\label{for:lensing:polar_integral_3}
	I(r) = \frac{1}{2\rmi}\oint\rmd z\ \frac{g}{z} = \pi \cdot g.
\end{align}
The denominator of \for{for:lensing:average_ellipticity} leads to $\int r\rmd r\ \pi P(r)$. Thus, using Eqs. \eqref{for:lensing:numerator} and \eqref{for:lensing:polar_integral_3}, we find
\begin{align}
	\langle\epsilon\rangle = g.
\end{align}
A similar calculation can be done for $|g|\geq 1$. Finally, the expected average of the observed ellipticities is 
\begin{align}\label{for:shear_estimator_2}
	\langle\epsilon\rangle = \left\{\begin{array}{ll}
		g     & \text{if $|g|\leq 1$,}\\
		1/g^* & \text{if $|g|\geq 1$.}
	\end{array}\right.
\end{align}
\forFull{for:shear_estimator_2} means that, whatever the intrinsic ellipticity distribution is, as soon as it is isotropic, then \textit{the mean of ellipticity samples yields locally an \textbf{unbiased} estimator of the reduced shear}\index{Shear estimator}. Cosmologists usually consider the regime where $|\kappa|\ll 1$. In this case, $g\approx\gamma$. The implication is that the convergence is not directly measurable, but the shear is, and we need to use an inversion to retrieve $\kappa$.

\index{Shape-shear-ellipticity relationship}Let us revise the notion of shape at the end of this section. Shape is a concept which is very ambiguous for a source galaxy. Even if the galaxy image is isolated from other objects, it is not straightforward to define the boundary since the surface brightness decreases continuously from the inner part. Also, even if the boundary is defined, the word ``shape'' in its common sense only correspond to the case of isophotes. Therefore, it would be better to replace the notion of shape with ellipticity. On the other hand, setting $\epsilon\src=0$, \for{for:lensing:intrinsic_observed_relation_2} yields $\epsilon=g$. This is exactly the transformation described by \for{for:lensing:circular_example_1} and \fig{fig:lensing:Shear_visualization}. Hence, the shear and the ellipticity are mathematically the same\footnote{Let $G=\{x|x\in\mathbb{C}, |x|\leq1\}$ be the set of ellipticities and shears whose norm is smaller than 1. Define $\ast$ as the ``lensing operator'': $x\ast y\equiv (x+y)/(1+x^*y)$. Then $(G,\ast)$ is a \textit{magma}: $\forall x,y\in G$, $x\ast y\in G$.}, and \fig{fig:lensing:Shear_visualization} can also be interpreted as the elliptical fit to images.

\subsection{Magnification effect}
\label{sect:lensing:extraction:muBias}

Apart from distortion, \acro{WL} also magnifies images. By measure theory, the magnification is given by the Jacobian matrix of the change of variables between the source space and the observed space, which is just the distortion matrix. More precisely, the magnification $\mu$ is defined as
\begin{align}\label{for:lensing:magnification}
	\mu \equiv \left\vert \frac{\partial\btheta}{\partial\bbeta} \right\vert = \det\mathcal{A}\inv = \frac{1}{(1-\kappa)^2 - |\gamma|^2}.
\end{align}
In the \acro{WL} regime, if $|\kappa|\ll 1$ and $|\gamma|\ll 1$, the dominant term in the denominator is $1-2\kappa$. Thus, we relate the magnification to the convergence by $\mu\approx 1+2\kappa$. 

The magnification effect makes angular separations larger if $\kappa>0$. Consequently, images become larger and more distant from each other. This decreases locally the number density of sources. Meanwhile, the brightness of sources is enhanced, so more galaxies will be detected. In the following, I will construct the mathematical formalism for this twofold effect.

According to the Liouville theorem\index{Liouville theorem}, the surface brightness density is invariant to the lensing effect, which means that
\begin{align}
	I(\btheta) = I\src(\bbeta).
\end{align}
This can be understood in the following way. If an image is magnified by lensing, its apparent angle becomes larger. Meanwhile, lensing converges light so some photons which would have escaped from the observer are deflected and contribute to the total flux. Therefore, the image size and the total flux increase at the same time, and the Liouville theorem tells us that both increase at the same rate.

Let $\rmd\Omega$ and $\rmd\Omega\src$ be a small surface fraction of the sky (solid angle) of an image in each space and $\rmd B$ and $\rmd B\src$ the corresponding brightness. The implication of the Liouville theorem is $\rmd B/\rmd\Omega = \rmd B\src/\rmd\Omega\src$. On the other hand, the ratio between two image sizes is the magnification, so
\begin{align}\label{for:lensing:brightness_relation}
	\mu = \frac{\rmd\Omega}{\rmd\Omega\src} = \frac{\rmd B}{\rmd B\src}.
\end{align}
Let $n(B,z)$ be the surface number density of observed galaxies with brightness $B$ at redshift $z$ and $n\src(B\src,z)$ the corresponding function in the $\bbeta$ space. Let $n(B,z)\rmd B\rmd\Omega\rmd z$ be the number counts of galaxies within solid angle $\rmd\Omega$ and redshift slice $\rmd z$ with brightness in $[B, B+\rmd B[$. Let $\rmd B\src$ be the corresponding brightness gap in the $\bbeta$ space and $\rmd\Omega\src$ the corresponding spanning area. Then, the number-count conservation leads to the equality
\begin{align}
	n(B,z)\rmd B\rmd\Omega\rmd z = n\src\big(B\src,z\big)\rmd B\src\rmd\Omega\src\rmd z.
\end{align}
From \for{for:lensing:brightness_relation}, the brightness $B$ in the lensed space will become $B/\mu$ in the $\bbeta$ space. By integrating between $B$ (resp. $B\src$) and infinity,
\begin{align}
	\int_B^\infty \rmd B\ n(B,z) = \frac{1}{\mu} \int_{B\src = B/\mu}^\infty \rmd B\src\ n\src\big(B\src,z\big).
\end{align}
Denote $n(>B,z)$ (resp. $n\src(>B\src,z)$) as the number density of galaxies with brightness larger than $B$ (resp. $B\src$) at redshift $z$. We find the relation of magnification effect \citep[see also][]{Broadhurst_etal_1995}:
\begin{align}\label{for:lensing:magnification_bias}
	n(>B, z) = \frac{1}{\mu}\ n\src\left(>\frac{B}{\mu},z\right).
\end{align}

\begin{figure}[tb]
	\centering
	\includegraphics[scale=0.65]{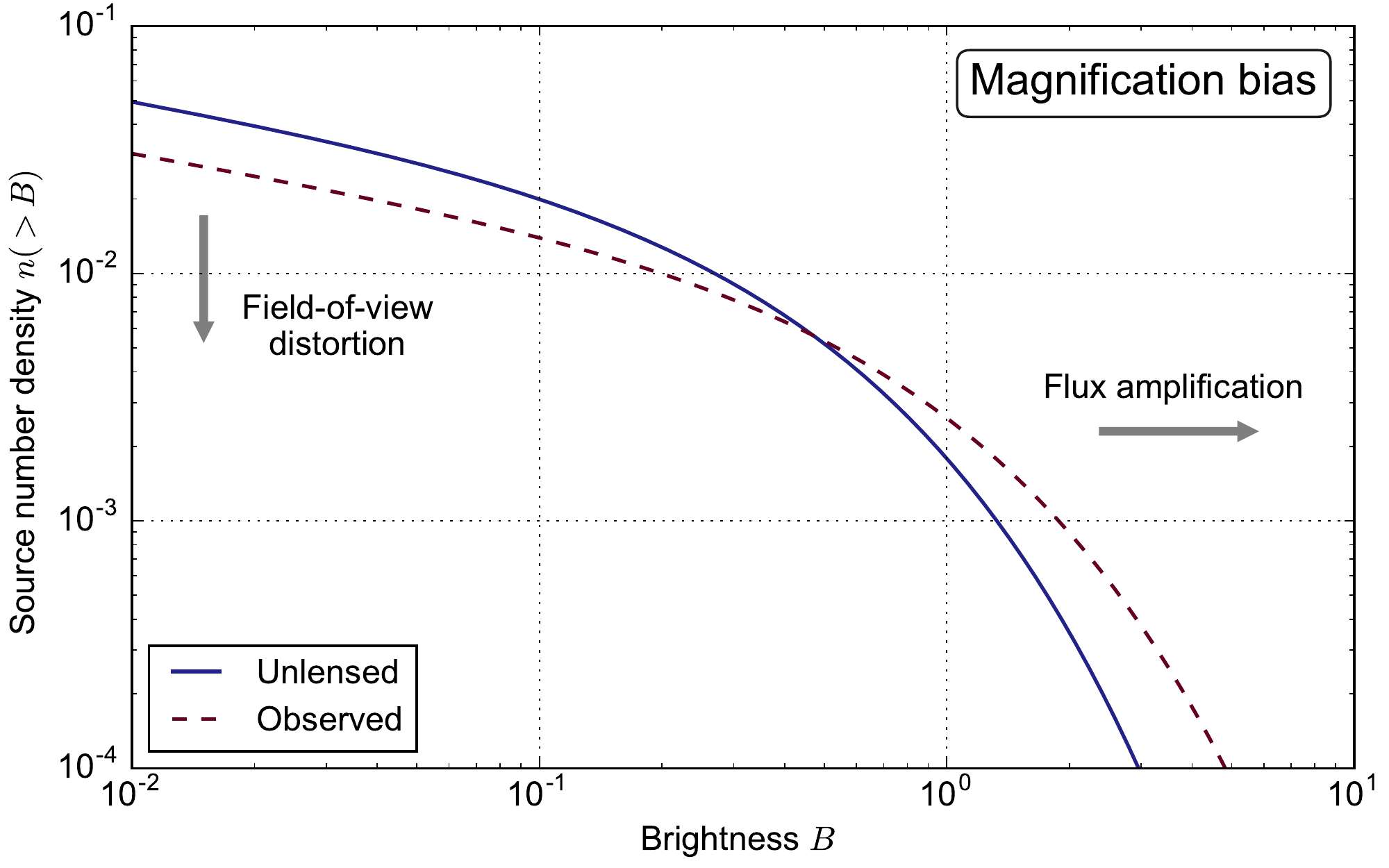}
	\caption{Illustration of the magnification bias. Weak lensing creates two effects on source number counts: the flux amplification and the field-of-view distortion. The combination of both results in a tilt of the density distribution. The number of luminous galaxies enhances and the number of faint ones depletes. The effect is exaggerated.}
	\label{fig:lensing:Magnification_bias}
\end{figure}

Historically, this effect is called \textit{magnification bias}\index{Magnification bias}\index{Magnification effect} since it biases the probability of quasar detection \citep{Turner_1980, Canizares_1981}. As said earlier, the source number counts with regard to brightness are subject to two effects. The preceding factor $1/\mu$ describes the field-of-view distortion: the stretching area increases angular separations to each other; and the inner factor represents the flux amplification. As illustrated by \fig{fig:lensing:Magnification_bias}, if $\mu>1$, the first factor diminishes the distribution function globally (number counts decrease); and the second effect shifts the curve toward the right (number counts increase). At the end of the day, the impact of the magnification on galaxy number density is ambiguous!

From \for{for:lensing:magnification_bias}, we can construct an estimator of $\mu$ as follows. By assuming $n\src(>B)\propto B^{-\alpha}$, if we can properly estimate the unlensed surface density $n\src(>B)$, e.g. counting galaxies from cluster-free areas, then the magnification $\mu$ will become an observable since
\begin{align}
	\frac{n(>B)}{n\src(>B)}=\mu^{\alpha-1}.
\end{align}
And the convergence $\kappa$ is also observable now! Thanks to \for{for:lensing:magnification}, the estimator of $\kappa$ is 
\begin{align}
	\kappa\approx \frac{1}{2(1-\alpha)}\left(1-\frac{n(>B)}{n\src(>B)}\right).
\end{align}
However, it is very difficult to measure the number density locally, either for the lensed or for unlensed density. A plausible way to overcome this difficulty is stacking. By stacking clusters of similar mass and redshift, the statistic can be enhanced. The stacked number density estimator provides a probe of the convergence profile, which can be combined with the tangential shear profile to constrain halo profiles more tightly \citep{Umetsu_etal_2014}.

\subsection{Second-order statistics}

Lensing scientists (or ``lensers'') have been interested in different statistical quantities to extract cosmological information from the large-scale structures (\acro{LSS}). The most studied among these is the second-order statistics, including the power spectrum\index{Power spectrum}. Following \citet{Schneider_2005}, by introducing the Fourier transform of $\kappa$:
\begin{align}
	\tilde{\kappa}(\vect{\ell}) = \int\rmd^2\btheta\ \rme^{-\rmi \vect{\ell}\btheta},
\end{align}
the \acro{2D} convergence power spectrum $P_\kappa(\ell)$ can be defined as
\begin{align}
	\langle \tilde{\kappa}(\vect{\ell})\tilde{\kappa}^*(\vect{\ell}')\rangle = (2\pi)^2 \delta(\vect{\ell}-\vect{\ell}') P_\kappa(\ell),
\end{align}
where $\delta$ is the Dirac delta function. Using Limber's approximation \citep{Limber_1953}, $P_\kappa(\ell)$ can be related to the matter power spectrum $P_\delta(\ell)$ by \citep{Kaiser_1992}
\begin{align}
	P_\kappa(\ell) = \frac{9H_0^4\OmegaM^2}{4\rmc^4}\int_0^{w_\maxx}\rmd w\ \frac{g^2(w)}{a^2(w)} P_\delta\left(k=\frac{\ell}{f_K(w)}, w\right),
\end{align}
where $g(w)$ is the lens efficiency given by \for{for:lensing:lens_efficiency}, and the projection integral is carried out over the comoving distance $w$ up to the limiting value $w_\maxx=w(a=0)$. Similarly, the shear power spectrum leads to 
\begin{align}
	\langle \tilde{\gamma}(\vect{\ell})\tilde{\gamma}^*(\vect{\ell}')\rangle = (2\pi)^2 \delta(\vect{\ell}-\vect{\ell}') P_\gamma(\ell).
\end{align}
Actually, $|\tilde{\gamma}|=|\tilde{\kappa}|$ (see further, Eq. \ref{for:lensing:psi_relation_3}), so $P_\gamma(\ell)=P_\kappa(\ell)$. Hence in the following, the focus will be put only on $P_\kappa(\ell)$.

Consider now the shear two-point-correlation function (\acro{2PCF})\index{Two-point-correlation function}. Since the shear has two components, its pair product can be separated into three parts, defined as
\begin{align}
	\xi_\pm(\theta) \equiv \langle\gamma_+\gamma_+\rangle(\theta) \pm \langle\gamma_\times\gamma_\times\rangle(\theta)\ \ \text{and}\ \ \xi_\times(\theta) \equiv \langle\gamma_+\gamma_\times\rangle(\theta).
\end{align}
These correlation functions should be invariant under parity transformation. Under the transformation, $\gamma_+\rightarrow\gamma_+$, $\gamma_\times\rightarrow-\gamma_\times$, $\xi_\times$ also changes its sign, so it vanishes and we usually leave it out. How do these quantities relate to $P_\kappa(\ell)$? Following \citet{Kaiser_1992}, if one transforms $\gamma$ into $\tilde{\gamma}$, the correlation functions become
\begin{align}
	\xi_+(\theta) = \int_0^{+\infty}\frac{\ell\rmd\ell}{2\pi}J_0(\ell\theta) P_\kappa(\ell)\ \ \ \text{and}\ \ \ \xi_-(\theta) = \int_0^{+\infty}\frac{\ell\rmd\ell}{2\pi}J_4(\ell\theta) P_\kappa(\ell),
\end{align}
where $J_n$ is the $n$-th order Bessel function of the first kind. By the orthonormality of the Bessel functions, these integrals are invertible, thus they become
\begin{align}\label{for:lensing:shear_xi_equality}
	P_\kappa(\ell) = 2\pi \int_0^{+\infty}\theta\rmd\theta\ J_0(\ell\theta)\xi_+(\theta) = 2\pi \int_0^{+\infty}\theta\rmd\theta\ J_4(\ell\theta)\xi_-(\theta).
\end{align}

\begin{figure}[tb]
	\centering
	\includegraphics[scale=0.65]{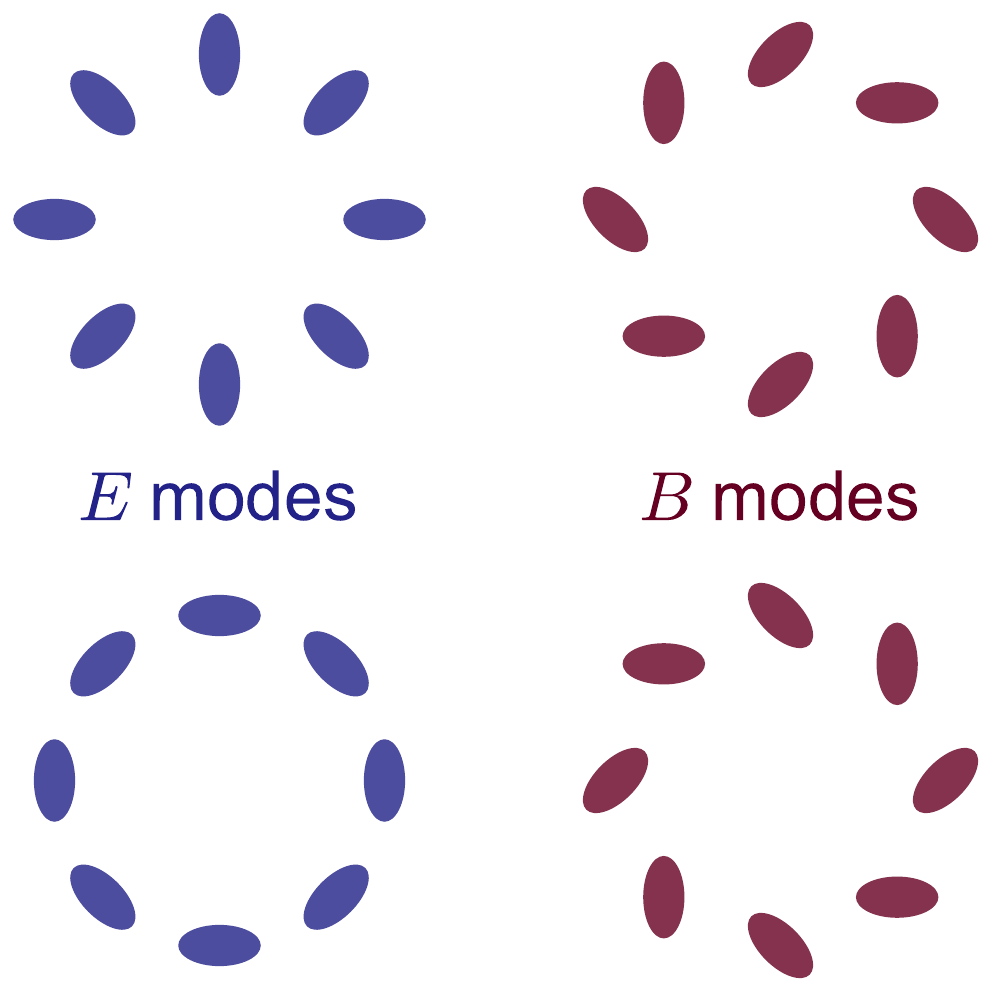}
	\caption{Illustration of typical patterns from weak lensing $E$- and $B$-modes.}
	\label{fig:lensing:EB_modes}
\end{figure}

In real life, the equality $\eqref{for:lensing:shear_xi_equality}$ might not be strictly satisfied due to observational noise and systematics. Following the similarity between the lensing shear and the \acro{CMB} polarization, the shear can be decomposed into two fields such that one of them satisfies \for{for:lensing:shear_xi_equality}, called \textit{$E$-modes}\index{$E$-modes}; and the other violates \for{for:lensing:shear_xi_equality}, called \textit{$B$-modes}\index{$B$-modes}. The characteristic patterns related to $E$- and $B$- modes are illustrated in \fig{fig:lensing:EB_modes}. To construct the $B$-mode shear, the most convenient way might be to introduce the ``imaginary'' part (in both usual and mathematical meaning) of the convergence, such that $\kappa=\kappa^E +\rmi\kappa^B$. \citet{Crittenden_etal_2002} and \citet{Schneider_etal_2002} showed that in this case,
the $E$- and $B$- mode power spectra can be written as
\begin{align}
	P_E(\ell) = \pi\int_0^{+\infty}\theta\rmd\theta\ \Big[J_0(\ell\theta)\xi_+(\theta) + J_4(\ell\theta)\xi_-(\theta)\Big], \label{for:lensing:E_modes}\\
	P_B(\ell) = \pi\int_0^{+\infty}\theta\rmd\theta\ \Big[J_0(\ell\theta)\xi_+(\theta) - J_4(\ell\theta)\xi_-(\theta)\Big]. \label{for:lensing:B_modes}
\end{align}
Unfortunately, Eqs. \eqref{for:lensing:E_modes} and \eqref{for:lensing:B_modes} extend to infinity. It is practical to introduce a filter function to reduce the integration range. Some examples can be found in the literature, including the top-hat filter \citep{Kaiser_1992}, the aperture mass \citep{Schneider_etal_1998}, the optimized ring statistic \citep{Fu_Kilbinger_2010}, and the Complete Orthogonal Sets of $E$-/$B$-mode Integrals (\acro{COSEBI}s, \citealt{Schneider_etal_2010}). Let $W(\theta)$ be the filter acting on the convergence field and $\widetilde{W}(\ell)$ its Fourier transform. Following \citet{Crittenden_etal_2002}, \citet{Schneider_etal_2002}, and \citet{Schneider_Kilbinger_2007}, one can construct 
\begin{align}
	T_+(\theta) = \int_0^{+\infty}\ell\rmd\ell\ J_0(\ell\theta)\widetilde{W}^2(\ell)\ \ \ \text{and}\ \ \ T_-(\theta) = \int_0^{+\infty}\ell\rmd\ell\ J_4(\ell\theta)\widetilde{W}^2(\ell),
\end{align}
which are implicitly linked together by a relation similar to \for{for:lensing:shear_xi_equality}. If $T_+$ vanishes outside an interval $[\theta_\minn, \theta_\maxx]$ and satisfies the following condition:
\begin{align}
	\int_{\theta_\minn}^{\theta_\maxx}\theta\rmd\theta\ T_+(\theta) = 0 = \int_{\theta_\minn}^{\theta_\maxx}\theta^3\rmd\theta\ T_+(\theta),
\end{align}
then $T_-$ also has non-zero values only on $[\theta_\minn, \theta_\maxx]$, and $E$- and $B$-mode information (denoted simply as $E$ and $B$) can be separated with finite integrations as
\begin{align}
	E &= \int_{\theta_\minn}^{\theta_\maxx}\theta\rmd\theta\ \Big[T_+(\theta)\xi_+(\theta) + T_-(\theta)\xi_-(\theta)\Big],\\ 
	B &= \int_{\theta_\minn}^{\theta_\maxx}\theta\rmd\theta\ \Big[T_+(\theta)\xi_+(\theta) - T_-(\theta)\xi_-(\theta)\Big]. 
\end{align}
More generally, the convergence power spectrum can be decomposed into various $E_n$ and $B_n$ by choosing successively $T_{n,\pm}$ with different finite supports. Finally, as detailed in \sect{sect:lensing:extraction:shape}, the correlations $\xi_\pm$ in the \acro{WL} regime can be estimated unbiasedly by counting pairs of $\langle\epsilon\epsilon^*\rangle$:
\begin{align}
	\hat{\xi}_+(\theta) = \frac{\sum_{ij}\omega_i\omega_j\Big(\epsilon_+(\btheta_i)\epsilon_+(\btheta_j) \pm \epsilon_\times(\btheta_i)\epsilon_\times(\btheta_j)\Big)}{\sum_{ij}\omega_i\omega_j}.
\end{align}
Note that the sum runs over all pairs $(i, j)$ fulfilling some binning conditions on $\theta = |\btheta_i-\btheta_j|$.

Recent surveys have successfully provided interesting cosmological constraints from the \acro{2PCF}. \citet{Kilbinger_etal_2013} processed the data from \acro{CFHTLenS} and \citet{Jee_etal_2013} studied the Deep Lens Survey (\acro{DLS}). Both results are consistent with the Wilkinson Microwave Anisotropy Probe (\acro{WMAP}). Some ongoing surveys such as \acro{KiDS} and \acro{DES} are also delivering early-stage results with the shear \acro{2PCF} \citep{Kuijken_etal_2015, Hildebrandt_etal_2016, TheDarkEnergySurveyCollaboration_etal_2015}.

\subsection{Higher-order statistics}

Higher-order statistics intrigue cosmologists for two reasons. First, as mentioned earlier, there is no simple solution for modelling the nonlinear matter power spectrum, which limits the amount of the information extracted from the \acro{2PCF}. Second, the \acro{2PCF}s only retain the Gaussian information. Unlike the \acro{CMB} which can almost be fully described by its Gaussian part, the \acro{WL} field is rich of non-Gaussianities. Cosmological information encrypted in the non-Gaussian part can be complementary to the power spectrum, and this can be extracted by higher-order statistics, some of which I display in the following.

\subsubsection{Bispectrum}

A common option of higher-order statistics is the three-point-correlation function (\acro{3PCF})\index{Three-point-correlation function}. As the \acro{2PCF} is related to the power spectrum, the \acro{3PCF} can be related to the bispectrum\index{Bispectrum}, and then to the matter distribution which depends on cosmology \citep{Schneider_Lombardi_2003, Takada_Jain_2003, Zaldarriaga_Scoccimarro_2003} (also \citealt{Takada_Jain_2003a, Takada_Jain_2003b, Schneider_etal_2005}). Similar to the power spectrum, the infinite integral problem also arises from observation. To overcome this obstacle, one can focus on third-order moments which summarize the local \acro{3PCF} information \citep{Kilbinger_Schneider_2005}. For the data-driven studies, \citet{Semboloni_etal_2011} took the data from the Hubble Space Telescope (\acro{HST}) Cosmological Evolution Survey (\acro{COSMOS}) and derived the first cosmological constraints with the third-order statistics. \citet{Fu_etal_2014} combined the \acro{2PCF} and \acro{3PCF}, and showed that parameter influence is improved compared to the \acro{2PCF}-only studies. For the even higher order, \citet{Takada_Jain_2002} addressed the kurtosis of the shear field, and found that its main contribution comes from massive halos with $M>\dix{14}\Msol$.

\subsubsection{Minkowski functionals}

Minkowski functionals\index{Minkowski functionals} (\acro{MF}s) capture the morphology of a considered region by extracting some characteristic quantities. For a given threshold $\nu$ and a smoothed \acro{2D} field $K(\btheta)$, \acro{MF}s are defined as
\begin{align}
	V_0(\nu)\equiv \int_{S_\nu}\rmd S,\ \ \ V_1(\nu)\equiv \frac{1}{4}\int_{\partial S_\nu}\rmd\ell,\ \ \ \text{and}\ \ \ V_2(\nu)\equiv \frac{1}{2\pi}\int_{\partial S_\nu}\rmd\ell\ \mathcal{K},
\end{align}
where $S_\nu\equiv \{\btheta| K(\btheta)>\nu\}$, $\partial S_\nu\equiv \{\btheta| K(\btheta)=\nu\}$, and $\mathcal{K}$ is the geodesic curvature along the contours. One can see at ease that $V_0(\nu)$ is the area above the threshold, and $V_1(\nu)$ counts for the boundary length. If $K(\btheta)$ is a smoothed Gaussian random field, \acro{MF}s can be computed analytically \citep{Sato_etal_2001}, whereas \citet{Taruya_etal_2002} provided predictions for a log-normal field. Among the recent studies, two series of work focus on application to \acro{WL}. For \acro{CFHTLenS} data, forecasts using the Fisher formalism \citep{Shirasaki_etal_2012, Shirasaki_Yoshida_2014} and cosmological constraints \citep{Kratochvil_etal_2012, Petri_etal_2013, Petri_etal_2015} were obtained with \acro{MF}s.

\subsubsection{Peak counts}
\index{Peak counts}

\begin{figure}[tb]
	\centering
	\includegraphics[scale=0.65]{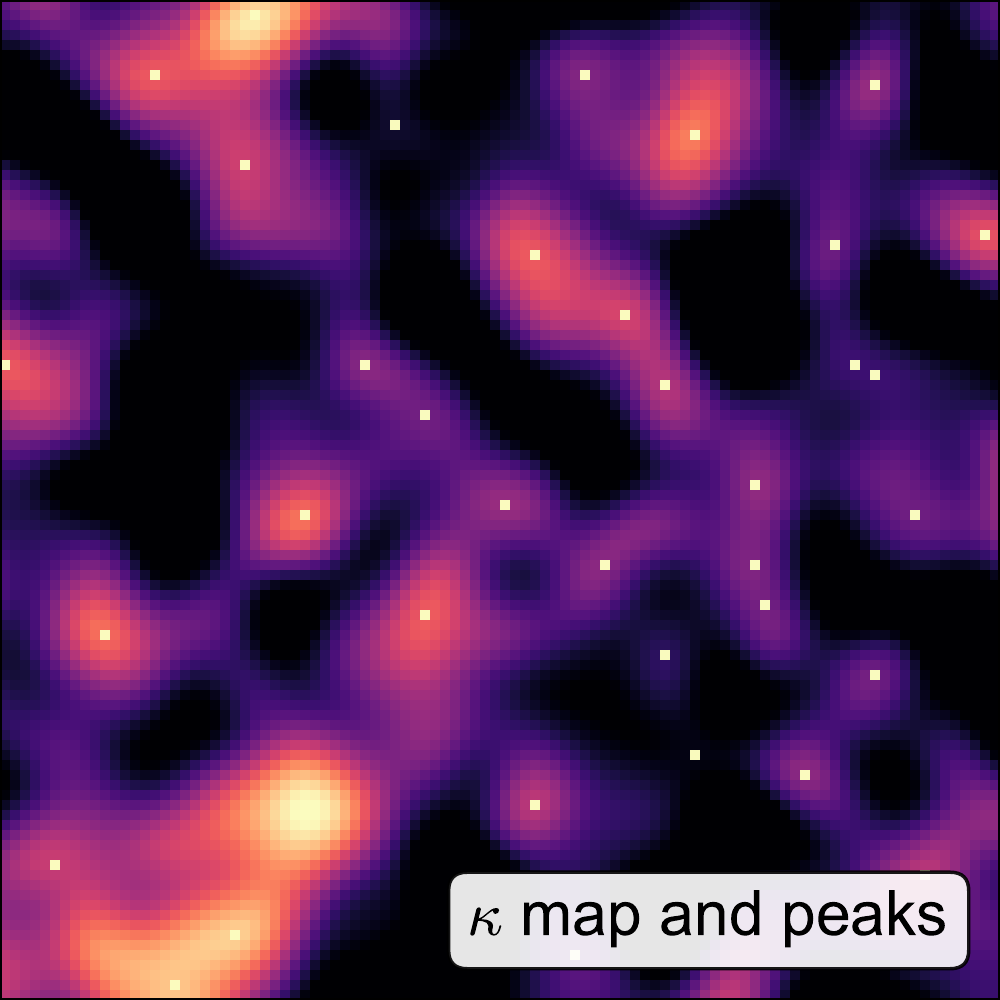}
	\caption{Peaks from a weak-lensing map.}
	\label{fig:lensing:Peak_selection}
\end{figure}

Recently, \textit{weak-lensing peak counts}\index{Peak counts} have become a popular research topic. Defined as the local maxima (\fig{fig:lensing:Peak_selection}) of the aperture mass (see \sect{sect:filtering:linear:aperture}) or convergence, peaks are tracers of massive structures, particularly galaxy clusters and \acro{DM} halos. The abundance of these objects is a cosmology-dependent statistic: whereas the \acro{2PCF} is modelled via the convergence and matter power spectra, peak counts are modelled via the halo mass function and the collapse theory. Unlike optical richness, the X-ray luminosity or temperature, or the Compton-$y$ parameter from the Sunyaev–Zel'dovich (\acro{SZ}) effect, the connection of \acro{WL} peak counts to the mass function is more direct since the mass-to-shear relation is simpler than the mass-to-light relation, which depends on baryonic matter \citep{Schneider_1996}. Besides, peaks are straightforward to identify in the surveys. These reasons make \acro{WL} peak counts a potential cosmological probe.

\section{Observational and modelling challenges}
\label{sect:lensing:challenges}

After introducing the basis of \acro{WL} and its measurement, I would like to turn to the practical side and discuss the observational challenges that cosmologists might encounter for \acro{WL} surveys. \acro{WL} is mainly observed in the optical and near-infrared bands, since source counts are highest in these wavelengths. The ``detector'' are cameras with photometers and spectrometers to estimate redshifts. Images need to be processed in order to separate individual galaxies and to determine the lensing signal.

\subsubsection{Shape measurement}

\begin{figure}[tb]
	\centering
	\includegraphics[width=0.7\textwidth]{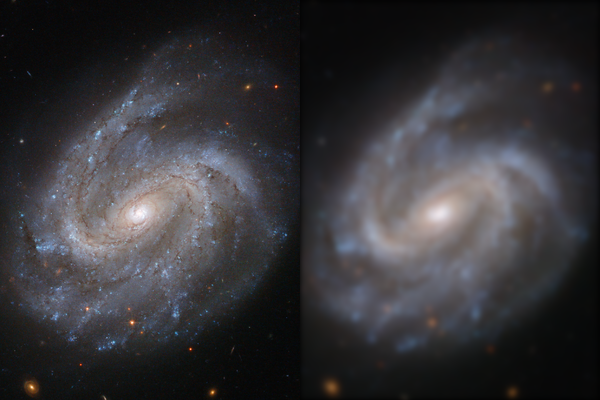}
	\caption{Left panel: a beautiful high-resolution image of NGC 201. Right panel: the same image blurred with a Gaussian \acro{PSF}. (Source: ESA/Hubble)}
	\label{fig:lensing:PSF}
\end{figure}

\index{Shape measurement}Measuring the galaxy shape up to a very high precision is important since \acro{WL} surveys aim to measure very small shape distortions. For example, Euclid, a stage IV \acro{WL} survey, aims to measure the ellipticity with an error $\leq2\dixx{-4}$ \citep{Laureijs_etal_2011}. However, such a task is difficult. One main reason is the point spread function (hereafter \acro{PSF}). The \acro{PSF} characterizes the fact that a distant point-like source always results in a blurred image in an observation. The origin of the blurring could be, for example, the optical system of the telescope, and the perturbations in Earth's atmosphere for a ground-based survey. 

In general, the \acro{PSF} varies with position, wavelength, and time. A straightforward way to handle it is to consider the images of stars as some local \acro{PSF} samples since stars are so small that they can be considered as point sources, and to interpolate over the field of view. Various sophisticated methods using Bayesian model fitting \citep{Miller_etal_2007} and super-resolution \citep{NgoleMboula_etal_2015} also exist. 

Besides the \acro{PSF}, the pixelization adds some additional uncertainty to shape estimation. Even though the resolution of ground-based telescopes reaches the order of sub-arcseconds, the size of distant faint galaxies is not much larger. The instrumental and photon noise adds another dimension to the complex shape measurement problem. Today, robust methods are strongly required as the data volume increases to derive precise cosmological constraints.

\subsubsection{Shape noise}

The term \textit{shape noise}\index{Shape noise} qualifies the randomness of galaxy intrinsic orientation. In some studies, one can read ``shot noise''\index{Shot noise} instead. This terminology comes from the consideration that the expected noise for the convergence is proportional to $1\sqrt{N_\gala}$ where $N_\gala$ is the number of averaged galaxies. However, this is misleading since (1) shape noise does not follow the Poisson law \citep{Poisson_1837}, and (2) the proportionality to $1\sqrt{N_\gala}$ is no longer valid if the convergence is obtained with a weighted average. Therefore, the term ``shot noise'' should be avoided.

Theoretically, the distribution of the shape noise is unknown. However, since lensing is in general weak, $\epsilon\approx\epsilon\src+g$ and the overall distribution on $\epsilon\src$ will be similar to the one of $\epsilon$. In practice, cosmologists assume a (truncated) normal distribution of the same variance $\sigma_{\epsilon_1}^2 = \sigma_{\epsilon_2}^2$ for both components. Throughout this work, I define the variance of the \textit{intrinsic ellipticity dispersion}\index{Ellipticity dispersion, intrinsic} $\sigma_\epsilon^2$ as the sum of variances of both components, i.e. $\sigma_\epsilon^2 \equiv \sigma_{\epsilon_1}^2 + \sigma_{\epsilon_2}^2$. Alternatively, some studies define $\sigma_\epsilon^2$ as the variance of one component, which results in a difference of factor $\sqrt{2}$. The value of $\sigma_\epsilon$ could potentially depend on cosmology on the galaxy evolution level. However, cosmologists usually account for it as an external parameter.

From \acro{CFHTLenS}, the dispersion is $\sigma_\epsilon\approx0.4$ \citep{Kilbinger_etal_2013}. From $N$-body simulations, only very few galaxies carry a convergence larger than 0.1. As a result, we are facing a problem where the signal is totally submerged in the noise. Therefore, filtering becomes an important step. I will discuss this topic in \chap{sect:filtering}.

\subsubsection{Mask effects}

\begin{figure}[tb]
	\centering
	\includegraphics[scale=0.65]{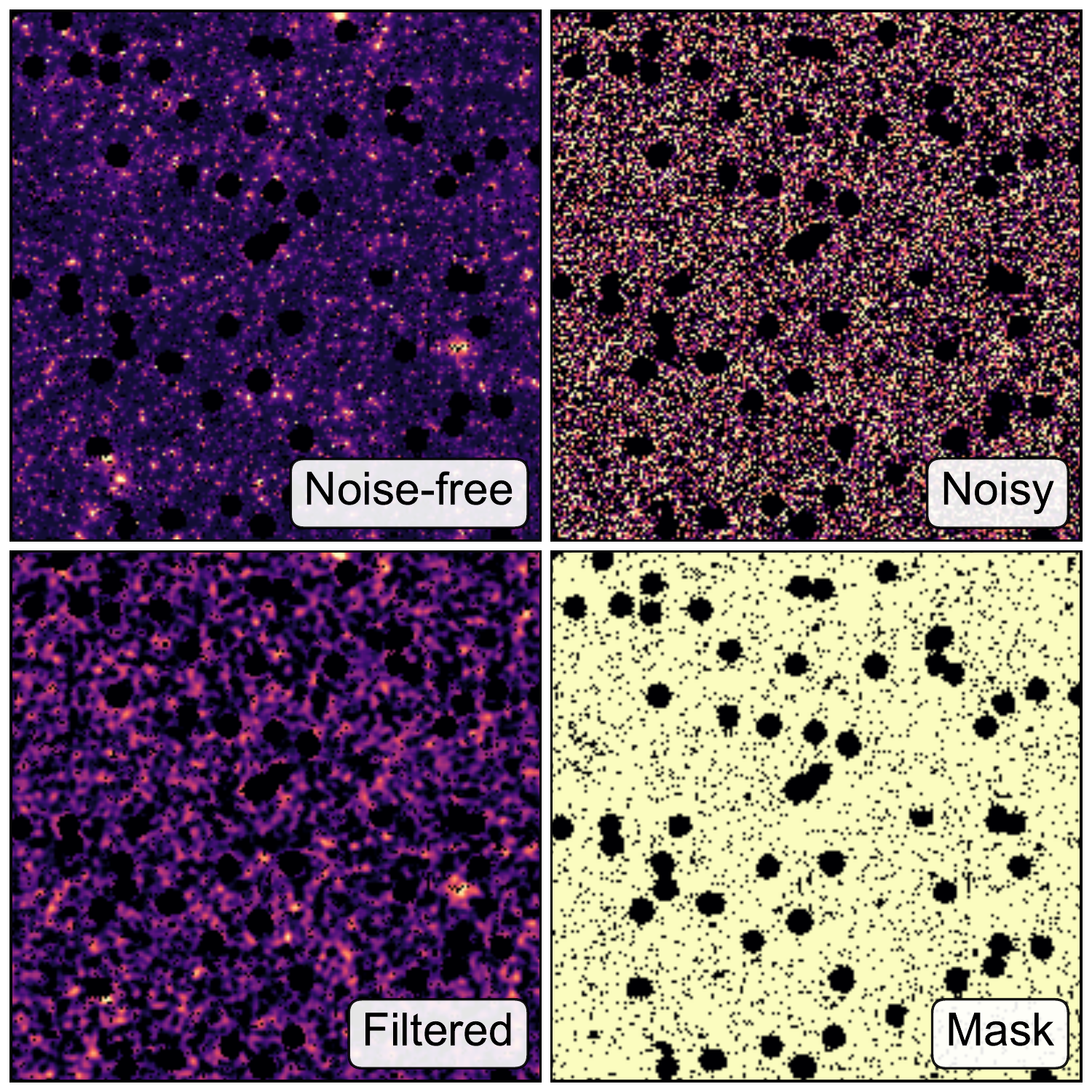}
	\caption{Example of lensing maps with a mask taken from \acro{CFHTLenS} data.}
	\label{fig:lensing:Maps_and_mask}
\end{figure}

\index{Mask effects}Like all cosmological surveys, real-life \acro{WL} data are partially masked. The origin of these masks is various. A very common type of mask is bright stars. Light from these objects diffracts in the telescope and creates spike-like artifacts. Normally, these spikes can be modelled by the \acro{PSF}. However, if they are too large and too luminous, nearby images will be contaminated. A usual way to handle these is to cut out a disk-like region centered at the star position (see \fig{fig:lensing:Maps_and_mask}).

Also, the charge-coupled devices (\acro{CCD}) of the camera do not perform perfectly. Individual abnormal pixels can create dotted masks, and depending on the geometry of electronic components, these masks can also be striped. 

Finally, because \acro{WL} probes in optical and infrared bands, the Milky Way becomes naturally the greatest mask for any full sky survey. The star and gas emission in the galactic plane makes approximately between one-third and one-half of the sky unavailable for the survey. Also, large foreground galaxies which do not contribute to lensing signals contaminate astronomical images. An optimistic size of the full-sky survey contains actually $\lesssim 25000$ deg$^2$.

\subsubsection{Inversion problem and mass-sheet degeneracy}

Even though we saw earlier that the magnification helps to estimate the convergence, this is intractable without stacking due to a low signal-to-noise ratio. Because of this reason, a $\kappa$ map should be determined from a $\gamma$ map (more exactly a $g$ map). As we have seen, $\kappa$ and $\gamma$ are related via \for{for:lensing:psi_relation_1}. A direct way to derive $\kappa$ is to ``invert'' this relation. Actually, if we go to Fourier space, derivatives become multiplications by wave numbers. Thus, \for{for:lensing:psi_relation_1} leads to
\begin{align}\label{for:lensing:psi_relation_2}
	\hat{\kappa} = \frac{\ell_1^2+\ell_2^2}{2}\hat{\psi},\ \ \hat{\gamma}_1 = \frac{\ell_1^2-\ell_2^2}{2}\hat{\psi}, \ \ \hat{\gamma}_2 = \ell_1\ell_2\hat{\psi}.
\end{align}
Readers can see that both $\hat{\gamma}_1$ and $\hat{\gamma}_2$ can be used to express $\hat{\kappa}$. To minimize the noise variance, we can set $\hat{\kappa} = (1-\alpha)\hat{\gamma}_1 + \alpha\hat{\gamma}_2$ and derive it with regard to $\alpha$ to find the optimum. The same result can be recovered by treating $\hat{\gamma}$ directly as $\hat{\gamma}_1+\rmi\hat{\gamma}_2$:
\begin{align}\label{for:lensing:psi_relation_3}
	\hat{\gamma} = \frac{(\ell_1+\rmi \ell_2)^2}{2}\hat{\psi},\ \ \text{so}\ \ \hat{\kappa} = \frac{\ell_1^2+\ell_2^2}{(\ell_1+\rmi \ell_2)^2}\hat{\gamma}.
\end{align}
The final results from both methods, called Kaiser-Squires inversion\index{Kaiser-Squires (\acro{KS}) inversion} \citep{Kaiser_Squires_1993}, are respectively
\begin{align}\label{for:lensing:KS_inversion}
	\hat{\kappa} = \frac{\ell_1^2-\ell_2^2}{\ell_1^2+\ell_2^2}\hat{\gamma}_1 + \frac{2\ell_1\ell_2}{\ell_1^2+\ell_2^2}\hat{\gamma}_2\ \ \text{and}\ \ \hat{\kappa} = \frac{\ell_1^2-\ell_2^2 - 2\rmi \ell_1\ell_2}{\ell_1^2+\ell_2^2}\hat{\gamma},
\end{align}
which are actually identical by simple algebra and \for{for:lensing:psi_relation_2}.

However in reality, only $g$ is directly measurable. If we use the weak-lensing approximation, i.e. $\gamma=(1-\kappa)g\approx g$, then $\kappa$ will be biased high. To correct this bias, one can iteratively apply \for{for:lensing:KS_inversion}, i.e. using the convergence from the previous iteration $\kappa\upp{t-1}$ to adjust the shear $\gamma\upp{t}=g/(1-\kappa\upp{t-1})$ and to yield a new estimation of the convergence $\kappa\upp{t}$ from \for{for:lensing:KS_inversion}. An alternative method using a similar technique is proposed by \citet{Seitz_Schneider_1995}.

Note that \for{for:lensing:KS_inversion} does not possess any solution when $\ell_1=\ell_2=0$. In real space, this wave mode corresponds to the constant term, say $\kappa_0$. It means that the global level of the convergence map is undetermined by this method. In principle, this degeneracy should not exist if we have the following idealistic lensing information: if $\kappa$ is known over the full sky at a fixed source depth $w$, then the mean of $\kappa$, which is $\kappa_0$, is necessarily 0. The reason is that \for{for:lensing:convergence_2} does not integrates over the mass density but the density contrast $\delta$, and the mean of the contrast over the sky is zero In reality, cosmologists possess only finite lensing samples which are located in the non-masked field of view and are not equally distant. These facts cause $\kappa_0\neq 0$. However, if the field is large $\kappa_0\approx0$ is a good approximation.

A related concept is proposed by \citet{Falco_etal_1985}. Consider, for a source catalogue and a gravitational lens, the following change of variables:
\begin{align}\label{for:lensing:lambda_transformation}
	1-\kappa'(\btheta) = \lambda(1-\kappa(\btheta))\ \ \ \text{or}\ \ \ \kappa'(\btheta) = \lambda\kappa(\btheta) + (1-\lambda),
\end{align}
where $\lambda$ is an arbitrary constant and $\kappa(\btheta)$ is the true convergence from this system. In the cluster lensing framework, this $\lambda$-transformation\index{$\lambda$-transformation} implies $\Sigma'(\btheta)=\lambda\Sigma(\btheta) + (1-\lambda)\Sigma_\crit$ from \for{for:lensing:kappa_halo_2}. Thus, if the profile is circular, from \for{for:lensing:tangential_shear_profile_2} we obtain $\gamma_+'(\btheta)=\lambda\gamma_+(\btheta)$. More generally, any distribution of $\Sigma$ can be considered as the convolution by a properly chosen kernel of a punctual mass (which is a ``circular'' profile), so $\gamma'(\btheta)=\lambda\gamma(\btheta)$ everywhere. Therefore, the reduced shear is invariant and the transformation \eqref{for:lensing:lambda_transformation} leads to a series of admissible solution $(\kappa(\btheta)$, $\gamma(\btheta))$ for the same observation. This is called the \textit{mass-sheet degeneracy}\index{Mass-sheet degeneracy}. For example, an Einstein ring could be an image of a strongly lensed background galaxy, or a lensing-free image of a huge ring-like structure.

Sometimes, the mass-sheet degeneracy also refers to $\kappa_0$, claiming that $\kappa_0$ and $\lambda$ are the same degree of freedom. This is true when the field of view is small. Actually, $\lambda$ is a local degeneracy since the $\lambda$-transformation is only applied to a circular area neighboring the cluster \citep{Falco_etal_1985}, whereas $\kappa_0$ is a global degeneracy which accounts for the whole field of view. Fundamentally, $\kappa_0$ and $\lambda$ are still different concepts.

\subsubsection{Source redshifts}

The lensing signal depends not only on the distance from the observer to the lens, but also on the one to the source. However, because of the broad kernel which acts in \for{for:lensing:convergence_2}, a very high precision on source redshifts is not necessarily required. Therefore, most surveys count on the photometric technique \citep{Connolly_etal_1995, Benitez_2000} to determine the source redshifts for a wide field, which is less precise but efficient, while the spectroscopic technique is only applied at a relatively restricted field size, to determine redshifts with high precision and to calibrate the photometric results. A very straightforward way to include the source information is to transform \for{for:lensing:convergence_2} into an integral over the redshift distribution, reasoning in a probabilistic way. An alternative is to split the source catalogue into different slices depending on their depth, and to perform a tomographic analysis. This is particularly interesting to study cosmological evolution for different epochs, which helps breaking some parameter degeneracies.

\begin{figure}[tb]
	\centering
	\includegraphics[scale=0.65]{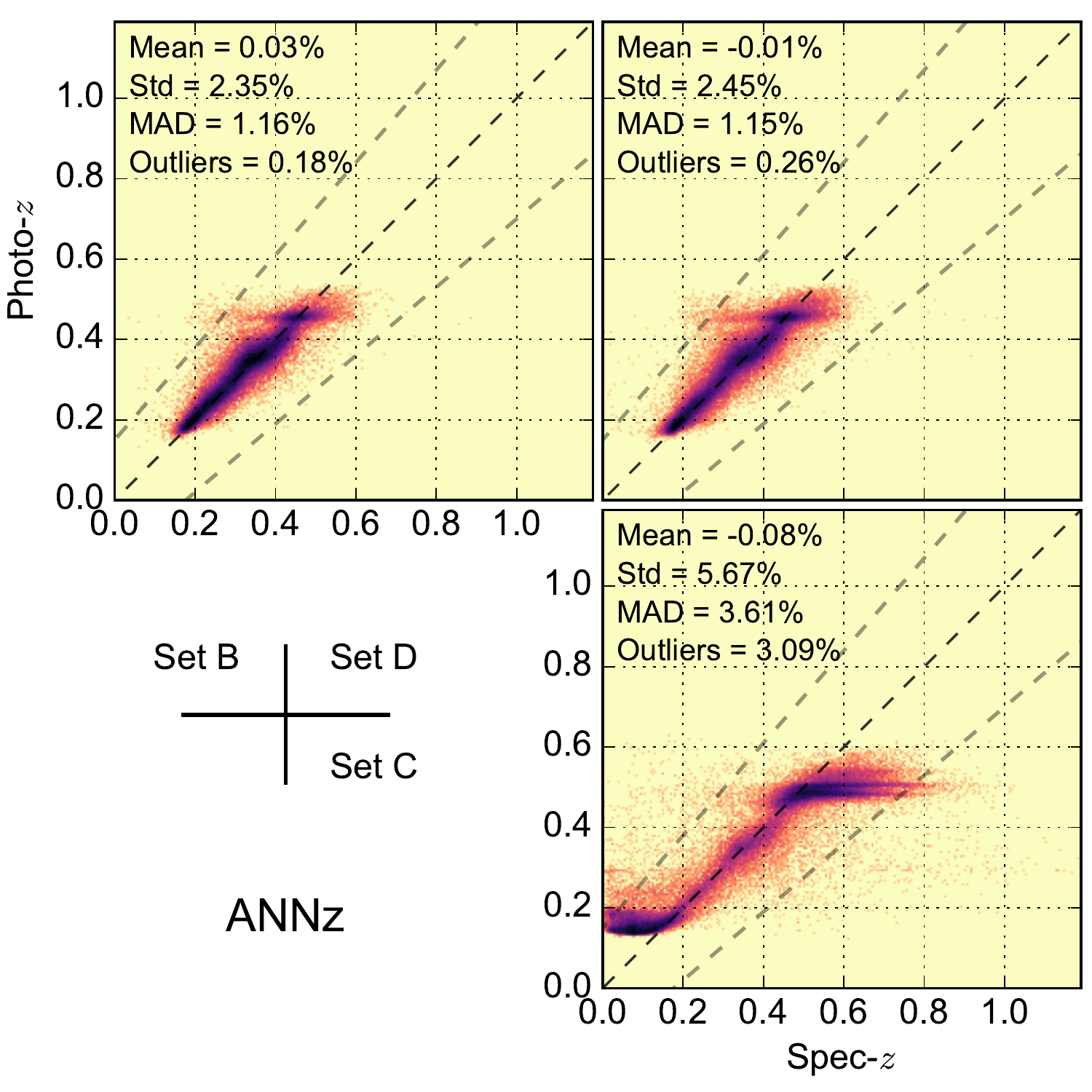}
	\caption{Comparison between the spectroscopic and photometric redshifts for three data sets constructed from Sloan Digital Sky Survey (\acro{SDSS}) data releases 12. All three panels show the scatter of the photometric redshift estimation from samples with the spectroscopic redshift measurement (considered as the truth). In the bottom right panel, we can see some catastrophic failures, where the estimation error is flagrant.}
	\label{fig:lensing:Spec_vs_photo}
\end{figure}

In this context, the focus of redshift studies for \acro{WL} has been put on modelling the errors on photometric redshifts (\acro{photo-$z$})\index{Photometric redshift} and understanding how these errors propagate to the lensing signal. An inevitable phenomenon for photometric redshifts are the ``catastrophic'' failures:\index{Catastrophic failure} the difference between the true and estimated values to be several times larger than the characteristic error. These failures are problematic if the proportion is large.

Recently, a technique taking advantage of the galaxy correlation function to derive the redshift distribution has been rediscovered and promoted \citep{Seldner_Peebles_1979, Roberts_Odell_1979, Newman_2008, Schmidt_etal_2013}. Such a technique has already been used by \citet{Hildebrandt_etal_2016} as an alternative to \acro{photo-$z$}.

\subsubsection{Intrinsic alignment}

\begin{figure}[tb]
	\centering
	\includegraphics[scale=0.65]{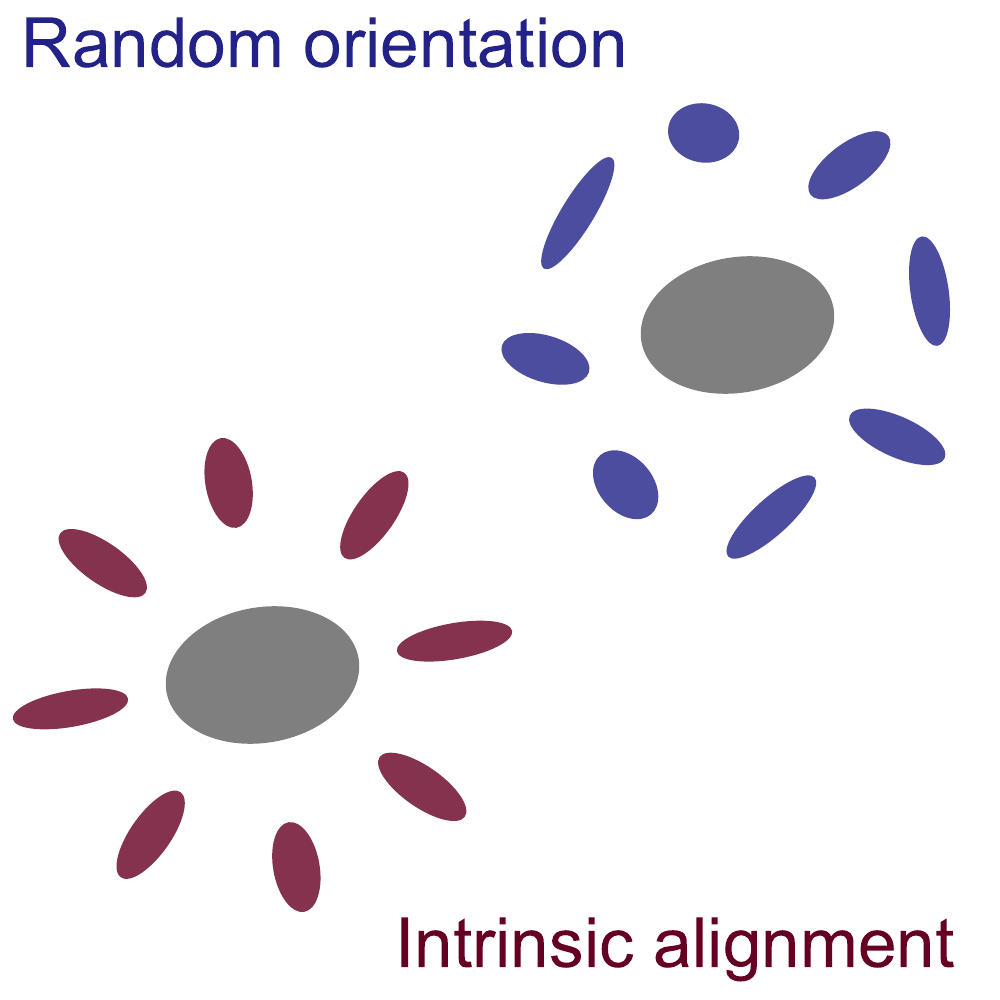}
	\caption{Illustration of intrinsic alignment. Galaxies can be influenced by the tidal force of nearby halos and aligned with the radial direction. The effect is exagerated in the figure.}
	\label{fig:lensing:Intrinsic_alignment}
\end{figure}

As we have seen in \sect{sect:lensing:extraction:shape}, the interpretation of the lensing signal and convergence maps is based on the isotropy hypothesis of galaxy orientation. The intrinsic galaxy orientation should not possess any privileged direction. This is however not true in reality. First, galaxies that are close to a halo are subject to a tidal force, and are constantly stretched in radial direction during their formation. Second, the angular momenta of the halo and the galaxy can be correlated. These physical processes break the isotropy hypothesis locally, which causes a phenomenon called \textit{intrinsic alignment} (\acro{IA})\index{Intrinsic alignment}. If the isotropy hypothesis is still applied, the presence of \acro{IA} creates spurious lensing signals \citep{Heavens_etal_2000, Bridle_Abdalla_2007}. In terms of second-order statistics, this contamination can be separated into two terms: an intrinsic-intrinsic correlation and a shear-intrinsic correlation. Removing these contaminations is challenging. Alternatively, some physically motivated models for the intrinsic shape of galaxies have been proposed \citep{Hirata_Seljak_2004, Bridle_King_2007}. This provides a solution for modelling using a forward approach. In any case, studies have already shown that neglecting or mis-modelling \acro{IA} could introduce important biases on \acro{WL} signals \citep[e.g.][]{Kirk_etal_2015}.

\subsubsection{Baryonic effects}

\begin{figure}[tb]
	\centering
	\includegraphics[width=\textwidth]{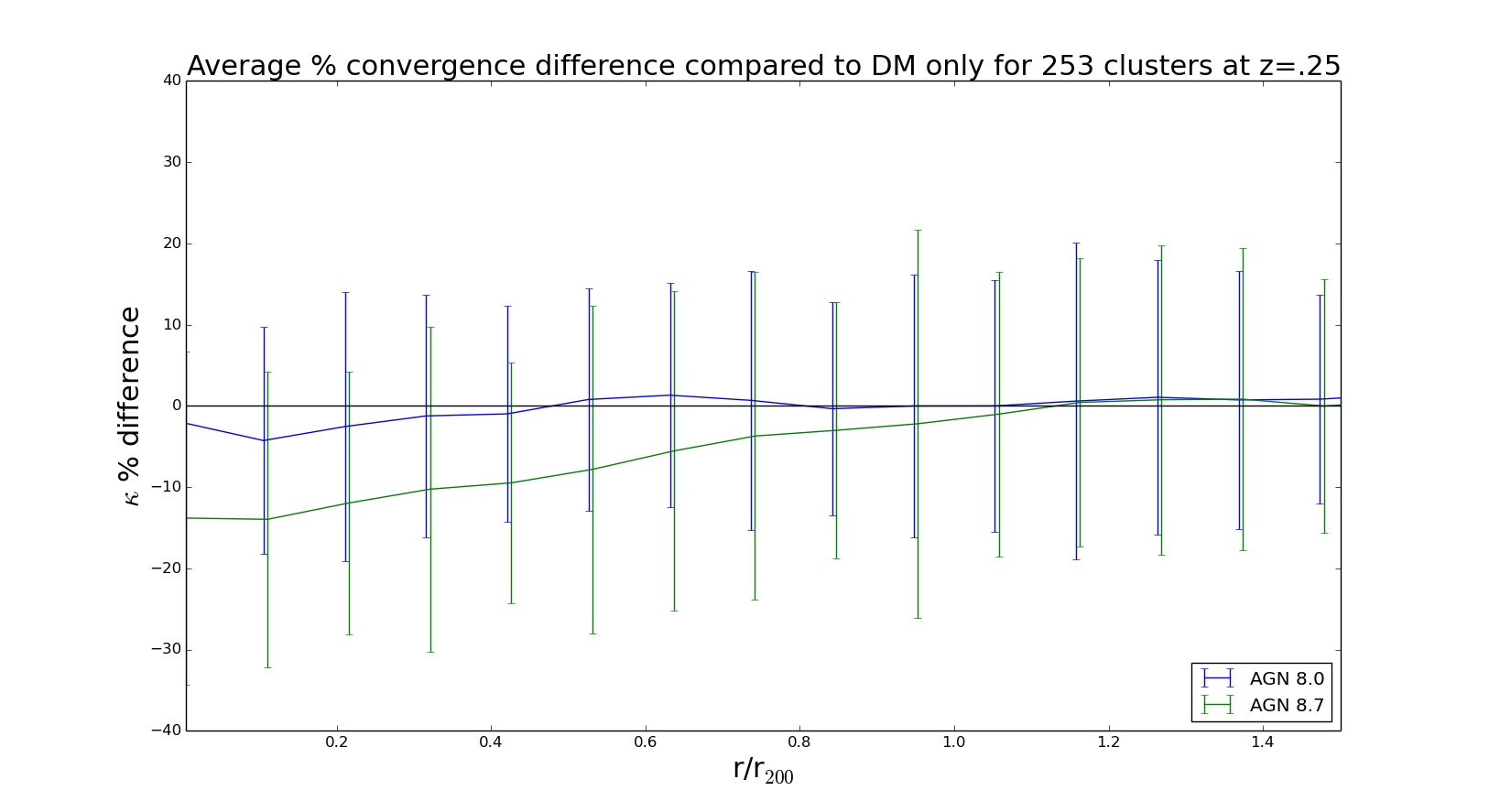}
	\caption{Preliminary result of the difference between profiles of \acro{DM}-only clusters and those with baryonic processes. In the extreme \acro{AGN} case (green), baryons can change the lensing signal by 10--15\% near the center of the clusters. (Source: Brandyn Lee)}
	\label{fig:lensing:Baryonic_effects}
\end{figure}

\index{Baryonic effects}At small scales, structure formation is influenced by baryonic physics. To solve the \textit{overcooling problem}, i.e. the star formation rate is higher than expected, the feedback from active galactic nuclei (\acro{AGN}) are required during galaxy formation \citep{Dubois_etal_2013, Okamoto_etal_2014}. Recent hydrodynamic simulations showed that including this feedback could change the \acro{DM} halo profiles \citep{Duffy_etal_2010, Martizzi_etal_2012}, which changes then the lensing signal at small scales. Most \acro{WL} studies until recently, especially those which focus on second-order statistics, do not account for baryonic effects since at large scales this is insignificant. However, if cosmologists desire to probe the power spectrum at high-$\ell$ or to study peak counts, including baryons would be indispensable. 

\subsubsection{Nonlinear spectrum}

\begin{figure}[tb]
	\centering
	\includegraphics[scale=0.65]{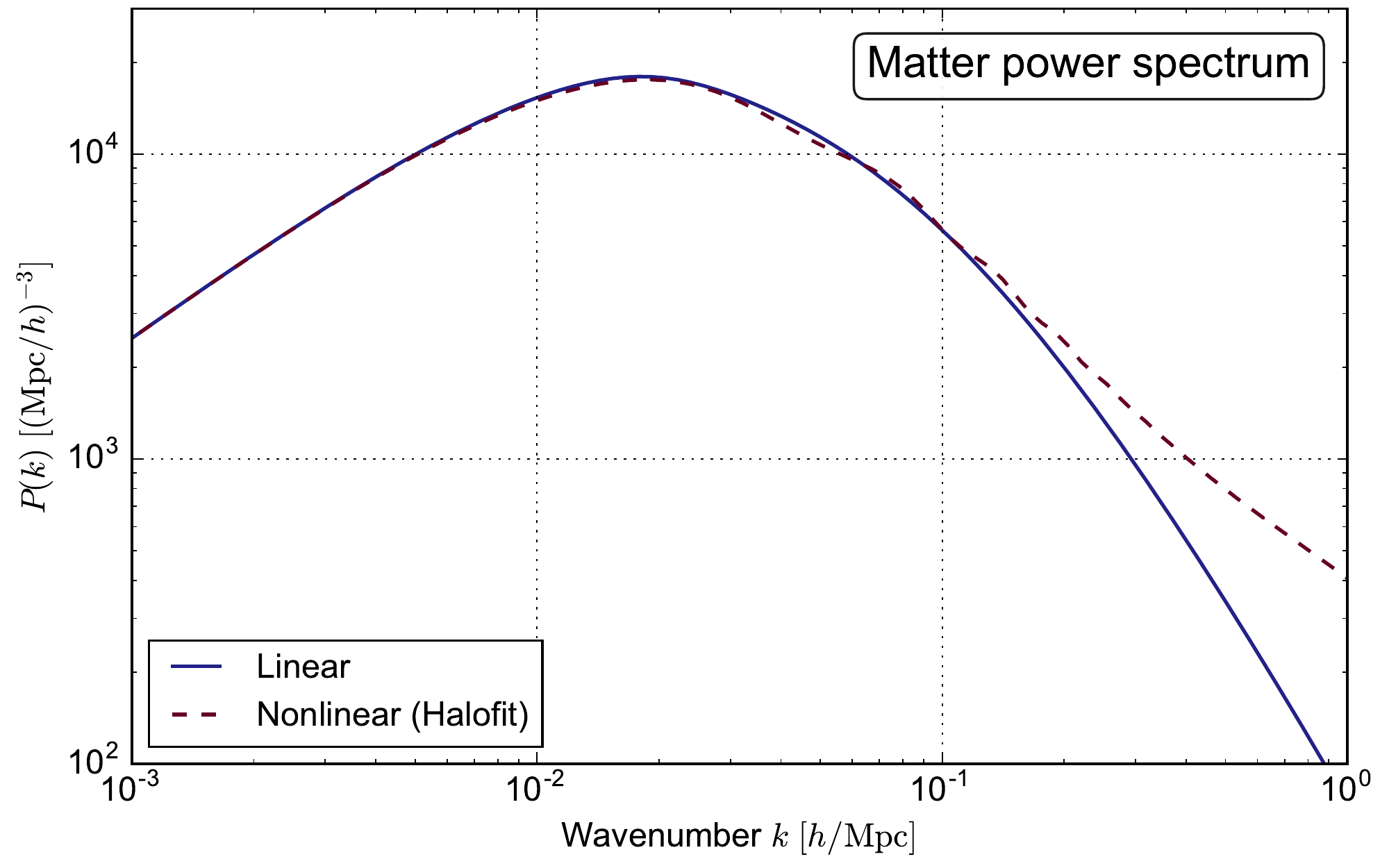}
	\caption{Comparison between the linear and the nonlinear power spectrum. The nonlinear spectrum is computed using the \textsc{Halofit} algorithm. We can see that at small scales the linear spectrum under estimates the power. Also, the spectrum oscillates for $k > \dix{-2}$. This is due to the baryonic acoustic oscillations (\acro{BAO}).}
	\label{fig:lensing:Nonlinear_power_spectrum}
\end{figure}

There exist several methods to extract cosmological information from \acro{WL}. The one which acquired the greatest focus to date is the power spectrum. This is mostly motivated by the fact that the
matter spectrum can be well modelled by theory on large scales. However on small scales, complex gravitational interactions make the matter distribution nonlinear and non-Gaussian, and invalidate the prediction from the linear theory. Until now, two families of corrections have been proposed, which are perturbation theory \citep[e.g.][]{Makino_etal_1992, Bernardeau_etal_2002} and effective field theory \citep[e.g.][]{Baumann_etal_2012, Carrasco_etal_2012}. Nevertheless, neither of them can provide a satisfactory result for scales below few $\Mpc/h$ \citep{Cooray_Sheth_2002, Smith_etal_2003, Carrasco_etal_2012}. For this reason, modelling using $N$-body simulations is an interesting option. This difficulty also provides a great motivation to explore \acro{WL} observables other than the power spectrum.

\section{Weak-lensing peak counts: state of the art}
\label{sect:lensing:state}

\subsubsection{A bit of history}

The idea of counting peaks in \acro{WL} was first proposed by \citet{Schneider_1996}. The original argument was that at signal-to-noise ratio (\acro{S/N}) of $\gtrsim 5$, peaks are presumably real halos, so we probe directly the high-mass end of the mass function to discriminate cosmologies. The detection was expected to be carried out by the aperture mass \citep{Kaiser_etal_1994, Schneider_1996}. The following studies of the same series also stayed in this ``lens-count'' approach: \citet{Kruse_Schneider_1999, Kruse_Schneider_2000} estimated the expected number of detections for different cosmological models, and provided an exponential model to describe the fast decreasing tail; \citet{Reblinsky_etal_1999} validated the previous works by $N$-body simulations. On the other hand, \citet{Jain_etal_2000}, \citet{Munshi_Jain_2000}, and \citet{Bernardeau_Valageas_2000} studied the full-$\kappa$ distribution (this is actually one of the Minkowski functionals!) modelled with appropriate noise properties and derived its cosmological dependency. 

Later, \citet{VanWaerbeke_2000} and \citet{Jain_VanWaerbeke_2000} combined both concepts and modelled directly the peak function without distinguishing between true and spurious peaks. Since then, studies of \acro{WL} peak counts can be roughly classified into two categories \citep{Lin_etal_2016a}. The first category is concerned with \textit{cluster-oriented purposes}\index{Cluster-oriented purpose} which stays close to the original philosophy of \citet{Schneider_1996} to establish explicitly the \textit{selection function}, i.e. the mass-convergence correspondence. These studies are usually interested in high \acro{S/N}, positional offsets, peak-height variations, and projection effects focusing on indicators such as purity and completeness. They may also cross-check with optical richness and X-ray data. The other category focuses on \textit{cosmology-oriented purposes}\index{Cosmology-oriented purpose} and adopts the idea of \citet{Jain_VanWaerbeke_2000}. These studies make the selection function implicit. They attempt to directly model \acro{WL} peak counts from cosmological models, including contributions from massive clusters, projections of \acro{LSS}, spurious signals, or a mixture of all of these cases \footnote{Of course, this is not a strict classification, since cluster detection and peak counts rely on the same peak-finding technique. Some work satisfy both purposes.}. This is precisely the focus of this thesis. 

In the following, I outline some remarkable series of works on \acro{WL} peak counts, organized by their respective working group. Before this, it should be appropriate to mention first \citet{Bartelmann_etal_2001, Bartelmann_etal_2002} who proposed to model the expected detection as the sum of the theoretical prediction and a Gaussian fluctuation, and who argued that cluster counts could be used to constrain the equation of state of the dark energy. These works influenced some of the following results.

\subsubsection{Tokyo}

One of the series is led by \citet{Hamana_etal_2004}. From $N$-body simulations, the authors studied the mass-convergence correspondence in detail, and extended the analytical cluster-count prediction model from \citet{Kruse_Schneider_2000} and \citet{Bartelmann_etal_2002} by emperically establishing the selection function. This model was improved by \citet{Hamana_etal_2012} as halo triaxiality has been taken into account. Modelling the lens geometry fluctuation, the projection from \acro{LSS}, and the shape noise, \citet{Hamana_etal_2015} applied the model from previous works to the Subaru Suprime-Cam data and obtained cosmological constraints. Recently, the focus of this group has been put on the ongoing \acro{HSC} survey. \cite{Osato_etal_2015} drew particular attention on the baryonic impact on the \acro{WL} power spectrum and peaks, providing forecast for \acro{HSC}. \citet{Shirasaki_etal_2016} and \citet{Higuchi_Shirasaki_2016} examined the sensitivity of cluster counts to $f(R)$ parameters by Fisher information \citep{Fisher_1922} and $N$-body simulations, respectively validating the feasibility of the probe. Later, \citet{Shirasaki_2016} tried to model peak counts analytically by combining different orders of moments using local-Gaussianized transformation. Finally, \citet{Shirasaki_etal_2016a} investigated Fisher information provided by different \acro{WL} observables, including peak counts.

\subsubsection{Heidelberg}

Another remarkable series is led by \citet{Maturi_etal_2005} focusing on optimizing cluster detection. \citet{Maturi_etal_2005} proposed a filter optimized for identifying \acro{NFW}-like halos. Note that \citeauthor{Schirmer_2004} (\citeyear{Schirmer_2004} \footnote{PhD thesis not available on NASA/ADS, but via \url{http://hss.ulb.uni-bonn.de/2004/0326/0326.htm}.}, see also \citealt{Hetterscheidt_etal_2005}) also proposed an optimized filter for the same issue. The difference is that \citet{Schirmer_2004} fitted an analytical function to the expected signal from an \acro{NFW} profile, while \citet{Maturi_etal_2005} argued that in order to detect clusters better, the projection from \acro{LSS} should be treated as noise and taken into account. The author proposed then to consider the new noise variance from the sum of the \acro{LSS} spectrum and the galaxy shape noise, and constructed a filter adapted to the \acro{NFW} signal and the total noise. \citet{Pace_etal_2007} compared this filter with other commonly used ones. Then, \citet{Maturi_etal_2007} applied this method to the \acro{GaBoDS} data in parallel with \citet{Schirmer_etal_2007}. 

Later, \citet{Maturi_etal_2010} proposed an analytical model to model peak counts. While \citet{Bartelmann_etal_2002} and \citet{Hamana_etal_2004} model the selection function either by a Gaussian fluctuation or $N$-body simulations, this new approach, more aligned with cosmology-oriented purposes, is directly based on the peak theory of the Gaussian random field \citep{Bardeen_etal_1986, Bond_Efstathiou_1987}. The definition of peaks has been changed, from the local maxima, pixels with values higher than their eight neighbors, to the \textit{contiguous areas} with values above a given threshold. The motivation is that the new definition simplifies the analytical expression. After that, \citet{Maturi_etal_2011} extended the previous work and yielded the Fisher forecast on $f_\mathrm{NL}$. Recently, \citet{Reischke_etal_2016} came out with a correction on very high \acro{S/N} for the model of \citet{Maturi_etal_2010}.

\subsubsection{Philadelphia-Bonn}

In this working group, \citet{Marian_Bernstein_2006} started with forecasting the constraining power of cluster counts. Then, using $N$-body simulations, \citet{Marian_etal_2009, Marian_etal_2010} examined the projection effect in the noise-free case, and \cite{Marian_etal_2011} provided a quick look on sensitivity of peaks on probing $f_\mathrm{NL}$. Many Fisher analyses have been performed by the authors. They focused on the mass function \citep{Smith_Marian_2011}, $\wZero$ and $\LCDM$ parameters \citep{Marian_etal_2012}, $f_\mathrm{NL}$ with \acro{2PCF} and peak counts \citep{Hilbert_etal_2012}, and $\wZero$ and $\LCDM$ parameters with peak abundance combined with $\gamma_+$ profiles and the peak-peak correlation \citep{Marian_etal_2013}. In particular, \citet{Marian_etal_2012} also proposed a hierarchical filtering algorithm for mass reconstruction. This actually coincides with the idea of the matched filter. As a remark, the authors of these studies claimed to aim for cosmology-oriented purposes \footnote{For example, we can read from \citet{Marian_etal_2009}: ``However, our goal here is not to establish the correspondence between the two-dimensional and three-dimensional masses of individual halos, but the cosmology dependence of the shear-peak abundance.''}. However, the methods and tools turned out to stay close to cluster-oriented approaches.

\subsubsection{Garching-Bonn-DES}

\citet{Dietrich_Hartlap_2010} provided the first realistic forecast on the $\OmegaM$-$\sigEig$ constraints. Modelling with a large amount of $N$-body simulations, they explored the joint constraining power of peak counts and the \acro{2PCF}, for obvious cosmology-oriented purposes. They showed that peaks alone constrain $\OmegaM$ and $\sigEig$ better than the \acro{2PCF}. \citet{Dietrich_etal_2012} studied the impact of the \acro{LSS} projection effect and the noise on peak positions. Later, \citet{Kacprzak_etal_2016} followed \citet{Dietrich_Hartlap_2010} and performed the $\OmegaM$-$\sigEig$ constraints with the \acro{DES} data.

\subsubsection{New York}

At the beginning, this series of studies followed a pure cluster-oriented approach. \citet{Wang_etal_2004} performed Fisher analyses with X-ray, \acro{SZ}, and \acro{WL}-selected clusters, while \citet{Fang_Haiman_2007} considered a similar problem with \acro{WL} clusters and \acro{2PCF}. However, the strategy turned into a cosmology-oriented focus later. \cite{Wang_etal_2009} investigated the properties of the full $\kappa$ \acro{PDF}. As mentioned earlier, the $\kappa$ \acro{PDF} is one of the \acro{MF}. Since then the group developped a remarkable series of studies on both peak counts and \acro{MF}. Similar to \citet{Dietrich_Hartlap_2010}, they modelled the observables with $N$-body simulations. First, \citet{Kratochvil_etal_2010} tested the sensitivity to cosmology of peak counts. Then, the authors focused on three parameters: $\OmegaM$, $\sigEig$, and $\wZero$. \citet{Yang_etal_2011} examined the number of halos associated to a peak and yielded Fisher forecasts. In this study they found a counterintuitive fact: the noise enhances the peak \acro{S/N}. Later, \citet{Yang_etal_2013} investigated the baryonic impact on the \acro{WL} peaks and power spectrum. They discovered that low peaks with \acro{S/N} between 1 and 3.5 not only are robust to baryonic effects, but also contain valuable cosmological information. After studying the magnification bias, \citetalias{Liu_etal_2014a} (\citeyear{Liu_etal_2014a}) concluded that this is negligeable for current small-coverage surveys such as \acro{CFHTLenS}, but needs to be taken into account for Euclid and \acro{LSST}. Meanwhile, \acro{MF} are studied by \citet{Kratochvil_etal_2012}, \citet{Petri_etal_2013}, and \citet{Petri_etal_2015}. \citetalias{Liu_etal_2015} (\citeyear{Liu_etal_2015}) applied the previous works to the \acro{CFHTLenS} data and obtained cosmological constraints, and \citet{Liu_Haiman_2016} examined the peak-peak correlation from the same data set. Finally, \citet{ZorrillaMatilla_etal_2016} compared the fast modelling method proposed by \citet{Lin_Kilbinger_2015} with $N$-body simulations and examined its systematics in detail. The authors found a good agreement and proposed several potential improvements. Note that this group is also involved in the preparation for \acro{LSST}, where \citet{Bard_etal_2013} dealt with the impact of measurement errors on peak counts, and \citet{Bard_etal_2016} studied masks.

\subsubsection{Beijing}

Like many other working groups, this series of studies also appeared to be interested in cluster counts first, and then turned to a cosmology-oriented method afterwards. The series started with \citet{Tang_Fan_2005} who tried to establish a formal selection function for \acro{NFW} lensing peaks. Then, \citet{Fan_2007} attempted to study the impact from \acro{IA} on \acro{WL} peaks. The author modelled \acro{IA} as an enhancement of the noise variance, thus proposed a correction to cluster counts. After that, \citet{Fan_etal_2010} chose to use an analytical prescription to model peak counts based on the Gaussian peak theory. In the end, this approach is similar to \citet{Maturi_etal_2010} except for two differences. The first one is the definition of peaks for which \citet{Fan_etal_2010} opted for the local maximum. The second one is the filter shape, chosen to be Gaussian. However, since the filter from \citet{Maturi_etal_2010} and the Gaussian kernel can be exchanged with each other and be implemented in both models without difficulty, this difference is not crucial. \citet{Jiao_etal_2011} compared the Gaussian kernel with the \textsc{MRLens} nonlinear filter (see \citealt{Starck_etal_2006}, \sect{sect:filtering:nonlinear:MRLens}) on information extraction. Afterwards, \citetalias{Liu_etal_2014} (\citeyear{Liu_etal_2014}) studied the mask effect. They proposed a correction for the model of \citet{Fan_etal_2010} to handle the bias generated by missing data. Recently, using the Canada-France-Hawaii Telescope (\acro{CFHT}) Stripe 82 data, \citet{Shan_etal_2014} gave the observed peak function and compared with the model, and \citetalias{Liu_etal_2015a} (\citeyear{Liu_etal_2015a}) derived the cosmological constraints on $\OmegaM$, $\sigEig$, and halo concentration parameters. Finally, \citetalias{Liu_etal_2016} (\citeyear{Liu_etal_2016}) constrain parameters from the $f(R)$ theory using the \acro{CFHTLenS} data.

\subsubsection{Saclay}

\begin{figure}[tb]
	\centering
	\includegraphics[width=0.8\textwidth]{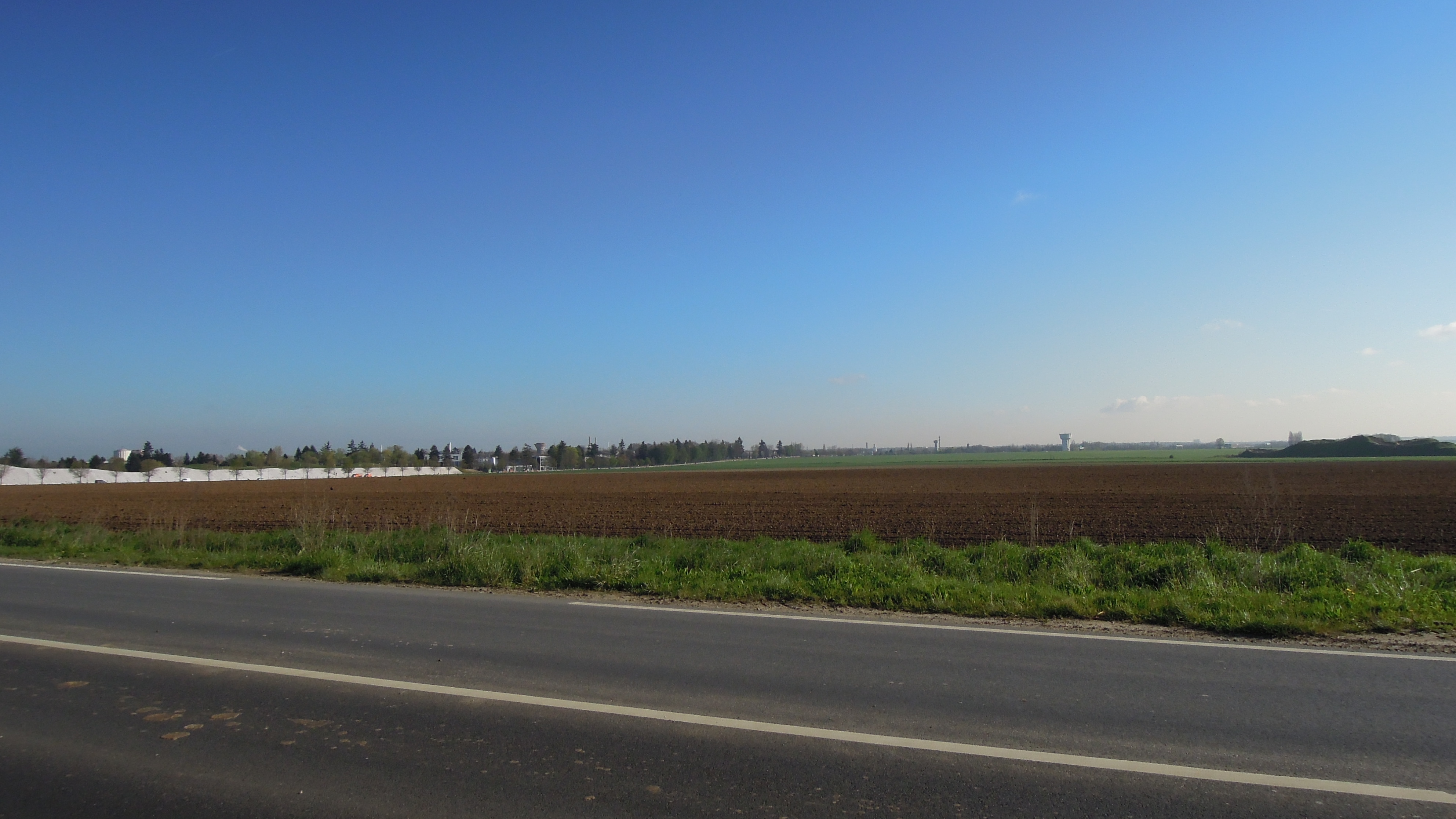}
	\caption{Saclay and its charming fields. (Are they Gaussian?)}
	\label{fig:lensing:Saclay}
\end{figure}

In this series, \citet{Pires_etal_2009a} compared different non-Gaussian observables by using different filters for convergence reconstruction, including the \textsc{MRLens} nonlinear method. They concluded that peak counts outperformed other higher-order statistics for discriminating between cosmological models. A similar study is carried out by \citet{Pires_etal_2012}, where they investigated the optimal scale for cosmological information extraction. Meanwhile, \citet{Berge_etal_2010} addressed a similar question with a different approach. The author performed the Fisher analysis with the power spectrum, bispectrum, and halo counts to explore the constraining power of these quantities.

Then, for full cosmology-oriented purposes, \citet[][hereafter \PaperI]{Lin_Kilbinger_2015} proposed a stochastic approach to model \acro{WL} peak counts (see \sect{sect:modelling:LK}). The model adopts the halo approach, and in a forward way computes the \acro{WL} peak counts from the considered mass function. This approach has the potential to include complex survey conditions in a straightforward way. The authors showed that the new model bypasses the costly $N$-body computation without losing consistency. Next, \citet[][hereafter \PaperII]{Lin_Kilbinger_2015} were interested in parameter constraint strategies. They discovered that the Gaussian likelihood is a good approximation. However, if the covariance is assumed to be independent from cosmology, the constraining power might be underestimated. The authors also showed that by using approximate Bayesian computation (\acro{ABC}, see \chap{sect:ABC}), cosmological constraints can be obtained without a likelihood function in a much faster way. Also, \citet[][hereafter \PaperIII]{Lin_etal_2016a} explored different filtering techniques, finding that using multiscale information from compensated filters extract more cosmological information from peak counts than the Gaussian function.

\subsubsection{Some other studies}

Further studies are presented below. \citet{White_etal_2002} investigated purity and completeness of cluster detection and claimed that noise and biases on mass were too large for cluster counts. Meanwhile, \citet{Weinberg_Kamionkowski_2002, Weinberg_Kamionkowski_2003} investigated the cluster abundance and analyzed its detection by \acro{WL} for different dark-energy models. \citet{Hennawi_Spergel_2005} proposed a tomographic matched filter adapted for cluster detection at different scales and redshift. This work involves comparisons between different filter sizes, filter shapes, and tomography binwidths, using purity and completeness as indicators. \citet{Jaroszynski_Kostrzewa_2010} studied the mass-convergence correspondence in the noiseless case. \citet{Li_2011} used wavelets to filter the convergence maps and studied the sensitivity of peaks to $\wZero$. The impact from the halo concentration has been addressed by \citet{King_Mead_2011}, while \citet{Schmidt_Rozo_2011} compared different filters mentioned in the literature. \citet{VanderPlas_etal_2012} applied inpainting to the convergence maps and discovered that spurious peaks were reduced by a factor of 3. \citet{Romano_etal_2012} provided a simple cluster-count forecast for the Euclid mission.

Some works yielded constraints from data. \citet{Dahle_2006} became the first to reconstruct the cluster mass function with \acro{WL} and to constrain $\Sigma_8$ from it. This study was also extended to the constraint on the neutrino mass by combining the \acro{WL} mass function with \acro{CMB}, supernovae of type Ia (\acro{SNIa}), and galaxy clustering \citep{Kristiansen_etal_2007}. 

Concerning Fisher analysis, readers find estimates on $\OmegaM$, $\sigEig$, $\wZero$ and $w_a^\mathrm{de}$ \citep{Shapiro_Dodelson_2007}, similar parameters with tomography \citep{Takada_Bridle_2007}, a joint analysis with halo abundance, primordial non-Gaussianity \citep{Dalal_etal_2008}, another joint analysis with galaxy and halo clustering \citep{Yoo_Seljak_2012}, estimates on $f(R)$ parameters \citep{Cardone_etal_2013}, on $M$-$c$ relation parameters \citep{Mainini_Romano_2014, Cardone_etal_2015}, and $\LCDM$ parameters using tomography \citep{Martinet_etal_2015}. 

Also, \acro{WL}-selected clusters have been investigated in several surveys, e.g. from the Subaru Suprime-Cam data \citep{Miyazaki_etal_2002}, the first Focal Reducer Spectrograph (\acro{FORS1}) field of the Very Large Telescope (\acro{VLT}) \citep{Hetterscheidt_etal_2005}, the \acro{DLS} data \citep{Wittman_etal_2006}, the \acro{CFHT} Legacy Survey Deep field (\acro{CFHTLS}-Deep) \citep{Gavazzi_Soucail_2007}, the Bonn lensing-optical-X-ray catalogue \citep{Dietrich_etal_2007}, the Garching-Bonn Deep Survey (\acro{GaBoDS}) \citep{Schirmer_etal_2007}, the \acro{CFHTLS}-Wide \citep{Shan_etal_2012}, the \acro{HSC} survey \citep{Miyazaki_etal_2015}, and the \acro{KiDS} data \citep{Viola_etal_2015}.

\subsubsection{Summary}

In this chapter, weak gravitational lensing has been presented considering both theoretical and observational aspects. 

Under the thin-lens approximation and the Born approximation, we see how weak lensing can be presented with two quantities, the convergence $\kappa$ and the shear $\gamma$, from a mass density distribution. We also derive the expected lensing signal for a particular case of cluster lensing.

Then, to reconstruct the lensing signal, an unbiased shear estimator is introduced. To maximize the extraction of cosmological information, we can not be satisfied with two-point statistics, and peak counts seem to be a good candidate to investigate.

Lensing faces numerous observational challenges, such as shape measurement, shape noise, masking, and source redshifts. Also, modelling all physical effects is not trivial. The mass-sheet degeneracy, intrinsic alignment, baryonic physics, and the nonlinear matter spectrum are further issues.

At the end, I provide a review on studies about weak-lensing peak counts, highlighting that methodologies vary as the purposes, cosmology-oriented or cluster-oriented, are different. In the next chapter, I would like to invite readers to address the core issue of the thesis: peak-count modelling.

\clearpage
\thispagestyle{empty}
\cleardoublepage


\chapter{Peak-count modelling}
\label{sect:modelling}
\fancyhead[LE]{\sf \nouppercase{\leftmark}}
\fancyhead[RO]{\sf \nouppercase{\rightmark}}

\subsubsection{Overview}

Although peak counts are straightforward to compute from a data-processing point of view, predicting them for a given cosmological model and a parameter set is more challenging. The aim of this chapter is to propose a solution to this. First, I will point out the difficulties related to peak-count modelling. Then, the focus will be put on a new method to model lensing peak counts and its pros and cons. The new model will be validated by $N$-body simulations. Finally, the last section will show a comparison with an analytical model. This chapter corresponds to \PaperI\ and some additional contents.

\section{Problematic of peak-count modelling}

The origin of the problem is that \textit{peak counts are not quantities that can be linked with a simple formula to any known analytical expression}\index{Peak-count modelling}. In the peak-count framework, this analytical support should be the halo mass function that cosmologists derive either from structure collapse theories or from $N$-body-simulation fits. Then, due to some complexity from halo geometry, the projection effect, shape noise, etc., the link between cosmology and observed lensing peaks is no longer trivial. How to properly model lensing peak counts taking into account realistic survey conditions? This is the question in which cosmologists are interested. Until now, three approaches have been proposed: (1) analytical models, (2) modelling using $N$-body simulations, and (3) a fast stochastic forward approach developed in this work.

\subsubsection{Drawbacks of analytical models}

The description of analytical solutions is as follows. Take a noisy convergence map. The idea is to consider galaxy shape noise as a Gaussian random field and lensing signal as a ``foreground''. In random field theory, the probability to have a local maximum of a given level can be well described as a function of the noise level and the value of the foreground. Therefore, constructing a \textit{peak function}\index{Peak function} (peak number density as a function of \acro{S/N} value) becomes possible. This approach has been adapted by \citet{Maturi_etal_2010, Maturi_etal_2011} and \citet{Fan_etal_2010}. The differences between two models are subtle. \citet{Fan_etal_2010} work directly on a convergence map while \citet{Maturi_etal_2010} extract peaks from a shear catalogue using aperture mass. \citet{Fan_etal_2010} define local maxima as peaks while \citet{Maturi_etal_2010} count \textit{contiguous areas exceeding a given threshold}. The remaining settings are similar.

In spite of providing explicit expressions for the peak function, analytical models have some harmful drawbacks. When realistic conditions are taken into account, the performance of these models are strongly limited. For example, how to model peak counts when a part of the data is masked? What about the bias from photometric redshifts (\acro{photo-$z$}) or the errors from shape measurement? The impact from these effects might not be minor for peak-count observables. A tentative approach for studying masking effects has been done by \citetalias{Liu_etal_2014} (\citeyear{Liu_etal_2014}). They improve on the model of \citet{Fan_etal_2010} with a two-regime strategy for dealing with masking effects. Nevertheless, modelling realistic survey settings and errors in general still remains unsolved for analytical solutions.

Another drawback is the difficulty to include additional cosmological or astrophysical features. The most intriguing topic among these is intrinsic alignment (\acro{IA}, see \sect{sect:lensing:challenges})\index{Intrinsic alignment}. For analytical models, the presence of \acro{IA} is interpreted as additional noise which is potentially non-Gaussian. This invalidates the peak function obtained from a Gaussian field and thus breaks down the models. An attempt to model the impact from \acro{IA} on peaks has been proposed by \citet{Fan_2007}. That work simply considers the Gaussian part of \acro{IA}, so that the observed variance becomes the sum of the original variance and the corrected one from \acro{IA}. However, this is strongly unsatisfactory since (1) \acro{IA} does not necessarily enhance the noise variance and (2) two totally opposite \acro{IA} variations might lead to the same effect in this consideration. Until now, the impact of \acro{IA} on \acro{WL} peak counts has not been studied yet, and no sophisticated solution for \acro{IA} features has been proposed for analytical peak-count models. Similar to \acro{IA}, the baryonic effects, influencing the massive core of clusters, could introduce biases to peak-count models. \citet{Osato_etal_2015} have examined this effect and have found that the constraint for $\wZero$ can be offset up to 0.061, which is $\sim 1\sigma$ for a \acro{HSC}-like 1400-deg$^2$ survey. Neglecting these extensions could result in large systematic biases. Taking them properly into account is crucial for peak studies.

The last disadvantage is the fact that analytical models count on external simulations to estimate statistical uncertainty. The propagation of errors from observational effects, such as \acro{photo-$z$} and masking, and scatters due to some stochastic processes, for example halo triaxiality and concentration dispersion, cannot be quantified by analytical calculations alone in general. In cosmological contexts, models of this kind require $N$-body simulations for estimating the covariance matrix of observables under the assumption of a Gaussian likelihood. At the end of the day, the total process from the theoretical basis to parameter constraints still requires considerable computational resources. As a result, analytical models seem to be neither accurate nor efficient.

\subsubsection{What about our old good friend, $N$-body simulations?}

Alternatively, a straightforward approach for modelling peak counts is to use $N$-body simulations. These runs simulate structure formation of the Universe throughout the time evolution, and provide simultaneously the prediction and the statistical variability of cosmological observables. Survey conditions and additional astrophysical features can be taken into account directly, since this is a forward modelling approach. However, $N$-body simulations have an embarrassing feature which is their large computation time. Depending on computational resources and simulation settings, the time cost for creating a $N$-body lightcone can vary between a day and several months. Repeating this process for a wide range of cosmological parameters is extremely expensive, especially when the number of parameters increases. This so-called \textit{curse of dimensionality}\index{Curse of dimensionality} makes the grid-point evaluation of likelihood using $N$-body simulations practically impossible.

Nowadays, different sampling techniques, such as Markov Chain Monte Carlo (\acro{MCMC}) and population Monte Carlo (\acro{PMC}), are widely used for accelerating the constraining process. With these techniques, the reasonable number of parameter sets to evaluate is no longer an exponential function of the parameter dimension. However, the time cost for $N$-body runs still remains very harmful. Cosmologists usually need to make a tradeoff between several factors: the dimension of parameters, the prior range, the resolution of the parameter space, statistical precisions, the resolution of simulations, and computing power. For example, the study of \citetalias{Liu_etal_2015} (\citeyear{Liu_etal_2015}) using $N$-body simulations adopts a strategy with a large \acro{3D} ($\OmegaM, \sigEig, \wZero$) prior but a poor resolution and a poor statistical precision (one run per parameter set).

The drawbacks described above motivates the development of a new model in this thesis work. I will present this new model in the next section.

\section{A new model to predict weak-lensing peak counts}
\label{sect:modelling:LK}

\subsection{Description}
\label{sect:modelling:LK:description}

The idea of the new model is to replace the time-costly $N$-body physical process by a ``shortcut'': an accelerated process which provides the same observables. A possible solution is semi-analytical computations under some assumptions. In this case, the new model preserves some characteristics of $N$-body runs, forward and probabilistic, and gets rid of the considerable time cost. This approach is called \textit{fast stochastic forward modelling}\index{Fast stochastic forward (\acro{FSF}) modelling} (\acro{FSF} modelling). Clearly, this principle is not reserved only for \acro{WL} peak counts. As long as the shortcut can be constructed, applying this fast modelling in other contexts is straightforward.

\begin{figure}[tb]
	\centering
	\includegraphics[width=0.65\textwidth]{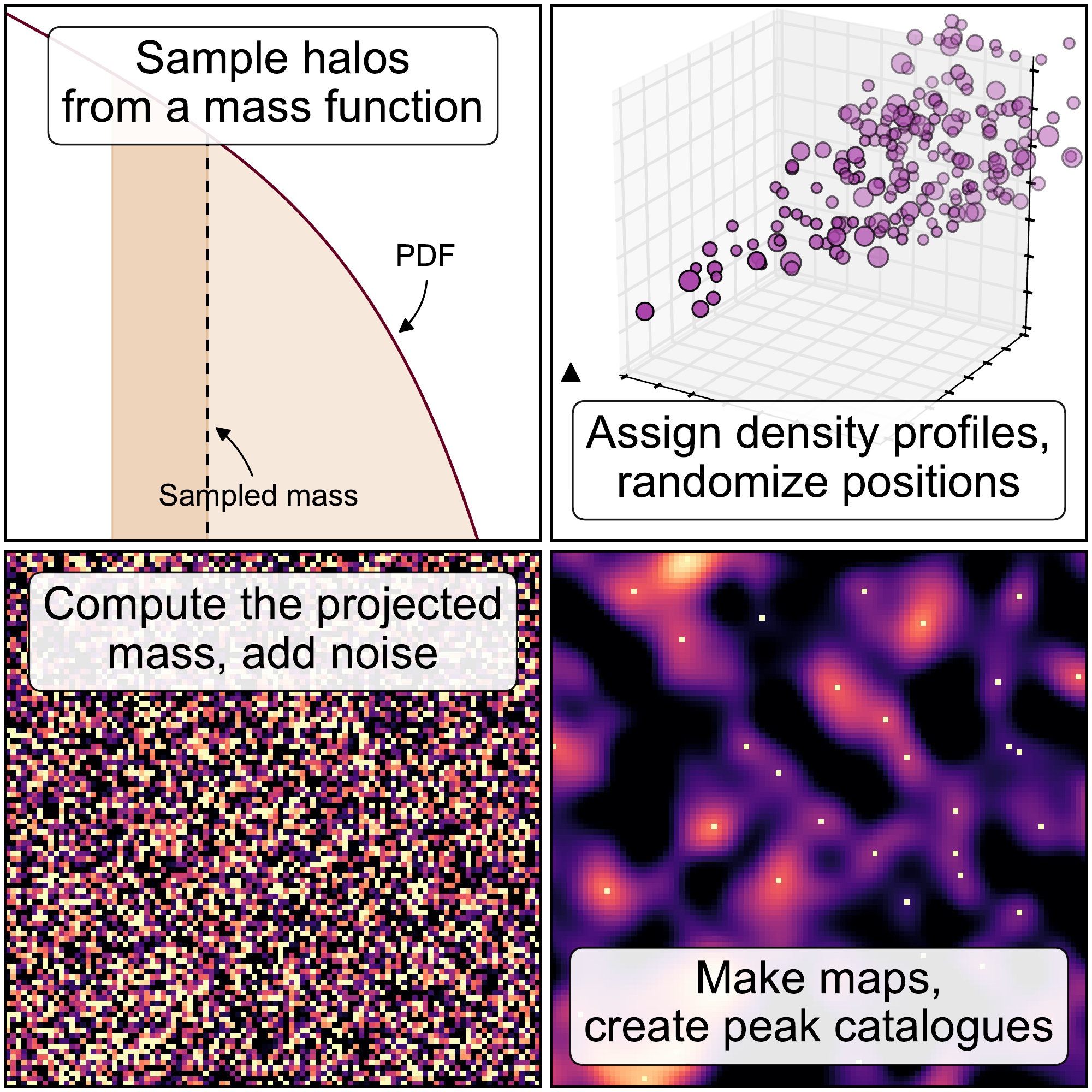}
	\caption{Illustration of various steps of our model to predict weak-lensing peak counts.}
	\label{fig:modelling:LK_model_1}
\end{figure}

The fast stochastic forward model to predict \acro{WL} peak counts developed in this thesis adopts steps as follows:
\begin{itemize}
	\item sample halos from a mass function;
	\item assign density profiles, randomize their position;
	\item compute the projected mass, add noise;
	\item make maps, create peak catalogues.
\end{itemize}
The two first steps are called ``fast simulations'' in \PaperI, because these processes correspond to the $N$-body physical process. Actually, a more intuitive name for fast simulations is the ``bubble model'' \index{Bubble model}. Since in most cases halos are considered to be spherical, a fast simulation box is actually a bubble chamber, or more precisely, a bubble lightcone with respect to a redshift distribution. The requirement for creating these is a mass function and a halo mass profile. Readers may notice that sampled halos might overlap in the \acro{3D} simulated boxes. We do not look for detecting or excluding these cases. The projected mass is computed by summing up contributions from each halo on the line of sight.

The model is implemented by our public code \textsc{\acro{Camelus}}\index{\textsc{\acro{Camelus}}} \footnote{Available at \url{http://github.com/Linc-tw/camelus/}.}, standing for \textit{Counts of Amplified Mass Elevations from Lensing with Ultrafast Simulation}. The cosmological computation in \textsc{\acro{Camelus}} is performed by \textsc{\acro{Nicaea}} \citep{Kilbinger_etal_2009} \footnote{Available at \url{http://www.cosmostat.org/software/nicaea/}.}.

\begin{figure}[tb]
	\centering
	\includegraphics[width=0.5\textwidth]{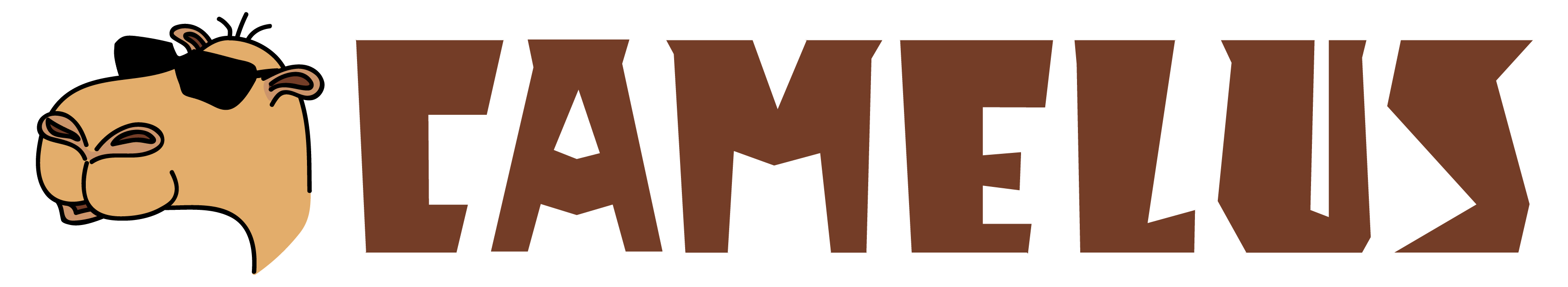}
	\caption{Logo of \textsc{\acro{Camelus}}.}
	\label{fig:modelling:Logo_Camelus_fig_name_horizontal}
\end{figure}

By construction of the fast simulations, one may find out that this model is based on two major hypotheses: 
\begin{itemize}
	\item diffuse, unbound matter, for example cosmological filaments, does not significantly contribute to peak counts;
	\item the spatial correlation of halos has a minor influence and can be ignored for peak counts.
\end{itemize}
Implicitly, this framework also admits the universality of the mass function and halo models. The first assumption is motivated by the fact that unbound mass is less concentrated than bound structures. The second one is supported by \citet{Marian_etal_2010}, who have proven that correlated structures affect number and height of peaks by only a few percent. Similarly, \citet{Kainulainen_Marra_2011} have found that assuming a stochastic distribution of halos can lead to accurate predictions of the convergence probability distribution function (\acro{PDF}). These studies motivate the use of halo one-point-only statistic in our model, while a detailed examination of above will be presented in \sect{sect:modelling:validation}.

Readers may understand now why fast simulations are good candidates to substitute the $N$-body physical process. Actually, most peaks with \acro{S/N} values (noted as $\nu$) lower than 3--4 are dominated by galaxy shape noise (see \fig{fig:modelling:peakHist_smallField}), while high peaks are basically true structures: either a single massive cluster or several halos on the same line of sight, and \citet{Yang_etal_2011} suggest that the first scenario is more plausible. They have found that $\sim$74\% of peaks with $\nu\geq 4.8$ are in the single-halo case. As a result, cutting off unbound objects and low-mass halos will not modify high-peak counts, and this helps drastically reducing the computation time. 

On the other hand, although low peaks are dominated by noise, \citet{Yang_etal_2013} have shown that they constrain cosmology better than high peaks due to the fact that cosmological variations are larger than statistical fluctuations. However, since our model introduces a lower mass cutoff during the sampling step, the missing low-mass halos might create a substantial impact on the regime of low peaks. Due to this reason, low-$\nu$ peaks are not included in the first consideration.

\subsection{Advantages}

What are the interests of an \acro{FSF} model like ours? What do we gain by taking the fast simulation? The improvements from it can be summarized by three characteristics: fast, flexible, and full \acro{PDF} information.
\vspace*{-2.5ex}

\paragraph{Fast} On a single-central-processing-unit (single-\acro{CPU}) machine, it only requires several seconds for creating a 25-deg$^2$ field. This is without using parallel computing such as message passing interface (\acro{MPI}) or graphics processing unit (\acro{GPU}) programming. The main cost is the computation of the projected mass, or nonlinear filtering method if it is used. 
\vspace*{-2.5ex}

\paragraph{Flexible} Including observational effects, such as masking and \acro{photo-$z$} errors, and additional features, such as \acro{IA} and baryonic feedbacks, is straightforward, since the model adopts a forward approach. A detailed discussion for possible extensions is presented in \sect{sect:modelling:extensions}.
\vspace*{-2.5ex}

\paragraph{Full PDF information} A stochastic model provides naturally an empirical \acro{PDF} of observables, which is supposed to contain all statistical information. This requires a large amount of model realizations, difficult to carry out for $N$-body runs. The fact that a \acro{FSF} model is fast makes the \acro{PDF} reconstruction feasible in practice.

With these three characteristics, constraining parameters becomes less restricted. For example, when the parameter dimension remains small, a grid-point evaluation of the likelihood in parameter space with high resolution will be feasible. Another example is the hypothesis of the constant covariance that almost all studies make for their Gaussian likelihood. A \acro{FSF} model can easily test the impact of the cosmology-dependent-covariance effect (\acro{CDC}, see \sect{sect:constraint:CDC}). It is also possible to apply non-parametric constraint methods such as $p$-value tests or approximate Bayesian computation (\acro{ABC}). A more detailed study will be presented in Chaps. \ref{sect:constraint} and \ref{sect:ABC}.

\section{Validation by $N$-body simulations}
\label{sect:modelling:validation}

\subsection{Description of $N$-body and ray-tracing simulations}
\label{sect:modelling:validation:NBody}

Putting calibration of systematics aside, here we attempt to validate our \acro{FSF} modelling by comparing it to $N$-body simulations. The $N$-body simulations used are the simulations from Dark Energy Survey (\acro{DES}) Blind Cosmology Challenge, called ``Aardvark''. They have been generated by \textsc{LGadget-2}, a \acro{DM}-only version of \textsc{\acro{Gadget}-2} \citep{Springel_2005}. The cosmological parameters for Aardvark form a \acro{WMAP}-like \acro{$\LCDM$} cosmology, with $\OmegaM = 0.23$, $\OmegaL = 0.77$, $\OmegaB = 0.047$, $\sigEig = 0.83$, $h = 0.73$, $n_\rms = 1.0$, and $\wZero = -1.0$. The \acro{DM} halos in Aardvark were identified using the \textsc{\acro{Rockstar}} friends-of-friends (\acro{FOF}) finder code \citep{Behroozi_etal_2013}. 

Ray-tracing simulations for Aardvark have been performed with \textsc{\acro{Calclens}} \citep{Becker_2013}. This is an algorithm yielding multi-plane lensing computation on a curved sky. Galaxies of Aardvark have been generated with \textsc{Addgals} (by M. Busha and R. Wechsler).

I use a \acro{HEALPix} \citep{Gorski_etal_2005} patch with \texttt{nside} =~2 of the halo catalogue, which is 860 deg$^2$ (large field), and a subpatch with \texttt{nside} =~8 of the galaxy catalogue and ray-tracing data, which is only 54 deg$^2$ (small field). In order to use as large field of view as possible, the map has been created in \acro{HEALPix} pixels. The \texttt{nside} of each pixel is 16384. Fast simulations from our model have also been generated in the same fields (see \sect{sect:modelling:validation:methodology}).

\subsection{Methodology}
\label{sect:modelling:validation:methodology}

\subsubsection{Comparison design}

In order to understand the impact from different hypotheses mentioned in \sect{sect:modelling:LK:description}, two intermediate steps have been included for comparison. This ends up with four different settings in total:
\begin{itemize}
	\item Case 1: full $N$-body runs,
	\item Case 2: replacing $N$-body halos with \acro{NFW} profiles of the same mass,
	\item Case 3: randomizing angular positions of halos from Case 2, 
	\item Case 4: our model.
\end{itemize}

Readers can notice that, first, halos of Case~1 are from the $N$-body process and recovered by a \acro{FOF} or spherical overdensity (\acro{SO}) finder, while halos of Cases~2, 3, and~4 follow \acro{NFW} profiles. Second, the position correlation between halos is conserved in Cases~1 and~2, while no correlation is retained for Cases~3 and~4. Last, Cases~1, 2, and~3 follow the same mass function as $N$-body simulations, while the one of Case~4 is an analytical model. At the end of the day, by comparing Cases~1 and~2, one studies the impact from unbound matter and halo asphericity. The comparison between Cases~2 and~3 should be interpreted as the impact of halo position correlations. And the difference between Cases~3 and~4, if there is any, corresponds to verify if the mass function from the $N$-body is well described by the analytical model. 

\subsubsection{Settings for Case 1}

This is tested in a simplified scenario, without masking, \acro{IA}, and baryonic physics. Also, only a slice of sources is considered. Such a configuration has widely been adopted in the literature for preliminary studies, including \citet{Hamana_etal_2004}, \citet{Fan_etal_2010}, and \citet{Yang_etal_2011}.

For Case~1, I take galaxies with redshifts between 0.9 and 1.1. Lensing in this case has already been described in \sect{sect:modelling:validation:NBody}. No inversion is needed since $\kappa$ has been directly obtained from ray-tracing. To make a grid map, I use the convergence values sampled by galaxies at this source slice, and carry out a bilinear interpolation. The ``truth'' map is then obtained by evaluating interpolated field values on \acro{HEALPix} pixel positions. In reality, this ``truth'' can never been recovered because of galaxy shape noise, thus the comparison should always be done between noisy fields. 

Shape noise can be modelled as a Gaussian random field $n(\btheta)$ added to convergence, giving the noisy field $\kappa_n(\btheta)$ as
\begin{align}
	\kappa_n(\btheta) = \kappa(\btheta) + n(\btheta).
\end{align}
For $n(\btheta)$, the noise level depends on the smoothing function. If we bin galaxies into map pixels, the smoothing function is just a top-hat filter based on the pixel support. In this case, the noise level $\sigma_\pix$ for a pixel of size $A_\pix$ is given by \citep{VanWaerbeke_2000}
\begin{align}
	\sigma_\pix^2 = \frac{\sigma_\epsilon^2}{2}\frac{1}{n_\gala A_\pix},
\end{align}
where $\sigma_\epsilon$ is intrinsic ellipticity dispersion and $n_\gala$ is galaxy number density. The intrinsic dispersion is set to $\sigma_\epsilon = 0.4$ which corresponds to a \acro{CFHTLenS}-like survey \citep{Fu_etal_2008}, and the value of $n_\gala$ will be discussed further in this section. 

Of course, galaxy binning is not necessary in general. The advantage of binning is that when the grid is Cartesian, the computation can be significantly sped up using the fast Fourier transform (\acro{FFT}). The speed is one of the priority of our algorithm, so this setting is always used. This is why the same spirit is kept even if the grid is not Cartesian here.

After adding random noise to pixels, the map is smoothed with a Gaussian kernel $W(\btheta)$ defined as
\begin{align}\label{for:modelling:Gaussian_kernel}
	W(\btheta) = \frac{1}{\pi\theta_\rmG^2} \exp\left(-\frac{\theta^2}{\theta_\rmG^2}\right),
\end{align}
with $\theta_\rmG = 1\ \arcmin$. One can then defined the smoothed convergence as
\begin{align}
	K_N(\btheta) \equiv (\kappa_n\ast W)(\btheta) = \int\rmd^2 \btheta'\ \kappa_n(\btheta-\btheta')W(\btheta'),
\end{align}
where $\kappa_n$ and $K_N$ denote noisy fields before and after smoothing respectively. In the validation study, this is performed by computing direct convolution. After smoothing, the noise level is expected to be (\citealt{VanWaerbeke_2000}; \sect{sect:filtering:linear:noise})
\begin{align}
	\sigma_\noise^2 = \frac{\sigma_\epsilon^2}{2}\frac{1}{2\pi n_\gala\theta_\rmG^2}.
\end{align}
The \acro{S/N} of each pixel is then defined as $\nu \equiv K_N / \sigma_\noise$. This quantity has been used for selecting peaks.

In practice, after obtaining the interpolated ``truth'' map, I add a random sampled value from $\mathcal{N}(0, \sigma_\pix^2)$ to each pixel as noise. Smoothing is done by computing a direct convolution, and peaks are identified as pixels such that the \acro{S/N} is larger than all eight neighbors. Some most outer pixels are abrogated to avoid the border effect. The width of this area corresponds approximately to $2.2\times\theta_\rmG$, which is three times the standard deviation of $W$. The field size for Case 1 is 54 deg$^2$.

\subsubsection{Settings for Case 4, a ``minimal pipeline''}

I would like to skip Cases~2 and 3 for the moment and come back to them later. For Case~4, peak counting is processed with a ``minimal pipeline''\index{Minimal pipeline}. I first generate fast simulation boxes. These boxes are actually composed of redshifts slices. In this validation test, slices are set to 10 equal bins from $z=0$ to 1. Halo masses are sampled from the mass function proposed by \citet[][Eq. \eqref{for:structure:massFct_J01}]{Jenkins_etal_2001}. The sampling mass range is $[M_\minn, M_\maxx]=[\dix{12}, \dix{17}]~\Msol/h$, and this implies that the total mass $M_\tot$ of each slice is not $\rho_\crit\cdot\OmegaM\cdot V$, but
\begin{align}
	M_\tot = V\int_{M_\minn}^{M_\maxx} \rmd\log M\ \cdot M\cdot\frac{\rmd n(z, \lessM)}{\rmd\log M},
\end{align}
where $V$ is the volume of the slice. Halo profiles are \acro{NFW} ones (inner slope $\alpha=1$) truncated at $r_\vir$. They are parametrized by mass and concentration. Here, I assume the mass-concentration relation as \for{for:structure:M_c_relation} and set $c_0=8$ and $\betaNFW=0.13$.

Sources are considered to be regularly spaced on a \acro{HEALPix} grid and fixed at a single redshift $z_\rms=1$. The \acro{HEALPix} grid corresponds either to the halo catalogue patch (large field) or to the galaxy catalogue patch (small field). In both cases, galaxies are located at the center of pixels with \texttt{nside} =~16384. This results in $A_\pix = 0.046~\arcmin^2$ and $n_\gala = 21.7~\arcmin\invSq$. This is also the value of $n_\gala$ for Case 1.

\begin{figure}[tb]
	\centering
	\includegraphics[scale=0.65]{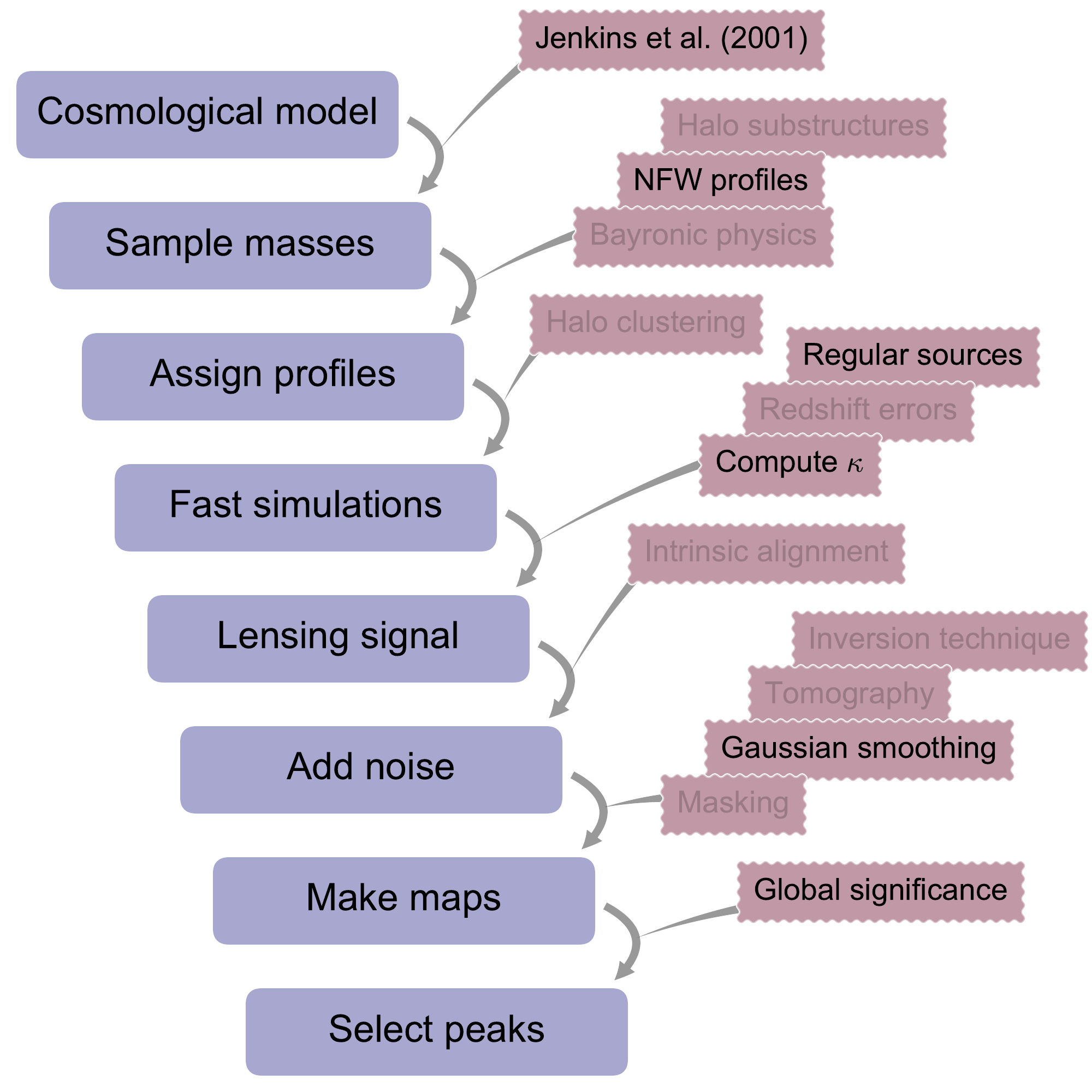}
	\caption{Diagram illustrating the processing pipeline for the validation test. Effects which are not considered are made faint. A more detailed description for all variations can be found in \sect{sect:modelling:extensions}.}
	\label{fig:modelling:LK_model_diagram_2}
\end{figure}

\begin{table}
	\centering
	\begin{tabular}{lcl}
		\hline\hline
		Parameter                               & Symbol            & Value\\
		\hline
		Lower sampling limit                    & $M_\minn$         & $\dix{12}~\Msol/h$\\
		Upper sampling limit                    & $M_\maxx$         & $\dix{17}~\Msol/h$\\
		Number of halo redshift bins            & -                 & 10\\
		\acro{NFW} inner slope                  & $\alpha$          & 1\\
		\acro{$M$-$c$} relation amplitude       & $c_0$             & 8\\
		\acro{$M$-$c$} relation power law index & $\betaNFW$        & 0.13\\
		Source redshift                         & $z_\rms$          & 1\\
		Intrinsic ellipticity dispersion        & $\sigma_\epsilon$ & 0.4\\
		Galaxy number density                   & $n_\gala$         & 21.7 arcmin$\invSq$\\
		Pixel size                              & $\theta_\pix$     & 0.215 arcmin\\
		Gaussian kernel size                    & $\theta_\rmG$     & 1 arcmin\\
		Noise level in a pixel                  & $\sigma_\pix$     & 0.283\\
		Noise level after smoothing             & $\sigma_\noise$   & 0.0242\\
		Small field size                        & -                 & 53.7 deg$^2$\\
		Large field size                        & -                 & 859 deg$^2$\\
		\hline\hline
	\end{tabular}
	\caption{List of parameter values adopted in this validation study.}
	\label{tab:modelling:parameters}
\end{table}

After setting lenses and sources, the lensing signal is calculated. For Case~4, this is done by summing up the projected mass along the line of sight following
\begin{align}\label{for:modelling:kappa_proj}
	\kappa_\proj(\btheta, w_\rms) \equiv \sum_{\halo\rms}\kappa_\halo(\btheta, w_\ell, w_\rms)
\end{align}
where $\kappa_\halo$ is given by Eqs. \eqref{for:lensing:G_kappa_TJ} and \eqref{for:lensing:kappa_halo_NFW}, and then I subtract the mean over the field at the end. The reason for this is that the projected mass from halos is always positive, while the true convergence can be negative in underdense regions. Recall that in \sect{sect:lensing:cluster:projection}, we have already seen that \for{for:modelling:kappa_proj} can be considered as the convergence only under an important condition: the ``empty space'' actually has the mass density of the background. Consequently, the calculation using \for{for:modelling:kappa_proj} without any correction leads to a $\kappa$ map from an universe in which the total mass is larger than what it should be for the input $\OmegaM$ value. In our model, instead of subtracting this additional mass, I simply subtract the mean of $\kappa_\proj$, i.e. considering $\kappa(\btheta) = \kappa_\proj(\btheta) - \overline{\kappa}_\proj$. The philosophy behind is to make a zero-mean convergence field an approximation of the true convergence field, which arises from a zero-mean density contrast on average. After obtaining the noiseless map, I add noise, smooth, and count peaks on \acro{S/N} as in Case~1.

\subsubsection{Settings for Cases 2 and 3}

For Cases~2 and~3, the processing is similar to Case~4 apart from fast simulations. In these two cases, I do not need to generate halos, but profiles are analytical, so the new lensing signal is computed by following Case~4. For Cases~2, 3, and~4, peaks are selected both from the large field and the small field, since these cases are not limited by the existing galaxy catalogue with ray-tracing information.

It is worth clarifying that $n_\gala = 21.7\ \arcmin\invSq$ is used for all cases. Thus, $n_\gala A_\pix$ is 1 automatically. Even if galaxy density of the sources in the redshift slice of [0.9, 1.1] from Case~1 is lower than $21.7\ \arcmin\invSq$, since the aim is to test different lens conditions under the same source configuration, the same noise level should be kept everywhere. I also consider that galaxies are distributed uniformly, so that $n_\gala$ is independent from position. As a result, here, $\nu$ is just a scaling of $K_N$. However, in general conditions, especially when masking is taken into account, $n_\gala$ becomes a local quantity which varies from one point to another. In this case, evaluating noise locally and counting peaks on true \acro{S/N} would be more authentic than scaling all $K_N$ to $\nu$ with the same factor.

For the small field, 8 independent noise maps are generated for analysis, and all cases with random processes (random position for Case~3 and fast simulation for Case~4) are carried out with 8 realizations. This ends up with averaging the result over either 8 or 64 peak histograms. For the large field, I generate 4 noise maps and 4 realizations instead, so the average is over 4 or 16 histograms. However, one should also notice that the large field is 16 times larger than the small field.

\subsection{Results}

\subsubsection{Small field, low-$\nu$ regime}

\begin{figure}[tb]
	\centering
	\includegraphics[width=0.8\textwidth]{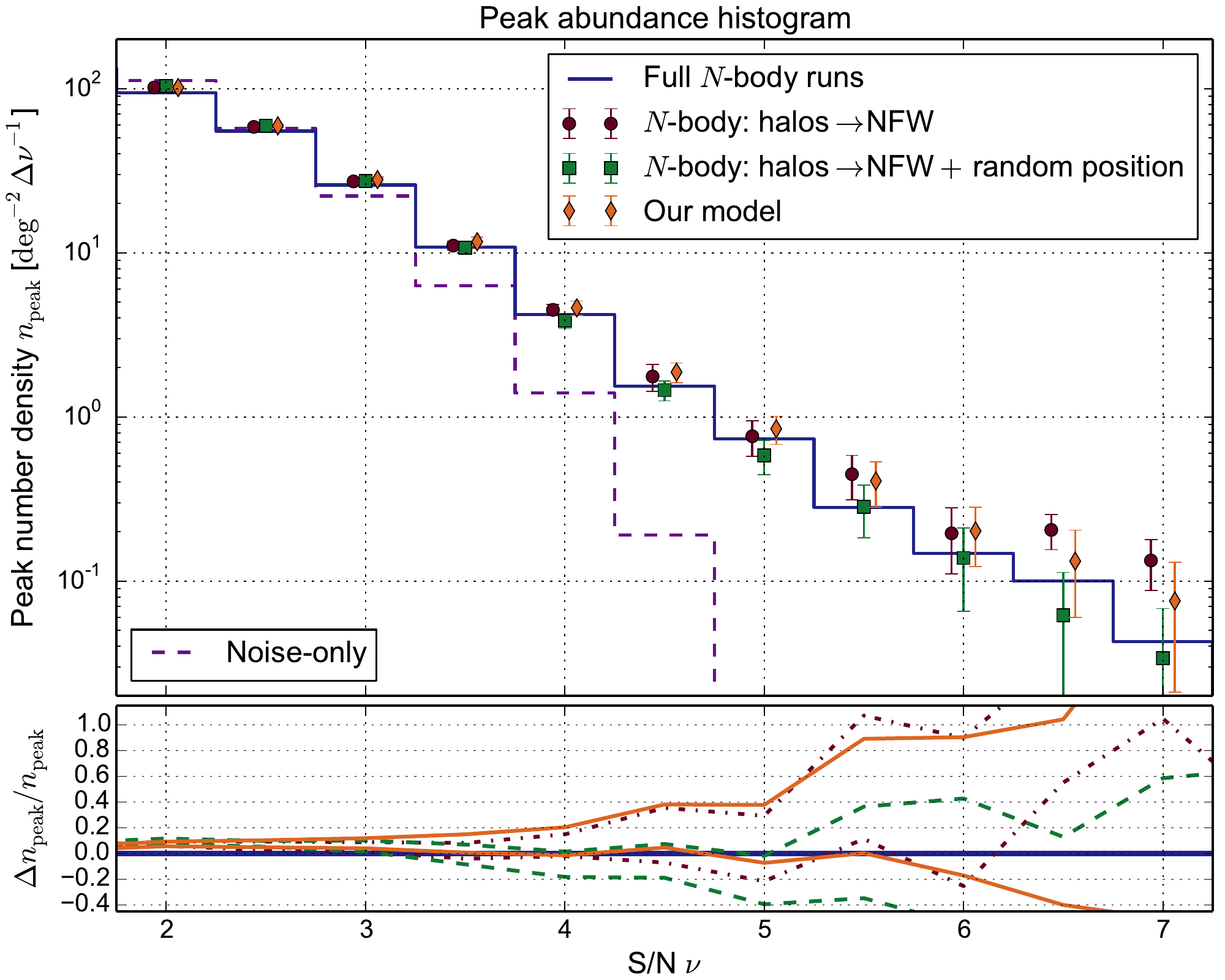}
	\caption{Comparison of the peak abundance from different cases evoked in \sect{sect:modelling:validation:methodology}. On the upper panel, blue solid lines: full $N$-body runs (Case 1); dark red circles: replacement of halos by \acro{NFW} profiles (Case 2); green squares: replacement of halos by \acro{NFW} profiles and randomization of halo angular positions (Case 3); orange diamonds: fast simulations, corresponding to our model (Case 4); purple dashed line: peaks from noise-only maps. On the lower panel, I display the upper and lower limits of error bars shifted with regard to Case 1. This refers to the standard deviation over 4 maps (dark red dash-dotted lines for Case 2) or 16 maps (green dashed lines for Case 3, orange solid lines for Case 4). The field of view is 54 deg$^2$.}
	\label{fig:modelling:peakHist_smallField}
\end{figure}

\figFull{fig:modelling:peakHist_smallField} shows the comparison between four cases on the small field. Only halos located in the 54-deg$^2$ subpatch are taken into account for analysis. The $x$-axis represents peak \acro{S/N}, denoted as~$\nu$. The $y$-axis stands for the peak function $n_\peak(\nu)$, which are histogram counts divided by the binwidth and the field area. We can discover that in the regime of low peaks with $\nu\leq 3.75$, $n_\peak(\nu)$ does not vary much between four cases. This is not surprising because noise is dominating in this range, as shown by the noise-only peak-count histogram (purple dashed lines in \fig{fig:modelling:peakHist_smallField}). 

The second observation from the low-$\nu$ regime is non-additivity. Combining peaks from cosmological structures and noise does not give the total count from the noisy field. This is because a pixel needs to ``cooperate'' with neighbors to become a local maximum and summing up two fields yields an ambiguous effect for this ``cooperation''. Peaks with $\nu\leq 2.75$ in \fig{fig:modelling:peakHist_smallField} show that appending structures to the noise field even decreases the peak counts.

The lower panel of \fig{fig:modelling:peakHist_smallField} gives the deviation of each case compared to Case~1. It shows that there exists a systematic overcount of 10\% in the low-$\nu$ regime. Several explanations for its origin are possible. First, it could come from the $\kappa$-mean subtraction. For example, if \acro{NFW} profiles truncated at $2r_\vir$ instead of $r_\vir$ are used, then $\overline{\kappa}_\proj$ would enhance and peaks will have a lower \acro{S/N}. Second, it could be the fact that a lower limit $M_\minn$ has been defined for halo mass sampling. Small halos might have an impact on low-peak counts. Finally, it could also be an effect caused by profiles, as I will discuss in the comparison Case 1-Case 2.

\subsubsection{Small field, high-$\nu$ regime, comparison Case 1-Case 2}

As the \acro{S/N} increases, the differences between cases become significant. We can observe that in the high-$\nu$ regime, replacing profiles enhances the peak counts while randomizing position introduces an opposite effect of a similar order of magnitude. Since peak counts at $\nu\gtrsim 4$ can barely contributed by unbound matter or small halos, the enhancement from Case~2 might be mainly caused by halo asphericity or triaxiality.

\begin{figure}[tb]
	\centering
	\includegraphics[width=0.8\textwidth]{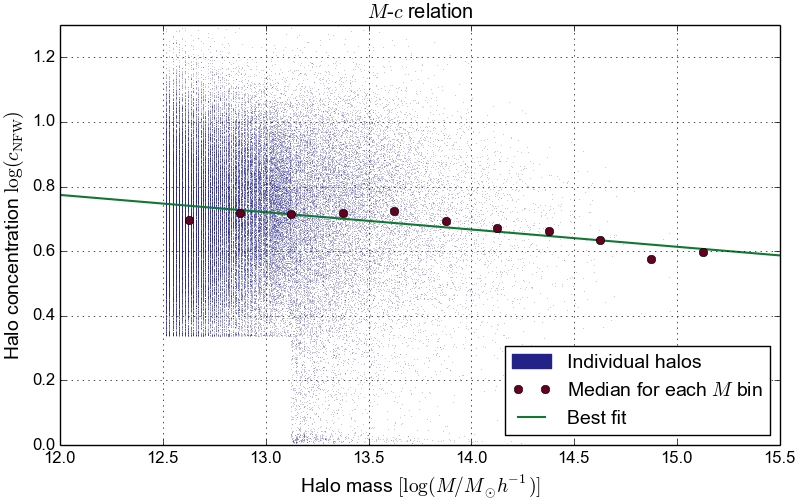}
	\caption{Scatter of the concentration for the Aardvark halos at redshift between 0.4 and~0.5. The concentration $c$ is derived from the ratio $r_\vir/r_\rms$, where $r_\rms$ is given by fitting an analytical \acro{NFW} profile to $N$-body particles in halos. For each mass bin in the logarithmic space, the value of the median of the concentration is drawn with dark red circles, and the best fit of the \acro{$M$-$c$} relation using \for{for:structure:M_c_relation} is presented by the green line.}
	\label{fig:modelling:M_c_relation}
\end{figure}

\begin{figure}[tb]
	\centering
	\includegraphics[width=0.8\textwidth]{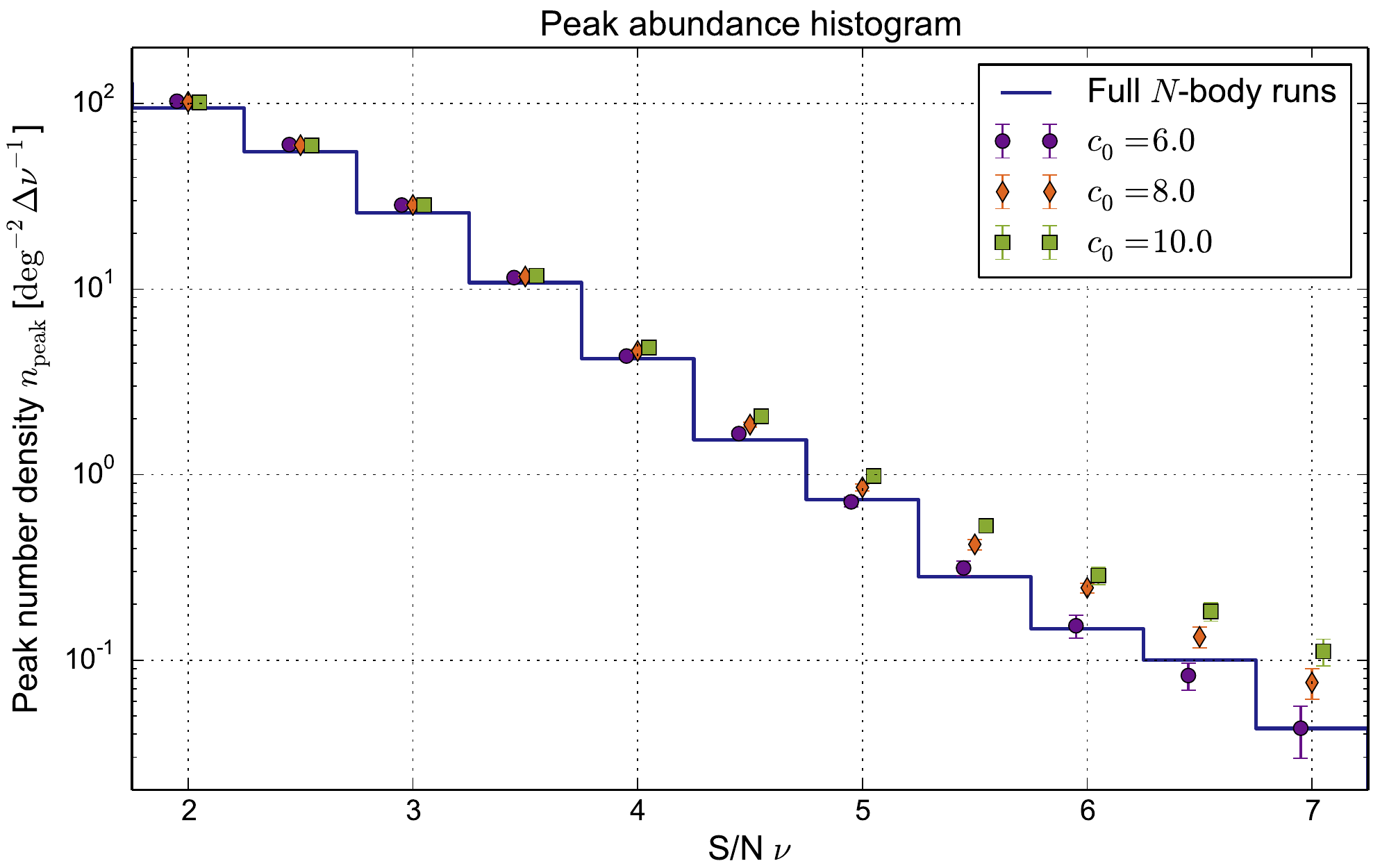}
	\caption{Peak abundance from our model for different $c_0$ values. This test has been carried out on the large field, so that orange diamonds are the same runs as Case 4 from \fig{fig:modelling:peakHist_largeField}, which will be mentioned later. This figure suggests that halo parameters and cosmological parameters can be highly degenerated together. The results from $N$-body runs (Case 1, blue solid lines), are only indicative, since they are derived from the small field and the fluctuation is large in principle.}
	\label{fig:modelling:peakHist_concentration}
\end{figure}

Another explanation comes from the mass-concentration relation (\acro{$M$-$c$} relation)\index{Mass-concentration relation}. Simulations have shown that for \acro{NFW} profiles, $M$ and $c$ are not linked with a tight relation. On the contrary, the scatter of $c(M)$ is rather large (e.g. \fig{fig:modelling:M_c_relation}). At a fixed $M$, the larger $c$ is, the higher the peak is observed. Thus, the modelling of the \acro{$M$-$c$} relation could have an impact on the comparison between Cases~1 and~2. \figFull{fig:modelling:peakHist_concentration} illustrates a simple test of dependency of our model on $c_0$. Note that the runs are done on the large field, whereas the blue lines still stand for Case~1 on the small field. While all the remaining settings are identical to Case~4, one can clearly observe different peak abundance from different $c_0$ values. Thus, it is indispensable to include halo parameters such as $c_0$ and $\betaNFW$ into the parameter set for future works focusing on constraints. In the validation test, the value of $c_0 = 8$ is suggested by fitting with Aardvark runs, assuming \for{for:structure:M_c_relation} as the \acro{$M$-$c$} relation. However, this fit suffers from some technical limits: a cutoff at $\log(c)\approx 0.32$ for $M\lesssim \dix{13.2}~\Msol/h$ and some excess at $c = 1$ for $M\gtrsim \dix{13.2}~\Msol/h$, as we can observe from \fig{fig:modelling:M_c_relation}.

One possible way to improve the bias induced by the \acro{$M$-$c$} relation is to take the individual concentration of $N$-body halos into account. Recall that in Case~2, only masses are retained from $N$-body runs. However, with this strategy, the model loses its universality. For future works, it could be interesting to set up a new intermediate case as current Case 2 with $N$-body concentrations. In this case, the impact from the \acro{$M$-$c$} relation can be tested.

\subsubsection{Small field, high-$\nu$ regime, comparison Case 2-Case 3}

By comparing Cases~2 and~3 from the upper panel of \fig{fig:modelling:peakHist_smallField}, I discover that position randomization decreases peak counts by 10\% to 50\% in the high-$\nu$ regime ($\nu\geq 3.75$). The explanation of this is that decorrelating angular positions breaks down the two-halo term, so that halos overlap less in projection on the field of view and decreases high-peak counts. Recall that \citet{Yang_etal_2011} have shown that high peaks ($\nu\geq 4.8$) are mainly contributed by one single halo. The same study has also shown that 12\% of total high-peak counts are derived by multiple halos. Apparently, these 12\% are not negligible, and this number agrees with the difference between Cases~2 and~3 within the bin of $4.75 \leq\nu\leq 5.25$, the dominant bin for peaks with $\nu\geq 4.8$.

\subsubsection{Small field, high-$\nu$ regime, comparison Case 3-Case 4}

\begin{figure}[tb]
	\centering
	\includegraphics[width=0.9\textwidth]{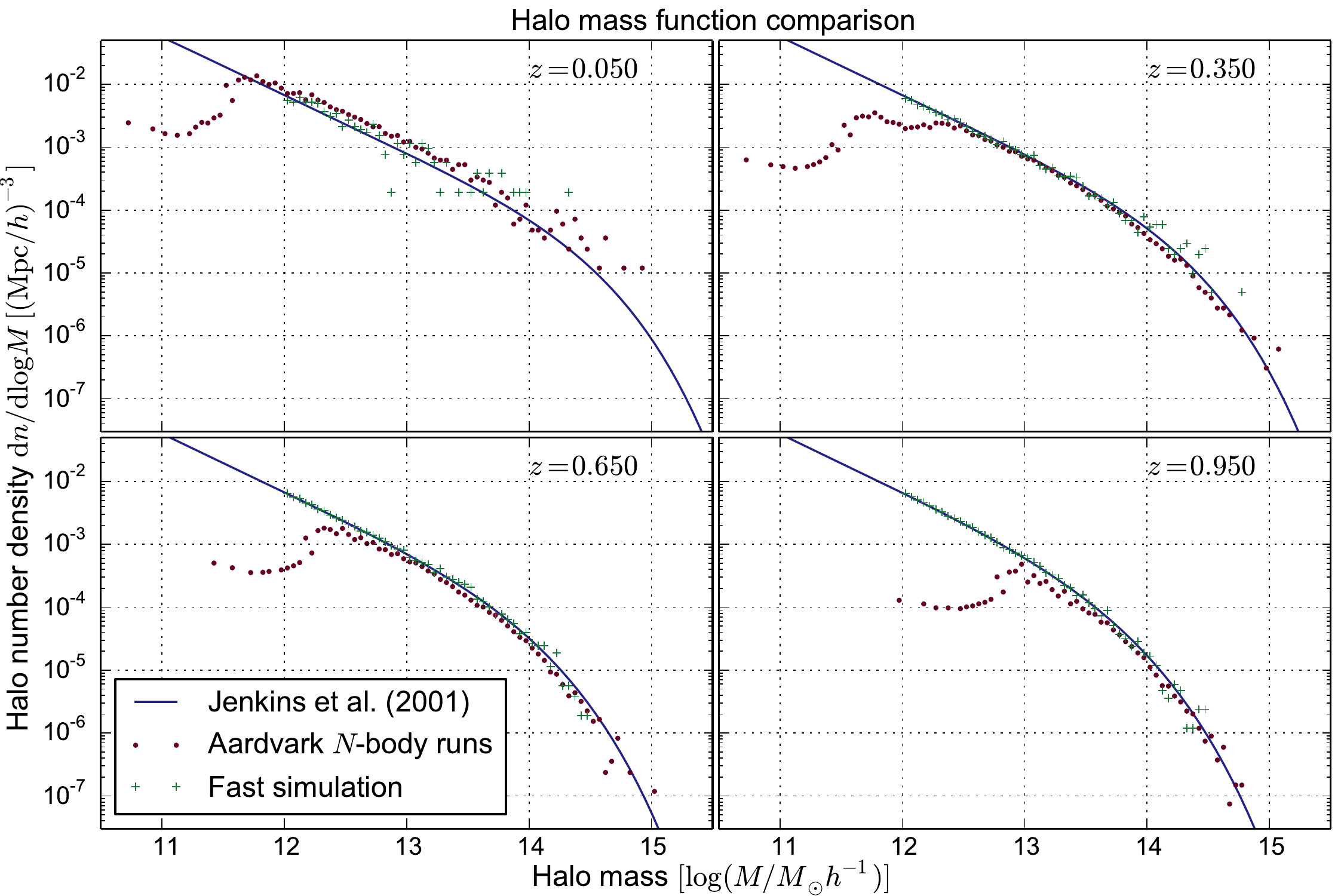}
	\caption{Mass function comparison between an analytical model and simulations at different redshifts. At low redshift, $N$-body runs count more halos than the analytical model. At high redshift, this effect is slightly reversed. In all cases, the low-mass part disagrees due to the mass resolution in the simulations.}
	\label{fig:modelling:massComp_aardvark_fast1}
\end{figure}

The last comparison from \fig{fig:modelling:peakHist_smallField}, between Cases~3 and~4, studies the impact of the mass function modelling. The result shows that more peaks have been found using the analytical model of \citet{Jenkins_etal_2001}, and the excess somehow compensates for the deficit from randomization. To explain this excess, we can take a look at \fig{fig:modelling:massComp_aardvark_fast1} which draws both mass functions at four different redshifts. While both mass functions agree well with each other, some discrepancies are still worth being pointed out.

First, at low redshift, $N$-body runs have more halos than the analytical model by a factor of a few. However, low redshift means poor lensing efficiency for halos, so they should not have strong impact. In addition, the number of peaks varies in the opposite way to the number of halos from Case~3 to Case~4. This is then not the origin of the excess.

\begin{figure}[tb]
	\centering
	\includegraphics[width=0.8\textwidth]{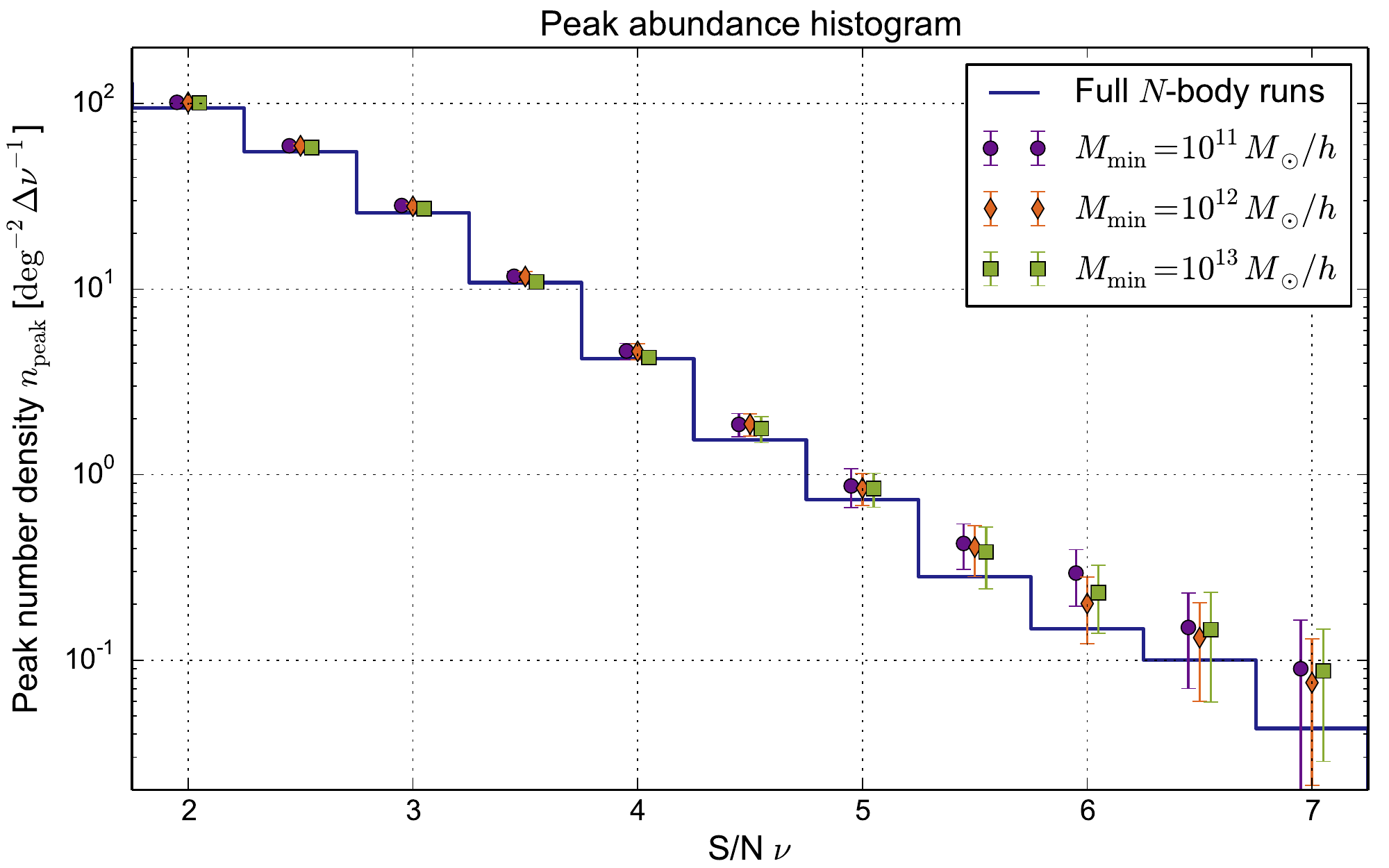}
	\caption{Peak abundance from our model for different $M_\minn$ values. This test has been carried out on the small field, because the lower $M_\minn$ is, the more halos present in fast simulations, which make the computation intractable. Orange diamonds are the same runs as Case 4 from \fig{fig:modelling:peakHist_smallField}, while blues lines stand for Case 1. Without quantitative comparisons, peak counts seem to be insensitive to low-mass halos.}
	\label{fig:modelling:peakHist_Mmin}
\end{figure}

The second difference is that at low $M$, the $N$-body mass function has an irregular tail for the reason of limited resolution, and depending on redshift, this irregular tail either contains more or less halos than fast simulation boxes. This hypothesis does not provide satisfactory explanation either because low-mass halos have a minor influence on high peak counts. To justify this, \fig{fig:modelling:peakHist_Mmin} shows the result from our model with three different values of $M_\minn$. Some hints suggest that deficits may occur for peaks with $3.25\leq\nu\leq 4.75$ when $M_\minn = \dix{13}~\Msol/h$, but for $M_\minn\lesssim\dix{12}~\Msol/h$, high-peak counts are not so sensitive to $M_\minn$. As a result, the irregular tail of the $N$-body mass function can be ignored.

The last difference is that at high redshift ($z=0.65$ and 0.95), the analytical model predicts more halos. Although the quantitative results between the halo excess ($\lesssim50$\% for high $M$ and high $z$) and the peak excess (50\%--100\% for $\nu\geq4.25$) do not match perfectly, this still remains the most probable explanation for the Case 3-Case 4 difference. Since the same massive halo can result in different \acro{S/N} due to the statistical fluctuation, there is no one-to-one relation between halo masses and peak heights. Therefore, predicting the quantitative impact of mass function on peak abundance is difficult. As a result, the difference between Cases~3 and~4 would be mainly the consequence of the mismatch between two mass functions.

\subsubsection{Large field}

\begin{figure}[tb]
	\centering
	\includegraphics[width=0.8\textwidth]{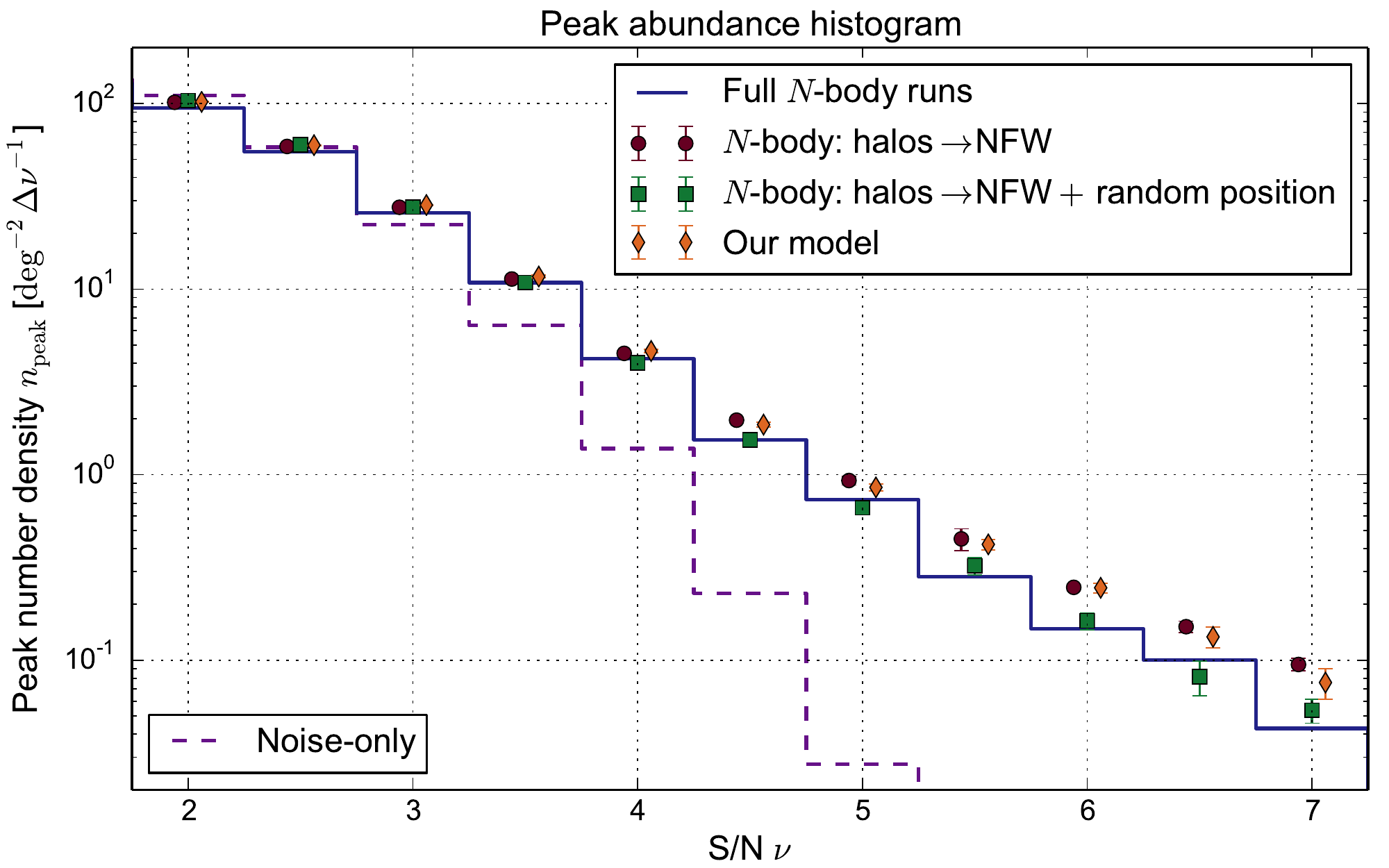}
	\caption{Similar to \fig{fig:modelling:peakHist_smallField}, comparison between four cases on the large field. Here, Case 1 is only indicative since it is still the result from the small field.}
	\label{fig:modelling:peakHist_largeField}
\end{figure}

The end of this section focuses on the same comparison on the large field. \figFull{fig:modelling:peakHist_largeField} shows Cases~2, 3, and~4 performed on the large field together with the Case~1 from the small field. The difference between sizes of fields is a factor of 16, so Case~1 (blue lines) is only indicative in \fig{fig:modelling:peakHist_largeField} since the uncertainty of blue lines is comparable to error bars in \fig{fig:modelling:peakHist_smallField}. Nevertheless, the result, with a weaker statistical uncertainty, confirms the all observations taken from \fig{fig:modelling:peakHist_smallField}. It suggests that Cases~1 and~3 are in a very good agreement. 

Overall, the model works very well for $\nu\leq4$. On the small field, systematic biases are smaller than the statistical uncertainty for all $\nu$. Although the validation study has shown that not all biases are not entirely understood, the variation from different cosmological predictions are found to be larger than biases (see \sect{sect:modelling:sensitivity}), hence are able to perform model selection. Also, in the next section, reader are going to see how our model can be improved. As far as the absolute bias is concerned, a larger set of $N$-body simulations would be required in the future.

\section{Model extensions}
\label{sect:modelling:extensions}

\begin{figure}[tb]
	\centering
	\includegraphics[scale=0.65]{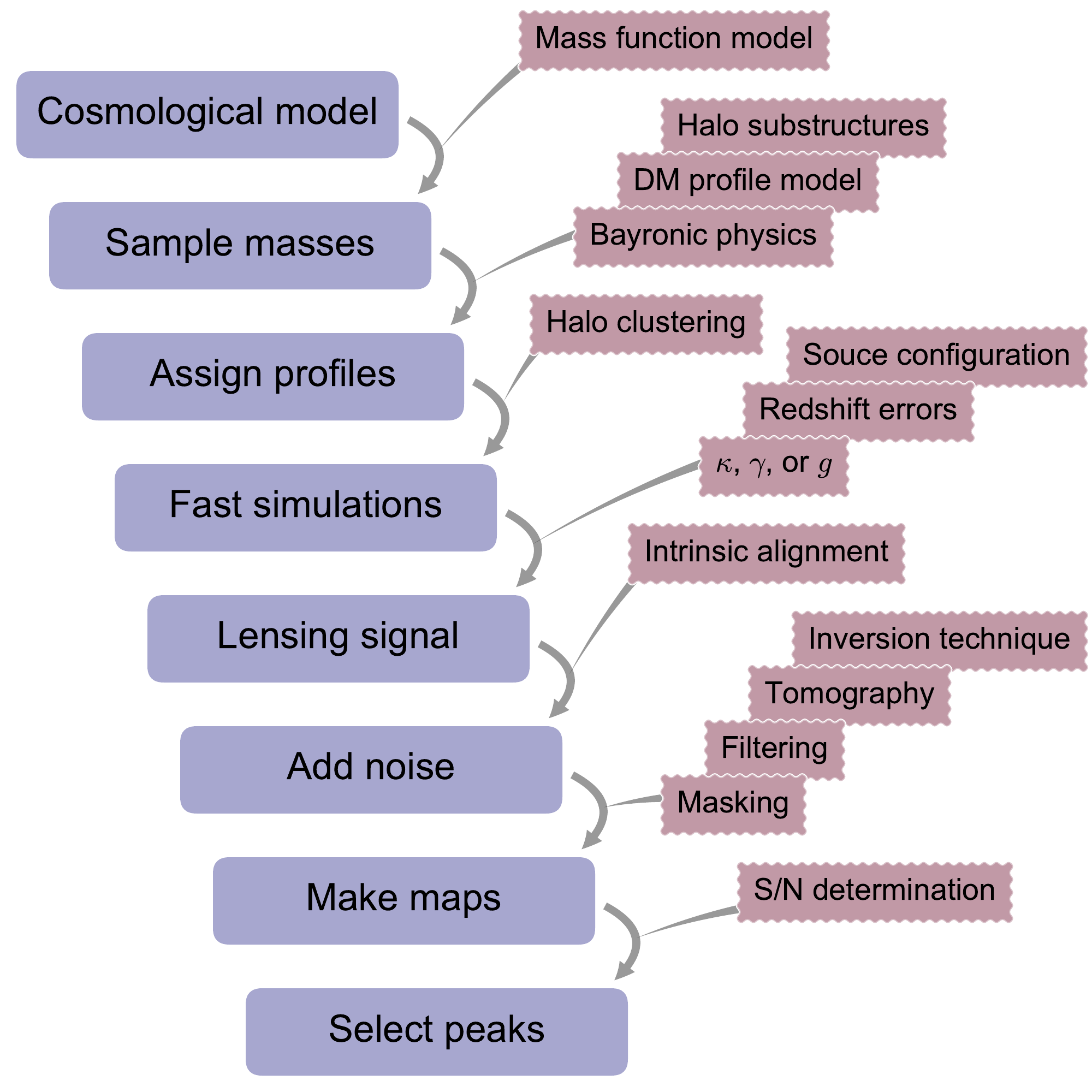}
	\caption{An indicative diagram listing possible extensions for our model. From each blue rectangle to another, red stamps are more physics to consider, observational effects, or processing techniques which would improve the understanding or the precision of \acro{WL} peak counts. Implicitly, this diagram implies the forward nature of our model.}
	\label{fig:modelling:LK_model_diagram_1}
\end{figure}

Thanks to the flexibility of our model, a wide range of options or extensions for different steps is available. These extensions could help us improving the physical modelling and extracting cosmological information in a more optimal way. Below is a non-exhaustive list of extensions that one may include in our model (see also \fig{fig:modelling:LK_model_diagram_1}).

\paragraph{Mass function} The mass function can be replaced by any distribution of our interests. Conversely, our model can also be used to discriminate mass function models. Nevertheless, this is not the aim of this thesis. In this thesis, the default input mass function is the one from \citet{Jenkins_etal_2001}.
\vspace*{-2.5ex}

\paragraph{Halo structures \& baryonic feedback} Throughout this work, the \acro{NFW} profiles (see \sect{sect:structure:profile:NFW}) have been chosen for halos. However, other density profiles such that the projected density is known are applicable. Therefore, providing an analytical expression of the projected mass would be the key for using the Einasto profile \citep{Einasto_1965} or elliptical profiles. The later one refers to studies related to halo triaxiality\index{Halo triaxiality} where the privileged direction of halos are taken into account. Beyond the dark-matter-only (\acro{DM}-only) profiles, those offered by baryonic feedback \citep{Yang_etal_2013, Osato_etal_2015} and substructures inside the halos could also be included in studies.
\vspace*{-2.5ex}

\paragraph{Halo clustering}\index{Halo clustering} It is possible not to break down the halo spatial correlation. The halo clustering information can be generated by using some fast algorithms such as \textsc{PTHalos} \citep{Scoccimarro_Sheth_2002}, \textsc{\acro{Pinocchio}} \citep[][see also \citealt{Heisenberg_etal_2011}]{Monaco_etal_2002}, and remapping \acro{LPT} \citep{Leclercq_etal_2013}. However, this is not the aim of the thesis. Zero angular correlation for halos is assumed throughout this work.
\vspace*{-2.5ex}

\paragraph{Galaxy redshift, redshift errors, tomography} One can either suppose a constant source redshift or generate sources from a distribution law. If a model for \acro{photo-$z$} or other redshift errors is provided, our forward model generates without any difficulties a series of realistic observed redshifts for galaxies. To better extract cosmological information, performing tomographic studies is also feasible.
\vspace*{-2.5ex}

\paragraph{Galaxy shape noise \& intrinsic alignment}\index{Shape noise} Another extension for our model is the noise model. When isotropy for galaxy shape orientation is assumed, Gaussian noise is added. Beyond Gaussian noise, some physically-motivated models for \acro{IA} can be added easily. In this case, one needs to identify pairs of galaxy and hosted halo. The galaxy orientation can be provided by an \acro{IA} model depending, on its type and the distance to the halo center.
\vspace*{-2.5ex}

\paragraph{Inversion \& filtering} A large number of options are available for our model to make a mass map. For example, using a $\kappa$-peak approach, one could compute directly the convergence signal without worrying about the inversion problem (see \sect{sect:lensing:challenges}), while a more realistic way would be simulating shears and using inversion techniques (\citealt{Kaiser_Squires_1993}, \citealt{Seitz_Schneider_1995}, etc.) to get convergence maps. Besides, one can also adopt an aperture-mass approach: convolving the shear field with a zero-mean filter.  By definition, the aperture mass is already a smoothing, while both $\kappa$-peak methods require in addition filtering techniques, linear or nonlinear ones, to reduce noise (see further \chap{sect:filtering}). For all three modelling approaches, taking masking into account is not necessarily trivial. Reducing the effective survey area, determining near-mask areas with a higher noise level (\citetalias{Liu_etal_2014} \citeyear{Liu_etal_2014}), or filling missing data with inpainting (method: \citealt{Pires_etal_2009}, data application: \citealt{Jullo_etal_2014}), are different options.
\vspace*{-2.5ex}

\paragraph{S/N determination} In most studies, the noise level in \acro{S/N} is a global value derived from the whole survey, which yields a global significance. However, in reality, depending on mask and the galaxy spatial distribution, the local noise level is not uniform. Taking the local significance into account would make the modelling more accurate (see also \chap{sect:filtering}).

As a summary, our model is capable to take into account a wide range of potential systematic sources: astrophysical ones (halo modelling, baryons, \acro{IA}), observational ones (masking, redshift errors), or the ones from data processing ($\kappa$-$\gamma$ inversion, inpainting). For the reason of time, not all topics are addressed in this thesis. The settings chosen will be detailed in each analysis.

\section{Sensitivity to cosmology}
\label{sect:modelling:sensitivity}

Before addressing a detailed study on parameter constraints in Chap. \ref{sect:constraint}, I will show here a first look at the sensitivity of the \acro{FSF} modelling to some cosmological parameters. Is our model sensitive to cosmology? Here, without performing constraints, some model realizations are shown by varying the matter density $\OmegaM$ and the matter fluctuation amplitude $\sigEig$. The aim is to give an idea of different behaviors of the model regarding to these two most lensing-sensitive parameters.

On the $\OmegaM$-$\sigEig$ plane, constraints from lensing second-order statistics tend to follow curved degeneracy lines, which results in ``banana-shaped''\index{Banana-shaped} contours. Thus, four groups of parameter sets have been proposed to be compared with the input cosmology of Aardvark $(\OmegaM^\inp, \sigEig^\inp) = (0.23, 0.83)$. With $\Delta\OmegaM = 0.03$ and $\Delta\sigEig = 0.05$, these four groups are successively $(\OmegaM^\inp\pm\Delta\OmegaM, \sigEig^\inp)$, $(\OmegaM^\inp, \sigEig^\inp\pm\Delta\sigEig)$, $(\OmegaM^\inp\pm\Delta\OmegaM, \sigEig^\inp\pm\Delta\sigEig)$, and $(\OmegaM^\inp\pm\Delta\OmegaM, \sigEig^\inp\mp\Delta\sigEig)$. Thus, the first two groups vary only one of the parameters at a time, and the two others are variations along the diagonal (45$^\circ$) and the anti-diagonal (135$^\circ$) directions, respectively. This makes nine different parameter sets in total, including $(\OmegaM^\inp, \sigEig^\inp)$.

\begin{figure}[tb]
	\centering
	\includegraphics[width=\textwidth]{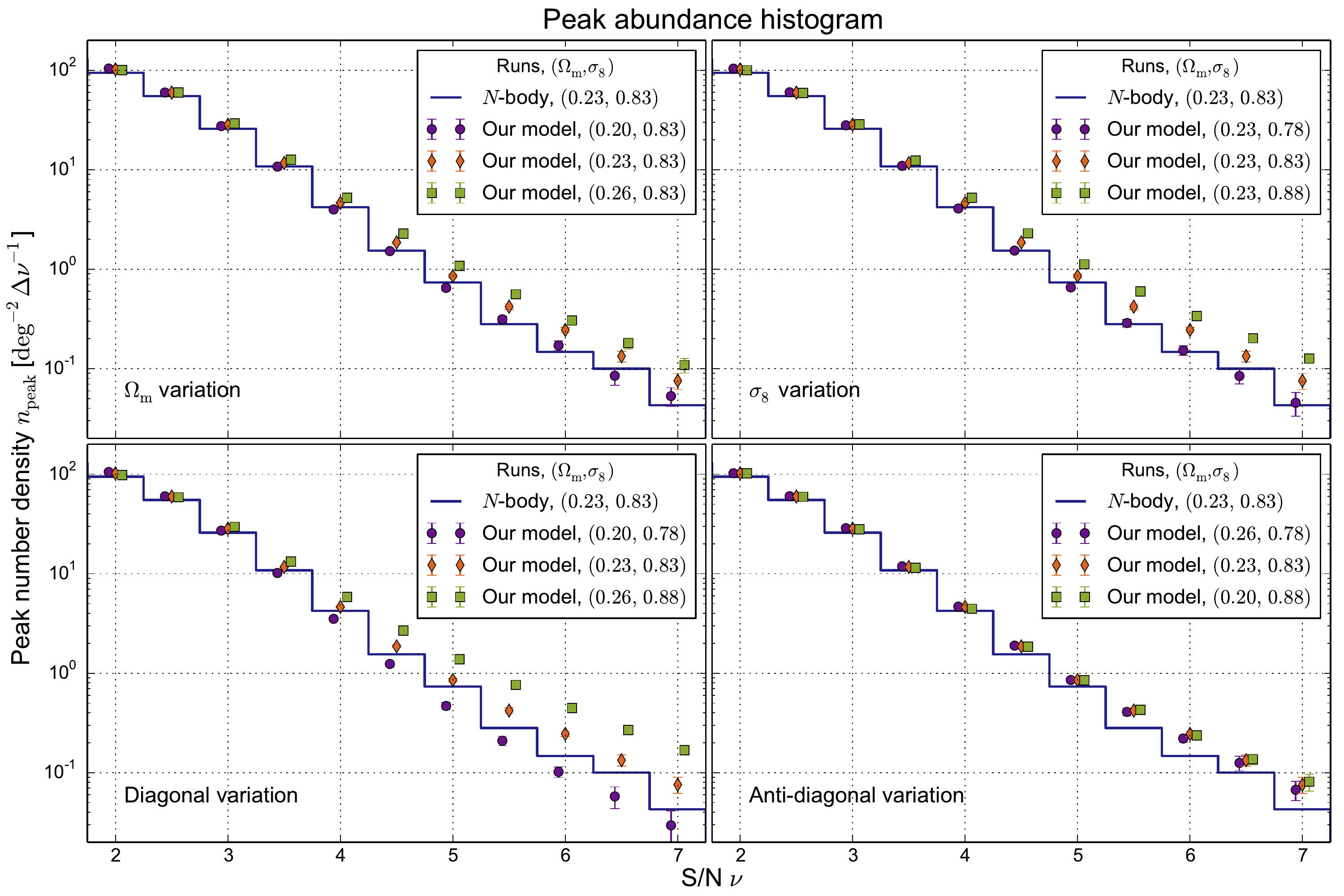}
	\caption{Peak abundance from our model for different cosmological parameter sets. The four panels display different directions of variation on the $\OmegaM$-$\sigEig$ plane. All orange diamonds are the same runs as in \fig{fig:modelling:peakHist_largeField}. This test shows that the degeneracy lines correspond approximately to the anti-diagonal direction.}
	\label{fig:modelling:peakHist_param}
\end{figure}

The four panels of \fig{fig:modelling:peakHist_param} show the comparison between models with different cosmologies. The variation of high-peak counts is neatly discernible when cosmology changes. Both bottom panels indicate that the degeneracy lines follow rather the anti-diagonal direction. This conforms to the a banana-shaped constraint contour suggested by other lensing observables. In addition, if the degeneracy slope of peak counts differs from the power spectrum, combining information from both would improve cosmological constraints \citep{Dietrich_Hartlap_2010}.

The blue lines in \fig{fig:modelling:peakHist_param} are peak counts from Case~1, which is considered as the ``truth''. Under this consideration, one may estimates that the bias from our model is on the order of 0.03 for $\OmegaM$, or equivalent to $\sim0.05$ for $\sigEig$. However, one should remind that this ``truth'' is only computed on the small field while all nine model runs are on the large one. Therefore, the ``truth'' is only indicative. Nevertheless, the model is not perfect and more studies are needed in order to understand the sources of biases.

\section{Comparison with an analytical model}

\subsection{Description of the FSL model}

The last part of this chapter focuses on the model from \citet*[][hereafter \acro{FSL} model]{Fan_etal_2010}\index{Fan-Shan-Liu (\acro{FSL}) model}. Based on an analytical approach, it computes directly the values of the peak function for various $\nu$ from the peak theory. Consider a smoothed Gaussian random field $K_N=K+N$ fluctuating around its mean value $K$ with a zero-mean random part $N$. In the context of \acro{WL} peaks, $N$ characterizes galaxy shape noise and $K$ is just the projected mass of foreground halos. Given a position, the joint probability of values of $K_N$ and its derivatives is well known. Now, if a point is a local maximum, the first derivatives should be zero and both eigenvalues of the second derivative matrix should be negative. These two conditions lead to determine a peak probability function which depends on the values and the derivatives of the foreground.

For the \acro{FSL} model, all is reasoned in terms of probability. The requirement of the \acro{FSL} model is also a mass function model and a halo profile. The mass function characterizes the halo population as a probability function of mass and redshift. In other words, it may be interpreted as, given a line of sight, the probability of crossing a halo of mass $M$ at redshift $z$. Now, if one ensures that the halo does not overlap with others on the line of sight and if the profile is known, then one knows with which probability the line of sight pierces the halo at a particular angular position. Knowing the relative position to the halo, one also obtains the smoothed convergence and the derivatives. And by combining the probabilities of the halo population, the relative position, the random fluctuation, and the condition of being peaks, one reaches eventually the peak function.

The basic assumption of the \acro{FSL} model is that halos do not overlap on the \acro{2D} sky. Actually, this assumption is not necessary because the probability of having two halos on the line of sight can be characterized by a term proportional to the product of two mass functions. However, this computation is very expensive, and since the non-overlapping hypothesis has been proven to be a good approximation \citep{Yang_etal_2011}, the following derivation only focuses on the first order of the \acro{FSL} model.

\subsection{Formalisms}
\label{sect:modelling:FSL:formalisms}

\subsubsection{Joint PDF from a Gaussian random field}

Consider a Gaussian random field $n$. Let $n$ be smoothed by a kernel function $W$ to become $N \equiv n\ast W$, assumed to have zero mean everywhere. The smoothed field $N$ is characterized by its moments $\sigma_i$ for $i=0,1,2,\ldots$, defined as \citep{VanWaerbeke_2000}
\begin{align}
	\sigma_i^2 \equiv \int \frac{\rmd^2 \vect{\ell}}{(2\pi)^2}\ \ell^{2i} \vert \widetilde{N}(\ell)\vert^2,
\end{align}
where $\widetilde{N}$ is the Fourier transform of $N$. Locally on the field, peak selection depends on the value and the first and the second derivatives of the considered position. In a \acro{2D} space, these are the value $N$, two first derivative components $N_{,1}$ and $N_{,2}$, and three effective terms from second derivatives $N_{,11}$, $N_{,22}$, and $N_{,12}$. Denoting $\bx = (N, N_{,1}, N_{,2}, N_{,11}, N_{,22}, N_{,12})$ as the considered random vector, the six-dimensional joint Gaussian \acro{PDF} is
\begin{align}\label{for:modelling:joint_pdf_1}
	P(N, N_{,1}, N_{,2}, N_{,11}, N_{,22}, N_{,12}) = \frac{1}{\sqrt{(2\pi)^6\det|\bC|}}\exp\left[-\frac{1}{2}\bx^T\bC\inv\bx\right],
\end{align}
where the covariance matrix $\bC$ is \citep{VanWaerbeke_2000}
\begin{align}
	\bC = \begin{pmatrix}
		\sigma_0^2    & 0            & 0             & -\sigma_1^2/2 & -\sigma_1^2/2 & 0\\
		0             & \sigma_1^2/2 & 0             & 0             & 0             & 0\\
		0             & 0            & \sigma_1^2/2  & 0             & 0             & 0\\
		-\sigma_1^2/2 & 0            & 0             & 3\sigma_2^2/8 & \sigma_2^2/8  & 0\\
		-\sigma_1^2/2 & 0            & 0             & \sigma_2^2/8  & 3\sigma_2^2/8 & 0\\
		0             & 0            & 0             & 0             & 0             & \sigma_2^2/8\\
	\end{pmatrix}.
\end{align}
After simple calculations, \for{for:modelling:joint_pdf_1} becomes
\small
\begin{align}
	P\Big(\nOne, \nTwo, \nThree, \nFour, \nFive&, \nSix\Big) \notag\\
	= \frac{2}{\pi^3\sqrt{1-\gamma_\ast^2}}\ &\exp\left[ -\frac{\nOne^2 +2\gamma_\ast\nOne\big(\nFour+\nFive\big) + \big(\nFour+\nFive\big)^2}{2\big(1-\gamma_\ast^2\big)} \right] \notag\\
	\times\ &\exp\left[ -\big(\nFour-\nFive\big)^2 -4\nSix\mbox{}^2 -\nTwo\mbox{}^2 -\nThree\mbox{}^2 \right], \label{for:modelling:joint_pdf_2}
\end{align}
\normalsize
where $\nOne\equiv N/\sigma_0$, $\nOne_{,i}\equiv N_{,i}/\sigma_1$, and $\nOne_{,ij}\equiv N_{,ij}/\sigma_2$ for $i,j\in\{1,2\}$, and $\gamma_\ast \equiv \sigma_1^2/\sigma_0\sigma_2$. 

\subsubsection{Peak density as a function of the foreground}

When the foreground clusters are present, $K\neq0$. One should not count peaks from $N$ but from $K_N=K+N$. To write down the joint probability for quantities related to $K_N$, one only needs to replace $N$ by $K_N-K$ in \for{for:modelling:joint_pdf_2}. Before doing so, it is useful to apply some changes of variable as follows. First, with the same logic as before, I use dimensionless quantities: 
\begin{align}
	\kOne\equiv K/\sigma_0,\ \ \kOne_{,i} &\equiv\partial_i K/\sigma_1,\ \ \kOne_{,ij} \equiv\partial_i\partial_j K/\sigma_2, \notag\\
	\kOne_N\equiv K_N/\sigma_0,\ \ \kOne_{N,i}&\equiv\partial_i K_N/\sigma_1,\ \ \kOne_{N,ij}\equiv\partial_i\partial_j K_N/\sigma_2. 
\end{align}
Further, since the zero-order moment $\sigma_0$ is nothing but $\sigma_\noise$, $K_N/\sigma_0$ is the \acro{S/N}. Therefore, one may set $\nu = \kOne_N$ to simplify the notation. Then, consider the second derivatives of $\kOne_N$. The matrix $-\kOne_{N,ij}$ is symmetric, so it has two real eigenvalues, say $\lambda_{N1}$ and $\lambda_{N2}$ with $\lambda_{N1}\geq\lambda_{N2}$. By defining
\begin{align}
	x_N \equiv \lambda_{N1}+\lambda_{N2}\ \ \text{and}\ \ e_N \equiv \frac{\lambda_{N1}-\lambda_{N2}}{2(\lambda_{N1}+\lambda_{N2})},
\end{align}
the second derivatives can be parametrized by these quantities as
\begin{align}
	\kOne_{N,11} &= -\frac{x_N}{2} \left(1+ 2e_N \cos 2\theta_N \right),\\
	\kOne_{N,22} &= -\frac{x_N}{2} \left(1- 2e_N \cos 2\theta_N \right),\\
	\kOne_{N,12} &= - x_N e_N \sin 2\theta_N.
\end{align}
where $\theta_N$ is a mixing angle in $[0, \pi]$. After the change of variables, taking into account the Jacobian matrix, the probability from \for{for:modelling:joint_pdf_2} becomes (see \citealt{Bardeen_etal_1986}, \citealt{Bond_Efstathiou_1987}, and \citealt{Fan_etal_2010})
\small
\begin{align}
	P\Big(&\nu, x_N, e_N, \theta_N, \knTwo, \knThree\Big) \notag\\
	&= \frac{4x_N^2e_N}{\pi^3\sqrt{1-\gamma_\ast^2}}\ 
	 \exp\left[ -\frac{1}{2}\big(\nu-\kOne\big)^2 -\frac{\big(x_N+\kFour+\kFive-\gamma_\ast\big(\nu-\kOne\big)\big)^2}{2\big(1-\gamma_\ast^2\big)} \right] \notag\\
	&\hspace{2cm}\times \exp\left[ -4x_N^2e_N^2 -4x_Ne_N\cos 2\theta_N\big(\kFour-\kFive\big) -\big(\kFour-\kFive\big)^2 \right] \notag\\
	&\hspace{2cm}\times \exp\left[ -8x_Ne_N\sin 2\theta_N\kSix -4\kSix\mbox{}^2 \right] \notag\\
	&\hspace{2cm}\times \exp\left[ -\big(\knTwo-\kTwo\big)^2 - \big(\knThree-\kThree\big)^2 \right]. \label{for:modelling:joint_pdf_3}
\end{align}
\normalsize
Here, \for{for:modelling:joint_pdf_3} is the probability for general cases. Two conditions need to be added for peak selection: the vanishing of the first derivatives and negative eigenvalues of $\kOne_{N,ij}$. The first one is ensured by adding $\delta(\knTwo)\delta(\knThree)$, where $\delta$ stands for the Dirac delta function, and the second is translated by the condition $\lambda_{N1}\geq\lambda_{N2}\geq 0$, or more precisely $\Theta(x_N)\Theta(e_N)\Theta(1-2e_N)\Theta(\theta_N)\Theta(\pi-\theta_N)$ with $\Theta$ being the Heaviside step function. Moreover, a surface term $(\sigma_2/\sigma_1)^2(x_N^2/4)(1-4e_N^2)$ should be introduced to transform \for{for:modelling:joint_pdf_3} from a one-point probability into a normalized quantity, $n_\peak^\los$, such that the dimension is equal to a surface density. The label \texttt{los} stands for line of sight \footnote{The ``number density for a line of sight'' does not make sense, so I avoid using this interpretation. It should rather be understood, as I said, as a probability which has a dimension of a surface density.}. At the end of the day, $n_\peak^\los$ is linked to ``foreground quantities'' $\kOne$, $\kOne_{,i}$, and $\kOne_{,ij}$ via 
\small
\begin{align}
	&n_\peak^\los\Big(\nu \Big\vert \kOne, \kTwo, \kThree, \kFour, \kFive, \kSix\Big) \notag\\
	&\hspace{0.5cm}= \int_0^{+\infty}\rmd x_N\int_0^{1/2}\rmd e_N\int_0^\pi\rmd\theta_N \int_\mathbb{R}\rmd\knTwo\ \delta(\knTwo) \int_\mathbb{R}\rmd\knThree\ \delta(\knThree) \notag\\
	&\hspace{1.5cm}\times P\left(\nu, x_N, e_N, \theta_N, \knTwo, \knThree\right) \cdot (\sigma_2/\sigma_1)^2 \left(x_N^2/4\right) \left(1-4e_N^2\right), 
\end{align}
\normalsize
which I write explicitly as
\small
\begin{align}
	&n_\peak^\los\Big(\nu \Big\vert \kOne, \kTwo, \kThree, \kFour, \kFive, \kSix\Big) \notag\\
	&\hspace{0.5cm}= \frac{1}{4\pi^2\theta_\ast^2\sqrt{1-\gamma_\ast^2}}\ \exp\left[ -\frac{1}{2}\big(\nu-\kOne\big)^2 -\kTwo\mbox{}^2 -\kThree\mbox{}^2 \right]\notag\\
	&\hspace{1.2cm}\times \int_0^{+\infty}\rmd x_N\ \exp\left[ -\frac{\big(x_N+\kFour+\kFive-\gamma_\ast\big(\nu-\kOne\big)\big)^2}{2\big(1-\gamma_\ast^2\big)} \right] \notag\\
	&\hspace{1.5cm}\times x_N^4 \exp\left[ -\big(\kFour-\kFive\big)^2 -4\kSix\mbox{}^2 \right]
	 \cdot \int_0^{1/2}\rmd e_N\ \exp\Big[ -4x_N^2e_N^2 \Big]\cdot 8e_N\left(1-4e_N^2\right) \notag\\
	&\hspace{1.8cm}\times \int_0^\pi\frac{\rmd\theta_N}{\pi}\ \exp\left[ -4x_Ne_N\cos 2\theta_N\big(\kFour-\kFive\big) -8x_Ne_N\sin 2\theta_N\kSix \right], \label{for:modelling:n_peak_los_2}
\end{align}
\normalsize
where $\theta_\ast^2 \equiv 2\sigma_1^2/\sigma_2^2$ is equivalent to a surface.

In principle, one can integrate \for{for:modelling:n_peak_los_2} numerically, since the number of dimension is only three. However, as we are going to see later, the total integral also contains \acro{S/N}, redshift, mass, and radius. Whatever the size of the cosmological parameter space is, this constraint requires already seven additional dimensions while performing model evaluation. In order to reduce the cost, \citet{Fan_etal_2010} have added some short cuts. Define $F(x_N)$ as the two last lines of \for{for:modelling:n_peak_los_2}, which is
\begin{align}
	F(x_N) &\equiv x_N^4 \exp\left[ -\big(\kFour-\kFive\big)^2 -4\kSix\mbox{}^2 \right] \notag\\
	&\times \int_0^{1/2}\rmd e_N\ \exp\Big[ -4x_N^2e_N^2 \Big]\cdot 8e_N\left(1-4e_N^2\right) \notag\\
	&\times \int_0^\pi\frac{\rmd\theta_N}{\pi}\ \exp\left[ -4x_Ne_N\cos 2\theta_N\big(\kFour-\kFive\big) -8x_Ne_N\sin 2\theta_N\kSix \right], \label{for:modelling:FSL_F_1}
\end{align}
If no foreground is present, this function has an analytical solution which is $F(x_N) = x_N^2-1+\exp(-x_N^2)$ \citep{Bond_Efstathiou_1987}. In the current scenario, if one considers that the foreground contains only one spherical halo on the line of sight, then \for{for:modelling:FSL_F_1} can be simplified. The spherical symmetry allows one to choose a new coordinate system such that $\kThree$ and $\kSix$ vanish. Therefore, $F(x_N)$ becomes $F(x_N, \kFour-\kFive)$. Inspired by the exact solution in the case without foreground, \citet{Fan_etal_2010} have proposed an approximate form for $F$:
\begin{align}\label{for:modelling:FSL_F_2}
	F\left(x_N, \kDiff\right) = x_N^2-1+\exp\left(-x_N^2\right) + \kDiff\mbox{}^2 \left[\exp\left(-\frac{x_N^3}{g\big(\kDiff\big)}\right)-1\right],
\end{align}
where $\kDiff\equiv \kFour-\kFive$ and $g$ is a function of $\kDiff$ whose values are to be determined by fitting. In this way, the integration over $e_N$ and $\theta_N$ only needs to be evaluated once. Let $\kSum\equiv \kFour+\kFive$. In the coordinate system where $\kThree$ and $\kSix$ vanish, one can rewrite \for{for:modelling:n_peak_los_2} as
\begin{align}
	n_\peak^\los\Big(\nu \Big\vert \kOne, \kTwo, \kSum&, \kDiff\Big) = \frac{1}{4\pi^2\theta_\ast^2\sqrt{1-\gamma_\ast^2}}\ \exp\left[ -\frac{1}{2}\big(\nu-\kOne\big)^2 -\kTwo\mbox{}^2 \right] \label{for:modelling:n_peak_los_3}\\
	&\times \int_0^{+\infty}\rmd x_N\ F\left(x_N, \kDiff\right)\exp\left[ -\frac{\big(x_N+\kSum-\gamma_\ast\big(\nu-\kOne\big)\big)^2}{2\big(1-\gamma_\ast^2\big)} \right],\notag
\end{align}
where $F(x_N)$ is of course given by \for{for:modelling:FSL_F_2}.

\subsubsection{Densities from two regimes: halo-covered and non-covered areas}

Now we need to compute, for a halo of mass $M$ at redshift $z$, the values of $\kOne$ and its derivatives. \citet{Fan_etal_2010} have considered \acro{NFW} profiles (\for{for:structure:NFW_profile}) and have set $\kappa = \kappa_\proj$ (Eqs. \eqref{for:lensing:G_kappa_TJ}, \eqref{for:lensing:kappa_halo_NFW}, and \eqref{for:modelling:kappa_proj}) without correction. The normalized smoothed convergence $\kOne$ is obtained via 
\begin{align}\label{for:modelling:kOne_d0}
	\kOne(\btheta) = \sigma_0\inv \int\rmd^2 \btheta'\ W(\btheta-\btheta')\kappa(\btheta').
\end{align}
And from \for{for:modelling:kOne_d0} we derive easily
\begin{align}
	\kOne_{,i}(\btheta) &= \sigma_1\inv \int\rmd^2 \btheta'\ \partial_iW(\btheta-\btheta')\kappa(\btheta'), \label{for:modelling:kOne_d1}\\
	\kOne_{,ij}(\btheta) &= \sigma_2\inv \int\rmd^2 \btheta'\ \partial_i\partial_j W(\btheta-\btheta')\kappa(\btheta'). \label{for:modelling:kOne:d2}
\end{align}
If one considers the kernel from \for{for:modelling:Gaussian_kernel}, which is the same as in \citet{Fan_etal_2010}, these quantities become
\small
\begin{align}
	\kOne(\theta_1, 0) &= \int \rmd^2 \btheta'\ 
	\frac{1}{\pi\theta_\rmG^2\sigma_0} \exp\left[-\frac{(\theta_1 - \theta_1')^2+(-\theta_2')^2}{\theta_\rmG^2}\right] \kappa(\btheta'), \label{for:modelling:kOne}\\
	\kTwo(\theta_1, 0) &= \int \rmd^2 \btheta'\ -\frac{2}{\theta_\rmG^2}(\theta_1-\theta_1') \cdot 
	\frac{1}{\pi\theta_\rmG^2\sigma_1} \exp\left[-\frac{(\theta_1 - \theta_1')^2+(-\theta_2')^2}{\theta_\rmG^2}\right] \kappa(\btheta'), \label{for:modelling:kTwo}\\
	\kSum(\theta_1, 0) &= \int \rmd^2 \btheta'\ 
	\frac{4}{\theta_\rmG^2}\Bigg[\frac{(\theta_1 - \theta_1')^2+(-\theta_2')^2}{\theta_\rmG^2} -1\Bigg] \cdot 
	\frac{1}{\pi\theta_\rmG^2\sigma_2} \exp\left[-\frac{(\theta_1 - \theta_1')^2+(-\theta_2')^2}{\theta_\rmG^2}\right] \kappa(\btheta'), \label{for:modelling:kSum}\\
	\kDiff(\theta_1, 0) &= \int \rmd^2 \btheta'\ 
	\frac{4}{\theta_\rmG^2}\Bigg[\frac{(\theta_1 - \theta_1')^2-(-\theta_2')^2}{\theta_\rmG^2}\Bigg] \cdot 
	\frac{1}{\pi\theta_\rmG^2\sigma_2} \exp\left[-\frac{(\theta_1 - \theta_1')^2+(-\theta_2')^2}{\theta_\rmG^2}\right] \kappa(\btheta'). \label{for:modelling:kDiff}
\end{align}
\normalsize
Only $\theta_2 = 0$ is taken into account because for a spherical profile, this condition leads to $\kThree = \kSix = 0$. Readers should be reminded that $\kappa$ in Eqs. \eqref{for:modelling:kOne}, \eqref{for:modelling:kTwo}, \eqref{for:modelling:kSum}, and \eqref{for:modelling:kDiff} depends on halo redshift and mass. Therefore, $\kOne$, $\kTwo$, $\kSum$, and $\kDiff$ are implicit functions of $z$, $M$, and $\theta_1$. Finally, for a field of view of size $\rmd^2\Omega$, the total number of peaks from halos is given by
\begin{align}
	n_\peak^\halo(\nu)\rmd^2\Omega &= \int\rmd z\ \frac{\rmd V(z)}{\rmd z}\int\rmd\log M\ \frac{\rmd n(z, \lessM)}{\rmd\log M} \int_0^{\theta_\vir}\rmd \theta_1\ 2\pi\theta_1 \notag\\
	&\times n_\peak^\los\Big(\nu\Big\vert \kOne(z ,M, \theta_1), \kTwo(z ,M, \theta_1), \kSum(z ,M, \theta_1), \kDiff(z ,M, \theta_1)\Big),  \label{for:modelling:n_peak_halo}
\end{align}
where 
\begin{align}
	\rmd V(z) = \rmd w\cdot w^2(z)\rmd^2\Omega = \frac{D_\rmH\cdot w^2(z)}{E(z)}\ \rmd z\rmd^2\Omega
\end{align}
is the comoving volume contained in a slice of redshift $[z, z+\rmd z[$ and $\rmd n(z, \lessM)/\rmd\log M$ is the halo volume density with log mass contained between $\log M$ and $\log M+\rmd \log M$ (defined as \for{for:structure:dn_over_dlogM}). The term $E(z)$ is given by \for{for:cosmology:E}. 

Equation \eqref{for:modelling:n_peak_halo} is straightforward to interpret. The integration over $z$ yields the light-cone volume for $\rmd^2\Omega$; the integration over $\log M$ gives the number of halos in the lightcone; the integration over $\theta_1$ shows the total covered surface by these halos; and finally $n_\peak^\halo$ represents the probability by unity of surface that a peak at level $\nu$ would be observed. However, this is not the end of the story. Equation \eqref{for:modelling:n_peak_halo} only takes into account those lines of sight where a halo is present. That is the reason for the label \texttt{halo}. From the sky area not covered by halos, peaks can still occur from random fluctuations. This foreground-free case is described by \for{for:modelling:n_peak_los_3} by setting all $K$-related quantities to zero, leading to
\begin{align}
	n_\peak^\los\Big(&\nu\Big\vert \kOne=\kTwo=\kSum=\kDiff=0\Big) \notag\\
	&= \frac{\exp(-\nu^2/2)}{4\pi^2\theta_\ast^2\sqrt{1-\gamma_\ast^2}}\int_0^{+\infty}\rmd x_N\  \left(x_N^2-1+\exp(-x_N^2)\right) \exp\left[ -\frac{(x_N-\gamma_\ast\nu)^2}{2\big(1-\gamma_\ast^2\big)} \right], 
\end{align}
and the peak density occurs from pure random fluctuation $n_\peak^\mathrm{ran}$ reads
\begin{align}
	n_\peak^\mathrm{ran}(\nu)\rmd^2\Omega &= \left(\rmd^2\Omega-\int\rmd z\ \frac{\rmd V(z)}{\rmd z}\int\rmd\log M\ \frac{\rmd n(z, \lessM)}{\rmd\log M} \int_0^{\theta_\vir}\rmd \theta_1\ 2\pi\theta_1\right) \notag\\
	&\times n_\peak^\los\Big(\nu\Big\vert \kOne=\kTwo=\kSum=\kDiff=0\Big). \label{for:modelling:n_peak_ran}
\end{align}
As a result, combining Eqs. \eqref{for:modelling:n_peak_halo} and \eqref{for:modelling:n_peak_ran}, the peak number density from the \acro{FSL} model is 
\begin{align}\label{for:modelling:n_peak_FSL}
	n_\peak(\nu) = n_\peak^\halo(\nu) + n_\peak^\mathrm{ran}(\nu).
\end{align}

\subsection{Results}
\label{sect:modelling:FSL:results}

\begin{figure}[tb]
	\centering
	\includegraphics[width=0.8\textwidth]{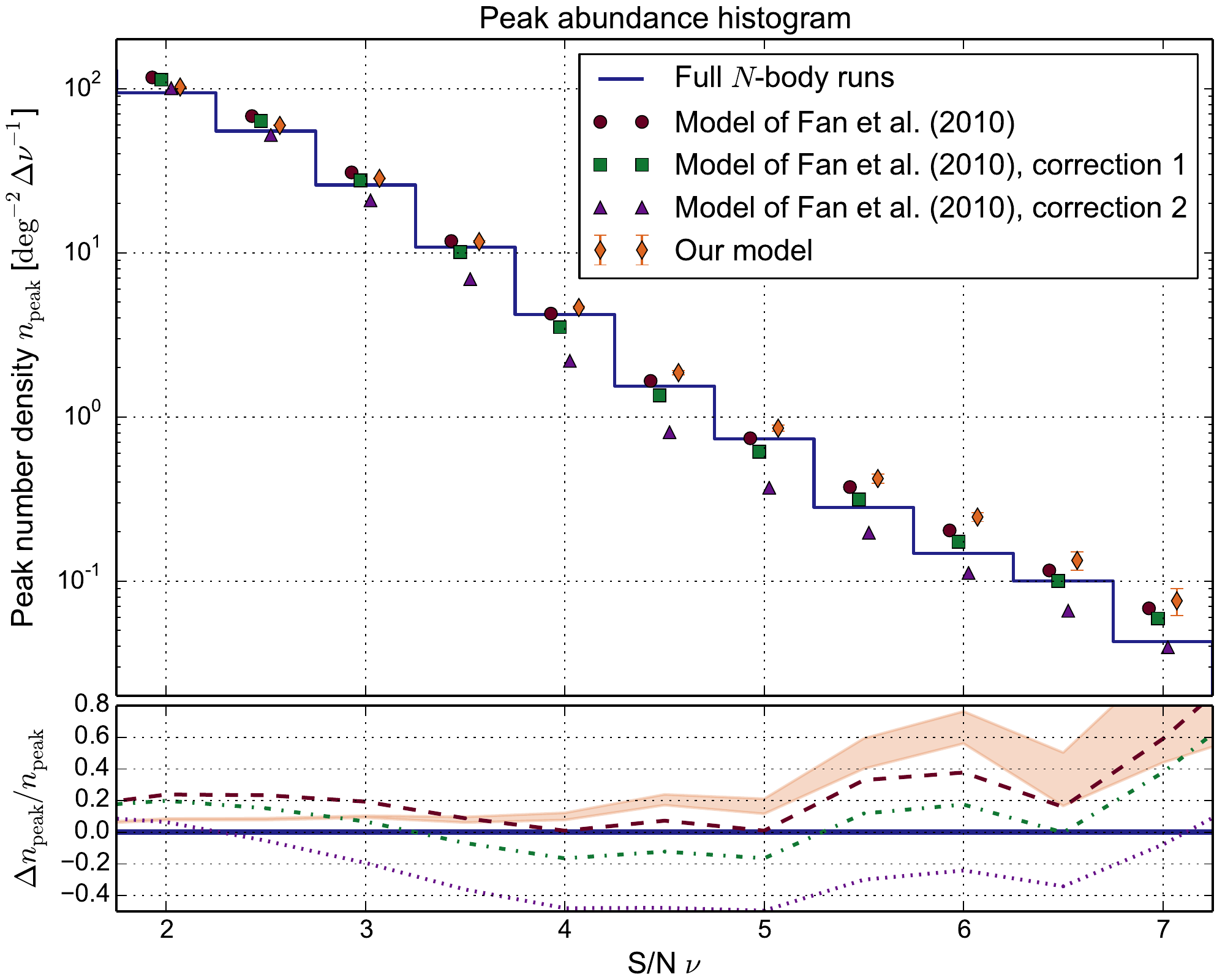}
	\caption{Comparison with an analytical model. On the upper panel, we show peak histograms from $N$-body runs (blue solid lines, Case 1 in \sect{sect:modelling:validation:methodology}), our model (orange diamonds, Case 4 in \sect{sect:modelling:validation:methodology}), the \acro{FSL} model (dark red circles), and two corrections (green squares and purple triangles) for the latter (see \sect{sect:modelling:FSL:results}). On the lower panel, the deviations of all cases with regard to $N$-body runs are shown. The orange band delineates the upper and lower limits of the error bars of our model, whereas for the other three cases, the lines indicate the mean. Results from $N$-body runs are obtained from the small field, while those from our model are performed on the large field.}
	\label{fig:modelling:peakHist_Fan_vs_ours}
\end{figure}

\figFull{fig:modelling:peakHist_Fan_vs_ours} shows the comparison between $N$-body runs, the \acro{FSL} model, and our model with the same input cosmology. Both models use the same noise level and smoothing scale $\theta_\rmG = 1~\arcmin$. The lower panel shows the relative difference of each model compared to the $N$-body runs. It is worth highlighting that the histogram of $N$-body runs (blue lines) is extracted from a patch of 54 deg$^2$. The cosmic-variance effect might be important. Nevertheless, both models show an overall good agreement with the Aardvark $N$-body simulations  (dark red circles and orange diamonds; the two other cases will be discussed later).

At first glance, \fig{fig:modelling:peakHist_Fan_vs_ours} seems to suggest that the \acro{FSL} model performs better. However, several problems can be revealed if by looking closer. As I mentioned above, the \acro{FSL} model is based on the assumption that halos do not overlap in projection. The direct consequence of this is that a cutoff at low redshift needs to be introduced. Otherwise, whatever the physical size of the halo is, the angular size would converge to the half of the sky ($2\pi\ \rad^2$) when $z\rightarrow 0$. The total halo-covered area would be much larger than the field of view in this case. Taking into account that low-redshift halos have poor lensing efficiency, this probably does not create any significant bias.

The other restriction related to the total halo-covered area is the lower mass limit. If $M_\minn$ is low, then halos will be numerous. Of course, most of them will be small and have a reduced angular size, but it seems that the decreasing rate of the angular size is dominated by the increasing rate of the population as $M_\minn$ decreases. As a result, by $M_\minn \lesssim 10^{13.3}~\Msol/h$, the halo-covered area will exceed the total field of view and break down the model, since it assumes no overlap in projection between halos. For the \acro{FSL} model in \fig{fig:modelling:peakHist_Fan_vs_ours}, $M_\minn$ is set to $\dix{13.6}~\Msol/h$. However, as indicated by \fig{fig:modelling:peakHist_Mmin}, this value is already large enough to create some deficits on middle-peak counts. Recall that our model has $M_\minn = \dix{12}~\Msol/h$. By increasing $M_\minn$, we expect that the orange diamonds would move closer to the blue solid lines; this however does not imply a better modelling! As a conclusion, the \acro{FSL} model would have faced the same augmentation if it manages to go lower than $M_\minn = 10^{13.3}~\Msol/h$. A more detailed analysis should be done in the future, especially for comparing with a larger $N$-body reference field.

\subsubsection{Revising the $\kappa_\proj$ formalism}

Even though high-peak counts are underestimated by the \acro{FSL} model, this result presents already an excess. What is the origin of these overcounts? Readers may have noticed that, contrary to our model, the \acro{FSL} model assumes $\kappa = \kappa_\proj$. Not subtracting anything from $\kappa_\proj$ leads systematically to an overestimation of peak counts. The space which is not occupied by halos is not empty (see \sect{sect:lensing:cluster:projection})! The projected mass accounted in Eqs. \eqref{for:modelling:n_peak_halo} and \eqref{for:modelling:n_peak_ran} are only carried out over $[M_\minn, M_\maxx[$, which is somewhat between 10\% and 40\% of the total mass. If this additional mass is not extracted, peak counts can be largely influenced.

Below I propose a correction for the \acro{FSL} model. Let us rewrite the matter density as
\begin{align}
	\rho\big(f_K(w)\btheta, w\big) = \sum_{\halo\rms}\rho_\halo\big(f_K(w)\btheta, w\big) + \big(1-\lambda(w)\big)\bar{\rho}(w),
\end{align}
where $\rho_\halo$ is halo density profiles and
\begin{align}\label{for:modelling:lambda}
	\lambda(w(z)) \equiv \frac{1}{\rho_\crit\OmegaM}\int_{M_\minn}^{M_\maxx}\rmd\log M\ \frac{\rmd n(z, \lessM)}{\rmd\log M}\cdot M,
\end{align}
The quantity $\lambda(w)$ is the proportion of the mass falling in $[M_\minn, M_\maxx[$ at the epoch corresponding to $w$ from the observer \footnote{No scaling factor should appear in \for{for:modelling:lambda}. The density that the mass function gives is per comoving volume. The matter density, usually denoted as $\rho(z)$, is per physical volume at $z$. In terms of comoving volume, it is per physical volume at $z=0$, which is $\rho_\crit\OmegaM$.}. With the factor $1-\lambda(w)$, the new mean density is now the correct one as $\bar{\rho}(w)=\rho_\crit\OmegaM(1+z(w))^3$. The resulting density contrast is
\begin{align}\label{for:modelling:delta_decomposition}
	\delta\big(f_K(w)\btheta, w\big) = \sum_{\halo\rms}\frac{\rho_\halo\big(f_K(w)\btheta, w\big)}{\bar{\rho}(w)} - \lambda(w).
\end{align}
Now, if we insert \for{for:modelling:delta_decomposition} into \for{for:lensing:convergence_2}, the definition of $\kappa$, we find
\begin{align}
	\kappa(\btheta, w) = \kappa_\proj(\btheta, w) - \kappa_1(w)
\end{align}
with $\kappa_\proj(\btheta, w)$ given by \for{for:modelling:kappa_proj} and 
\begin{align}\label{for:modelling:kappa_1}
	\kappa_1(w) \equiv \frac{3H^2_0 \OmegaM}{2\rmc^2} \int_0^w \rmd w'\ \frac{f_K(w-w')f_K(w')}{f_K(w)} \frac{\lambda(w')}{a(w')}.
\end{align}

For $z_\rms$ fixed at 1, $\kappa_1$ is a constant. The numerical result implies that $\kappa_1 = 0.0034$ for $M_\minn=\dix{13.6}~\Msol/h$ in the Aardvark cosmology, which is 0.14$\sigma$ for a Gaussian smoothing of $\theta_\rmG=1~\arcmin$. In \fig{fig:modelling:peakHist_Fan_vs_ours}, the result taking $\kappa_1$ into account is labelled as \texttt{correction1}. Readers can see that the $\kappa_1$-correction improves neatly the model. The difference between both deviations from $N$-body runs (dark red dashed and green dash-dotted lines from the lower panel) tends to zero as \acro{S/N} decreases. This conforms to \for{for:modelling:n_peak_FSL} since $n_\peak^\mathrm{ran}$, which does not change, is more dominant in this regime. Computing the $\kappa_1$-correction for $M_\minn=10^{12}~\Msol/h$, the result is shown as \texttt{correction2} in \fig{fig:modelling:peakHist_Fan_vs_ours}. Of course, $\kappa_\proj$ from $M \in [\dix{12}, \dix{13.6}]~\Msol/h$ is not taken into account in this case. However, on one hand, the comparison between Corrections 1 and 2 indicates the variation of $\kappa_1$ that one expects in each bin, showing that ignoring this difference is not appropriate; on the other hand, the comparison between Correction 2 and our model reveals the order of magnitude of the contribution from $\kappa_\proj$ for $M \in [\dix{12}, \dix{13.6}]~\Msol/h$, which suggests that $M_\minn=\dix{13.6}~\Msol/h$ is not pertinent.

A similar examination has been done for our model. For a setting conforming to \tab{tab:modelling:parameters}, we have found $\kappa_1 = 0.013$, and $\vert\kappa_1 - \overline{\kappa}_\proj\vert \lesssim 10^{-4}$ for all realizations from our model. This displays an excellent agreement between subtracting $\overline{\kappa}_\proj$ and the $\lambda(w)$ modelling. The zero-mean convergence is therefore a good approximation. Nevertheless, for our model, applying \for{for:modelling:kappa_1} to the computation of the projected mass is feasible. For example, a table of $\lambda(w)$ can be precomputed for a wide range of $w$, then instead of calculating $\kappa_\proj$ for each galaxy and subtracting the mean after the fact, one may take directly $\kappa_\proj-\kappa_1$ as the lensing signal. This improvement will also solve the problem that $\overline{\kappa}_\proj$ depends on the cutoff size of the truncated profiles. The same improvement technique is also applicable on the \acro{FSL} model.

Regarding the complexity of the realistic conditions, the \acro{FSL} model seems to be limited. Modelling non-Gaussian features from \acro{photo-$z$} errors, \acro{IA}, and non-uniformly-distributed galaxies with a simple Gaussian random field would be a great challenge.  Meanwhile, on the road to parameter constraints, estimating the covariance matrix would be impossible without the help of $N$-body simulations for analytical models like this one. This is however not the case of our model.

\subsubsection{Summary}

In this chapter, I have introduced a fast stochastic forward model to predict lensing peak counts. It performs fast simulations, which substitute the $N$-body physical process with a stochastic one. The advantages of the new model can be summarized to three characteristics: fast, flexible, and full \acro{PDF} information.

Adopting progressively different steps of the new model, intermediate cases permit to test two major hypotheses that the model makes: neglecting unbound matter and breaking halo correlation. In spite of some biases, the model agrees well with $N$-body simulations. Most importantly, it has been shown that the cosmological sensitivity is larger than the biases of the model.

The origin of some systematic biases are known, including halo correlation and halo concentration. Thanks to the flexibility of the new model, many options of improvement are possible. However, some other biases are not entirely understood. To study this, a larger $N$-body simulation set and more realistic probe conditions would be required for the future.

The new model has been compared to the \acro{FSL} analytical model. The deviation of both from $N$-body runs are similar. I have also proposed a correction which improves the \acro{FSL} model. However, even with the correction, the potential of the \acro{FSL} model is limited by its basic assumption: non-overlapping halos in projection.

In the next chapter, I am going to demonstrate a tremendous asset of our fast stochastic model: the availability of the full \acro{PDF} information, and how this can provide various constraint strategies for cosmological parameters.

\clearpage
\thispagestyle{empty}
\cleardoublepage


\chapter{Parameter constraint strategies}
\label{sect:constraint}
\fancyhead[LE]{\sf \nouppercase{\leftmark}}
\fancyhead[RO]{\sf \nouppercase{\rightmark}}

\subsubsection{Overview}

As the link between cosmological models and peak-count observables is established, we are now able to perform model selection. Comparing cosmological models or constraining parameters requires sizing the scatter and uncertainty, and this can be done by exploring the full \acro{PDF} information provided by our model. In this chapter, I am going to propose several constraining techniques. After introducing the methodology, the impact from the true covariance on the Gaussian likelihood is studied. Various alternatives to the Gaussian likelihood are then presented: the copula likelihood, the true likelihood, and model selection using $p$-values. This chapter corresponds to Sects. 2, 3, 4, and 5 of \PaperII.

\section{Methodology}
\label{sect:constraint:methodology}

\subsection{Likelihood formalism}
\label{sect:constraint:methodology:likelihood}

After the sensitivity test shown in \fig{fig:modelling:peakHist_param}, this chapter aims to present parameter constraints derived from our peak-count model. This requires a parameter estimator which is usually the likelihood function $\Like$. Hereafter, $\bx$ denotes a data vector, $\bpi$ a parameter set, and in order to avoid confusions, $P$ is always a probability function in the space of $\bx$ and $\mathcal{P}$ the one in the parameter space.

From Bayes' theorem \citep{Bayes_1763}, the conditional probability between $\bx$ and $\bpi$ is
\begin{align}\label{for:constraint:Bayes}
	\mathcal{P}(\bpi|\bx)P(\bx) = P(\bx|\bpi) \mathcal{P}(\bpi).
\end{align}
Let $\bx^\obs$ be the observation. By setting the Bayesian evidence to unity, $P(\bx=\bx^\obs) = 1$, one can rewrite \for{for:constraint:Bayes} as
\begin{align}\label{for:constraint:likelihood}
	\mathcal{P}(\bpi|\bx^\obs) = \Like(\bpi|\bx^\obs) \mathcal{P}(\bpi),
\end{align}
where $\Like(\bpi|\bx^\obs)\equiv P(\bx^\obs|\bpi)$ is the \textit{likelihood function}\index{Likelihood function}, $\mathcal{P}(\cdot)$ the prior, and $\mathcal{P}(\cdot|\bx^\obs)$ the posterior. In this chapter, I will use $\Like(\bpi)$ instead of $\Like(\bpi|\bx^\obs)$ to simplify the notation.

\subsection{Analysis design}
\label{sect:constraint:methodology:design}

The study from this chapter focuses only on constraints of $\bpi =~(\OmegaM, \sigEig)$. The other cosmological parameters are fixed, including $\OmegaB=0.047$, $h=0.78$, $n_\rms=0.95$, and $\wZero=-1$. The dark energy density, which in this case corresponds to a cosmological constant, is noted by $\OmegaL$ and set to $1-\OmegaM$ to match a flat universe.

\begin{figure}[tb]
	\centering
	\includegraphics[width=0.5\textwidth]{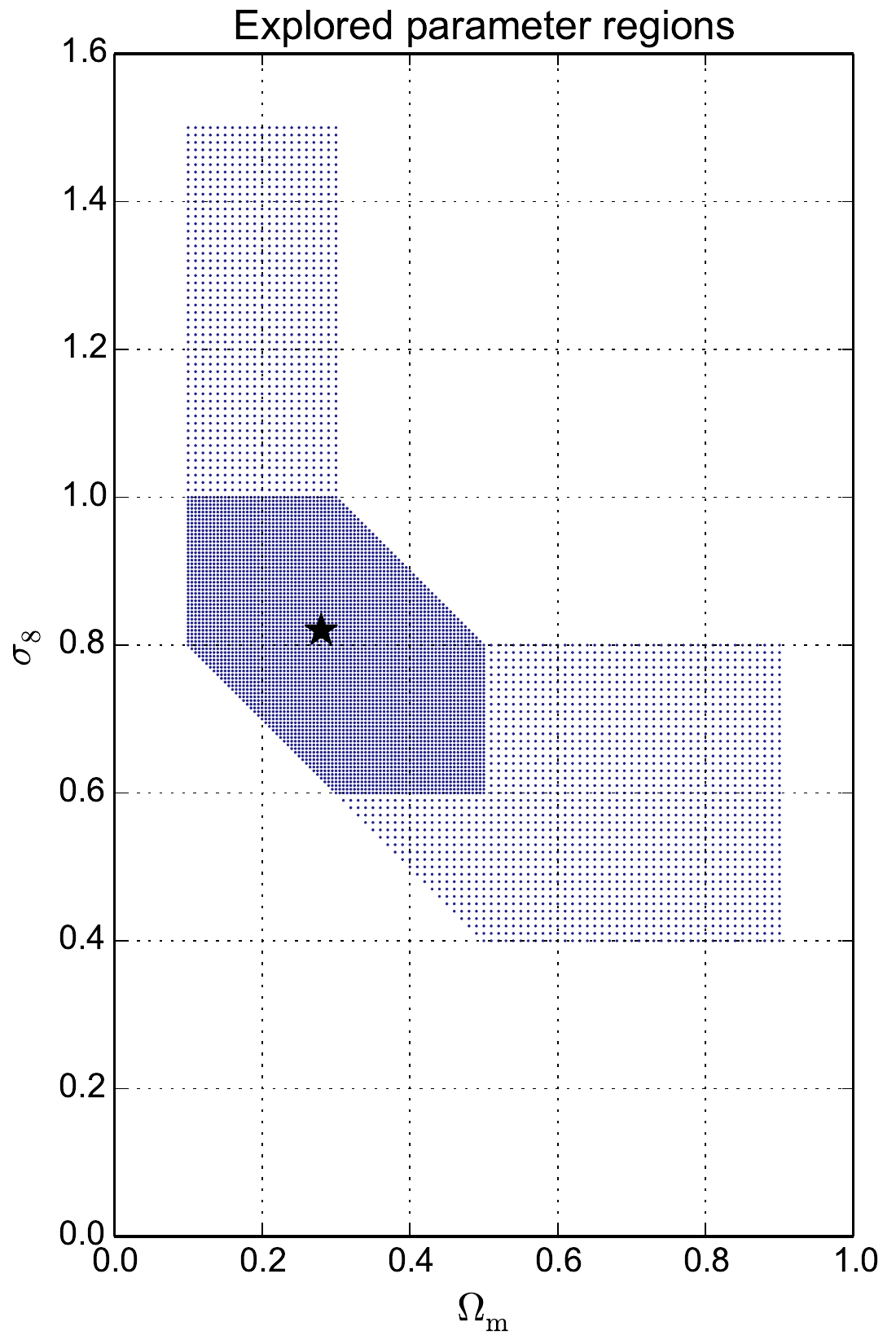}
    \caption{Location of 7821 points on which the likelihoods are evaluated. In the central condensed area, the interval between two grid points is 0.005, while in both wing zones it is 0.01. The black star displays $(\OmegaM^\inp, \sigEig^\inp) = (0.28, 0.82)$.}
	\label{fig:constraint:Explored_regions}
\end{figure}

On the $\OmegaM$-$\sigEig$ plane, a region where the posterior density is potentially high is explored. A ``grid-point evaluation''\index{Grid-point evaluation} of the likelihood is carried out. The resolution of the grid is $\Delta\OmegaM=\Delta\sigEig=0.005$ in the center zone, and $\Delta\OmegaM=\Delta\sigEig=0.01$ in the outer area, as shown by \fig{fig:constraint:Explored_regions}. This results in a total of 7821 points in the parameter space to evaluate. For each parameter vector, $N=1000$ realizations of a 25-deg$^2$ field are carried out. This is the data set from which statistical properties such as mean and covariance are evaluated. The setting for our peak model is similar to the description in \sect{sect:modelling:validation:methodology} with \acro{NFW} profiles, regular sources, $\kappa$ peaks, Gaussian smoothing, and global significance (as sketched in \fig{fig:modelling:LK_model_diagram_2}). The input parameter values are detailed in \tab{tab:constraint:parameters}.

\begin{table}[tb]
	\centering
	\begin{tabular}{lcl}
		\hline\hline
		Parameter                        & Symbol            & Value\\
		\hline
		Lower sampling limit             & $M_\minn$         & $\dix{12}~\Msol/h$\\
		Upper sampling limit             & $M_\maxx$         & $\dix{17}~\Msol/h$\\
		Number of halo redshift bins     & -                 & 10\\
		NFW inner slope                  & $\alpha$          & 1\\
		$M$-$c$ relation amplitude       & $c_0$             & 11\\
		$M$-$c$ relation power law index & $\betaNFW$        & 0.13\\
		Source redshift                  & $z_\rms$          & 1\\
		Intrinsic ellipticity dispersion & $\sigma_\epsilon$ & 0.4\\
		Galaxy number density            & $n_\gala$         & 25 arcmin$\invSq$\\
		Pixel size                       & $\theta_\pix$     & 0.2 arcmin\\
		Gaussian kernel size             & $\theta_\rmG$     & 1 arcmin\\
		Noise level in a pixel           & $\sigma_\pix$     & 0.283\\
		Noise level after smoothing      & $\sigma_\noise$   & 0.0226\\
		Effective field area             & -                 & 25 deg$^2$\\
		\hline\hline
	\end{tabular}
	\caption{List of parameter values adopted in the study of this chapter.}
	\label{tab:constraint:parameters}
\end{table}

Apart from the data set mentioned earlier, a single model realization under an input cosmology, $\bpi^\inp = (\OmegaM^\inp, \sigEig^\inp)$, is also performed. This serves as the mock observation. In other words, denoting $P(\bx|\bpi)$ as the \acro{PDF} to obtain a data vector $\bx$ from our stochastic model under the cosmology $\bpi$, the observed data vector $\bx^\obs$ for this chapter is just a sample point drawn from $P(\cdot|\bpi^\inp)$. This study is not focusing on accuracy, but precision of the model, so that model biases and the random fluctuation related to generating a mock input can be neglected. The input parameters are set to $\OmegaM^\inp =~0.28$ and $\sigEig^\inp =~0.82$, which corresponds to a \acro{WMAP}-9-like cosmology.

\subsection{Data vector definitions}
\label{sect:constraint:methodology:vector}

Peak-count information can be presented not only as histograms. For example, \citet{Dietrich_Hartlap_2010} have used quantities related to the inverse peak cumulative distribution function (\acro{CDF}) to define observables, i.e. the data vector. Which definition can actually extract better the cosmological information? In order to answer to this question, all constraining techniques are applied on three different types of observables derived from the same data set. These three definitions of the data vector are (1) the abundance of peaks found in each \acro{S/N} bin (binned peak function, noted as $\bx^\abd$); (2) the \acro{S/N} values at some given percentiles of the peak \acro{CDF} (inverse peak \acro{CDF}, noted as $\bx^\pct$); and (3) similar to the second type, but with a cutoff for peaks below a certain threshold value (noted as $\bx^\cut$). Mathematically, the $i$-th component of the two last types of observables can be denoted as $x_i$, thereby satisfying
\begin{align}\label{for:constraint:percentile}
	u_i = \int_{\nu_\minn}^{x_i} n_\peak(\nu)\rmd\nu,
\end{align}
where $n_\peak(\nu)$ is the peak function derived from an observation or a model realization, $\nu_\minn$ a cutoff, and $u_i$ a specific percentile. The vector $\bx^\pct$ corresponds to the case $\nu_\minn=-\infty$. 

\begin{figure}[tb]
	\centering
	\includegraphics[width=0.8\textwidth]{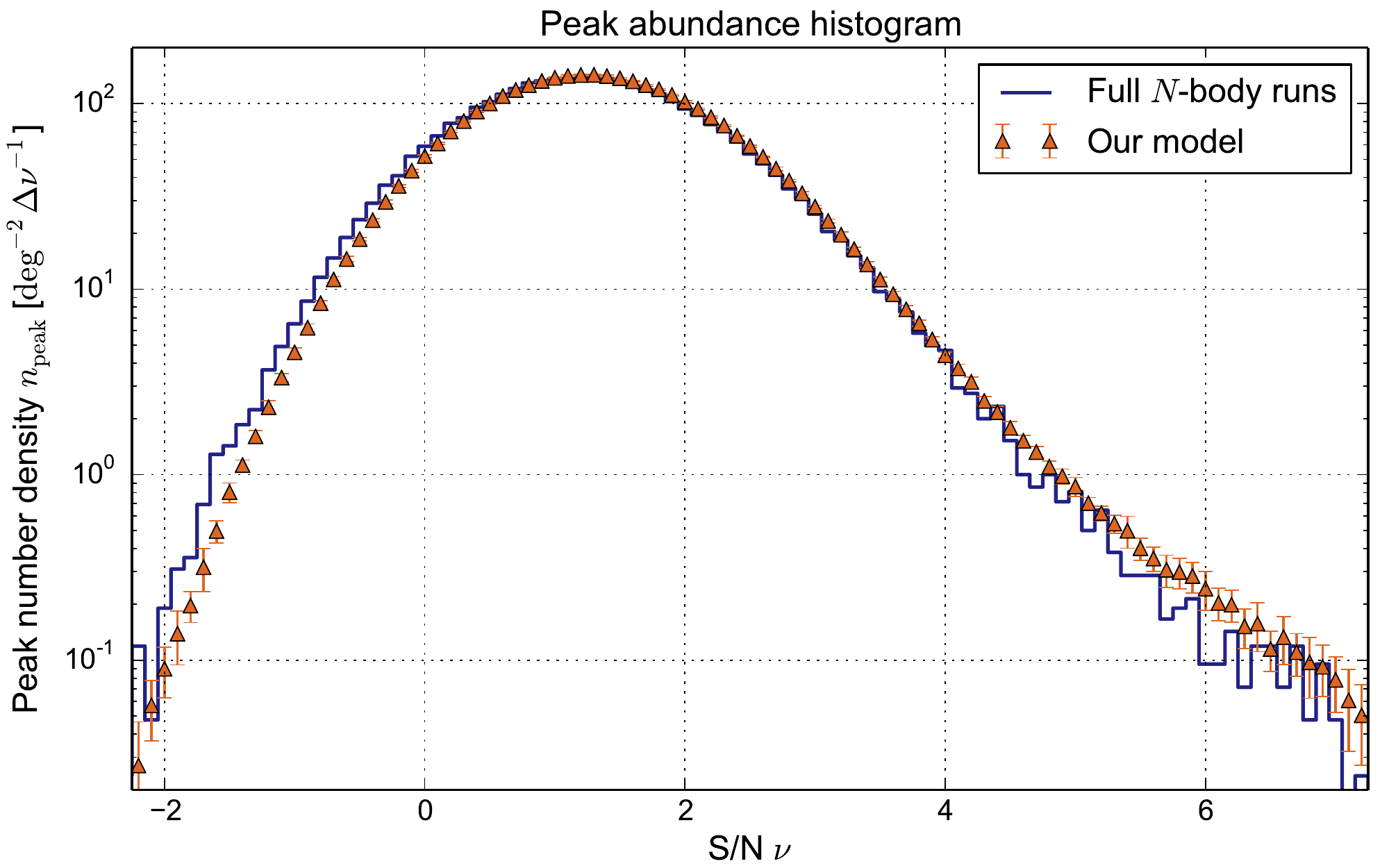}
    \caption{Similar to \fig{fig:modelling:peakHist_largeField}. Only $N$-body runs and our model are retained. The binwidth has been changed and the $x$ range has been extended to -2. We can see that the number of negative peaks is systematically underestimated.}
	\label{fig:constraint:peakHist_negative}
\end{figure}

The motivation for these choices is as follows. The \acro{CDF}-based data vectors are free from an artificial binwidth choice, thus avoid systematics derived from binning. However, their components might be highly correlated, while diagonal terms are probably more dominant in the covariance from \acro{PDF}-based data vectors. On the other hand, our modelling technique focuses on massive halos, which are overdensity features. The prediction on local maxima with negative \acro{S/N} has been found out to be incorrect, as shown by \fig{fig:constraint:peakHist_negative}, which will further impact on values of $\bx^\pct$. As a result, with real data, using directly $\bx^\pct$ might introduce substantial bias. For this reason, $\bx^\pct$ is included only for giving an idea about how much information we can extract from a data vector derived from the full inverse \acro{CDF}, and $\bx^\cut$ is a more realistic choice for future studies.

In \sect{sect:modelling:LK:description}, it is mentioned that although low-$\nu$ peaks might have better constraining power, it is not recommended to use for our model due to the lower mass cutoff $M_\minn$ during mass sampling. For the same reason, I focus here on peaks with $\nu\geq~3$ and define $\bx^\abd$, $\bx^\pct$, and $\bx^\cut$ as indicated in \tab{tab:constraint:data_vector}. Starting from choosing logarithmic separated percentiles for $\bx^\cut$ (with $\nu_\minn=3$), I determine the average of the corresponding $x_i$ over the~1000 realizations under $\bpi^\inp$. Then, with these $x_i$, I define $u_i$ for $\bx^\pct$ and bins for $\bx^\abd$ such that all three vectors represent approximately the same information. Of course, this coherence is only connected via the \acro{CDF} under $\bpi^\inp$ and might break down under another cosmology, but for the current studying framework, this choice can be considered reasonable.

\begin{table}[tb]
	\centering
	\begin{tabular}{cc@{\hspace*{0.6em}}c@{\hspace*{0.6em}}c@{\hspace*{0.6em}}c@{\hspace*{0.6em}}c}
		\hline\hline
		Label                 & \texttt{abd}\\
		Bins on $\nu$         & \small[3.0, 3.8[ & \small[3.8, 4.5[ & \small[4.5, 5.3[ & \small[5.3, 6.2[ & \small[6.2, $+\infty$[\\
		$x_i$ for $\bpi^\inp$ & 330   & 91    & 39    & 18    & 15\\
		\hline
		Label                 & \texttt{pct}\\
		$\nu_\minn$           & $-\infty$\\
		$u_i$                 & 0.969 & 0.986 & 0.994 & 0.997 & 0.999\\
		$x_i$ for $\bpi^\inp$ & 3.5   & 4.1   & 4.9   & 5.7   & 7.0\\
		\hline
		Label                 & \texttt{cut}\\
		$\nu_\minn$           & 3\\
		$u_i$                 & 0.5   & 0.776 & 0.9   & 0.955 & 0.98\\
		$x_i$ for $\bpi^\inp$ & 3.5   & 4.1   & 4.9   & 5.7   & 6.7\\
		\hline\hline
	\end{tabular}
	\caption{Definition of $\bx^\abd$, $\bx^\pct$, and $\bx^\cut$. The $x_i$ corresponds to the number of peaks for the first case and \acro{S/N} values for two others. Parameters such as $\nu_\minn$ and $u_i$ are used in \for{for:constraint:percentile}. As an indication, their values for the input cosmology $\bpi^\inp$ are also given. They were calculated by averaging over 1000 realizations.}
	\label{tab:constraint:data_vector}
\end{table}

\tab{tab:constraint:correlation_matrix} shows the correlation matrix for each of the data vectors under $\bpi^\inp$. It reveals that the inter-component correlation for $\bx^\abd$ is quite weak, and strong for $\bx^\pct$ and $\bx^\cut$. This suggests that the covariance should be included in likelihood analyses. The highest absolute value of off-diagonal terms does not exceed 17\% in the case of $\bx^\abd$. A similar result can be reproduced when binning peaks differently for $\bx^\abd$. 

\begin{table}[tb]
	\centering
	\begin{tabular}{cc}
		\texttt{abd} & 
		$\begin{pmatrix}
			1     & -0.05 & -0.09 & -0.08 & -0.16\\
			-0.05 & 1     & -0.05 & -0.01 & -0.12\\
			-0.09 & -0.05 & 1     & -0.04 & -0.11\\
			-0.08 & -0.01 & -0.04 & 1     & -0.06\\
			-0.16 & -0.12 & -0.11 & -0.06 & 1
		\end{pmatrix}$\\[9ex]
		\texttt{pct} & 
		$\begin{pmatrix}
			1    & 0.62 & 0.29 & 0.15 & 0.11\\
			0.62 & 1    & 0.58 & 0.36 & 0.25\\
			0.29 & 0.58 & 1    & 0.66 & 0.43\\
			0.15 & 0.36 & 0.66 & 1    & 0.59\\
			0.11 & 0.25 & 0.43 & 0.59 & 1
		\end{pmatrix}$\\[9ex]
		\texttt{cut} & 
		$\begin{pmatrix}
			1    & 0.58 & 0.31 & 0.20 & 0.15\\
			0.58 & 1    & 0.61 & 0.39 & 0.28\\
			0.31 & 0.61 & 1    & 0.65 & 0.47\\
			0.20 & 0.39 & 0.65 & 1    & 0.70\\
			0.15 & 0.28 & 0.47 & 0.70 & 1
		\end{pmatrix}$
	\end{tabular}
	\caption{Correlation matrices of $\bx^\abd$, $\bx^\pct$, and $\bx^\cut$ in the input cosmology. For $\bx^\abd$, the peak abundance is weakly correlated between bins.}
	\label{tab:constraint:correlation_matrix}
\end{table}

\subsection{How to qualify a constraint?}

Performing comparison between estimators requires quantitative criteria on constraint contours. In this respect, two indicators have been found in literature.\index{Constraint indicator} Inspired by \citet{Jain_Seljak_1997} and \citet{Maoli_etal_2001}, the first quantity is the error on
\begin{align}\label{for:constraint:Sigma_8}
	\Sigma_8 \equiv \sigEig\left( \frac{\OmegaM}{\Omega_0} \right)^\alpha.
\end{align}
In some papers, the pivot value $\Omega_0$ is set to 0.3. Here, the convention of $\Omega_0=0.27$, adopted by \citetalias{Liu_etal_2015} (\citeyear{Liu_etal_2015}) and \citetalias{Liu_etal_2015a} (\citeyear{Liu_etal_2015a}), is used. This error, noted as $\Delta\Sigma_8$, is the ``thickness'' of the banana-shaped contour on the $\OmegaM$-$\sigEig$ plane tilted by the slope $\alpha$. The value for the slope is usually taken from the best-fit result with the linear relation $\log\Sigma_8 = \log\sigEig + \alpha\log(\OmegaM/0.27)$. Then, assuming this $\alpha$ value, one can transform the estimator into a function of $\OmegaM$ and $\Sigma_8$, marginalize over $\OmegaM$, and derive $\Delta\Sigma_8$.

\begin{figure}[tb]
	\centering
	\includegraphics[width=\textwidth]{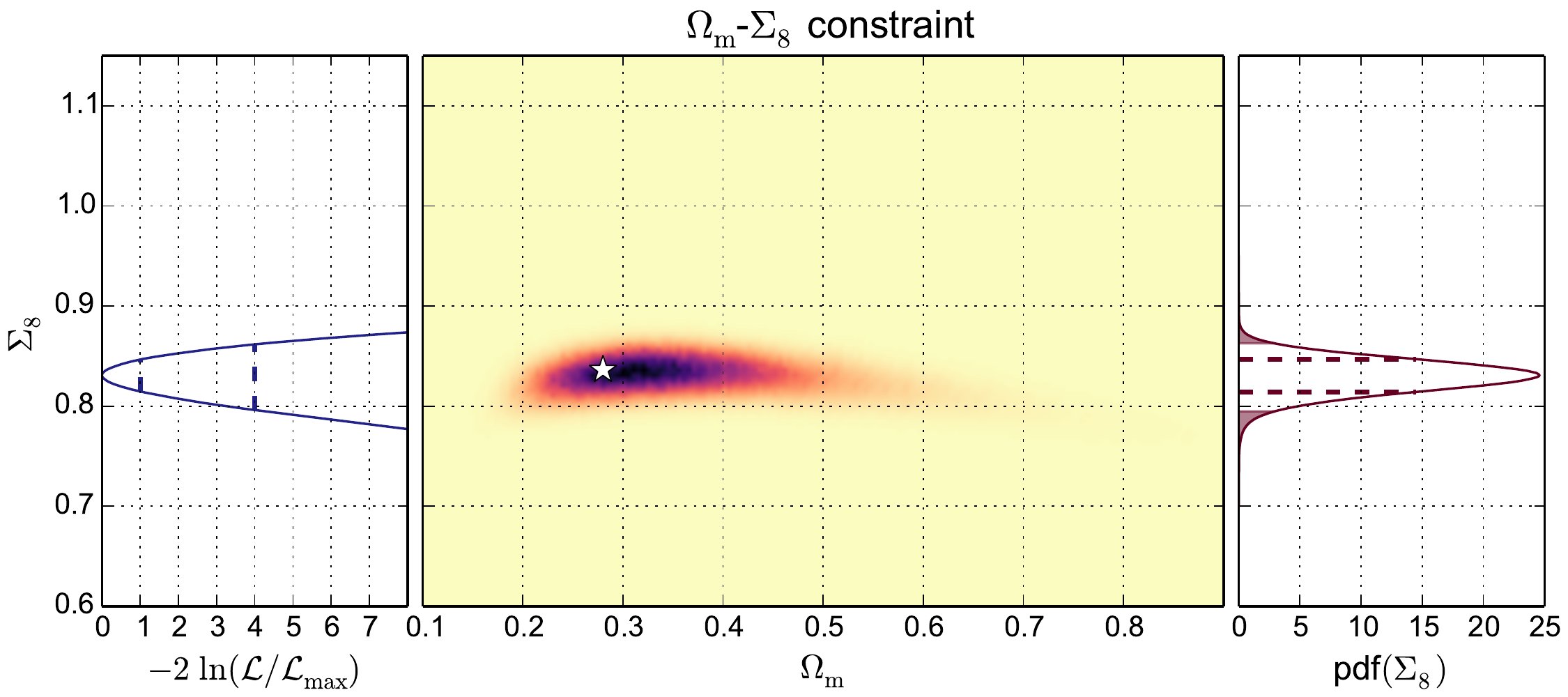}
	\caption{Middle panel: the likelihood value using $\bx^\abd$ on the $\OmegaM$-$\Sigma_8$ plane. The white star represents the input cosmology $\bpi^\inp$. Since $\log\sigEig$ and $\log\OmegaM$ form an approximately linear degeneracy, the quantity $\Sigma_8 \equiv \sigEig (\OmegaM/0.27)^\alpha$ allows us to characterize the thickness. Right panel: the marginalized \acro{PDF} of $\Sigma_8$. The dashed lines give the 1-$\sigma$ interval (68.3\%), while the borders of the shaded areas represent 2-$\sigma$ limits (95.4\%). Left panel: the log-value of the marginalized likelihood ratio. Dashed lines in the left panel give the corresponding value for 1- and 2-$\sigma$ significance levels, respectively.}
	\label{fig:constraint:pdfSigma8_abd5_cg_both}
\end{figure}

In this work, both frequentist and Bayesian analyses are performed. The frequentist approach proceeds what we call \textit{likelihood-ratio test}\index{Likelihood-ratio test} \citep[see e.g. Theorem 10.3.3 from][]{Casella_Berger_2002}. In a general mathematical context, this defines the \textit{$i$-$\sigma$ confidence region}\index{Confidence region} as
\begin{align}\label{for:constraint:confidence_region}
	\text{set of all}\ \ \bpi\ \ \text{such that}\ \ 1-p_i \geq \int_0^{-2\ln(\Like(\bpi)/\Like_\maxx)} \chi^2_f(t)\rmd t,
\end{align}
where $\chi^2_f$ is the chi-squared distribution with $f$ degrees of freedom ($f$ is the dimension of $\bpi$) and $p_i$ is the $p$-value corresponding to the $i$-$\sigma$ significance, defined as
\begin{align}\label{for:constraint:p_i}
	1-p_i \equiv \frac{1}{\sqrt{2\pi}}\int_{-i}^{+i} \exp(-t^2/2)\rmd t.
\end{align}
For example, $1 - p_1 \approx 68.3\%$ and $1 - p_2 \approx 95.4\%$. The interpretation for the confidence region is that \textit{the expected result from the considered model for any parameter inside the region should be at least better than the current observed data at the $i$-$\sigma$ level}, where better refers to more probable. Here, $f$ is just one since $\alpha$ has been fixed and the only parameter is $\Sigma_8$. The quantity $\Delta\Sigma_8$ is defined as the width of the 1-$\sigma$ interval. In the case of $f=1$, this is exactly the interval of $\Sigma_8$ such that $-2\ln(\Like(\Sigma_8)/\Like_\maxx) \leq 1$. The left panel of \fig{fig:constraint:pdfSigma8_abd5_cg_both} shows an example of the likelihood-ratio test and the derived confidence interval on $\Sigma_8$.

For Bayesian inference, one seeks for the most probable region based on the likelihood function. This is called \textit{$i$-$\sigma$ credible region}\index{Credible region} and is defined as
\begin{align}\label{for:constraint:credible_region}
	\text{set of all}\ \ \bpi\ \ \text{such that}\ \ 1 - p_i \geq \int \rmd^f\bpi'\ \Like(\bpi') \Theta(\Like(\bpi')-\Like(\bpi)).
\end{align}
As a result, $\Delta\Sigma_8$ is just the width of the 1-$\sigma$ credible interval, as shown on the right panel of \fig{fig:constraint:pdfSigma8_abd5_cg_both}. I would like to highlight that the term \textit{confidence} (region, contour, interval, etc.) always refers to a frequentist analysis, while the \textit{ad hoc} word for a Bayesian approach is \textit{credible}.

The second quantitative indicator for constraint contours is the figure of merit (\acro{FoM})\index{Figure of merit (\acro{FoM})} proposed by \citet{Dietrich_Hartlap_2010}. Analogous to the one from \citet{Albrecht_etal_2006}, this is the inverse of the area of the 2-$\sigma$ region on the $\OmegaM$-$\sigEig$ plane. Depending on the scenario, it can be either a confidence region (\for{for:constraint:confidence_region}) or a credible region (\for{for:constraint:credible_region}). By defining $\Delta\Sigma_8$ and \acro{FoM}, we can now analyze quantitatively the difference between estimators.

\section{Gaussian likelihood and cosmology-dependent covariance}
\label{sect:constraint:CDC}

\subsection{Formalisms}
\label{sect:constraint:CDC:formalisms}

\index{Cosmology-dependent covariance (\acro{CDC})}In this section, I show the parameter constraint contours from our model, and I further examine the so-called \textit{cosmology-dependent-covariance effect} (\acro{CDC} effect) on the Gaussian likelihood.

Following \sect{sect:constraint:methodology:likelihood}, the shape of the likelihood function is usually assumed to be a multivariate Gaussian, that is\index{Likelihood, Gaussian}
\begin{align}\label{for:constraint:Gaussian_likelihood}
	\Like_\rmg(\bpi) = \frac{1}{\sqrt{(2\pi)^d|\det\bC(\bpi)|}} \exp\left[-\frac{1}{2}\Delta\bx^T(\bpi) \cdot \bC\inv(\bpi) \cdot \Delta\bx(\bpi)\right],
\end{align}
where the label \texttt{g} stands for Gaussian, $d$ denotes the dimension of $\bx$, $\Delta\bx(\bpi) \equiv \bx^\obs-\bx^\model(\bpi)$, the difference between the observation and the model prediction $\bx^\model(\bpi)$, and $\bC(\bpi)$ is the covariance matrix for $\bx^\model(\bpi)$.

In general, estimating $\bC(\bpi)$ could be difficult or expensive. As I have mentioned in \chap{sect:modelling}, most studies call a relatively large amount of $N$-body simulations or resampling techniques such as bootstrap and jackknife to estimate the covariance. In order to make the estimation tractable, a common approximation is to assume that the covariance is independent from the cosmological parameters, so that $\bC(\bpi)=\bC(\bpi^\inp)$ is constant and the log-likelihood, defined as $L\equiv-2\ln\Like$, can be written as
\begin{align}\label{for:constraint:L_cg}
	L_\cg(\bpi) = \text{\acro{cst}} + \Delta\bx^T(\bpi)\cdot \widehat{\bC\inv}(\bpi^\inp)\cdot \Delta\bx(\bpi),
\end{align}
where $\widehat{\bC\inv}(\bpi^\inp)$ is the inverse covariance estimated at a chosen parameter $\bpi^\inp$, and the constant comes from the determinant in \for{for:constraint:Gaussian_likelihood} which is not important since we are always interested in the likelihood ratio. The remaining $\bpi$-dependent term in \for{for:constraint:L_cg} is called ``chi-squared term'' ($\chi^2$ term). The label \texttt{cg} is an abbreviation of constant-covariance Gaussian. If the covariance is estimated with resampling techniques, one would rather replace $\bC(\bpi^\inp)$ by $\bC(\bpi^\obs)$, but the idea remains the same, which is treating it as an external parameter.

Does this constant-covariance hypothesis introduce an impact on parameter constraints? To examine this, two other estimators are proposed. The first one still neglects the determinant term in \for{for:constraint:Gaussian_likelihood}, but considers the $\chi^2$ term parameter-dependent. In other words, this semi-varying-covariance Gaussian log-likelihood (labelled \texttt{svg}) is
\begin{align}
	L_\svg(\bpi) = \text{\acro{cst}} + \Delta\bx^T(\bpi)\cdot \widehat{\bC\inv}(\bpi)\cdot \Delta\bx(\bpi).
\end{align}
The second log-likelihood is a fully varying-covariance and Gaussian (labelled \texttt{vg}) function, mathematically given by
\begin{align}
	L_\vg(\bpi) = \text{\acro{cst}} + \ln\left\vert \det\widehat{\bC}(\bpi) \right\vert + \Delta\bx^T(\bpi)\cdot \widehat{\bC\inv}(\bpi)\cdot \Delta\bx(\bpi).
\end{align}

The likelihoods above are evaluated using the $N=1000$ model realization set described in \sect{sect:constraint:methodology:design}. The model prediction is the mean over $N$ samples for each $\bpi$. Denote $\bx^{(k)} = \left(x_1^{(k)}, \ldots, x_d^{(k)}\right)$ as the data vector of $k$-th realization, the unbiased estimators of the mean, the covariance matrix, and the inverse matrix \citep{Hartlap_etal_2007} can be written respectively as
\begin{align}
	x^\model_i &= \frac{1}{N}\sum_{k=1}^N x_i^{(k)}, \label{for:constraint:mean}\\
	\widehat{C}_{ij} &= \frac{1}{N-1}\sum_{k=1}^N \left(x_i^{(k)} - x^\model_i\right) \left(x_j^{(k)} - x^\model_j\right), \label{for:constraint:covariance}\\
	\widehat{\bC\inv} &= \frac{N-d-2}{N-1}\ \widehat{\bC}\inv. \label{for:constraint:inverse_covariance}
\end{align}

\subsection{The chi-squared term}

\begin{figure}[tb]
	\centering
	\includegraphics[width=0.49\textwidth]{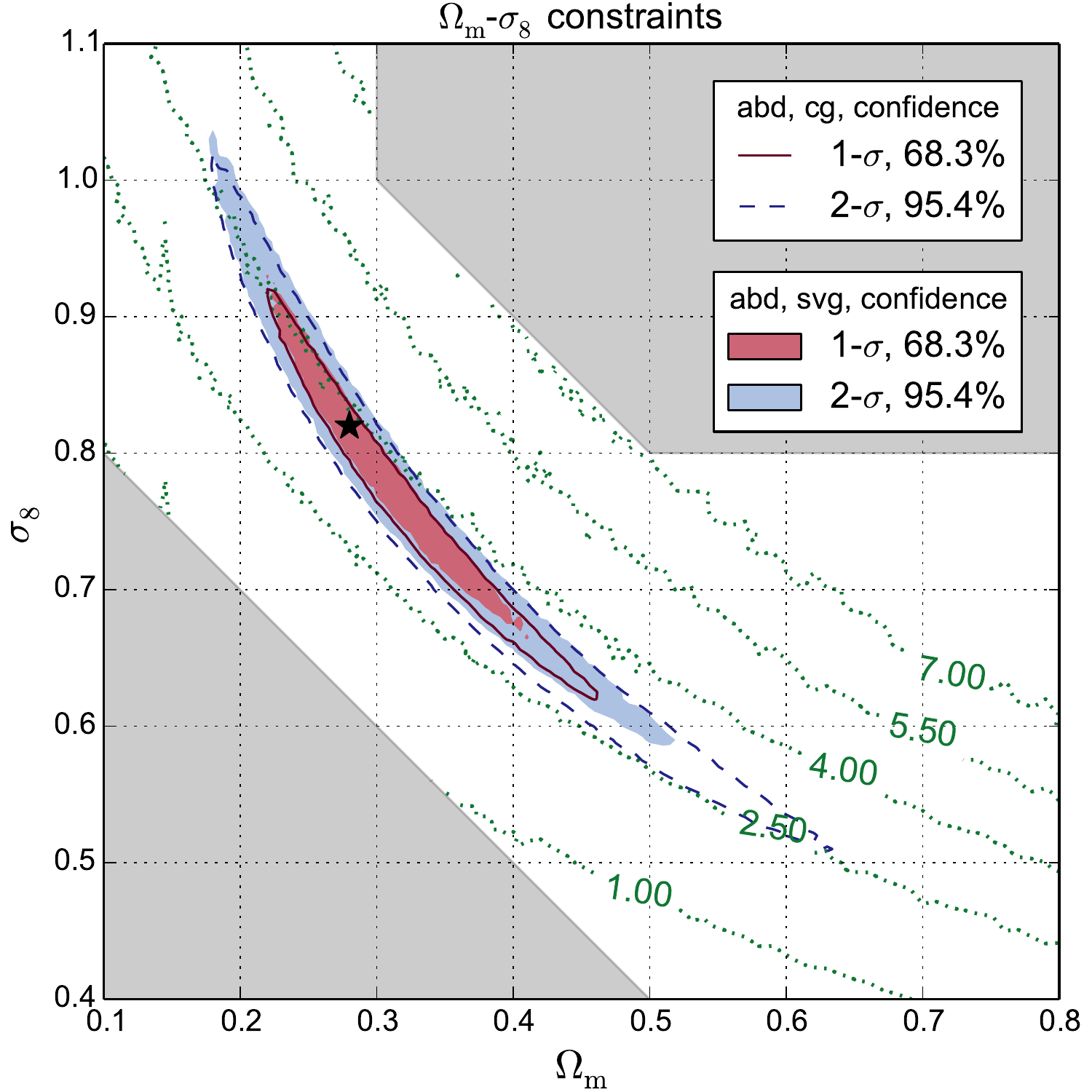}\hfill
	\includegraphics[width=0.49\textwidth]{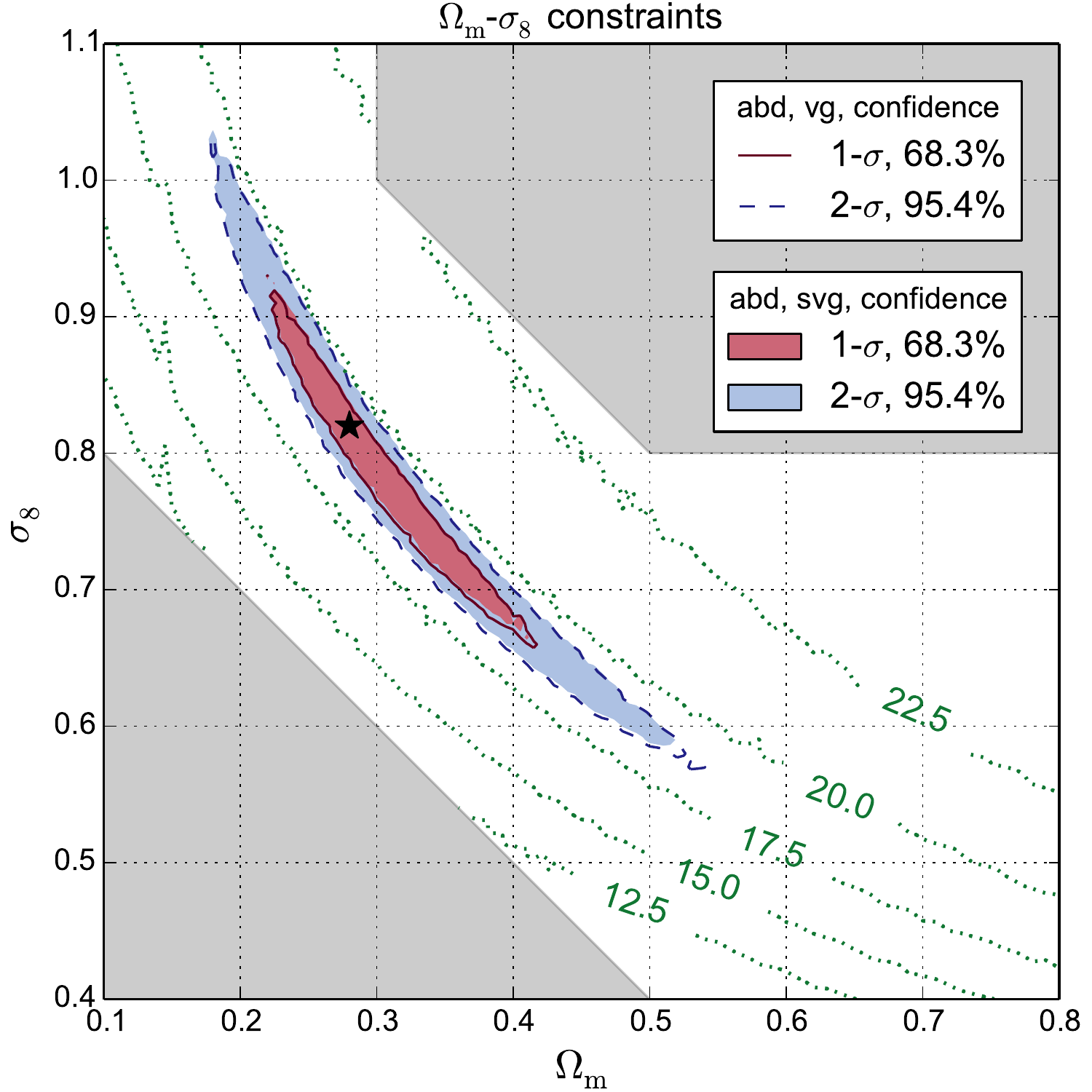}
	\caption{Confidence regions derived from $L_\cg$, $L_\svg$, and $L_\vg$ with $\bx^\abd$. The solid and dashed contours represent $L_\cg$ in the left panel and $L_\vg$ in the right panel, while the colored areas are from $L_\svg$. The black star stands for $\bpi^\inp$ and gray areas represent the non-explored parameter space. The green dotted lines are different isolines, the variance $\widehat{C}_{55}$ of the bin with highest \acro{S/N} in the left panel and $\ln(\det\widehat{C})$ for the right panel. The contour area is reduced by 22\% when taking the \acro{CDC} effect into account. The parameter-dependent determinant term does not contribute significantly.}
	\label{fig:constraint:contour_Gaussian_abd}
\end{figure}

\begin{figure}[!tb]
	\centering
	\includegraphics[width=0.49\textwidth]{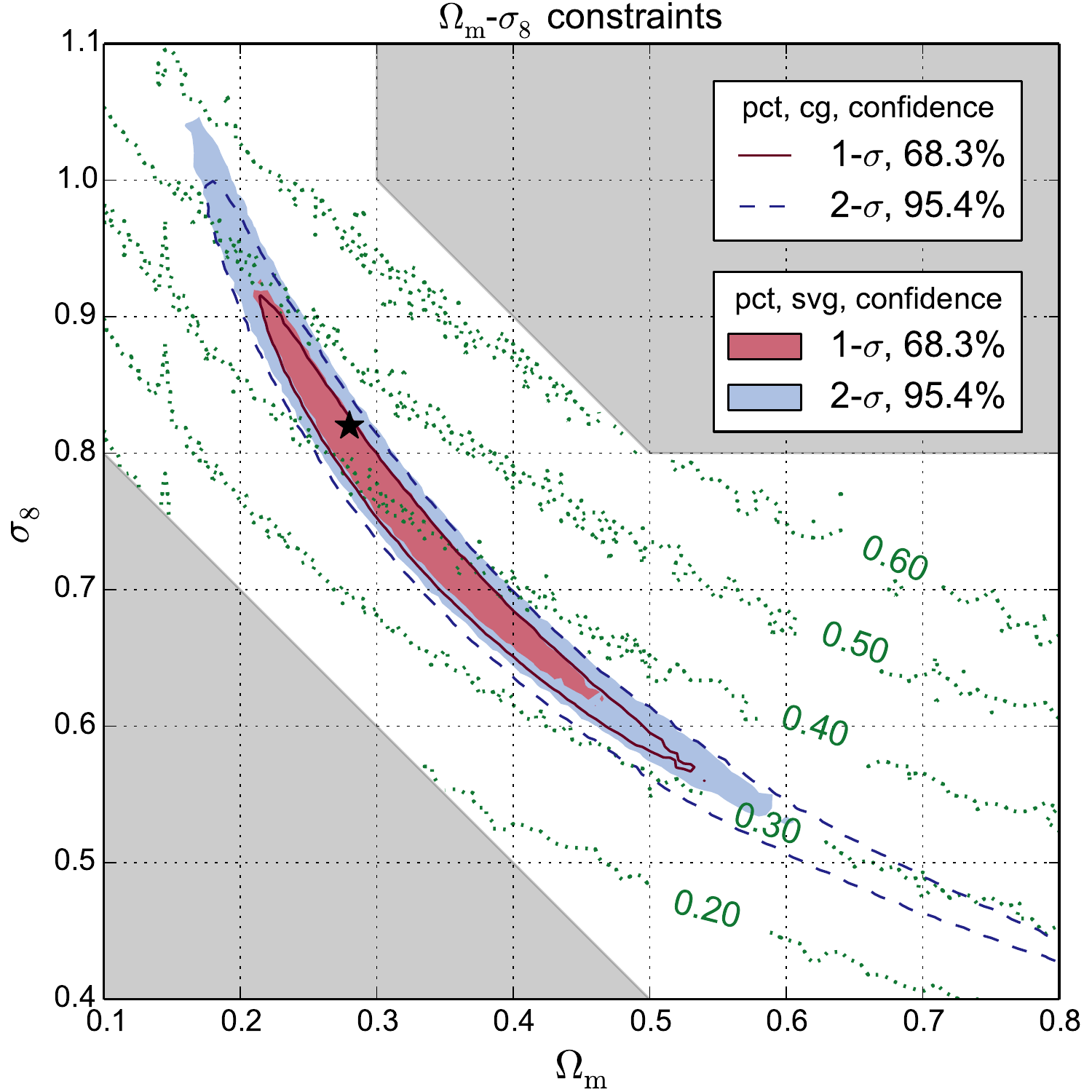}\hfill
	\includegraphics[width=0.49\textwidth]{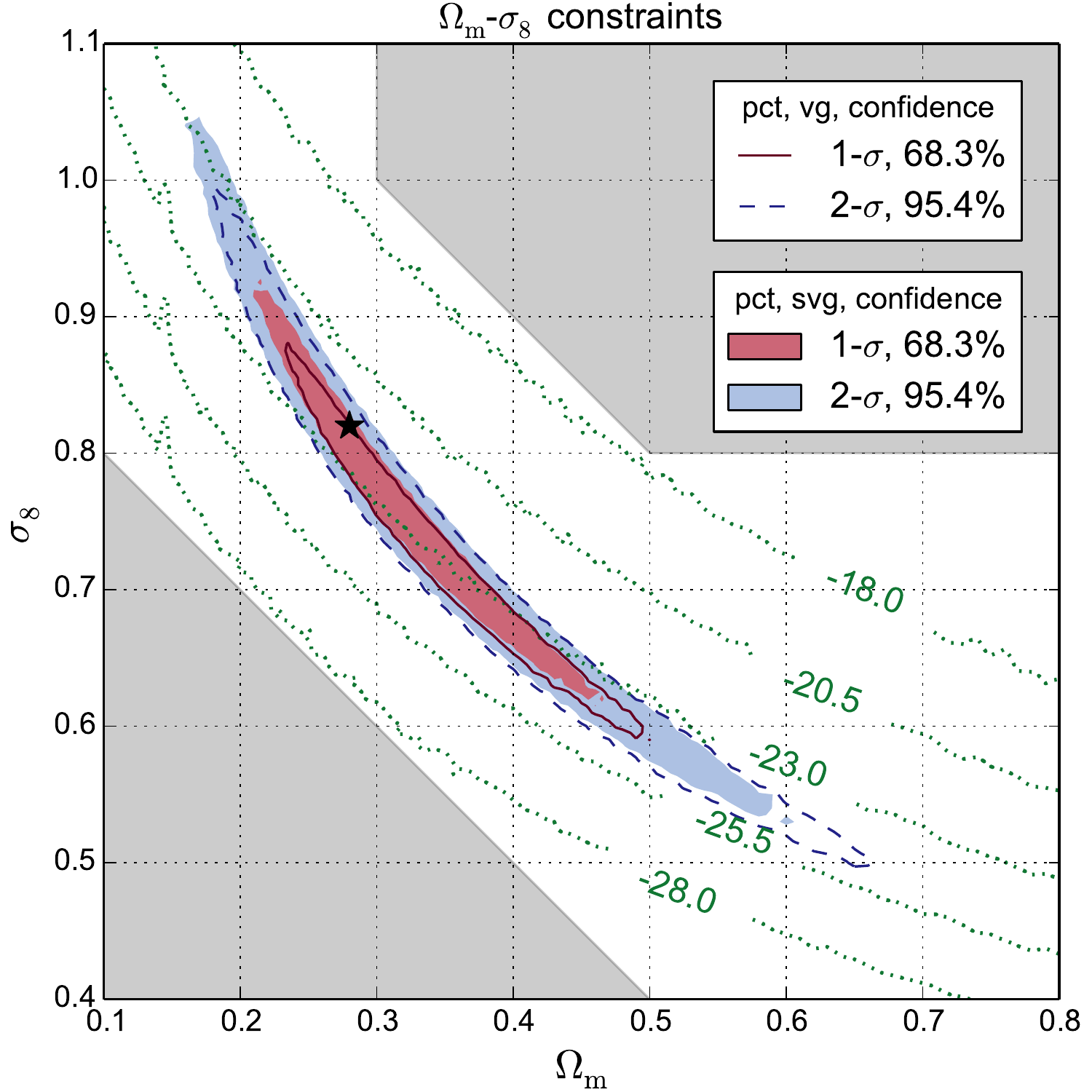}\\
	\includegraphics[width=0.49\textwidth]{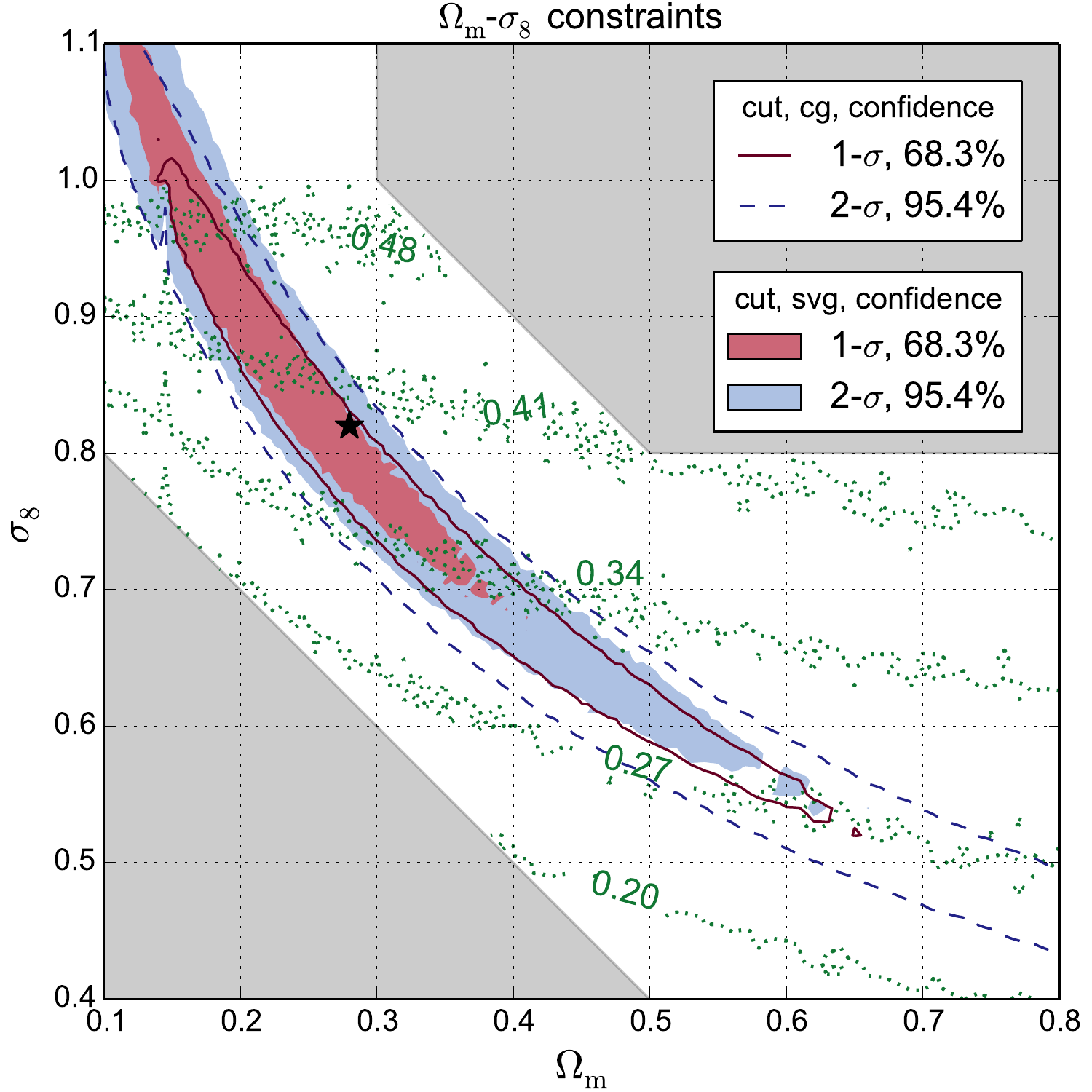}\hfill
	\includegraphics[width=0.49\textwidth]{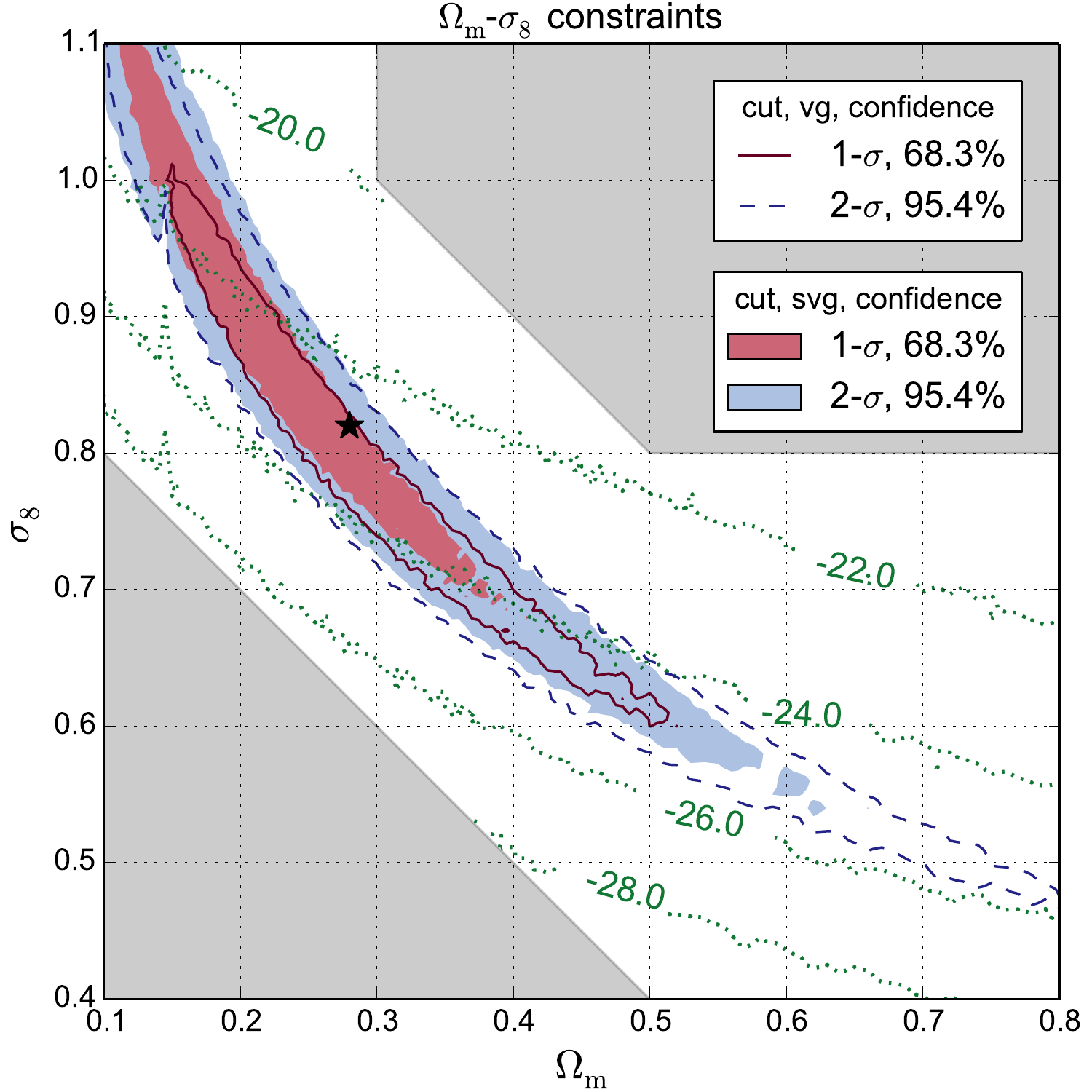}
	\caption{Similar to \fig{fig:constraint:contour_Gaussian_abd}. Confidence regions with $\bx^\pct$ and $\bx^\cut$. Both upper panels are drawn with $\bx^\pct$ and both lower panels with $\bx^\cut$. Both left panels are the comparison between $L_\cg$ and $L_\svg$, and both right panels between $L_\svg$ and $L_\vg$.}
	\label{fig:constraint:contour_Gaussian_pct}
\end{figure}

What is the impact of the \acro{CDC} effect on the $\chi^2$ term? The left panel of \fig{fig:constraint:contour_Gaussian_abd} shows the comparison between confidence regions derived from $L_\cg$ and $L_\svg$. The data vector is $\bx^\abd$. One can observe that the constraint contours have been shifted. As shown in \tab{tab:constraint:correlation_matrix}, the off-diagonal correlation coefficients are weak. Therefore, the variation in diagonal terms of $\widehat{\bC}$ plays a major role in the size of confidence regions. In the same figure, the isolines for $\widehat{C}_{55}$ are also drawn. These isolines cross the $\OmegaM$-$\sigEig$ degeneracy lines from $L_\cg$ and thus shrink the confidence region. More detailed examinations reveal that the isolines for $\widehat{C}_{11}$ and $\widehat{C}_{22}$ are noisy and that those for $\widehat{C}_{33}$ and $\widehat{C}_{44}$ coincide well with the original degeneracy direction. 

\tab{tab:constraint:indicator_confidence} shows the values of both indicators for different likelihoods. One can observe that using $L_\svg$ significantly improves the constraints by 24\% in terms of \acro{FoM} compared to $L_\cg$. Regarding $\Delta\Sigma_8$, the improvement is weak. As a result, using varying covariance matrices breaks down the degeneracy partially and shrinks the contour length, but does not effectively reduce the thickness. 

The same constraints derived from two other data vectors $\bx^\pct$ and $\bx^\cut$ can be found in the left panel of \fig{fig:constraint:contour_Gaussian_pct}. A similar \acro{CDC} effect exists for both. One can remark that $\bx^\pct$ has less constraining power than $\bx^\abd$. Also, $\bx^\cut$ is outperformed by both other data vectors. This difference is due to the cutoff value $\nu_\minn$. Introducing a cutoff at $\nu_\minn=3$ decreases the total number of peaks and amplifies the fluctuation of high-peak values in the \acro{CDF}. When we use percentiles to define the data vector, the distribution of each component of $\bx^\cut$ becomes wider than the one of the corresponding component of $\bx^\pct$, and this greater scatter in the \acro{CDF} enlarges the contours. However, the cutoff also introduces a tilt for the contours. \tab{tab:constraint:best_fit_confidence} shows the best fit alpha for the different cases. The difference in the tilt could be a useful tool for improving the constraining power, as demonstrated by \citet{Dietrich_Hartlap_2010}. Nevertheless, I do not explore any joint analysis since \citet{Dietrich_Hartlap_2010} consider peak counts and power spectrum while here, $\bx^\abd$ and $\bx^\cut$ contain basically the same information.

\begin{table}[tb]
	\centering
	\begin{tabular}{c|cc|cc|cc}
		\hline\hline
		          & \multicolumn{2}{c|}{$\bx^\abd$} & \multicolumn{2}{c|}{$\bx^\pct$} & \multicolumn{2}{c}{$\bx^\cut$}\\
		\hline
		          & $\Delta\Sigma_8$ & FoM & $\Delta\Sigma_8$ & FoM & $\Delta\Sigma_8$ & FoM\\
		\hline
		$L_\cg$   & 0.032            & 46  & 0.037            & 31  & 0.065            & 13 \\
		$L_\svg$  & 0.031            & 57  & 0.032            & 42  & 0.054            & 21 \\
		$L_\vg$   & 0.031            & 56  & 0.032            & 43  & 0.052            & 18 \\
		$L_\cc$   & 0.032            & 43  & 0.038            & 33  & 0.056            & 13 \\
		$L_\vc$   & 0.033            & 52  & 0.034            & 39  & 0.058            & 16 \\
		$L_\true$ & 0.033            & 54  & 0.035            & 39  & 0.058            & 17 \\
		$p$-value & 0.035            & 39  & 0.037            & 27  & 0.067            & 12 \\
		\hline\hline
	\end{tabular}
    \caption{Values of $\Delta\Sigma_8$ and the figure of merit (\acro{FoM}) for confidence regions from different analysis approaches. $L_\cg$, $L_\svg$, and $L_\vg$ are introduced in \sect{sect:constraint:CDC:formalisms}, $L_\cc$ and $L_\vc$ in \sect{sect:constraint:copula:likelihood}, and $L_\true$ and $p$-value in \sect{sect:constraint:nonParam}. In each case, $\bx^\abd$, $\bx^\pct$, or $\bx^\cut$ are taken respectively as data vector as indicated in the table rows.}
	\label{tab:constraint:indicator_confidence}
\end{table}

\begin{table}[tb]
	\centering
	\renewcommand{\arraystretch}{1.2}
	\begin{tabular}{c|cc|cc|cc}
		\hline\hline
		              & \multicolumn{2}{c|}{$\bx^\abd$} & \multicolumn{2}{c|}{$\bx^\pct$} & \multicolumn{2}{c}{$\bx^\cut$}\\
		\hline
		              & $\Sigma_8$ $^{+1\sigma}_{-1\sigma}$ & $\alpha$ & $\Sigma_8$ $^{+1\sigma}_{-1\sigma}$ & $\alpha$ & $\Sigma_8$ $^{+1\sigma}_{-1\sigma}$ & $\alpha$\\
		\hline
		$L_\cg$   & 0.831$^{+0.016}_{-0.016}$ & 0.54 & 0.822$^{+0.018}_{-0.019}$ & 0.54 & 0.800$^{+0.030}_{-0.035}$ & 0.45\\
		$L_\svg$  & 0.831$^{+0.016}_{-0.015}$ & 0.52 & 0.820$^{+0.015}_{-0.016}$ & 0.51 & 0.800$^{+0.031}_{-0.023}$ & 0.40\\
		$L_\vg$   & 0.829$^{+0.015}_{-0.015}$ & 0.52 & 0.819$^{+0.015}_{-0.016}$ & 0.52 & 0.800$^{+0.024}_{-0.028}$ & 0.42\\
		$L_\cc$   & 0.830$^{+0.016}_{-0.016}$ & 0.54 & 0.825$^{+0.018}_{-0.020}$ & 0.54 & 0.807$^{+0.025}_{-0.031}$ & 0.46\\
		$L_\vc$   & 0.829$^{+0.016}_{-0.016}$ & 0.52 & 0.823$^{+0.016}_{-0.019}$ & 0.53 & 0.798$^{+0.029}_{-0.029}$ & 0.44\\
		$L_\true$ & 0.828$^{+0.018}_{-0.015}$ & 0.53 & 0.823$^{+0.015}_{-0.020}$ & 0.53 & 0.800$^{+0.028}_{-0.030}$ & 0.44\\
		$p$-value & 0.835$^{+0.016}_{-0.019}$ & 0.54 & 0.823$^{+0.018}_{-0.018}$ & 0.54 & 0.798$^{+0.032}_{-0.034}$ & 0.45\\
		\hline\hline
	\end{tabular}
	\renewcommand{\arraystretch}{1.0}
	\caption{Best fits of $(\Sigma_8, \alpha)$ for confidence regions from different analysis approaches. $L_\cg$, $L_\svg$, and $L_\vg$ are introduced in \sect{sect:constraint:CDC:formalisms}, $L_\cc$ and $L_\vc$ in \sect{sect:constraint:copula:likelihood}, and $L_\true$ and $p$-value in \sect{sect:constraint:nonParam}. Readers should note that the values for $\Sigma_8$ are indicative since no real observational data are used.}
	\label{tab:constraint:best_fit_confidence}
\end{table}

\subsection{The determinant term}

The right panel of \fig{fig:constraint:contour_Gaussian_abd} shows the comparison between $L_\svg$ and $L_\vg$ with the data vector $\bx^\abd$. The difference between contours is minor. This shows that adding the determinant term does not result in significant changes of the parameter constraints. The isolines from $\ln(\det \widehat{\bC})$ illustrate well this phenomenon. The gradients, this time, are perpendicular to the degeneracy lines. Meanwhile, including the determinant makes the contours slightly larger, but almost negligibly so. Quantitatively, the total improvement in the contour area compared to $L_\cg$ is 22\%.

However, a different change is seen for $\bx^\pct$ and $\bx^\cut$ (\fig{fig:constraint:contour_Gaussian_pct}). Adding the determinant to the likelihood computed from these observables induces a shift of contours toward the higher $\OmegaM$ area. In the case of $\bx^\cut$, this shift compensates for the contour offset from the varying $\chi^2$ term, but does not improve either $\Delta\Sigma_8$ or \acro{FoM} significantly, as shown in \tab{tab:constraint:indicator_confidence}. As a result, using the Gaussian likelihood, the total \acro{CDC} effect can be summed up as an improvement of at least 14\% in terms of thickness and 38\% in terms of area.

\begin{table}[tb]
	\centering
	\begin{tabular}{c|cc|cc|cc}
		\hline\hline
		          & \multicolumn{2}{c|}{$\bx^\abd$} & \multicolumn{2}{c|}{$\bx^\pct$} & \multicolumn{2}{c}{$\bx^\cut$}\\
		\hline
		          & $\Delta\Sigma_8$ & FoM & $\Delta\Sigma_8$ & FoM & $\Delta\Sigma_8$ & FoM\\
		\hline
		$L_\cg$   & 0.033            & 43  & 0.038            & 31  & 0.066            & 15 \\
		$L_\svg$  & 0.031            & 53  & 0.033            & 41  & 0.056            & 20 \\
		$L_\vg$   & 0.031            & 53  & 0.032            & 40  & 0.055            & 18 \\
		$L_\cc$   & 0.033            & 40  & 0.040            & 30  & 0.071            & 14 \\
		$L_\vc$   & 0.033            & 47  & 0.035            & 36  & 0.060            & 16 \\
		$L_\true$ & 0.034            & 49  & 0.036            & 36  & 0.061            & 17 \\
		ABC       & 0.056            & 31  & 0.044            & 33  & 0.068            & 16 \\
		\hline\hline
	\end{tabular}
	\caption{Similar to \tab{tab:constraint:indicator_confidence}, but for credible regions. The row \acro{ABC} represents the results from \sect{sect:ABC:constraints:results}.}
	\label{tab:constraint:indicator_credible}
\end{table}

\begin{table}[tb]
	\centering
	\renewcommand{\arraystretch}{1.2}
	\begin{tabular}{c|cc|cc|cc}
		\hline\hline
		          & \multicolumn{2}{c|}{$\bx^\abd$} & \multicolumn{2}{c|}{$\bx^\pct$} & \multicolumn{2}{c}{$\bx^\cut$}\\
		\hline
		          & $\Sigma_8$ $^{+1\sigma}_{-1\sigma}$ & $\alpha$ & $\Sigma_8$ $^{+1\sigma}_{-1\sigma}$ & $\alpha$ & $\Sigma_8$ $^{+1\sigma}_{-1\sigma}$ & $\alpha$\\
		\hline
		$L_\cg$   & 0.831$^{+0.017}_{-0.016}$ & 0.54 & 0.822$^{+0.018}_{-0.020}$ & 0.54 & 0.800$^{+0.030}_{-0.036}$ & 0.45\\
		$L_\svg$  & 0.831$^{+0.016}_{-0.015}$ & 0.52 & 0.820$^{+0.016}_{-0.017}$ & 0.51 & 0.800$^{+0.032}_{-0.024}$ & 0.40\\
		$L_\vg$   & 0.829$^{+0.015}_{-0.015}$ & 0.52 & 0.819$^{+0.015}_{-0.017}$ & 0.52 & 0.800$^{+0.025}_{-0.029}$ & 0.42\\
		$L_\cc$   & 0.830$^{+0.017}_{-0.017}$ & 0.54 & 0.825$^{+0.018}_{-0.022}$ & 0.54 & 0.807$^{+0.030}_{-0.041}$ & 0.46\\
		$L_\vc$   & 0.829$^{+0.016}_{-0.016}$ & 0.52 & 0.823$^{+0.016}_{-0.019}$ & 0.53 & 0.798$^{+0.030}_{-0.030}$ & 0.44\\
		$L_\true$ & 0.828$^{+0.019}_{-0.015}$ & 0.53 & 0.823$^{+0.015}_{-0.021}$ & 0.53 & 0.800$^{+0.030}_{-0.032}$ & 0.44\\
		ABC       & 0.819$^{+0.030}_{-0.025}$ & 0.50 & 0.817$^{+0.022}_{-0.022}$ & 0.51 & 0.799$^{+0.034}_{-0.034}$ & 0.42\\
		\hline\hline
	\end{tabular}
	\renewcommand{\arraystretch}{1.0}
	\caption{Similar to \tab{tab:constraint:best_fit_confidence}, but for credible regions. The row \acro{ABC} represents the results from \sect{sect:ABC:constraints:results}. The values for $\Sigma_8$ are only indicative.}
	\label{tab:constraint:best_fit_credible}
\end{table}

Overall, the results from Bayesian inference are very similar to the likelihood-ratio test. Thus, only $\Delta\Sigma_8$ and \acro{FoM} are shown in \tab{tab:constraint:indicator_credible} and best fits in \tab{tab:constraint:best_fit_credible}. I would like to recall that a similar analysis has been done by \citet{Eifler_etal_2009} on shear covariances. This work agrees with their conclusions: a relatively large impact from the $\chi^2$ term and negligible change from the determinant term. However, the total \acro{CDC} effect is more significant in the peak-count framework than for the power spectrum.

\section{Copula analysis}

\subsection{What is copula?}

\index{Copula}Consider a multivariate joint distribution $P(x_1, \ldots, x_d)$. In general, $P$ could be far from Gaussian so that imposing the Gaussian likelihood could induce biases. The idea of the copula technique is to evaluate the likelihood in a new data space in which the Gaussian approximation is better. Using a change of variables, individual marginalized distributions of $P$ can be approximated to Gaussian ones. This is achieved by a series of successive \acro{1D}, axis-wise transformations. In general, the multivariate Gaussianity of the transformed distribution is still \textbf{not} guaranteed. However, in some cases, this transformation tunes the distribution and makes it ``more Gaussian'', so that evaluating the likelihood in the tuned space is more realistic \citep{Benabed_etal_2009, Sato_etal_2011}.

From Sklar’s theorem \citep{Sklar_1959}, any multivariate distribution $P(x_1, \ldots, x_d)$ can be decomposed into the copula density multiplied by marginalized distributions. A comprehensible and elegant demonstration is given by \citet{Rueschendorf_2009}. Readers are also encouraged to follow \citet{Scherrer_etal_2010} for detailed physical interpretations and \citet{Sato_etal_2011} for a very pedagogical derivation of the Gaussian copula transform.

Let $\bx = (x_1, \ldots, x_d)$ and $P(\bx)$ a multivariate distribution. For all $i$, the marginalized distribution is naturally
\begin{align}
	P_i(x_i) \equiv \int P(\bx) \prod_{j\neq i} \rmd x_j,
\end{align}
and its \acro{CDF} leads
\begin{align}
	F_i(x_i) \equiv \int_{-\infty}^{x_i} P_i(x') \rmd x'.
\end{align}
We can also define the $d$-dimensional cumulative distribution of $P$ as
\begin{align}
	F(\bx) \equiv \int_{-\infty}^{x_1}\cdots\int_{-\infty}^{x_d} P(x_1', \ldots, x_d') \rmd x_1'\cdots\rmd x_d'.
\end{align}
Now, Sklar's theorem indicates that there exists an unique $d$-dimensional distribution function $C$ defined on $[0, 1]^d$ with uniform marginals such that
\begin{align}
	F(\bx) = C(F_1(x_1), \ldots, F_d(x_d)).
\end{align}
This function $C$ is called the \textit{copula}. In other words, by defining $u_i\equiv\ F_i(x_i)$ and $x_i=F\inv_i(u_i)$, the copula is defined as
\begin{align}
	C(\vect{u}) \equiv F(F\inv_1(u_1), \ldots, F\inv_d(u_d)).
\end{align}
We can compute the \textit{copula density}, $c(\vect{u})$, given by
\begin{align}
	c(\vect{u}) \equiv \frac{\partial C(\vect{u})}{\partial u_1 \cdots \partial u_d} = \frac{\partial F(F\inv_1(u_1), \ldots, F\inv_d(u_d))}{\partial x_1 \cdots \partial x_d} \cdot \frac{\partial x_1}{\partial u_1}\cdots\frac{\partial x_d}{\partial u_d},
\end{align}
and this results in
\begin{align}\label{for:constraint:copula_transform_1}
	P(\bx) = c(\vect{u}) P_1(x_1) \cdots P_d(x_d),
\end{align}
which means that \textit{any multivariate distribution can be decomposed into the product of the copula density and marginalized distributions}.

On the other hand, let $q_i \equiv \Phi_i\inv(u_i)$, where $\Phi_i$ is the \acro{CDF} of the normal distribution with the same mean $\mu_i$ and variance $\sigma_i^2$ as the law $P_i$. Mathematically, this leads 
\begin{align}
	\Phi_i(q_i) &\equiv \int_{-\infty}^{q_i} \phi_i(q') \rmd q',\\ 
	\phi_i(q_i) &\equiv \frac{1}{\sqrt{2\pi\sigma_i^2}} \ \exp\left[ -\frac{(q_i-\mu_i)^2}{2\sigma^2_i} \right]. 
\end{align}
The relation between $\vect{q}$ and $\bx$ is just a change of variables, so there exists a new joint \acro{PDF} in the $\vect{q}$-space which corresponds to $P$ in $\bx$-space, say $P'$. Then, we have $P'(\vect{q}) = P(\bx)$. The marginal \acro{PDF} and \acro{CDF} of $P'$ are nothing but $\phi_i$ and $\Phi_i$, respectively. Thus, applying \for{for:constraint:copula_transform_1} to $P'$ and $\phi_i$, one gets
\begin{align}\label{for:constraint:copula_transform_2}
	P'(\vect{q}) = c(\vect{u}) \phi_1(q_1) \cdots \phi_d(q_d).
\end{align}
By uniqueness of $c$, the copula densities in Eqs. \eqref{for:constraint:copula_transform_1} and \eqref{for:constraint:copula_transform_2} turn out to be the same. Thus, we obtain
\begin{align}
	P(\bx) = P'(\vect{q}) \frac{P_1(x_1) \cdots P_d(x_d)}{\phi_1(q_1) \cdots \phi_d(q_d)}.
\end{align}
Readers should note that the marginals of $P'$ are identical to the ones of a multivariate Gaussian distribution $\phi$ with mean $\vect{\mu}$ and covariance $\bC$, where $\bC$ is the covariance matrix of $\bx$. The \acro{PDF} of this Gaussian distribution is
\begin{align}
	\phi(\vect{q}) &\equiv \frac{1}{\sqrt{(2\pi)^d|\det\bC|}} \exp\left[ -\frac{1}{2}\sum_{i=1}^d\sum_{j=1}^d(q_i-\mu_i) C_{ij}\inv (q_j-\mu_j)\right].
\end{align}
Finally, by approximating $P'$ to $\phi$, one gets the \textit{Gaussian copula transform}:
\begin{align}\label{for:constraint:copula_transform_4}
	P(\bx) = \phi(\vect{q}) \frac{P_1(x_1) \cdots P_d(x_d)}{\phi_1(q_1) \cdots \phi_d(q_d)}.
\end{align}

Why is it better to calculate the likelihood in this way? In the usual case, people often approximate the unknown shape of $P(\bx)$ to a normal distribution: $P(\bx) \approx \phi(\bx)$. Here, by applying the Gaussian copula transform, we carry out this approximation into the new space of $\vect{q}$: $P'(\vect{q}) \approx \phi(\vect{q})$. Since $q_i = \Phi\inv_i(F_i(x_i))$, at least the marginals of $P'(\vect{q})$ are strictly Gaussian. And \for{for:constraint:copula_transform_4} gives the corresponding value in $\bx$-space while taking the Gaussian approximation $P'(\vect{q}) \approx \phi(\vect{q})$ in $\vect{q}$-space. This is the reason that the copula transform makes the distribution ``more Gaussian''.

\begin{figure}[tb]
	\centering
	\includegraphics[scale=0.65]{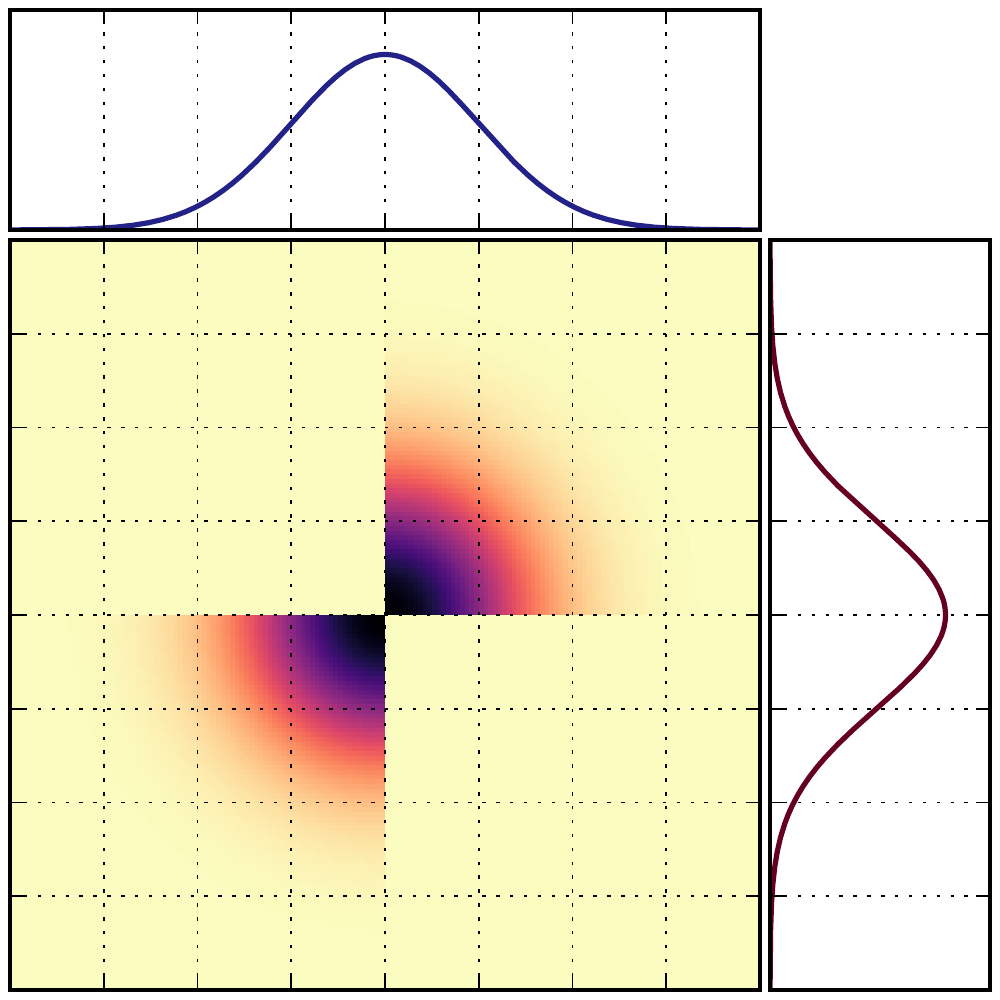}
	\caption{An example that the copula transform is insensitive. The multivariate distribution is not very ``Gaussian'', but its two marginals are both perfectly Gaussian, so the copula will not do adjustment.}
	\label{fig:constraint:Copula_counterexample}
\end{figure}

However, in some specific cases, the copula has no effect at all. Consider the following example:
\begin{align}
	f(x,y) = 2\phi(x,y)\Theta(xy),\ \ \text{where}\ \ \phi(x,y) = \frac{1}{\sqrt{(2\pi)^2}}\ \exp\left(-\frac{x^2+y^2}{2}\right).
\end{align}
Here, $\phi$ is just the \acro{2D} standard normal distribution and $\Theta$ is the Heaviside step function. The value of $f$ is 2 times $\phi_2$ if $x$ and $y$ have the same sign and 0 otherwise. The marginal \acro{PDF} of $f$ and $\phi_2$ turn out to be the same (see \fig{fig:constraint:Copula_counterexample}). As a result, the Gaussian copula transform does nothing for $f$. The copula likelihood of $f$ is exactly the original Gaussian likelihood, while $f$ remains ``highly'' non Gaussian. This example proves that the copula is not the ultimate solution. However, if we do not have any prior knowledge, then the result with the copula transformation should be at least as good as the usual Gaussian likelihood.

\subsection{Copula likelihood}
\label{sect:constraint:copula:likelihood}

We can now write down the copula likelihood. By applying \for{for:constraint:copula_transform_4} to $P(\bx^\obs|\bpi)$, $x_i$ becomes $x_i^\obs$ and $q_i$ becomes $q_i^\obs$ similarly. The \textit{copula likelihood}\index{Likelihood, copula} $\Like_\rmc$ is then
\begin{align}
	\Like_\rmc(\bpi) &= \frac{1}{\sqrt{(2\pi)^d|\det\bC|}} \times \exp\left[ -\frac{1}{2}\sum_{i=1}^d\sum_{j=1}^d \left(q^\obs_i-x_i^\model\right) C_{ij}\inv \left(q^\obs_j-x_j^\model\right)\right] \notag\\
	&\times \prod_{i=1}^d \left[\frac{1}{\sqrt{2\pi\sigma_i^2}} \ \exp\Bigg( -\frac{\big(q^\obs_i-x_i^\model\big)^2}{2\sigma^2_i} \Bigg)\right]\inv \times \prod_{i=1}^d P_i(x_i^\obs), 
\end{align}
where $\mu_i$, the mean of $q_i$, is also the mean of $x_i$ thus is replaced by $x_i^\model$. By detailing the dependency on $\bpi$ for all quantities and $\widehat{\mbox{\hspace*{0.8em}}}$ for estimated terms, the varying-covariance copula log-likelihood (labelled \texttt{vc}) is 
\begin{align}
	L_\vc(\bpi) = \text{\acro{cst}} &+ \ln\left\vert \det\widehat{\bC}(\bpi) \right\vert + \sum_{i=1}^d \sum_{j=1}^d \left[q_i^\obs(\bpi) - x^\model_i(\bpi)\right] \widehat{C\inv_{ij}}(\bpi) \left[q_j^\obs(\bpi) - x^\model_j(\bpi)\right] \notag\\
	&- 2 \sum_{i=1}^d \ln\hat{\sigma}_i(\bpi) - \sum_{i=1}^d \left(\frac{q_i^\obs(\bpi) - x^\model_i(\bpi)}{\hat{\sigma}_i(\bpi)}\right)^2 - 2 \sum_{i=1}^d \ln\hat{P}_i(x^\obs_i|\bpi). \label{for:constraint:L_vc}
\end{align}
Here, $\hat{P}_i(x^\obs_i|\bpi)$ should be understood as a one-point evaluation of $\hat{P}_i(\cdot|\bpi)$ on $x^\obs_i$, and $\hat{P}_i(\cdot|\bpi)$ is the $i$-th marginal \acro{PDF} in the $\bx$ space, which is $\bpi$-dependent. This can be computed by applying kernel density estimation (\acro{KDE}, see \sect{sect:constraint:nonParam:true} and \for{for:constraint:KDE}) on the realization set. The quantities $x^\model_i(\bpi)$, $\hat{\sigma}_i(\bpi)$, and $\widehat{C\inv_{ij}}(\bpi)$ are estimated following Eqs. \eqref{for:constraint:mean}, \eqref{for:constraint:covariance}, and \eqref{for:constraint:inverse_covariance}. Finally, $q^\obs_i(\bpi) = \Phi_i\inv(\hat{F}_i(x^\obs_i|\bpi))$, where $\hat{F}_i(\cdot|\bpi)$ is the \acro{CDF} corresponding to $\hat{P}_i(\cdot|\bpi)$ and $\Phi_i$ also depends implicitly on $\bpi$ via $\mu_i$ and $\hat{\sigma}_i$.

Similar to \sect{sect:constraint:CDC:formalisms}, we can also study the \acro{CDC} effect on the copula likelihood. Obtained straightforwardly from \for{for:constraint:L_vc}, the constant-covariance copula log-likelihoods (labelled \texttt{cc}) is
\begin{align}
	L_\cc(\bpi) = \text{\acro{cst}} &+ \sum_{i=1}^d \sum_{j=1}^d \left[q_i^\obs(\bpi) - x^\model_i(\bpi)\right] \widehat{C\inv_{ij}} \left[q_j^\obs(\bpi) - x^\model_j(\bpi)\right] \notag\\
	&- \sum_{i=1}^d \left(\frac{q_i^\obs(\bpi) - x^\model_i(\bpi)}{\hat{\sigma}_i}\right)^2 - 2 \sum_{i=1}^d \ln\hat{P}_i\left(x^\obs_i - x^\model_i(\bpi)\right), 
\end{align}
where $\hat{P}_i(\cdot)$ denotes the zero-mean marginal \acro{PDF}, assuming that the distribution of each $x_i$ around its mean value is cosmology-independent. This \acro{PDF}, $\hat{\sigma}_i$, and $\widehat{C\inv_{ij}}$ are estimated with the 1000 realizations under $\bpi^\inp$.

\subsection{Results}

\begin{figure}[tb]
	\centering
	\includegraphics[width=0.49\textwidth]{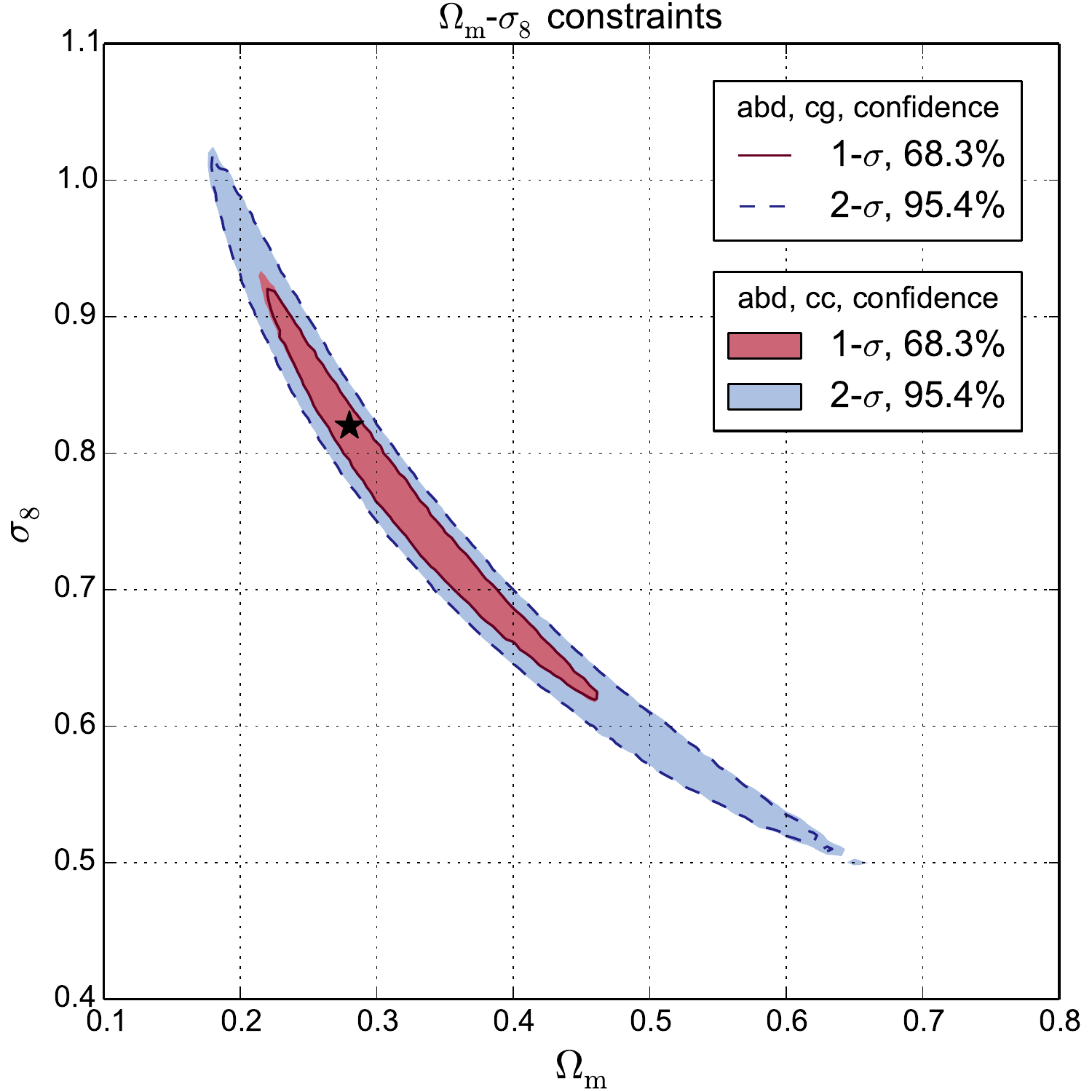}\hfill
	\includegraphics[width=0.49\textwidth]{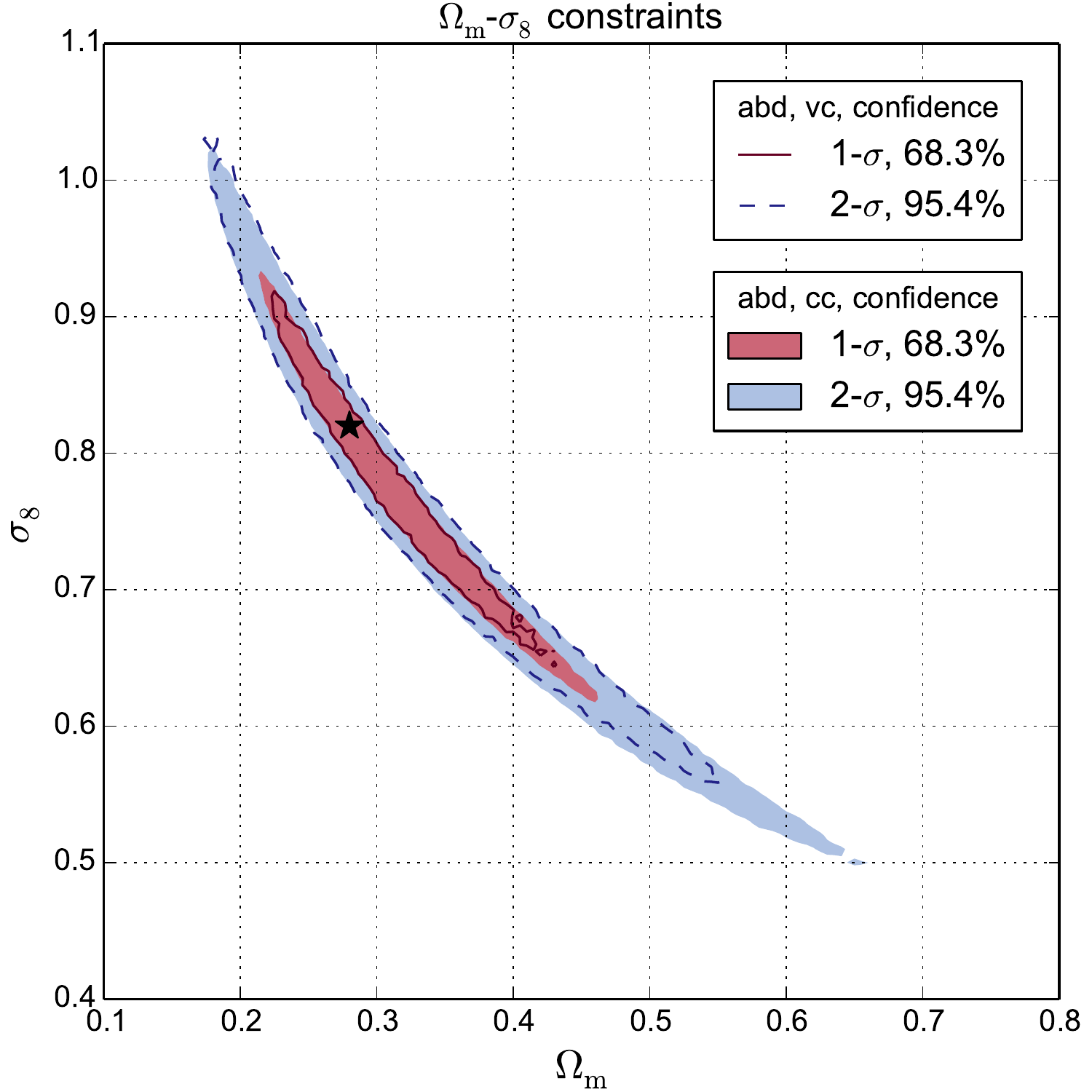}
	\caption{Confidence regions derived from copula analyses. Left panel: comparison between $L_\cg$ (solid and dashed contours) and $L_\cc$ (colored areas). Right panel: comparison between $L_\cc$ (colored areas) and $L_\vc$ (solid and dashed contours). The evolution tendency from $L_\cc$ to $L_\vc$ is similar to the evolution from $L_\cg$ to $L_\vg$. The data vector is $\bx^\abd$.}
	\label{fig:constraint:contour_copula}
\end{figure}

The comparisons between different likelihoods are shown in \fig{fig:constraint:contour_copula}. The left panel focuses on the impact from the copula transformation using $\bx^\abd$. It illustrates the confidence contour evolution from $L_\cg$ to $L_\cc$. It turns out that the Gaussian likelihood is a very good approximation for lensing peak-counts. Presented in \tab{tab:constraint:indicator_confidence}, quantitative results reveal that $L_\cg$ provides slightly more optimistic $\OmegaM$-$\sigEig$ constraints. I would like to underline that the effect of the copula transform is ambiguous, and both tighter or wider constraints are possible. This has already shown by \citet{Sato_etal_2011}, who found that the Gaussian likelihood underestimates the constraint power for low $\ell$ of the lensing power spectrum and overestimates it for high $\ell$.

The right panel of \fig{fig:constraint:contour_copula} focuses on the \acro{CDC} effect, comparing confidence regions between $L_\cc$ and $L_\vc$ using $\bx^\abd$. When the \acro{CDC} is taken into account for the copula transform, the parameter constraints are submitted compared to a similar change to the Gaussian likelihood. Tighter constraints are obtained from $L_\vc$ than from $L_\cc$. The results from $\bx^\pct$ and $\bx^\cut$, as well as those on credible regions, lead to the same conclusion.

In summary, the copula likelihood with a fully varying covariance, $L_\vc$, is closer to the truth than the usual Gaussian likelihood and is easy to compute. Taking this weaker approximation into account, the constraining power could be improved by at least 10\% in terms of \acro{FoM} compared to $L_\cg$, the usual estimator.

\section{Toward non-parametric estimation}
\label{sect:constraint:nonParam}

\subsection{True likelihood}
\label{sect:constraint:nonParam:true}

\index{Likelihood, true}This section presents two ways to obtain parameter constraints with even less restrictive assumptions. The first one is to use the ``true'' likelihood: if our model provides the \acro{PDF} of $\bx$, why not use it directly? The original definition of likelihood is $\Like(\bpi)=P(\bx^\obs|\bpi)$, so we can write the true log-likelihood as
\begin{align}
	L_\true(\bpi) = -2 \ln\hat{P}(\bx^\obs|\bpi),
\end{align}
where $\hat{P}$ is an estimation of the \acro{PDF}. 

In this work, $\hat{P}$ is estimated with the realization set described in \sect{sect:constraint:methodology:design}, using kernel density estimation (\acro{KDE}).\index{Kernel density estimation (\acro{KDE})} Noting $\bx\upp{k}$ for $k=1,\ldots,N$, the $N$ samples of the data vector under $\bpi$, the multivariate density of $\bx$ is 
\begin{align}\label{for:constraint:KDE}
	\hat{P}(\bx|\bpi) &= \frac{1}{N} \sum_{k=1}^N W_{\vect{H}}\left(\bx-\bx\upp{k}\right)
\end{align}
with
\begin{align}\label{for:constraint:KDE_kernel}
	W_{\vect{H}}(\bx)&= \frac{1}{\sqrt{(2\pi)^d|\det\vect{H}|}}\exp\left[-\frac{1}{2} \bx^T \vect{H}\inv \bx\right]
\end{align}
and the bandwidth matrix $\vect{H}$ is given by Silverman's rule \citep{Silverman_1986} as
\begin{align}\label{for:constraint:KDE_bandwidth}
	\sqrt{H_{ij}} = \left\{\begin{array}{ll}
		\displaystyle
		\left[\frac{4}{(d+2)N}\right]^{\textstyle \frac{1}{d+4}}\hat{\sigma}_i & \text{if $i=j$,}\\[3ex]
		0 & \text{otherwise,}
	\end{array}\right.
\end{align}
where $d$ is the dimension of the data vector. 

\begin{figure}[tb]
	\centering
	\includegraphics[width=0.49\textwidth]{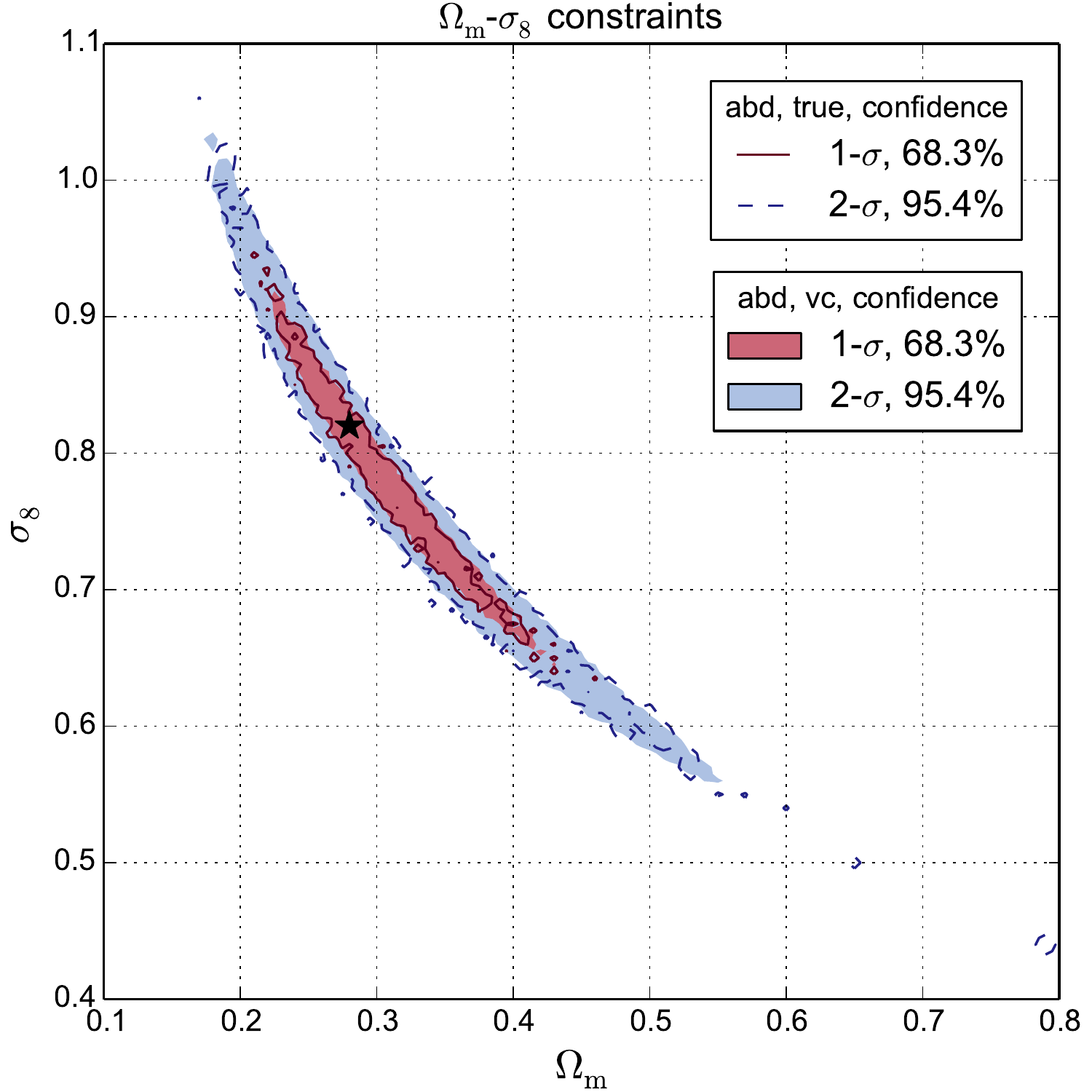}\hfill
	\includegraphics[width=0.49\textwidth]{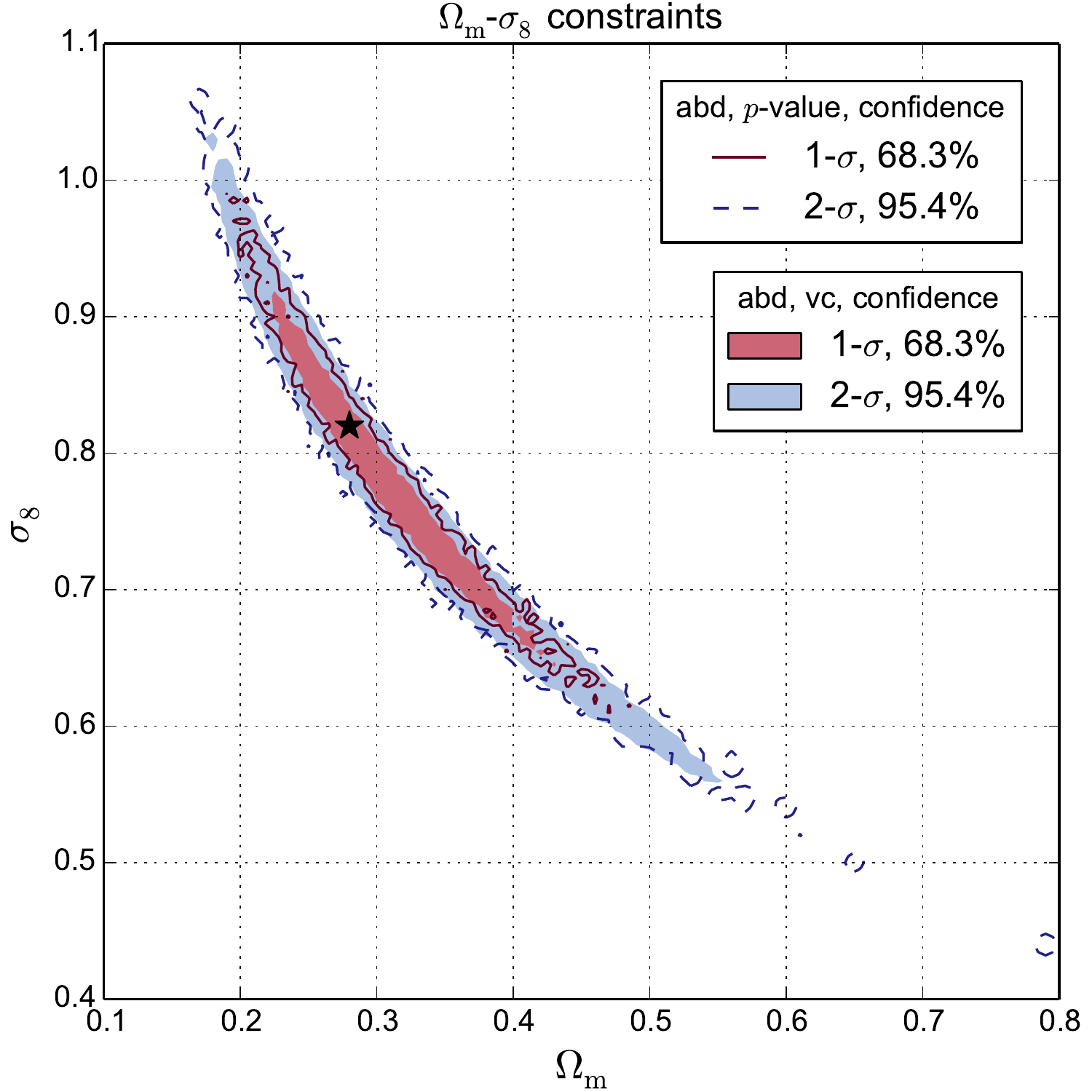}
	\caption{Left panel: confidence regions derived from $L_\vc$ (colored areas) and $L_\true$ (solid and dashed contours). Right panel: confidence regions derived from $L_\vc$ (colored areas) and $p$-value analysis (solid and dashed contours). The contours from $L_\true$ and $p$-value analysis are noisy due to a relatively low $N$. We can observe that $L_\vc$ and $L_\true$ yield very similar results. The data vector is $\bx^\abd$.}
	\label{fig:constraint:contour_nonParam}
\end{figure}

The left panel of \fig{fig:constraint:contour_nonParam} shows the confidence contours from $L_\true$ with $\bx^\abd$. Readers can notice that this constraint is very noisy. This is due to a relatively low number of realizations which fails to stabilize the probability estimation and prevents us from making definite conclusions. However, when we put the constraint from $L_\true$ and the one from $L_\vc$ together, both results agree well and thus suggest that we may substitute $L_\true$ with $L_\vc$, which bypasses the drawback of noisy estimation. Quantitative results from $\Delta\Sigma_8$ and \acro{FoM} (\tab{tab:constraint:indicator_confidence}) also suggest the same agreement. For constraints with $\bx^\pct$, $\bx^\cut$, and credible regions, I have found similar results and conclusions.

\subsection{$p$-value analysis}
\label{sect:constraint:nonParam:pValue}

\begin{figure}[tb]
	\centering
	\includegraphics[width=0.49\textwidth]{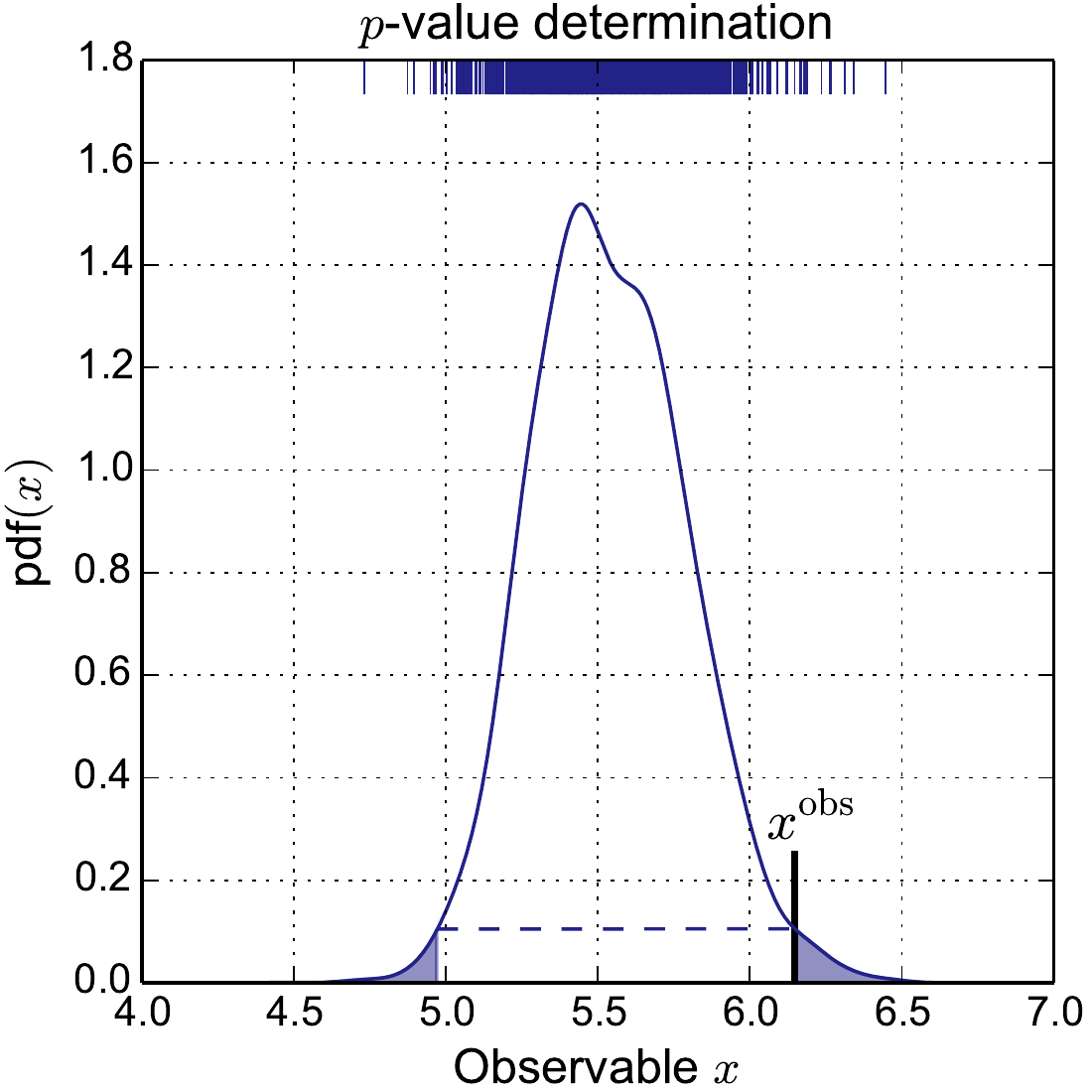}\hfill
	\includegraphics[width=0.49\textwidth]{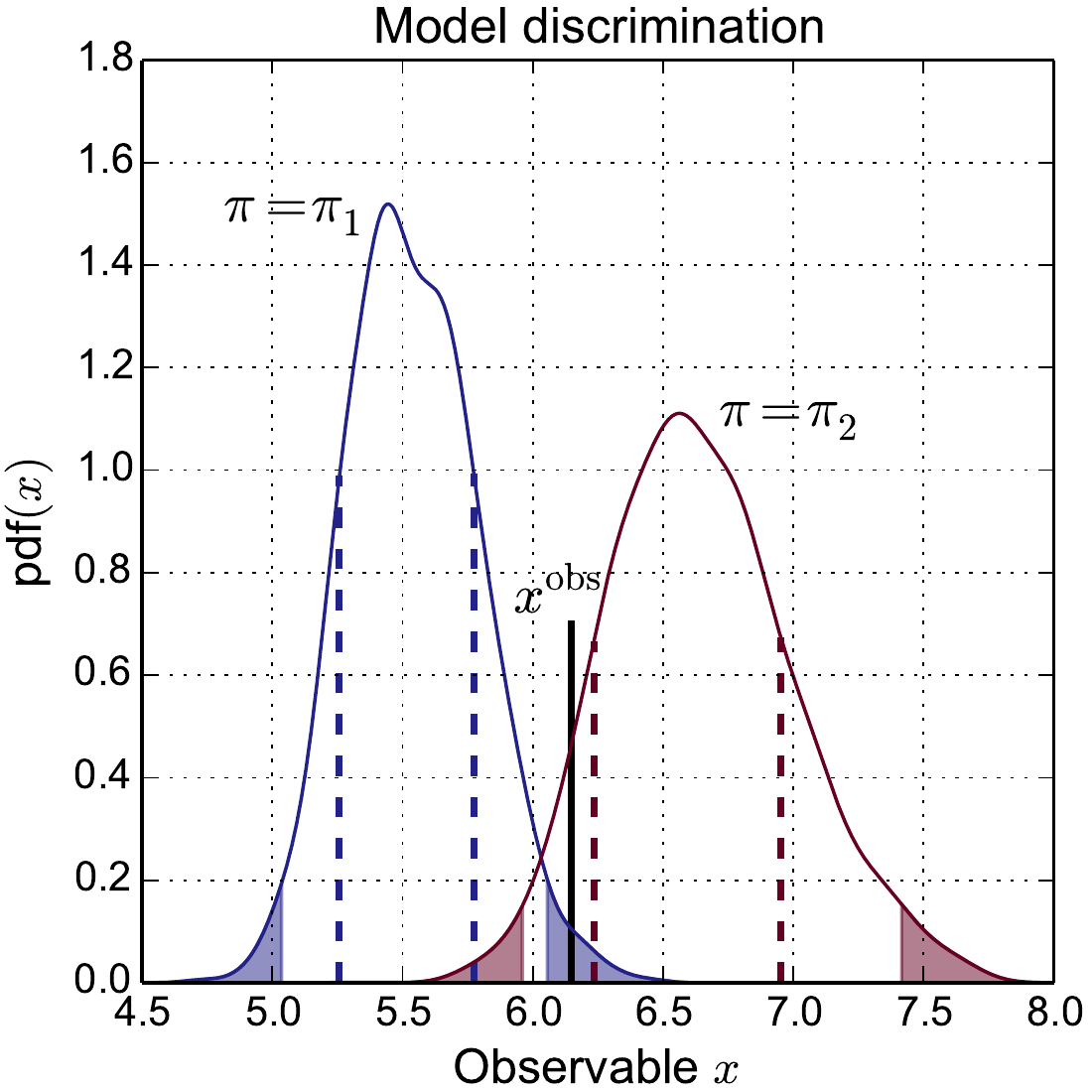}
	\caption{Examples for the $p$-value determination and for model selection. The $x$-axis indicates a one-dimensional data vector, and the $y$-axis is the \acro{PDF}, obtained from \acro{KDE} using the $N = 1000$ realizations. In the left panel, values of realizations are shown as bars in the rug plot at the top of the panel. The shaded area is the corresponding $p$-value for given observational data $x^\obs$. In the right panel, the dashed lines show 1-$\sigma$ intervals for two models (denoted by $\pi_1$ and $\pi_2$), and the shaded areas are intervals beyond the 2-$\sigma$ level. In this example, the model $\pi_1$ is excluded at more than 2-$\sigma$, whereas the significance of the model $\pi_2$ is between 1 and 2-$\sigma$.}
	\label{fig:constraint:p_value_determination}
\end{figure}
 
\index{$p$-value}Another non-parametric technique is the $p$-value analysis. Without passing by the likelihood, it gives directly the significance level of the observation for a model. For a observation $\bx^\obs$, the $p$-value associated to a parameter vector $\bpi$ is defined as 
\begin{align}\label{for:constraint:p_value}
	1-p(\bpi) \equiv \int \rmd^d\bx\ \hat{P}(\bx|\bpi)\times \Theta\left(\hat{P}(\bx|\bpi)- \hat{P}(\bx^\obs|\bpi)\right),
\end{align}
where $\Theta$ denotes the Heaviside step function. The integral extends over the region where $\bx$ is more probable than $\bx^\obs$, as shown by the left panel of \fig{fig:constraint:p_value_determination}. Thus, the proper interpretation of \for{for:constraint:p_value} is that if we generated $N$ universes, then at least $(1-p)N$ of them should have an observational result ``better'' than $\bx^\obs$, and ``better'' refers to more probable. 

As in \sect{sect:constraint:nonParam:true}, \acro{KDE} is performed to estimate the multivariate \acro{PDF} and \for{for:constraint:p_value} is numerically integrated to obtain the $p$-value. Monte Carlo integration is used for evaluating the five-dimensional integrals in this case (consequence of the curse of dimensionality). To reach $i$-$\sigma$ significance level, the $p$-value should reach at least $p_i$, which is given by \for{for:constraint:p_i}. And as the right panel \fig{fig:constraint:p_value_determination} shows, this provides a straightforward way to distinguish different cosmological models.

The result from $p$-value analysis with $\bx^\abd$ is presented in the right panel of \fig{fig:constraint:contour_nonParam} and \tab{tab:constraint:indicator_confidence}. The contours seem to be larger than $L_\true$, and are also quite noisy. This time, the ``noise'' is caused by a relative low number of realizations and/or fluctuation from the Monte Carlo integration. One way to reduce this noise is to reduce the data dimension $d$. If $d$ is small, on one hand, a low number of realizations can be ``sufficient'' for reconstructing the \acro{PDF}; on the other hand, the grid-point evaluation would become tractable. For these reasons, the same analysis is done with some \acro{2D} vectors which are obtained by combining components of $\bx^\abd$, and I observe that in some cases the constraint contours are less noisy without deteriorating the constraining power.

Another subtlety from the true likelihood and $p$-value analysis is the \acro{PDF} evaluation. While $L_\true(\bpi)$, which is also $P(\bx^\obs|\bpi)$, is a one-point evaluation at $\bx=\bx^\obs$, determining $p$-value requires a more global information, at least about the region where $P(\bx) < P(\bx^\obs)$. This makes evaluating $p$-values expensive. Furthermore, we should recall that \acro{KDE} is a biased estimator \citep{Zambom_Dias_2012}. What is the bias level compared to the Gaussian likelihood? How to correct it? Those are questions to answer to for further studies. Some further ideas, for example the Voronoi-Delaunay tessellation \citep[see e.g.,][]{Schaap_2007, Guio_Achilleos_2009}, could be an alternative to the \acro{KDE} technique.

\subsubsection{Summary}

I have displayed results of parameter constraints on $\OmegaM$-$\sigEig$ provided by our model, by using successively the Gaussian likelihood, the copula likelihood, the true likelihood, and the $p$-value. The different comparisons have shown that taking the true covariance into account improves the constraining power, and the copula likelihood is a good approximation to the true one.

The results also suggest that the \acro{PDF}-based data vector, $\bx^\abd$, retains the most cosmological information among three candidates. The full \acro{CDF}-based data vector $\bx^\pct$ also performs well, but could potentially cause serious biases due to our modelling strategy. And the adapted \acro{CDF}-based data vector $\bx^\cut$ loses information thus carries out larger contours.

The size of the data dimension could affect the performance of the density estimation. In order to find a balance between efficiency and precision, data vector choice, data compression, and density estimation need to be considered jointly, especially for non-parametric estimators. 

Recent results from \acro{CFHTLenS} (\citetalias{Liu_etal_2015} \citeyear{Liu_etal_2015}) and Stripe-82 (\citetalias{Liu_etal_2015a} \citeyear{Liu_etal_2015a}) obtained $\Delta\Sigma_8\sim0.1$, about 2--3 times larger than this study. However, I would like to highlight that redshift errors are not taken into account here and that the simulated galaxy density used in this work is much higher. Also, the source redshift has been set to $z_\rms = 1$, which is higher than the median redshift of both surveys ($\sim 0.75$). All these factors contribute to our smaller error bars.

In a more complete analysis, the number of parameters could be large. The grid-point evaluation will not be available anymore. Instead, sophisticated sampling techniques, such as Markov Chain Monte Carlo (\acro{MCMC}) or population Monte Carlo (\acro{PMC}) are usually suggested. With a similar spirit, I am going to present a robust and efficient technique in the next chapter, called approximate Bayesian computation (\acro{ABC}), which is notably suitable to our \acro{FSF} model.

\clearpage
\thispagestyle{empty}
\cleardoublepage


\chapter{Approximate Bayesian computation}
\label{sect:ABC}
\fancyhead[LE]{\sf \nouppercase{\leftmark}}
\fancyhead[RO]{\sf \nouppercase{\rightmark}}

\subsubsection{Overview}

In this chapter, I will show how to obtain parameter constraints with a method called approximate Bayesian computation. This technique, not so well known by the cosmological community, has a great potential. I will first introduce the concept and describe the philosophy of this method, and then show and comment on the constraint results carried out by a sophisticated iterative algorithm. This chapter corresponds to Sect. 6 of \PaperII.

\section{Introduction to ABC}

\subsection{State of the art}

The concept of approximate Bayesian computation (hereafter \acro{ABC}) has first been mentioned by \citet{Rubin_1984} to resolve the problem of the intractable likelihood estimation. For models that have some stochastic characteristics, i.e. from which one can sample, the technique provides an efficient estimation of the posterior. The basic idea of \acro{ABC} is combining the sampling with an accept-reject criteria in such a way that the process probes the likelihood without evaluating it.

After \citet{Rubin_1984}, the first \acro{ABC}-like algorithm was proposed by \citet{Tavare_etal_1997} who studied population genetics. Their algorithm bypasses the likelihood evaluation to estimate the posterior in a particular scenario where values of the data vector are discrete. The generalization for a continuous data space was first proposed by \citet{Pritchard_etal_1999}. Since then, statistical studies on \acro{ABC} have become prosperous. Plenty of variations, e.g. Markov Chain Monte Carlo \acro{ABC} \citep{Marjoram_etal_2003}, population Monte Carlo \acro{ABC} \citep{Beaumont_etal_2009}, lazy \acro{ABC} \citep{Prangle_2016}, weighted \acro{ABC} \citep{Killedar_etal_2015}, or Hamiltonian \acro{ABC} \citep{Meeds_etal_2015}, have been proposed.

The technique has been widely utilized, especially in biology-related domains \citep[e.g.][]{Beaumont_etal_2009, Berger_etal_2010, Csillery_etal_2010, Drovandi_Pettitt_2011}. However, applications of \acro{ABC} for astronomical purposes are still few. These are morphological transformation of galaxies \citep{Cameron_Pettitt_2012}, cosmological parameter inference using type Ia supernovae \citep{Weyant_etal_2013, Jennings_etal_2016a}, constraints of the disk formation of the Milky Way \citep{Robin_etal_2014}, cosmological constraints using weak-lensing peak counts (\PaperII, \PaperIII, \citealt{Peel_etal_2016}), strong lensing properties of galaxy clusters \citep{Killedar_etal_2015}, and the constraints on the halo occupation distribution (\acro{HOD}) parameters \citep{Hahn_etal_2016}. Apart from \PaperII, three other software packages which allow to perform \acro{ABC} in a general cosmological context have been released with their respective accompanied papers \citep{Ishida_etal_2015, Akeret_etal_2015, Jennings_Madigan_2016}.

\subsection{The discrete case}

\begin{algorithm}[tb]
	\begin{algorithmic}
		\REQUIRE
			\STATE observation $\bx^\obs$
			\STATE stochastic model $P(\cdot|\bpi)$
			\STATE number of particles $Q$
			\STATE prior $\mathcal{P}(\cdot)$
		\bigskip
		\FOR{$i = 1$ to $Q$}
			\REPEAT
				\STATE generate $\bpi_i$ from $\mathcal{P}(\cdot)$
				\STATE generate $\bx$ from $P(\cdot|\bpi_i)$
			\UNTIL{$\bx=\bx^\obs$\ \ (one-sample test)}
		\ENDFOR
	\end{algorithmic}
	\caption{Discrete \acro{ABC}}
	\label{algo:ABC:discrete}
\end{algorithm}

How does \acro{ABC} bypass the likelihood? The philosophy behind it can be easily illustrated in the case of discrete data as follows. Let $\bx$ be a data vector and $\bpi$ a parameter vector. Denote $\mathcal{P}(\cdot)$ as a prior function, $\mathcal{P}(\cdot|\bx^\obs)$ as the posterior given the observation $\bx^\obs$, and $P(\cdot|\bpi)$ as the probability of the model of the data given $\bpi$. I would like to remind readers that $P$ is a \acro{PDF} in the data space and $\mathcal{P}$ in the parameter space. Consider now the accept-reject process from Algo. \ref{algo:ABC:discrete}. Each time, we take a parameter candidate $\bpi_i$ from the prior, sample a $\bx$ from the model under $\bpi_i$, and accept $\bpi_i$ if $\bx$ and $\bx^\obs$ are identical. In the case of rejection, the parameter should be resampled. We call \textit{particles}\index{Particle} all accepted candidates of parameters $\bpi_{i=1,\ldots,Q}$ from this algorithm. The number of particles is then $Q$. Implicitly, each particle is an independent and identically distributed (\acro{iid}) sample drawn from $\mathcal{P}_\ABC(\cdot|\bx^\obs)$, given by
\begin{align}
  	\mathcal{P}_\ABC(\bpi|\bx^\obs) &= \sum_{\bx} \delta_{\bx, \bx^\obs} P(\bx|\bpi) \mathcal{P}(\bpi) \notag\\
  	&= P(\bx^\obs|\bpi) \mathcal{P}(\bpi) \notag\\
  	&= \mathcal{P}(\bpi|\bx^\obs),
  	\label{for:ABC:post_discrete}
\end{align}
where the last equality is given by \for{for:constraint:likelihood} and $\delta_{\bx, \bx^\obs}$ is Kronecker's delta. We can see that the posterior from the \acro{ABC} process, $\mathcal{P}_\ABC(\cdot|\bx^\obs)$, is nothing but the true posterior $\mathcal{P}(\cdot|\bx^\obs)$. Therefore, one is able to reconstruct the posterior without computing the likelihood.

Two important concepts worth being highlighted in Algo. \ref{algo:ABC:discrete}. First, if the criterion $\bx=\bx^\obs$ is not matched, the candidate $\bpi_i$ must be dropped and resampled from the prior. The reason is trivial: otherwise, all candidates would eventually be accepted and the \acro{ABC} process would sample from the prior. Second, it is sufficient to perform a \textit{one-sample test}\index{One-sample test}: a single draw of $\bx$ from $P(\cdot|\bpi_i)$ for each candidate $\bpi_i$. Why is that? It is easy to understand when $\bpi$ has also discrete values. In the doubly discrete case, a specific parameter value can be presented by more than one candidate, so asymptotically, having an $N$-sample test instead is equivalent to setting number of particles $N$ times larger. In a continuous parameter space, the ``noise'' related to an $N$-sample test, which is actually the statistical fluctuation related to the likelihood evaluated at a specific point, will be reduced when we reconstruct the global posterior. Recalling that particles are \acro{iid} samples from the \acro{ABC} posterior, the true ``noise'' here is just the same as any reconstruction of an underlying distribution with a limit number of samples.

Another good reason not to use more than one sample is the difficulty to define the accepting criterion. The new criteria might change the probability for accepting $\bpi_i$, which is just $\delta_{\bx, \bx^\obs} P(\bx|\bpi_i)$ in the one-sample case. Since this simple form of probability leads to an exact reconstruction of the posterior, the process is already optimal.

\subsection{The continuous case}

How does \acro{ABC} work in a continuous data space? Since the strict equality of $\bx=\bx^\obs$ is unlikely to happen in the continuous case, one should introduce a \textit{tolerance level} $\epsilon$ below which two data vectors are considered identical. However, real-world data may have very complicated representations and/or a very high dimension, and often need to be reduced to a vector of suitable size. This data compression problem, present in all statistical analyses, is emphasized in the \acro{ABC} context by adopting the formalism of the \textit{summary statistic}\index{Summary statistic} $s$. Instead of denoting $\bx$ as a data vector, in the rest of this chapter, $\bx$ is rather considered as a general data set, and then reduced into a vector $s(\bx)$ via $s$. Finally, we need a \textit{distance} $D$ to compare in a quantitative way two different data. This could be the Euclidean distance, a covariance-weighted function, or any other customized measure. The distance $D$ is obviously defined in the summary space.

\begin{algorithm}[tb]
	\begin{algorithmic}
		\REQUIRE
			\STATE requirements of Algo. \ref{algo:ABC:discrete}
			\STATE summary statistic $s(\cdot)$
			\STATE distance $D(\cdot, \cdot)$
			\STATE tolerance level $\epsilon$
		\bigskip
		\FOR{$i = 1$ to $Q$}
			\REPEAT
				\STATE generate $\bpi_i$ from $\mathcal{P}(\cdot)$
				\STATE generate $\bx$ from $P(\cdot|\bpi_i)$
			\UNTIL{$D\left(s(\bx), s(\bx^\obs)\right)\leq\epsilon$\ \ (one-sample test)}
		\ENDFOR
	\end{algorithmic}
	\caption{Continuous \acro{ABC}}
	\label{algo:ABC:continuous}
\end{algorithm}

\begin{figure}[tb]
	\centering
	\includegraphics[scale=0.65]{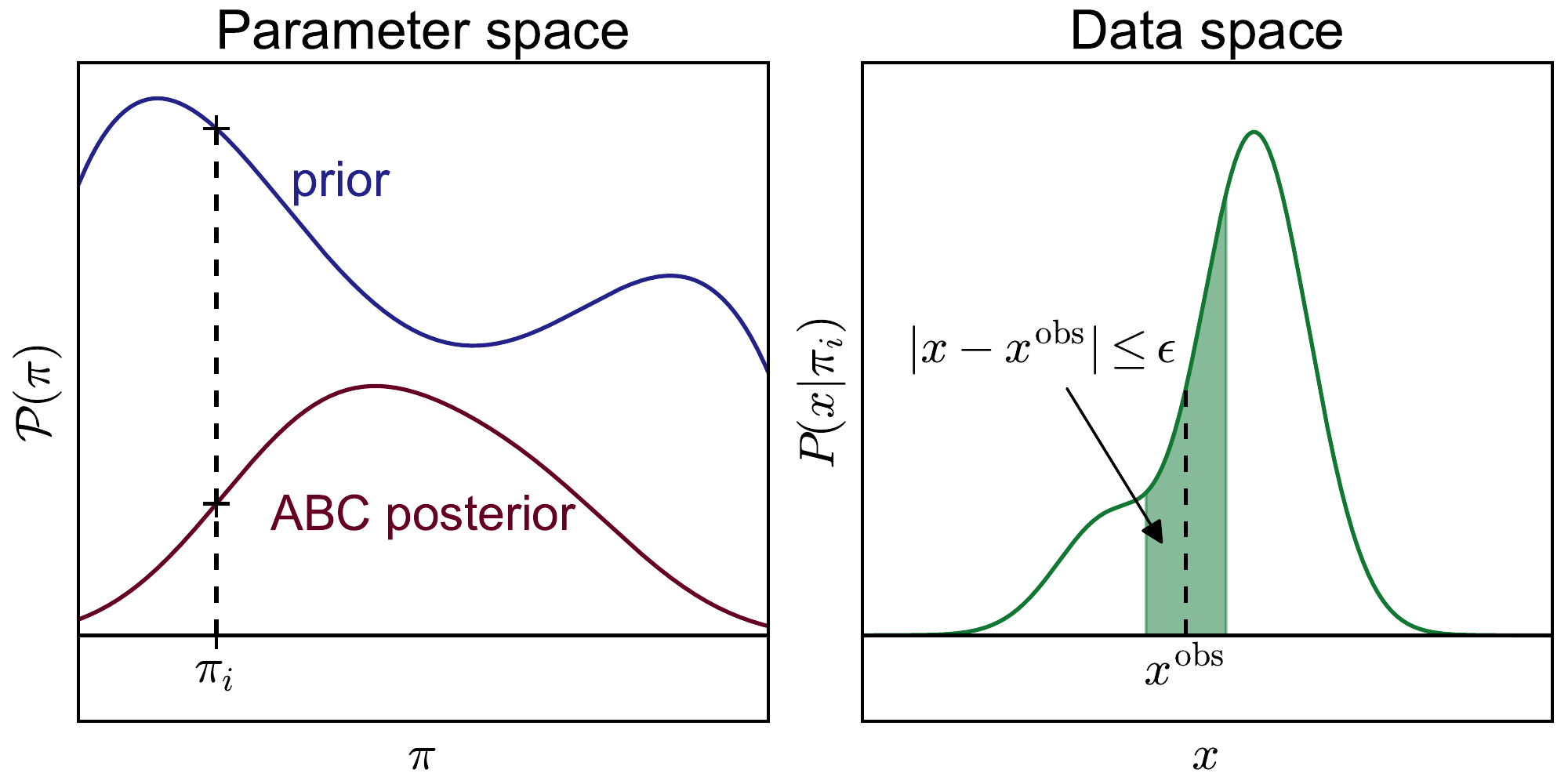}
	\caption{Illustration of the \acro{ABC} posterior. Left panel: the parameter space with the prior in blue and the \acro{ABC} posterior in dark red. Right panel: the data space with the \acro{PDF} of $x$ under a specific parameter $\pi_i$, in green. Here, $|x-x^\obs|$ replaces $D(s(\bx),s(\bx^\obs))$ to illustrate the difference between $x$ and $x^\obs$. If the total integral under the green curve is one, then the green area is the accepting probability, which is also the ratio of the cross on the red curse to the cross on the blue curve.}
	\label{fig:ABC:ABC_toy_model_1}
\end{figure}

The continuous \acro{ABC} algorithm is displayed in Algo. \ref{algo:ABC:continuous}. Similar to Algo. \ref{algo:ABC:discrete}, it generates $\bpi_i$ from the prior and does a one-sample test, this time with the criterion $D(s(\bx), s(\bx^\obs))\leq\epsilon$. In this case, the \acro{ABC} posterior from which the particles are sampled, denoted as $\mathcal{P}_{\ABC, \epsilon}(\cdot|\bx^\obs)$, leads to
\begin{align}\label{for:ABC:post_continuous}
	\mathcal{P}_{\ABC, \epsilon}(\bpi|\bx^\obs) = A_\epsilon(\bpi) \mathcal{P}(\bpi),
\end{align}
where $A_\epsilon(\bpi)$ is the probability that a candidate $\bpi$ passes the one-sample test within the error~$\epsilon$:
\begin{align}\label{for:ABC:accept_1}
	A_\epsilon(\bpi) \equiv \int \rmd \bx\ \mathbbm{1}_{D(s(\bx), s(\bx^\obs))\leq\epsilon}(\bx)P(\bx|\bpi).
\end{align}
The Kronecker delta from \for{for:ABC:post_discrete} has been replaced with the indicator function $\mathbbm{1}$ of the set of points $\bx$ that satisfy the tolerance criterion. Eqs. \eqref{for:ABC:post_continuous} and \eqref{for:ABC:accept_1} are visualized in \fig{fig:ABC:ABC_toy_model_1}, where $\mathcal{P}(\bpi)$ is represented by the blue curve, $\mathcal{P}_{\ABC, \epsilon}(\bpi|\bx^\obs)$ the red curve, and $P(\bx|\bpi)$ the green curve, so that the ratio of the red curve to the blue curve equals to the ratio of the green area to the total integral. From \fig{fig:ABC:ABC_toy_model_1}, it is straightforward to verify that 
\begin{align}
	\lim_{\epsilon\rightarrow 0} \frac{A_\epsilon(\bpi)}{2\epsilon} = P(\bx^\obs|\bpi) = \mathcal{L}(\bpi|\bx^\obs).
\end{align}
Therefore, the basic assumption of \acro{ABC} is to take a small value of $\epsilon$ and to consider the \acro{ABC} posterior as a good approximation of the underlying one, such that
\begin{align}\label{for:ABC:approximation}
	\mathcal{P}_{\ABC, \epsilon}(\bpi|\bx^\obs) \approx \mathcal{P}(\bpi|\bx^\obs).
\end{align}
As a result, the error can be separated into two parts: one from the approximation \eqref{for:ABC:approximation} and the other from the estimation of the desired integral \eqref{for:ABC:post_continuous}. For the latter, one-sample tests of $A_\epsilon(\bpi)$ correspond to a Monte Carlo estimation of the \acro{ABC} posterior, which is unbiased. This ensures the use of the one-sample test.

At the end of the day, \acro{ABC} can be summarized as a process which, \textit{instead of evaluating the likelihood, samples directly from a posterior which is considered close to the true one; and by reconstructing from samples, one obtains an estimate of the true posterior}.

\section{Parameter constraints with the PMC ABC algorithm}
\label{sect:ABC:constraints}

\subsection{Choice of the requirements}

In Algo. \ref{algo:ABC:continuous}, some of the requirements to be chosen by the user can play an important role to the result. First of all, choosing a summary statistic is not obvious. On the one hand, if the considered data vector is too large, the analysis will be slow and inefficient; on the other hand, if data compression is too violent, then we might lose useful information and possibly create biases. A \textit{sufficient} statistic which contains all useful information might be difficult to define. Second, similar to the summary statistic, the choice of the distance function can also be very various. Finally, what about the tolerance level? If $\epsilon$ is large, \for{for:ABC:accept_1} becomes a bad estimate. If $\epsilon$ is too small, $A_\epsilon(\bpi)$ is close to 0 and sampling becomes extremely difficult and inefficient. How to find a trade-off?

A study about the impact of the various choices necessary for \acro{ABC} has been done by \citet{McKinley_etal_2009}. The authors found that (1) doing an $N$-sample test instead of a single sample does not seem to improve the posterior estimation (similar conclusion are found by \citealt{Bornn_etal_2014}; (2) the specific choice of the tolerance level does not seem to be important; and (3) the choice of the summary statistic and the distance is crucial. Therefore, exploring a sufficient summary statistic to represent the whole data set becomes an essential concern for the ABC technique. There exist several methods to select ideal summaries and to compare between different ones, which can be divided into three classes: best subset selection, projection techniques, and regularization approaches \citep[see e.g.][]{Blum_etal_2013}. However, users still need to propose a collection of summary candidates. Usually, motivated by computational efficiency, the summary statistic has a low dimension and a simple form in practice. It seems that a simple summary can still produce reliable constraints \citep{Weyant_etal_2013}.

As far as the tolerance level is concerned, one of the solutions is population Monte Carlo (\acro{PMC}, \citealt{Iba_2001}; for applications to cosmology, see \citealt{Wraith_etal_2009}). \acro{PMC} is an iterative importance sampling process. Its combination with \acro{ABC} has first been proposed by \citet{Beaumont_etal_2009}. Meanwhile, based on the theoretical work of \citet{DelMoral_etal_2006}, \citet{Sisson_etal_2007} offered a sequential Monte Carlo (\acro{SMC}) sampler for application to \acro{ABC}. After the correction of a bias published by \citet{Sisson_etal_2009}, this \acro{SMC} \acro{ABC} approach is basically identical to the \acro{PMC} \acro{ABC}. For this reason, both names can be found in the literature. This iterative solution is the algorithm which has been used for this study.

\subsection{PMC ABC algorithm}

\begin{algorithm}[tb]
	\begin{algorithmic}
		\REQUIRE
			\STATE requirements of Algo. \ref{algo:ABC:discrete}
			\STATE summary statistic $s(\cdot)$
			\STATE distance $D(\cdot, \cdot)$
			\STATE transition kernel $W_{\bC}(\cdot)$
			\STATE decreasing sequence of tolerance levels $\epsilon\upp{0} \geq \epsilon\upp{1} \geq \ldots \geq \epsilon\upp{T}$
		\bigskip
		\STATE set $t = 0$
		\FOR{$i = 1$ to $Q$}
			\REPEAT 
				\STATE generate $\bpi_i\upp{0}$ from $\rho(\cdot)$ and $\bx$ from $P\left(\cdot|\bpi_i\upp{0}\right)$
			\UNTIL{$D \left(s(\bx), s(\bx^\obs)\right) \leq \epsilon\upp{0}$}
			\STATE set $\omega_i\upp{0} = 1/Q$
		\ENDFOR
		\STATE set $\bC\upp{0} = \cov\left(\bpi_i\upp{0}, \omega_i\upp{0}\right)$ 
		\bigskip
		\FOR{$t = 1$ to $T$}
			\FOR{$i = 1$ to $Q$}
				\REPEAT 
					\STATE generate $j$ from $\{1,\ldots, Q\}$ with weights {\footnotesize $\left\{\omega_1\upp{t-1}, \ldots, \omega_Q\upp{t-1}\right\}$}
					\STATE generate $\bpi_i\upp{t}$ from {\footnotesize $\mathcal{N}\left(\bpi\upp{t-1}_j, \bC\upp{t-1}\right)$} and $\bx$ from $P\left(\cdot|\bpi_i\upp{t}\right)$
				\UNTIL{$D\left(s(\bx), s(\bx^\obs)\right) \leq \epsilon\upp{t}$}
				\STATE set $\omega_i\upp{t} \propto$ {\scriptsize $\displaystyle\frac{\mathcal{P}\left(\bpi_i^{(t)}\right)}{\sum_{j=1}^Q \omega_j\upp{t-1} W_{\bC\upp{t-1}}\left(\bpi\upp{t}_i-\bpi\upp{t-1}_j\right)}$}
			\ENDFOR
			\STATE set $\bC\upp{t} = \cov\left(\bpi_i\upp{t}, \omega_i\upp{t}\right)$ 
		\ENDFOR
	\end{algorithmic}
	\caption{Population Monte Carlo \acro{ABC}}
	\label{algo:ABC:PMC}
\end{algorithm}

The basic idea of \acro{PMC}\index{Population Monte Carlo (\acro{PMC})} is to generate samples from a \textit{proposal function}\index{Proposal function} which depends on the previous iteration. In the case of the first iteration, the proposal is just the prior. Given a decreasing sequence of $T+1$ tolerance levels $\epsilon\upp{0} \geq \epsilon\upp{1} \geq \ldots \geq \epsilon\upp{T}$, we start \acro{ABC} by sampling candidate parameters from the prior. We attribute an equal weight to each particle, and compute the covariance matrix. Then, for all later iterations, candidates are taken from a normal law centered on a particle, randomly selected with regard to the weights. The normal law uses the covariance matrix from the previous iteration. At the end of each iteration, the weights are updated using a transition kernel $W_{\bC}$, where $\bC$ is an input covariance matrix (weighted) since the kernel is usually Gaussian. The covariance matrix is also updated by taking new weights into account. The samples from the final iteration are considered as the result from the \acro{PMC} \acro{ABC} algorithm, presented in Algo.~\ref{algo:ABC:PMC}.

In this way, we can actually write the analytical expression for the \acro{ABC} posterior. The proposal function described above is actually a Gaussian smoothing of the previous posterior. Therefore, the proposal function for the $t$-th iteration $\widetilde{\mathcal{P}}_t(\bpi)$ is given by 
\begin{align}
	\widetilde{\mathcal{P}}_0(\bpi) &= \mathcal{P}(\bpi), \label{for:ABC:proposal_1}\\
	\widetilde{\mathcal{P}}_t(\bpi) &= \int\rmd\bpi'\ W_{\bC\upp{t-1}}(\bpi-\bpi') \omega_{t-1}(\bpi')\mathcal{P}_{t-1}(\bpi'|\bx^\obs),\ \ \text{for $1\leq t\leq T$}, \label{for:ABC:proposal_2}
\end{align}
where $\mathcal{P}_t(\bpi|\bx^\obs)$ is denoted as the \acro{ABC} posterior after the $t$-th iteration, $\omega_t(\bpi)$ represents the weight function, and $\bC\upp{t}$ is the weighted covariance matrix derived from $\mathcal{P}_t(\bpi|\bx^\obs)$ and $\omega_t(\bpi)$. This weight function itself depends on the posterior of the previous iteration, so the proposal $\widetilde{\mathcal{P}}_t(\bpi)$ involves $\omega_{t-1}(\bpi)$ and $\mathcal{P}_{t-2}(\bpi|\bx^\obs)$. Note that in practice, posteriors, proposals, weights, and covariances are computed from discrete samples; here however, we are interested in the expectation values of the sampling process, so all quantities are computed from continuous functions. Considering Eqs. \eqref{for:ABC:proposal_1} and \eqref{for:ABC:proposal_2}, the $t$-th posterior can then be established via
\begin{align}
	A_{\epsilon_t}(\bpi) &= \int\rmd\bx\ \mathbbm{1}_{D(s(\bx), s(\bx^\obs))\leq\epsilon_t}(\bx)P(\bx|\bpi),\\ 
	\mathcal{P}_t(\bpi|\bx^\obs) &= A_{\epsilon_t}(\bpi) \widetilde{\mathcal{P}}_t(\bpi). 
\end{align}
The expression is more difficult to interpret now, especially for \for{for:ABC:proposal_2} where the weight function is involved. Qualitatively, each $A_{\epsilon_t}$ sharpens the distribution on points where the likelihood is high; and the kernel $W_{\bC}$ tends to smooth the posterior. As $\epsilon_t$ decreases, Bayesian inference tends to dominate over smoothing, thereby yields a converging result.

\begin{algorithm}[tb]
	\begin{algorithmic}
		\REQUIRE
			\STATE requirements of Algo. \ref{algo:ABC:discrete}
			\STATE summary statistic $s(\cdot)$
			\STATE distance $D(\cdot, \cdot)$
			\STATE transition kernel $W_{\bC}(\cdot)$
			\STATE shutoff parameter $r_\mathrm{stop}$
		\bigskip
		\STATE set $t = 0$
		\FOR{$i = 1$ to $Q$}
			\STATE generate $\bpi_i\upp{0}$ from $\mathcal{P}(\cdot)$ and $\bx$ from $P\left(\cdot|\bpi_i\upp{0}\right)$
			\STATE set $\delta_i\upp{0} = D \left(s(\bx), s(\bx^\obs)\right)$ and $\omega_i\upp{0} = 1/Q$
		\ENDFOR
		\STATE set $\epsilon\upp{1} = \median\left(\delta_i\upp{0}\right)$ and $\bC\upp{0} = \cov\left(\bpi_i\upp{0}, \omega_i\upp{0}\right)$ 
		\bigskip
		\WHILE{success rate $\geq r_\mathrm{stop}$}
			\STATE $t \leftarrow t + 1$
			\FOR{$i = 1$ to $Q$}
				\REPEAT 
					\STATE generate $j$ from $\{1,\ldots, Q\}$ with weights {\footnotesize $\left\{\omega_1\upp{t-1}, \ldots, \omega_Q\upp{t-1}\right\}$}
					\STATE generate $\bpi_i\upp{t}$ from {\footnotesize $\mathcal{N}\left(\bpi\upp{t-1}_j, \bC\upp{t-1}\right)$} and $\bx$ from $P\left(\cdot|\bpi_i\upp{t}\right)$
					\STATE set $\delta_i\upp{t} = D\left(s(\bx), s(\bx^\obs)\right)$
				\UNTIL{$\delta_i\upp{t} \leq \epsilon\upp{t}$}
				\STATE set $\omega_i\upp{t} \propto$ {\scriptsize $\displaystyle\frac{\mathcal{P}\left(\bpi_i^{(t)}\right)}{\sum_{j=1}^Q \omega_j\upp{t-1} W_{\bC\upp{t-1}}\left(\bpi\upp{t}_i-\bpi\upp{t-1}_j\right)}$}
			\ENDFOR
			\STATE set $\epsilon\upp{t+1} = \median\left(\delta_i\upp{t}\right)$ and $\bC\upp{t} = \cov\left(\bpi_i\upp{t}, \omega_i\upp{t}\right)$ 
		\ENDWHILE
	\end{algorithmic}
	\caption{Modified \acro{PMC} \acro{ABC} used in this work}
	\label{algo:ABC:modified}
\end{algorithm}

Instead of selecting an arbitrary tolerance sequence, I propose a modified version of the \acro{PMC} \acro{ABC} algorithm by introducing a \textit{shutoff parameter}\index{Shutoff parameter} $r_\mathrm{stop}$ (Algo. \ref{algo:ABC:modified}). The algorithm continues until the success rate $r\upp{t}$ of the one-sample test is smaller than $r_\mathrm{stop}$. And, inspired by \citet{Weyant_etal_2013}, each $\epsilon\upp{t}$ is determined by the median of the differences to the observation for all particles of the previous iteration. I also set $\epsilon\upp{0}$ which accepts everything from the prior at $t=0$. This is the \acro{ABC} algorithm used in this thesis work.

\subsection{Settings}

In this chapter, parameter constraints using \acro{ABC} are studied only with two free parameters $\OmegaM$ and $\sigEig$. The values of other cosmological parameters are detailed in \sect{sect:constraint:methodology:design}. The stochastic model $P(\cdot|\bpi)$ for Algo. \ref{algo:ABC:modified} is our \acro{FSF} peak-count model mentioned in \chap{sect:modelling}. The settings for the model are identical to the previous chapter, detailed in \tab{tab:constraint:parameters}. The same mock observation is also used.

The study has been done with three summary statistics following \chap{sect:constraint}. They have been chosen as $s(\bx)=\bx^\abd$, $\bx^\pct$, and $\bx^\cut$, as defined in \sect{sect:constraint:methodology:vector}. The corresponding distances are defined as
\begin{align}
	D(\bx^\mathrm{type}, \vect{y}^\mathrm{type}) \equiv \sqrt{\sum_{i=1}^5 \frac{(x_i^\mathrm{type} - y_i^\mathrm{type})^2}{C_{ii}}},
\end{align}
where \texttt{type} is \texttt{abd}, \texttt{pct}, or \texttt{cut}. It is simply a weighted Euclidean distance by neglecting the off-diagonal terms of the covariance matrix. In other words, the cross-correlations between bins are not taken into account. The weighted covariance used in the transition kernel and the normal law is explicitly
\begin{align}
	C_{ij} = \frac{\left(\sum_{k=1}^Q \omega_k\right) \cdot\sum_{k=1}^Q \omega_k (\pi_{k,i} - \bar{\pi}_i) (\pi_{k,j} - \bar{\pi}_j)}{\left(\sum_{k=1}^Q \omega_k\right)^2-\sum_{k=1}^Q \omega_k^2},
\end{align}
where $\pi_{k,i}$ stands for the $i$-th component of the $k$-th particle and $\omega_k$ is associated weight. The inverse is estimated via \for{for:constraint:inverse_covariance}. The prior is flat over a SIM-card-shaped region (see \fig{fig:ABC:ABC_evolution_abd5}). I have set $Q=250$ and $r_\mathrm{stop}=0.05$. Finally, the Gaussian transition kernel can be explicitly written in \for{for:constraint:KDE_kernel}.

\subsection{Results}
\label{sect:ABC:constraints:results}

\begin{figure}[tb]
	\centering
	\includegraphics[width=\textwidth]{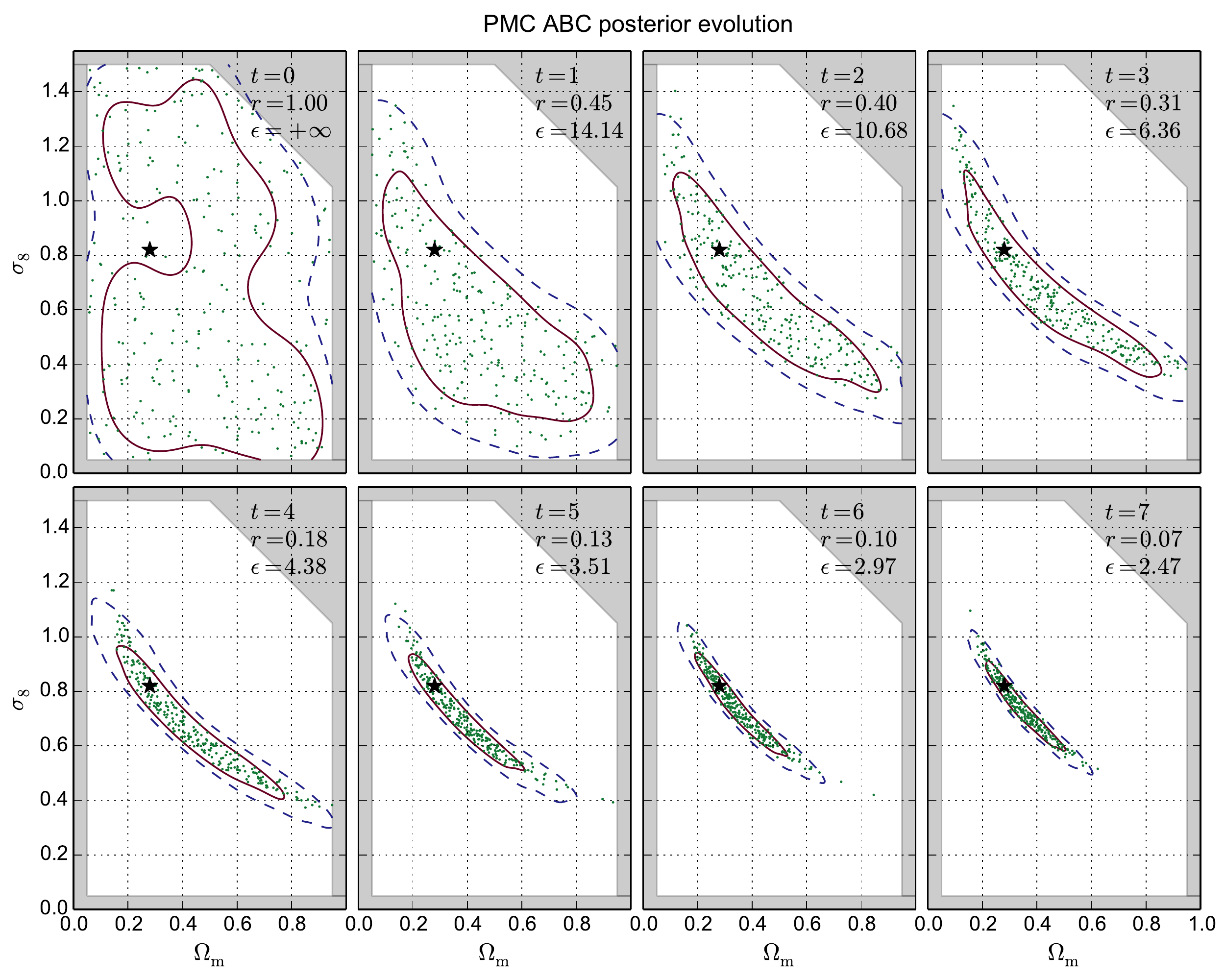}
	\caption{Evolution of the $\OmegaM$-$\sigEig$ posterior from the \acro{PMC} \acro{ABC} algorithm. Other cosmological parameters are fixed. This figure shows the first eight iterations with their success rate and tolerance level. The prior is presented as the white area, $Q=250$ particles as green dots, and the input parameter for the observation as the black star. The 1-$\sigma$ and 2-$\sigma$ regions are displayed by red solid lines and blue dashed lines, respectively. This run actually extends to $t=8$ and $t=9$, which satisfy respectively to $r_\mathrm{stop}=0.05$ and 0.03.}
	\label{fig:ABC:ABC_evolution_abd5}
\end{figure}

Let us first put the focus on the data vector $\bx^\abd$. For this run, the process stops at the end of the iteration $t=8$. \figFull{fig:ABC:ABC_evolution_abd5} shows \acro{ABC} posteriors at iterations $0\leq t\leq7$. These credible contours are constructed by kernel density estimation (\acro{KDE}) using \for{for:constraint:KDE}. Contrary to the proposal function, particle weights are neglected for posterior estimation. When $t=0$, we can see that the particles (green points) are distributed uniformly in the prior region (white area), as $\epsilon\upp{0}=+\infty$ implies. Because of stochasticity, the input parameter (black star) is even excluded at 1-$\sigma$. Gradually, particles move and the usual $\OmegaM$-$\sigEig$ degeneracy direction appears. The credible contours stabilize when $r\upp{t}$ and $\epsilon\upp{t}$ decrease, as indicated in \fig{fig:ABC:ABC_evolution_abd5}.

\begin{figure}[tb]
	\centering
	\includegraphics[width=0.49\textwidth]{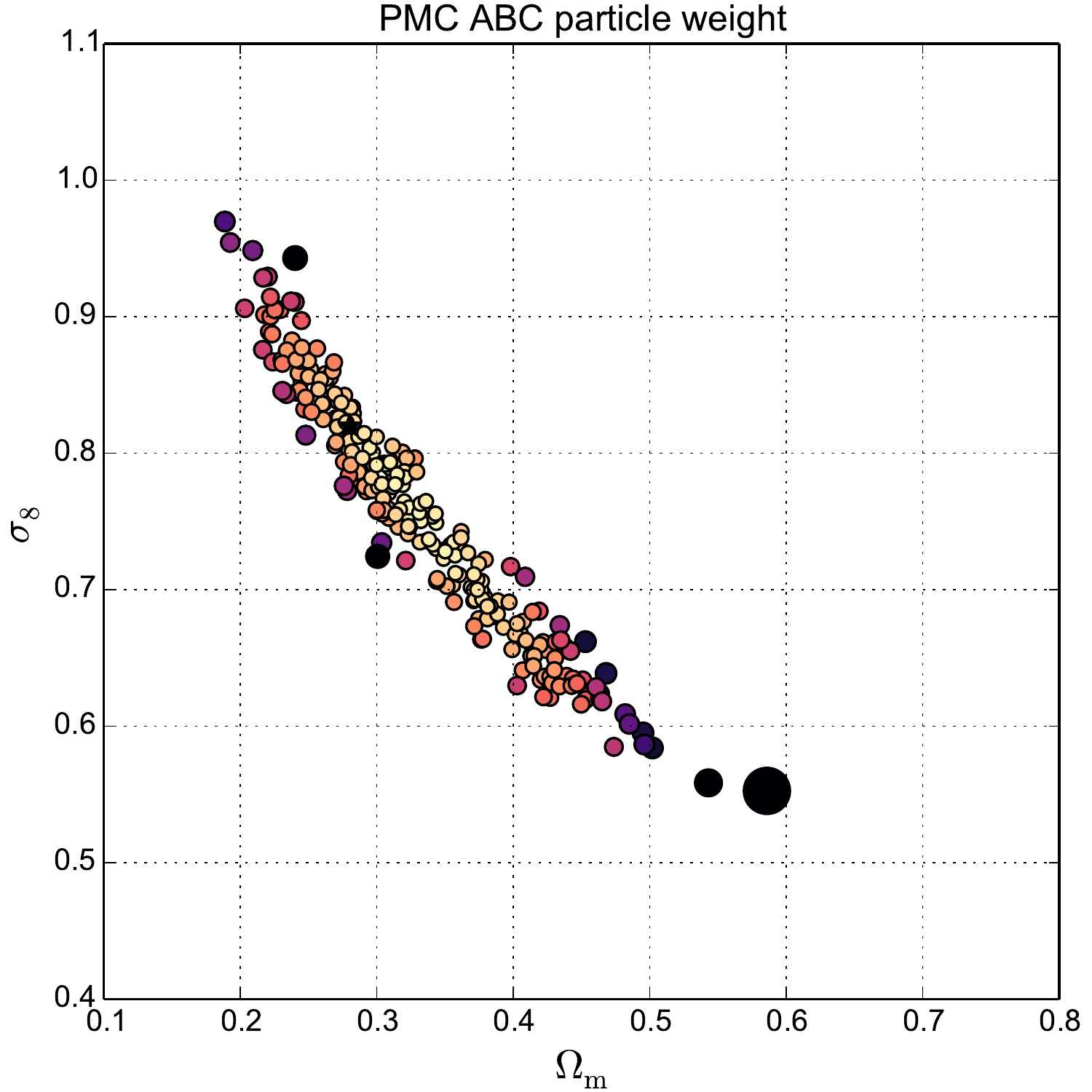}\hfill
	\includegraphics[width=0.49\textwidth]{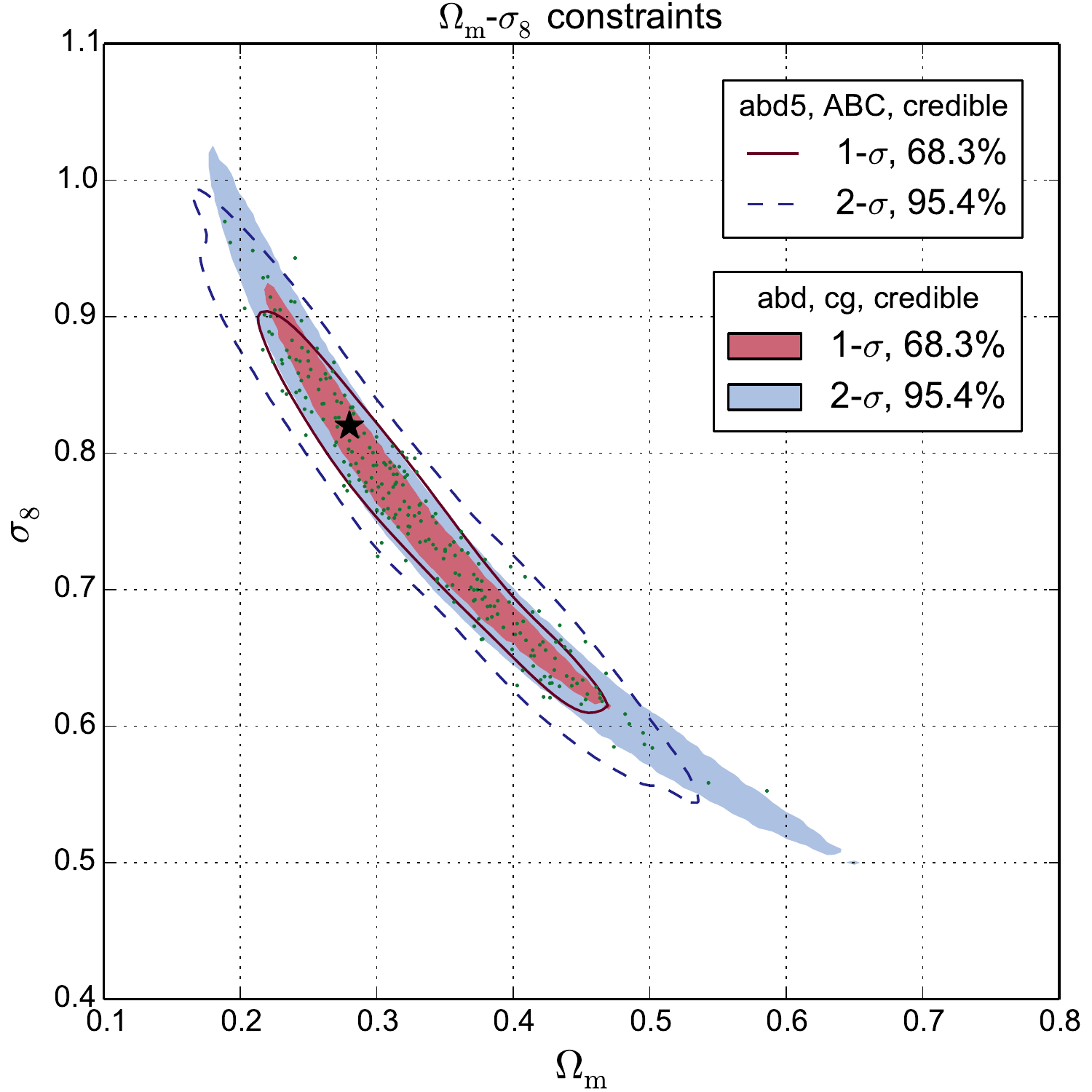}
	\caption{Results from \acro{PMC} \acro{ABC} with $\bx^\abd$. Left panel: particles at $t=8$ with their weights, illustrated in both size and color. Right panel: comparison between constraints from $L_\cg$ and \acro{ABC}; the legend is similar to \fig{fig:ABC:ABC_evolution_abd5}.}
	\label{fig:ABC:contour_abd5_cg_ABC_credible}
\end{figure}

The stabilization of the contours can be explained by the weight of particles. The left panel of \fig{fig:ABC:contour_abd5_cg_ABC_credible} shows particles from the final result with their weights, visualized by both size and color. We observe that isolated particles are more weighted than those in the high-density zone, as constructed by Algo. \ref{algo:ABC:modified}. This fact smooths the posterior, and makes the proposal for the next iteration a flatter function, avoiding undersampling the tails of the posterior and overestimating the constraining power.

The posterior given by the final result is presented on the right panel of \fig{fig:ABC:contour_abd5_cg_ABC_credible}. This result is compared with the constraint contours obtained with $L_\cg$ from \for{for:constraint:L_cg}. We can observe a nice agreement between these two constraint techniques, which validates the performance of \acro{PMC} \acro{ABC}. However, at both wings of the contour, the \acro{ABC} contour is less curved than the $L_\cg$ constraint, while by eyes, particles still seem to follow the banana-shaped degeneracy lines. The reason for this discrepancy is \acro{KDE}. First, \acro{KDE} smooths and gives systematically a biased estimation. This bias can be more visible at low-probability regions, which is the case of the wings of the contour. Second, the \acro{KDE} bandwidth matrix is chosen to be diagonal (\for{for:constraint:KDE_bandwidth}), so the cross-section of the Gaussian smoothing kernel is an ellipse whose long axis is aligned either with $\OmegaM$-axis or $\sigEig$-axis, not adapted to the degeneracy direction. Nevertheless, with this drawback, \acro{KDE} still offers a satisfactory result. The values of indicators and best-fits are presented in Tables \ref{tab:constraint:indicator_credible} and \ref{tab:constraint:best_fit_credible}. These results are also very consistent with other techniques studied in \chap{sect:constraint}.

The most important advantage of \acro{ABC} is efficiency. Start with estimating the time cost in terms of the number of model realizations. Let $N$ be the number of realizations under an input parameter set for estimation of the covariance matrix (for likelihoods) or the \acro{PDF} (for copula, the true likelihood, and $p$-value analysis). For parameter constraints by grid-point evaluation or \acro{MCMC}, the total time cost is $Q\times N$ where $Q$ is the number of parameter sets that are tested. For \acro{PMC} with likelihoods, we need to add the number of iterations $T$, so that the time cost becomes $T\times Q\times N$. For instance, for analyses of \chap{sect:constraint}, $Q=7821$ and $N=1000$.

What is then the time cost for \acro{PMC} \acro{ABC}? In this case, the number of realizations varies between iterations. It is the total number of candidates times one, becuase of the one-sample test. The number of candidates for the $t$-th iteration is $Q/r\upp{t}$. Thus, the total cost is $\left(\sum_t 1/r\upp{t}\right)\times Q\times1$. For the run with $\bx^\abd$, $Q=250$ and $\sum_t 1/r\upp{t}\approx 69$ (for $0\leq t\leq8$). So, we end up with $7821\times1000$ for analyses of \chap{sect:constraint} and $69\times250\times1$ for \acro{PMC} \acro{ABC}. The difference is about two orders of magnitude. If we set a more strict shutoff value $r_\mathrm{stop}=0.03$ for \acro{PMC} \acro{ABC} (in this case, $\sum_t 1/r\upp{t}\approx 102$ for the same run, $0\leq t\leq9$), then the difference is still neatly an order of magnitude. This is more likely the gain of the \acro{PMC} \acro{ABC} algorithm for future studies which seek to constrain parameters in a high-dimensional space, where grid-point evaluation is not feasible and $Q$ will take between 10,000 and 100,000 for both \acro{MCMC} and \acro{PMC}.

\begin{figure}[tb]
	\centering
	\includegraphics[width=0.49\textwidth]{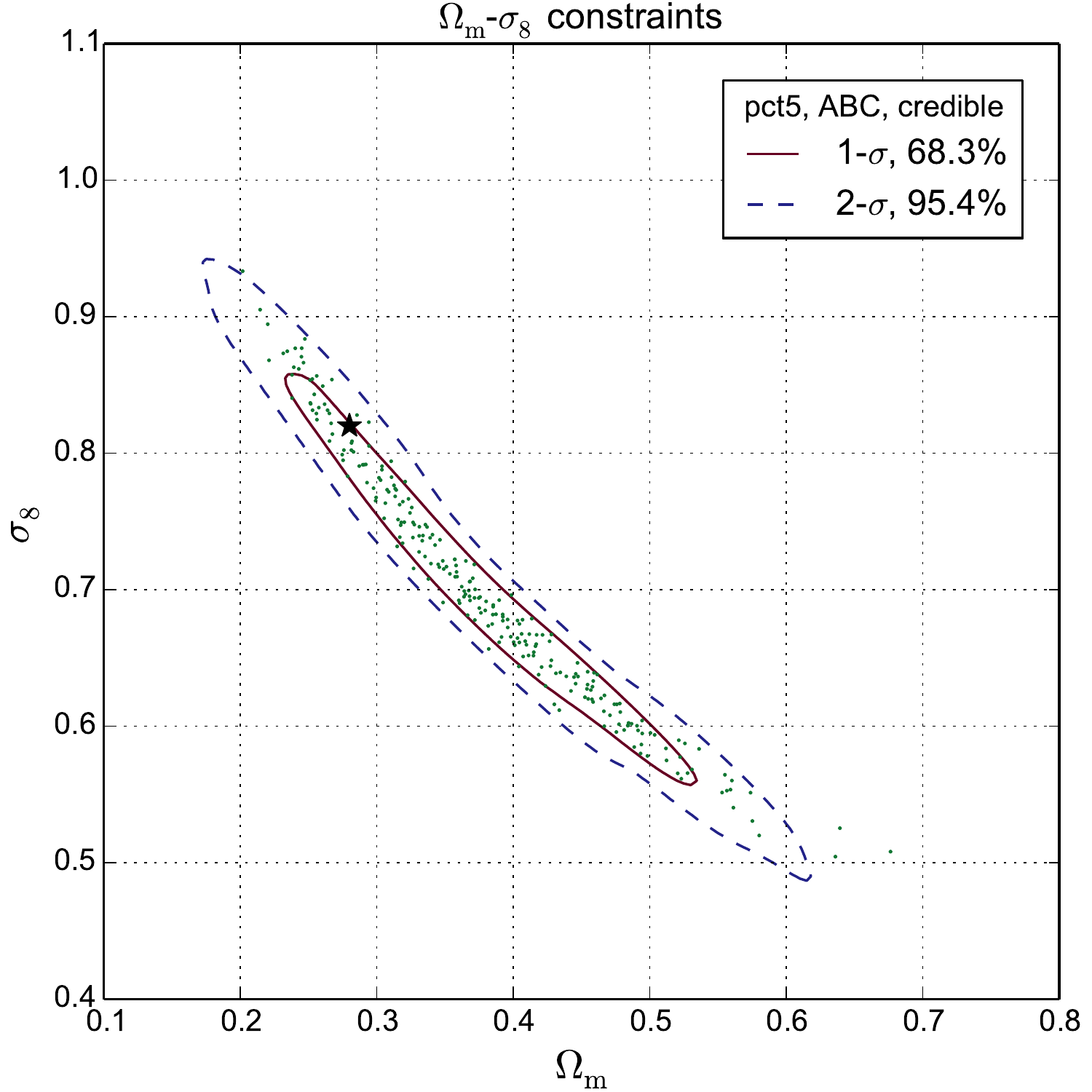}\hfill
	\includegraphics[width=0.49\textwidth]{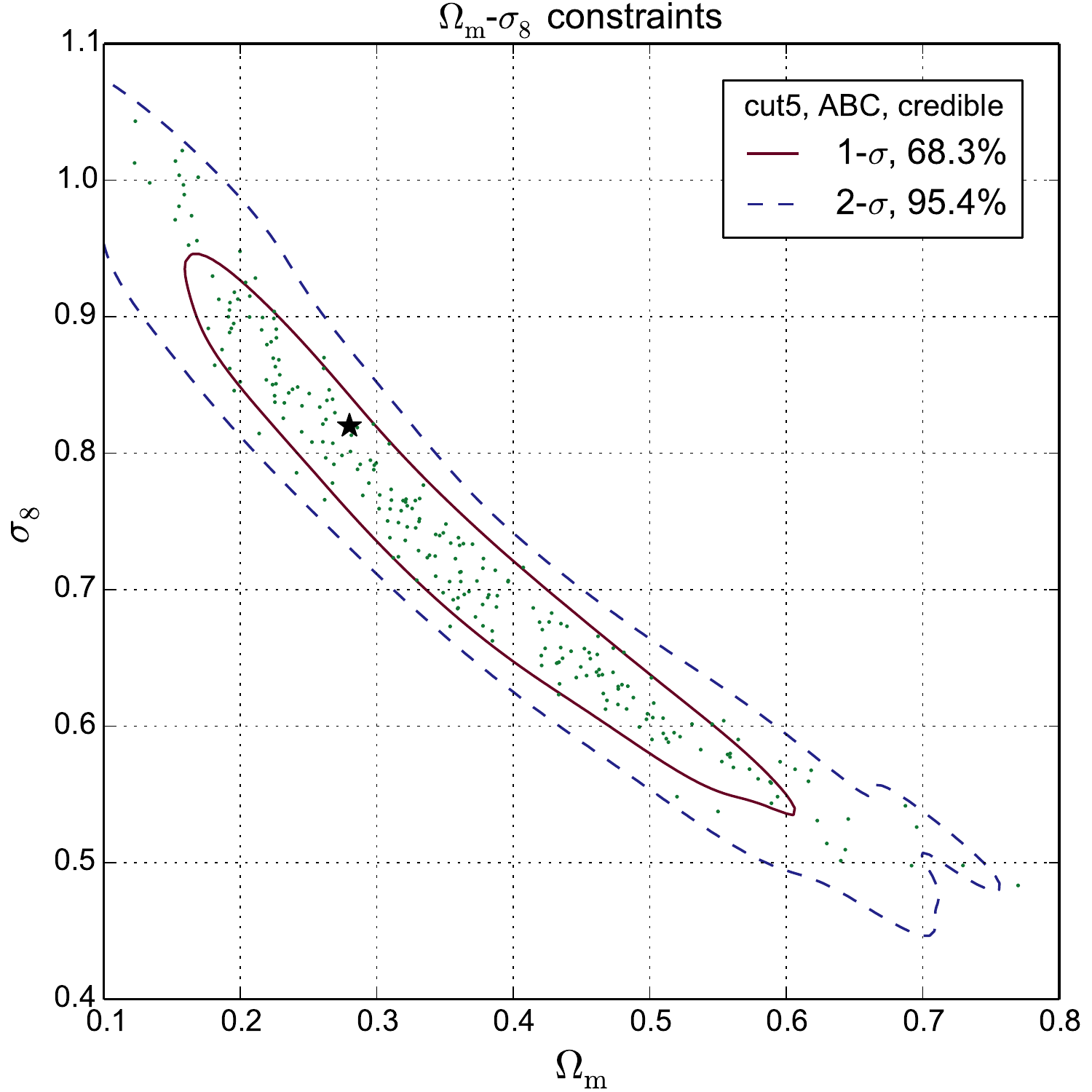}
	\caption{Credible regions from \acro{PMC} \acro{ABC} with $\bx^\pct$ and $\bx^\cut$. These contours correspond respectively to $t=9$ and $t=7$. The legend is similar to \fig{fig:ABC:ABC_evolution_abd5}. The wings of contours are also less curved in these cases.}
	\label{fig:ABC:ABC_contour_zoom_pct5_p250_t9}
\end{figure}

The results for runs with $\bx^\pct$ and $\bx^\cut$ are shown in \fig{fig:ABC:ABC_contour_zoom_pct5_p250_t9}, \tab{tab:constraint:indicator_credible}, and \tab{tab:constraint:best_fit_credible}. The conclusion taken from these results is similar to the interpretations above.

\subsubsection{Summary}

In this chapter, I present a new parameter constraining method called approximate Bayesian computation (\acro{ABC}). It bypasses likelihood evaluation and generates samples directly from an approximate posterior close to the true one. By reconstructing from samples, one obtains an estimate of the true posterior.

The \acro{ABC} process requires a stochastic model. The result may depend on the metric which defines the difference between two data sets. This metric involves a distance and the notion of summary statistic which can be interpreted as defining a data vector.

A fashionable way to perform \acro{ABC} is to combine it with population Monte Carlo (\acro{PMC}). Thanks to its iterative process, the \acro{PMC} \acro{ABC} algorithm decreases the tolerance level and refines the result. In this work, I proposed a modified version with shutoff parameter to decrease the number of arbitrary choices.

The result from \acro{PMC} \acro{ABC} agrees well with those from other constraining methods. Only a small number of particles is required. In terms of computational costs, it is two orders of magnitude lower than classical constraining methods using likelihoods. Robust and efficient, \acro{PMC} \acro{ABC} seems to be the most adapted solution for stochastic processes, including our model and $N$-body simulations.

Chapters \ref{sect:modelling}, \ref{sect:constraint}, and \ref{sect:ABC} have established together a ``pipeline'' from theoretical inputs to parameter constraints. Before applying this pipeline to the real data, I will take a closer look on filtering techniques in the next chapter.

\clearpage
\thispagestyle{empty}
\cleardoublepage


\chapter{Filtering technique comparisons}
\label{sect:filtering}
\fancyhead[LE]{\sf \nouppercase{\leftmark}}
\fancyhead[RO]{\sf \nouppercase{\rightmark}}

\subsubsection{Overview}

This chapter addresses two questions: the best filter choice and extraction of multiscale information. First, I will argue why some optimal filters proposed in the literature might not be necessarily optimal for cosmological purposes. An alternative standard will be established. Then, several filters, linear or nonlinear, will be compared. Two strategies for dealing with multiscale information will be introduced and tested. All analyses are processed with a realistic version of the fast model, with two parameter constraints methods: the copula likelihood and \acro{ABC}. This chapter corresponds to \PaperIII.

\section{Optimality of the filtering techniques}

\subsection{What are optimal filters optimal for?}

In weak lensing, galaxy shape noise usually dominates the lensing signal. Filters are then applied to observations to enhance the peak \acro{S/N}. Among different filtering methods, some optimization studies, such as \citet{Maturi_etal_2005} and \citet{Hennawi_Spergel_2005}, have been done for peak detections. What objectives are these filters optimal for? They actually refer to different meanings of optimum. On the one hand, \citet{Maturi_etal_2005} modelled large-scale structures as ``noise'' with respect to clusters. Following this reasoning, given a halo density profile on a given scale, the ideal shape for the smoothing kernel can be obtained. On the other hand, \citet{Hennawi_Spergel_2005} constructed a \textit{tomographic matched filter} algorithm. Given a kernel shape, this algorithm was able to determine the most probable position and redshift of presumed clusters.

Recall the vocabulary defined in \sect{sect:lensing:state}: \textit{cluster-oriented purposes}\index{Cluster-oriented purpose} means that cluster detection is put forward, even though these detections may be used to probe cosmology later; \textit{cosmology-oriented purposes} seek to constrain directly cosmology, bypassing cluster identification. With these concepts, readers can find out that both methods mentioned above aim for cluster-oriented purposes, and these might not be optimal if we are interested in cosmological constraints.

Why does the optimal cluster detection not imply the maximal cosmological information extraction from \acro{WL} peaks? First, a peak is not necessarily created by a single halo \citep{Yang_etal_2011, Liu_Haiman_2016}. Optimizing multiple-halo peaks with single-halo detection approaches is counter-effective. Second, large-scale structures could contribute to peak counts and increase cosmological implications from peaks. As we go to regimes with lower \acro{S/N}, these effects could become non-negligible. Furthermore, cluster-oriented studies usually focus on purity and completeness as performance indicators \citep{Hennawi_Spergel_2005, Pires_etal_2012, Leonard_etal_2014}. This duality makes the comparisons between filters ill-defined. Even if we only focus on one single detection method, the choice between high purity and high completeness is still ambiguous. What strategy extracts more cosmological information? What is its link to parameter constraints? These subtle questions remain unsolved.

For peak studies aiming for cosmology-oriented purposes, a more proper way to find the filtering optimality is to examine directly the cosmological results. This means that instead of looking at purity and completeness, we should compare directly the resulting constraints contours (width, size, etc.) between different filters. Under this standard, no filtering technique can claim to be optimal in an ``obvious'' way --- especially when we want to explore multiscale information. The only justification we can give is to compare between some potential candidates, and find out the one which outperforms a necessarily non-exhaustive list of other filters. This idea will be strictly followed in this chapter.

\subsection{Separated and combined strategies}
\label{sect:filtering:optimality:strategies}

Compared to the studies in previous chapters, an improvement has been put forward by exploring multiscale information instead of one single filter scale. There exist two strategies to deal with multiple scales, namely the separated and combined strategies. 

The \textit{separated strategy}\index{Separated strategy} (followed implicitly by \citealt{Maturi_etal_2005}, \citeyear{Maturi_etal_2010}; see also \citetalias{Liu_etal_2015} \citeyear{Liu_etal_2015}) applies a series of filters at different scales. This results in a series of filtered weak-lensing maps, and cosmological constraints are derived from the ensemble of these maps. For weak-lensing peaks, we can for example concatenate peak-count histograms to do parameter inference.

The \textit{combined strategy}\index{Combined strategy} (followed e.g. by \citealt{Hennawi_Spergel_2005, Marian_etal_2012}), sometimes called mass mapping\index{Mass mapping}, yields only one map. This means that the significances from different scales have been compared and summarized into a representative value. Both \citet{Hennawi_Spergel_2005} and \citet{Marian_etal_2012} use a likelihood approach to choose, for each position, the most adaptive filtering scale, so that only one filtered map is provided, from which we can estimate peak abundance and derive constraints.

Apart from linear filters, there also exist various nonlinear reconstruction techniques. For example, \citet{Bartelmann_etal_1996} proposed to minimize the error on shear and magnification together. Other techniques are sparsity-based methods such as \MRLens\ \citep{Starck_etal_2006}, \textsc{\acro{FASTLens}} \citep{Pires_etal_2009a}, and \textsc{\acro{Glimpse}} \citep{Leonard_etal_2014}. These approaches aim to map the projected mass through a minimization process. Therefore, the optimality of filters is a complex issue involving several aspects such as linearity, either separated or combined strategy for multiscale information, and the choice of scales. The aim of this chapter is to address this problem and to answer to following questions:
\begin{itemize}
	\item For a given kernel shape, with the separated strategy, what are the preferable characteristic scales?
	\item Which can extract more cosmological information, the compensated or non-compensated filters?
	\item Which can extract more cosmological information, the separated or combined strategy?
	\item How do nonlinear filters perform?
\end{itemize}

\section{Linear filters}

\subsection{Convergence filters and noise level}
\label{sect:filtering:linear:noise}

In this section, I will re-establish the formalism for filtering. Consider a noisy convergence map to filter. Let $\btheta$ be angular coordinates. Let $\theta_{\ker}$ be the size of the filtering kernel and $x=\theta/\theta_{\ker}$. The filter function will always be described by $W(x)$ with the dimensionless position $x$. A very common choice of $W$ is the Gaussian smoothing kernel:
\begin{align}
	W(x) \propto \exp\left(-x^2\right).
\end{align}

In principle, we assume that both components of the galaxy intrinsic ellipticity $\epsilon=\epsilon_1+\rmi\epsilon_2$ follow the same Gaussian distribution. By noting $\sigma_\epsilon^2=\sigma_{\epsilon_1}^2+\sigma_{\epsilon_2}^2$ as the sum of both variances, the noise for the smoothed convergence is also Gaussian. The new variance depends on the kernel shape and the expected galaxy number density $n_\gala$, which results in the global noise level \citep[see e.g.][]{VanWaerbeke_2000}:
\begin{align}\label{for:filtering:global_noise}
	\sigma_\noise^2 = \frac{\sigma_\epsilon^2}{2n_\gala}\cdot\frac{\parallel W\parallel_2^2}{\parallel W\parallel_1^2},
\end{align}
where $\parallel W\parallel_p$ stands for the $p$-norm of $W$. Equation \eqref{for:filtering:global_noise} is valid for unnormalized kernels because the normalization factors from the denominator and the numerator cancel each other. For example, if $W$ is a Gaussian with width $\theta_{\ker}$, then $\parallel W\parallel_2^2/\parallel W\parallel_1^2 = 1/2\pi\theta_{\ker}^2$. 

Equation \ref{for:filtering:global_noise} is the global noise level since $n_\gala$ is the global density. This is the true noise only if sources are distributed regularly. In realistic conditions, random fluctuations, mask effects, and clustering of source galaxies lead to irregular distributions, which results in a non-constant noise level. To properly take this into account, we can transform norms into discrete sums. Thus, the local noise level is
\begin{align}\label{for:filtering:local_noise}
	\sigma_\noise^2(\btheta) = \frac{\sigma_\epsilon^2}{2}\cdot\frac{\sum_i W^2(\btheta_i-\btheta)}{\left(\sum_i |W(\btheta_i-\btheta)|\right)^2},
\end{align}
where $\btheta_i$ is the position of the $i$-th galaxy, and $i$ runs over non-masked galaxies under the kernel $W$. This is nothing but the variance of a weighted sum of independent Gaussian random variables each with variance $\sigma_\epsilon^2/2$.

\begin{figure}[tb]
	\centering
	\includegraphics[width=0.8\textwidth]{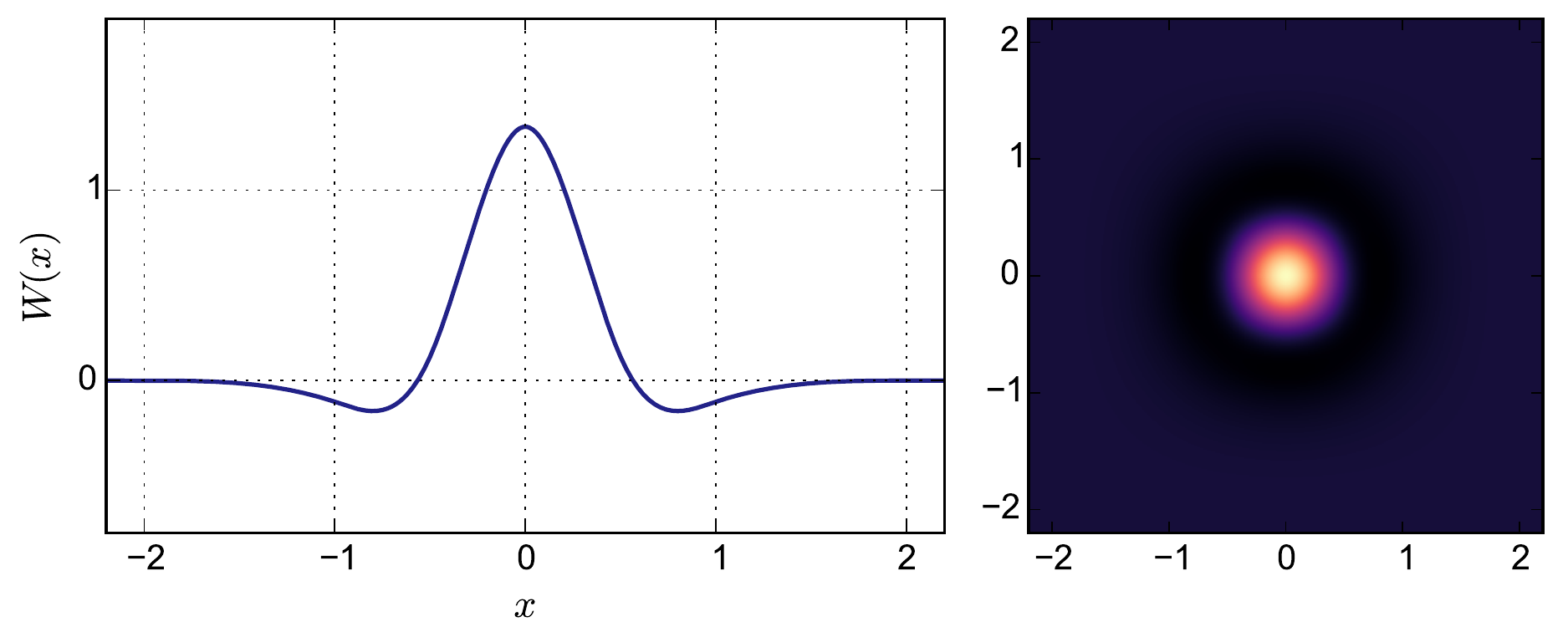}
	\caption{Left panel: the profile of the \acro{2D} starlet. It has a finite support [-2, 2]. Right panel: the bird-eye view of the \acro{2D} starlet.}
	\label{fig:filtering:Starlet}
\end{figure}

In this work, the \acro{2D} \textit{starlet} function \citep{Starck_etal_2002} is also studied. It is defined as 
\begin{align}\label{for:filtering:starlet}
	W(x, y) = 4\phi(2x)\phi(2y) - \phi(x)\phi(y),
\end{align}
where $x=\theta_1/\theta_\mathrm{ker}$, $y=\theta_2/\theta_\mathrm{ker}$, and $\phi$ is the B-spline of order 3, given by
\begin{align}
	\phi(x) = \frac{1}{12}\left(|x-2|^3 - 4|x-1|^3 + 6|x|^3 - 4|x+1|^3 + |x+2|^3\right).
\end{align}
Because of the property the B-spline, the starlet is a compensated function with compact support in $[-2,2]\times[-2,2]$. It does not conserve circular symmetry, but its isolines tend to be round (\fig{fig:filtering:Starlet}). For the starlet, $\parallel W\parallel_1^2\ \approx 0.979^2$ and $\parallel W\parallel_2^2 = 5(2/5+5/63)^2 - 2(1/3+1/5+1/21+1/48)^2\ \approx 0.652^2$ can be solved analytically.

\subsection{Aperture mass}
\label{sect:filtering:linear:aperture}

The \textit{aperture mass}\index{Aperture mass} $M_\ap$ \citep{Kaiser_etal_1994, Schneider_1996} was motivated by the fact that the convergence is not directly observable, and that shear inversion generates a free constant $\kappa_0$. To avoid this, one may convolve directly the shear with another filter and expect that this is equivalent to convergence filtering. This is possible only if the filter is circular. In this case, the convolution product is then the aperture mass. 

More precisely, we look for all pairs of filters $(U, Q)$ such that (1) $U$ is circularly symmetric, (2) $U$ is a compensated function, and (3) $U\ast\kappa = Q\ast\gamma_+$. With these conditions, convolving $\gamma_+$ with $Q$ results in a filtered convergence map that is not affected by either $\kappa_0$ or inversion systematics. 

To satisfy the third condition, $Q$ has to be related to $U$ by
\begin{align}\label{for:filtering:Q_U_relation}
	Q(\theta) \equiv \frac{2}{\theta^2}\int_0^\theta\rmd\theta'\ \theta'U(\theta')-U(\theta).
\end{align}
Equation \eqref{for:filtering:Q_U_relation} is directly derived from \for{for:lensing:tangential_shear_profile_1}. Then, $M_\ap$ is given by
\begin{align}
	M_\ap(\btheta) \equiv \int\rmd^2\btheta'\ U(\btheta)\kappa(\btheta-\btheta') = \int\rmd^2\btheta'\ Q(\btheta)\gamma_\rmt(\btheta-\btheta').
\end{align}
Here, we are particularly interested in the $Q$ function proposed by \citet{Schirmer_etal_2004} and \citet{Hetterscheidt_etal_2005}, given by
\begin{align}\label{for:filtering:Q_tanh}
	Q(x) \propto \frac{\tanh(x/x_\rmc)}{(x/x_\rmc)\left(1+\exp(a-bx)+\exp(-c+dx)\right)},
\end{align}
with $a=6$, $b=150$, $c=47$, $d=50$ to have a cutoff around $x=1$. This filter shape has been motivated by the tangential shear pattern generated by \acro{NFW} halo profiles. Also, we set $x_\rmc=0.1$ as suggested by \citet{Hetterscheidt_etal_2005}. 

For the noise level of a map filtered with $M_\ap$, Eqs. \eqref{for:filtering:global_noise} and \eqref{for:filtering:local_noise} are still valid. We only need to substitute $W$ with $Q$ \citep{Schneider_1996}.

\section{A sparsity-based nonlinear filter}

In this section, a nonlinear filtering technique using the sparsity of signals is introduced.

\subsection{What is sparsity?}
\label{sect:filtering:nonlinear:sparsity}

In signal processing, a signal is sparse in a specific representation if most of the information is contained in only a few coefficients. This means that either only a finite number of coefficients is non zero, or the coefficients decrease fast when rank-ordered.

A straightforward example is the family of sine functions. In real space, sine functions are not sparse. However, they are sparse in Fourier space since they become the Dirac delta functions. More generally, periodic signals are sparse in the Fourier space.

Why is this interesting? Because white noise is not sparse in any representation. Therefore, if the information of the signal can be compressed into a few strong coefficients, it can easily be separated from the noise. This concept of sparsity has been widely used in the signal processing domain for applications such as denoising, inpainting, deconvolution, inverse problems, or other optimization problems \citep{Daubechies_etal_2004, Candes_Tao_2006, Elad_Aharon_2006, Candes_etal_2008, Fadili_etal_2009}. Examples can also be found for studying astrophysical signals \citep{Lambert_etal_2006, Pires_etal_2009, Bourguignon_etal_2011, Carrillo_etal_2012, Bobin_etal_2014, NgoleMboula_etal_2015, Lanusse_etal_2016}.

\subsection{Wavelet transform}
\label{sect:filtering:nonlinear:wavelet}

From the previous section, one can see that the sparsity of a signal depends on its representation basis. In which basis is the weak lensing signal sparse? A promising candidate is the wavelet transform which decomposes the signal into a family of scaled and translated functions. Wavelet functions are all functions $\psi$ which satisfy the \textit{admissibility condition}:
\begin{align}
	\int_0^{+\infty}\frac{\rmd k}{k}\ |\tilde{\psi}(k)|^2 < +\infty.
\end{align}
For example, the starlet is one of the wavelet functions. One of the properties implied by this condition is $\int\psi=0$, which restricts $\psi$ to a compensated function. In other words, one can consider wavelet functions as highly localized functions with a zero mean. Such a function $\psi$ is called the mother wavelet, which can generate a family of daughter wavelets such as
\begin{align}
	\psi_{a,b}(x) = \frac{1}{\sqrt{a}}\psi\left( \frac{x-b}{a} \right),
\end{align}
which are scaled and translated versions of the mother $\psi$.

The wavelet transform \citep[see e.g. Chaps. 2 and 3 of][]{Starck_etal_2002} refers to the decomposition of an input image into several images of the same size each associated to a specific scale. Due to the property of wavelet functions, each resulting image gives the details of the original one at different scales. If we stack all the images, we recover the original signal.

In the peak-count scenario, peaks which are generated by massive clusters are considered as signals. Like clusters, these signals are local point-like features, and therefore have a sparse representation in the wavelet domain. As described in \sect{sect:filtering:nonlinear:sparsity}, white noise is not sparse. So one simple way to reduce the noise is to transform the input image into the wavelet domain, set a relatively high threshold $\lambda$, cut out weak coefficients smaller than $\lambda$, and reconstruct the clean image by stacking the thresholded images. This introduces non-linearity. In this chapter, we use the \acro{2D} starlet function as the mother wavelet, given by \for{for:filtering:starlet}, which satisfies the admissibility condition.

\subsection{The MRLens filter}
\label{sect:filtering:nonlinear:MRLens}

In this study, the nonlinear filtering technique \textit{MultiResolution tools for gravitational Lensing} \citep[\MRLens,][]{Starck_etal_2006} is applied to lensing maps. \MRLens\ is an iterative filtering based on a Bayesian framework that uses a multiscale entropy prior and the false discovery rate (\acro{FDR}, \citealt{Benjamini_Hochberg_1995}) which allows to derive robust detection levels in wavelet space.

More precisely, \MRLens\ first applies a wavelet transform to a noisy map. The mother wavelet is chosen to be the starlet function. Then, in the wavelet domain, it determines the threshold by \acro{FDR}. The denoising problem is regularized using a multiscale entropy prior only on the non-significant wavelet coefficients. Readers are welcome to read \citet{Starck_etal_2006} for a detailed description of the method.

Note that, whereas \citet{Pires_etal_2009a} selected peaks from different scales separately before the final reconstruction, here peaks are counted on the final reconstructed map. Actually, the methodology of \citet{Pires_etal_2009a} is close to filtering with a lower cutoff in the histogram defined by \acro{FDR}, thus similar to starlet filtering. With the vocabulary defined in \sect{sect:filtering:optimality:strategies}, \citet{Pires_etal_2009a} followed the separated strategy and here we attempt the combined strategy. This choice provides a comparison between cosmological information extracted with two strategies, by comparing starlet filtering to the \MRLens\ case.

\section{Methodology}

\subsection{Filter comparisons}
\label{sect:filtering:methodology:comparisons}

\begin{table}[tb]
	\centering
	\begin{tabular}{rccccc}
		\hline\hline
		Kernel                   & \multicolumn{5}{l}{Gaussian}\\
		$\theta_{\ker}$ [arcmin] & 1.2   & 2.4   & 4.8    & & \\
		$\sigma_\noise$          & 0.027 & 0.014 & 0.0068 & & \\
		\hline
		Kernel                   & \multicolumn{5}{l}{Starlet}\\
		$\theta_{\ker}$ [arcmin] & 2     & 4     & 8      & 12     & 16\\
		$\sigma_\noise$          & 0.027 & 0.014 & 0.0068 & 0.0045 & 0.0034\\
		\hline
		Kernel                   & \multicolumn{5}{l}{$M_\ap$ $\tanh$}\\
		$\theta_{\ker}$ [arcmin] & 2.125 & 4.25  & 8.5    & & \\
		$\sigma_\noise$          & 0.027 & 0.014 & 0.0068 & & \\
		\hline
	\end{tabular}
	\caption{List of kernel sizes $\theta_{\ker}$. These are chosen based on $\sigma_\noise$ such that the corresponding values are similar. The quantity $\sigma_\noise$ is computed using \for{for:filtering:global_noise} with $n_\gala=12~\arcmin\invSq$ and $\sigma_\epsilon=0.4$.}
	\label{tab:filtering:scales}
\end{table}

\begin{figure}[tb]
	\centering
	\includegraphics[width=0.8\textwidth]{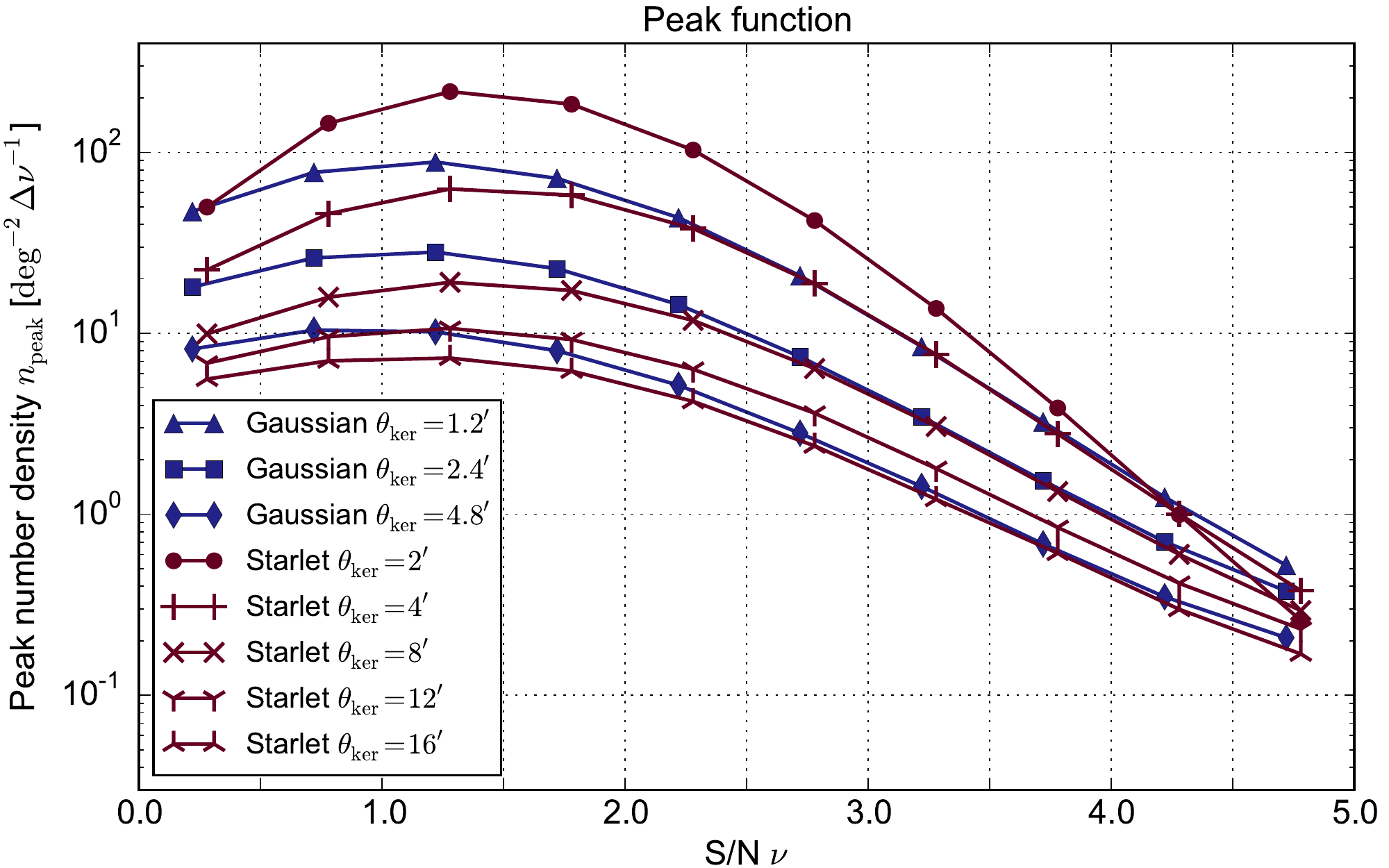}
	\caption{Peak function for different kernel sizes for an input cosmology $(\OmegaM, \sigEig, \wZero)=(0.28, 0.82, -0.96)$. The number counts are the mean over 400 realizations of 36 deg$^2$. Focusing on the range $2.5\leq\nu\leq4.5$, we find that the average number counts of Gaussian filtering with $\theta_{\ker}=$ 1.2, 2.4, and 4.8 arcmin correspond respectively to starlet filtering with $\theta_{\ker}=$ 4, 8, and 16 arcmin.}
	\label{fig:peakHist_gauss_vs_star_thesis}
\end{figure}

Comparing different filtering techniques for peak counts is the subject of this analysis. This comparison contains (1) the Gaussian kernel, (2) the starlet function, (3) the aperture mass with the hyperbolic tangent function, and (4) the nonlinear filtering technique \MRLens. 

The linear filters are parametrized with a single parameter, which is the size of the kernel $\theta_{\ker}$. In order to compare in a pertinent way bewteen multiscale information captured by different filter shapes, two rules are proposed. The first is to choose $\theta_{\ker}$ such that the 2-norms have the same value if kernels are normalized (by their respective 1-norms). The reason for this is that if the ratio of the 2-norm to the 1-norm is identical, then the comparison is based on the same global noise level (Eq. \ref{for:filtering:global_noise}). \tab{tab:filtering:scales} shows various values of $\theta_{\ker}$ that are used in this study and the corresponding $\sigma_\noise$ for different linear filters. For the Gaussian filter with $\theta_{\ker}=$ 1.2, 2.4, and 4.8 arcmin, the corresponding scales are $\theta_{\ker}=$ 2, 4, and 8 arcmin for the starlet function.

The second way is to calculate peak-count histograms, and set $\theta_{\ker}$ such that peak abundance is similar. \figFull{fig:peakHist_gauss_vs_star_thesis} shows an example for the Gaussian and starlet kernels with $\theta_{\ker}$ taken from \tab{tab:filtering:scales}. We can observe that, for Gaussian filtering with $\theta_{\ker}=$ 1.2, 2.4, and 4.8 arcmin, the correspondence for starlet filtering based on peak abundance is $\theta_{\ker}=$ 4, 8, and 16 arcmin, if we focus on peaks with $2.5\leq\nu\leq4.5$.

The \MRLens\ filter handles different scales by the wavelet transform mentioned in \sect{sect:filtering:nonlinear:wavelet}. The most important parameter is the \acro{FDR} $\alpha$. It is set to $\alpha=0.05$ for this analysis.

\subsection{Settings for the pipeline: from the mass function to peak catalogues}
\label{sect:filtering:methodology:pipeline}

\begin{figure}[tb]
	\centering
	\includegraphics[scale=0.65]{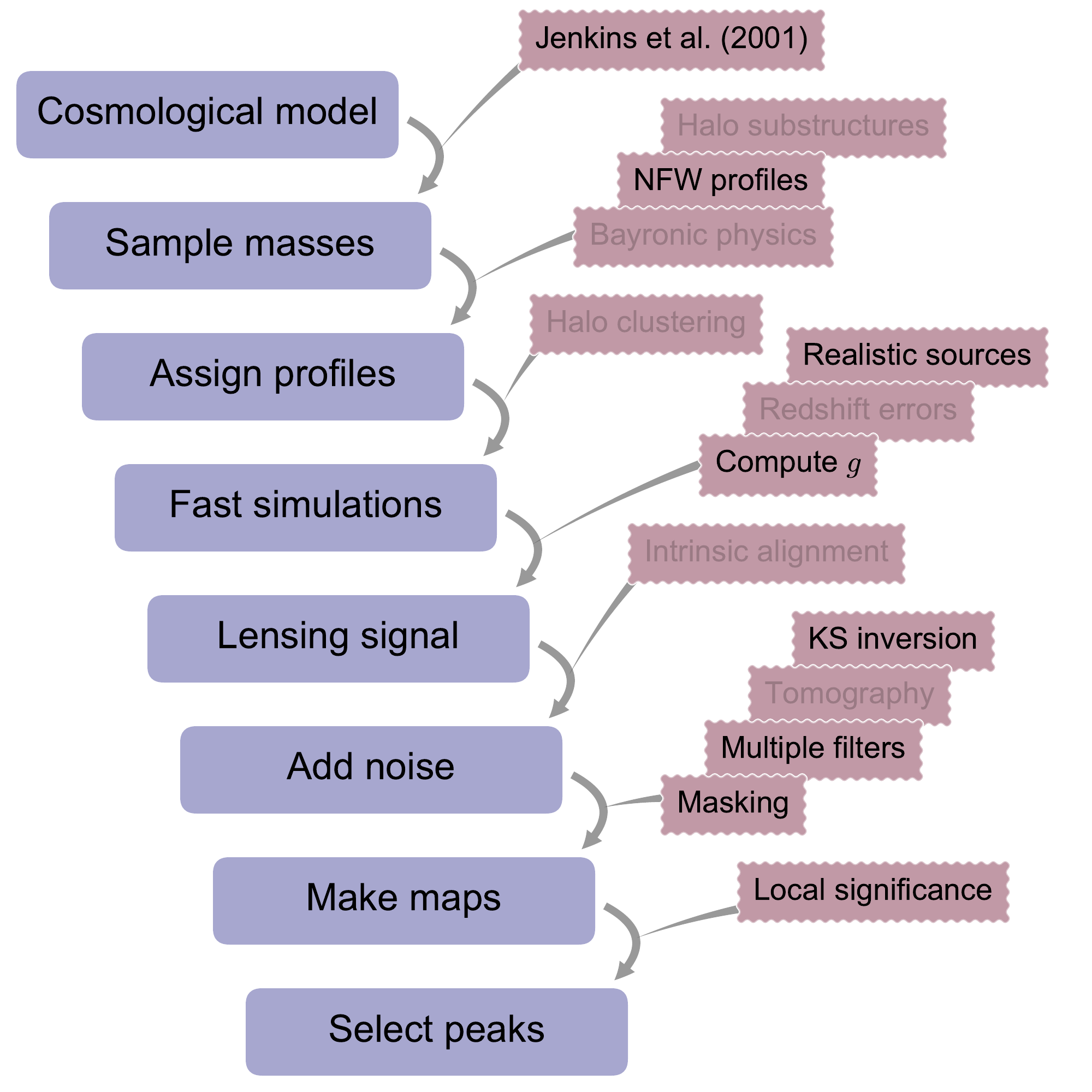}
	\caption{Diagram illustrating the processing pipeline for the filtering comparison. Effects which are not considered are made faint. A more detailed description for all variations can be found in \sect{sect:modelling:extensions}.}
	\label{fig:modelling:LK_model_diagram_3}
\end{figure}

\begin{table}[tb]
	\centering
	\begin{tabular}{lcl}
		\hline\hline
		Parameter                               & Symbol            & Value\\
		\hline
		Lower sampling limit                    & $M_\minn$         & $5\dixx{12}~\Msol/h$\\
		Upper sampling limit                    & $M_\maxx$         & $\dix{17}~\Msol/h$\\
		Maximum halo redshift                   & -                 & 3\\
		Number of halo redshift bins            & -                 & 60\\
		\acro{NFW} inner slope                  & $\alpha$          & 1\\
		\acro{$M$-$c$} relation amplitude       & $c_0$             & 11\\
		\acro{$M$-$c$} relation power law index & $\betaNFW$        & 0.13\\
		Intrinsic ellipticity dispersion        & $\sigma_\epsilon$ & 0.4\\
		Galaxy number density                   & $n_\gala$         & 12 arcmin$\invSq$\\
		Pixel size                              & $\theta_\pix$     & 0.8 arcmin\\
		Effective field area                    & -                 & 36 deg$^2$\\
		Threshold for filling factor            & $\lambda$         & 0.5\\
		\hline\hline
	\end{tabular}
	\caption{List of parameter values adopted in the study of this chapter.}
	\label{tab:filtering:parameters}
\end{table}

Compared to previous analyses, more realistic observational features are accounted for now. We apply a redshift distribution for source galaxies, include masks, construct the convergence from the reduced shear instead of computing it directly, test different filters, determine the noise level locally, and include the equation of state of dark energy for the constraints. A summary is illustrated by \fig{fig:modelling:LK_model_diagram_3}. Details are described below.

\subsubsection{Fast simulations}

For fast simulations, the settings remain similar to previous chapters. Halos are sampled from the model of \citet[][Eq. \ref{for:structure:massFct_J01}]{Jenkins_etal_2001}. The sampling range is set to $M=[5\dixx{12}, 10^{17}]\ \Msol/h$. This is done for 60 equal redshift bins from 0 to 3, on a field adequately larger than 36 deg$^2$ so that border effects are properly eliminated. Halo correlation is neglected. The truncated \acro{NFW} profiles (Eq. \ref{for:structure:TJ_profile}) are assigned to each halo. The \acro{$M$-$c$} relation is given by \for{for:structure:M_c_relation}. 

\subsubsection{Source catalogues}

Instead of sources at a fixed plane, a realistic redshift distribution is considered here. This is assumed to be a gamma distribution following \citet{Efstathiou_etal_1991}:
\begin{align}
	p(z) = \frac{z^2}{2z^3_0}\exp\left(-\frac{z}{z_0}\right),
\end{align}
where $z_0=0.5$ is the pivot redshift value. The positions of sources are random, but masking has been taken into account. The masked area of the W1 field of \acro{CFHTLenS} has been taken as the characteristic mask and is applied to all simulations. The source number density is set to $n_\gala=12\ \arcmin\invSq$, which corresponds to a \acro{CFHTLenS}-like survey \citep{Heymans_etal_2012}. The intrinsic ellipticity dispersion is $\sigma_\epsilon=0.4$, which is also close to the \acro{CFHTLenS} survey \citep{Kilbinger_etal_2013}.

\subsubsection{Ray-tracing}

For each galaxy, we compute $\kappa_\proj$ and $\gamma_\proj$ following 
\begin{align}
	\kappa_\proj(\btheta) = \sum\kappa_\halo(\btheta, w_\ell, w_\rms)\ \ \ \text{and}\ \ \ \gamma_\proj(\btheta) = \sum\gamma_\halo(\btheta, w_\ell, w_\rms),
\end{align}
where the sums run over all lens-source pairs and $\kappa_\halo$ and $\gamma_\halo$ are given respectively by Eqs. \eqref{for:lensing:kappa_halo_NFW} and \eqref{for:lensing:gamma_halo_NFW}. The observed ellipticity is computed as $\epsilon\upp{\rmo}=(\epsilon\src+g_\proj)/(1+g_\proj^*\epsilon\src)$, where $g_\proj\equiv\gamma_\proj/(1-(\kappa_\proj-\overline{\kappa}_\proj))$ is the reduced shear and $\epsilon\src$ is the intrinsic ellipticity, taken from a random generator. The subtraction from $\kappa_\proj$ of its mean over the field is required by the fact that it is always positive and can not be considered as the true convergence, as already evoked earlier.

\subsubsection{Map making}

Except for the aperture mass, galaxies are first binned into map pixels for the reason of efficiency. The mean of $\epsilon\upp{\rmo}$ is taken as the pixel's value. The pixel size is set to 0.8 arcmin. This results in regularly spaced data so that the algorithm can be accelerated. Then, the linear \acro{KS} inversion is applied before filtering. The iterative correction for the reduced shear is not used. By applying exactly the same processing to both observation and model prediction, we expect the systematics related to inversion (e.g. boundary effects, missing data, and negative mass density) to be similar so that the comparison is unbiased. 

For the aperture mass, the pixel's value is evaluated by convolving directly the lensing catalogue with the $Q$ filter (Eq. \ref{for:filtering:Q_tanh}), successively placed at the center of each pixel \citep[see also][]{Marian_etal_2012, Martinet_etal_2015}. The choice of filter sizes is detailed in \sect{sect:filtering:methodology:comparisons}.

\subsubsection{Peak selection}

For linear filters (the Gaussian, the starlet, the aperture mass), peaks are now selected based on their local noise level, determined by \for{for:filtering:local_noise}. In this case, the \acro{S/N} of a peak is a local maximum with
\begin{align}
	\nu(\btheta) \equiv \left\{\begin{array}{ll}
		(\kappa\ast W)(\btheta)/\sigma_\noise(\btheta) & \text{if Gaussian or starlet,}\\
		M_\ap(\btheta)/\sigma_\noise(\btheta) & \text{if aperture mass.} 
	\end{array}\right.
\end{align}

For the nonlinear filter, the notion of noise level does not easily apply. Actually, to determine the significance of a rare event from any distribution, instead of using the empirical standard deviation, it is more rigorous to obtain first the $p$-value and find how much $\sigma$ this value is associated with if the distribution was Gaussian. However, even if we compute the standard deviation instead, this process is still too expensive computationally for our purpose. Therefore, we simply select peaks on $\kappa$. 

Because of masks, only pixels satisfying a specific condition are examined. This criterion is based on the filling factor $f(\btheta)$ (\citealt{VanWaerbeke_etal_2013}; \citetalias{Liu_etal_2015a} \citeyear{Liu_etal_2015a}). A local maximum is selected as a peak only if $f(\btheta)\geq \lambda \bar{f}$, where $\bar{f}$ is the mean of $f$ over the map. We set $\lambda=0.5$. For analyses using binning, the filling factor is simply defined as the number of galaxies $N(\btheta)$ inside the pixel at $\btheta$. For the aperture mass, it is the $Q$-weighted sum of the number counts. In other words,
\begin{align}
	f(\btheta) \equiv \left\{\begin{array}{ll}
		N(\btheta)          & \text{if galaxies are binned,}\\
		\sum_i Q(\btheta_i) & \text{for the aperture mass,} 
	\end{array}\right.
\end{align}
where $\btheta_i$ is the position of the $i$-th galaxy. Also, we only count pixels in the inner area to avoid border effects. The size of this effective field is 36 deg$^2$.

\subsection{Data vector definitions}

The data vector $\bx$, for linear filters, is defined as the concatenation of \acro{S/N} histograms from various scales. In \chap{sect:constraint}, we have seen that the number counts from histograms are the most appropriate form to derive cosmological information from peak counts. After testing several values of $\nu_\minn$, we only keep peaks above $\nu_\minn=1$ for each kernel size. This choice maximizes the figure of merit of parameter constraints. Thus, the statement from \citet{Yang_etal_2013} that ignoring peaks with $\nu\leq 3$ corresponds to a loss of cosmological information is reconfirmed. For each scale, peaks are binned with width of $\Delta\nu=0.5$ up to $\nu=5$, and the last bin is $[5,+\infty[$.

For the nonlinear filter, peaks are binned directly by their $\kappa$ values into [0.02, 0.03, 0.04, 0.06, 0.10, 0.16, $+\infty$[. This configuration is chosen such that the average count per bin is large enough to assume a Gaussian fluctuation.

\subsection{Sampling in the parameter space}

The free parameters to constrain in this study are $(\OmegaM, \sigEig, \wZero)$. The values of other cosmological parameters are $h=0.78$, $\OmegaB=0.047$, and $n_\rms=0.95$. The Universe is assumed to be flat.

The mock observation is generated by a realization of our model, using a particular set $(\OmegaM, \sigEig, \wZero)=(0.28, 0.82, -0.96)$ as input parameters. In this way, we only focus on the precision of our model. Simulations runs are proceeded in two different ways. The first one consists of interpolating the likelihood, from which we draw credible regions from Bayesian inference, and the second is \acro{ABC}.

\begin{figure}[tb]
	\centering
	\includegraphics[width=0.5\textwidth]{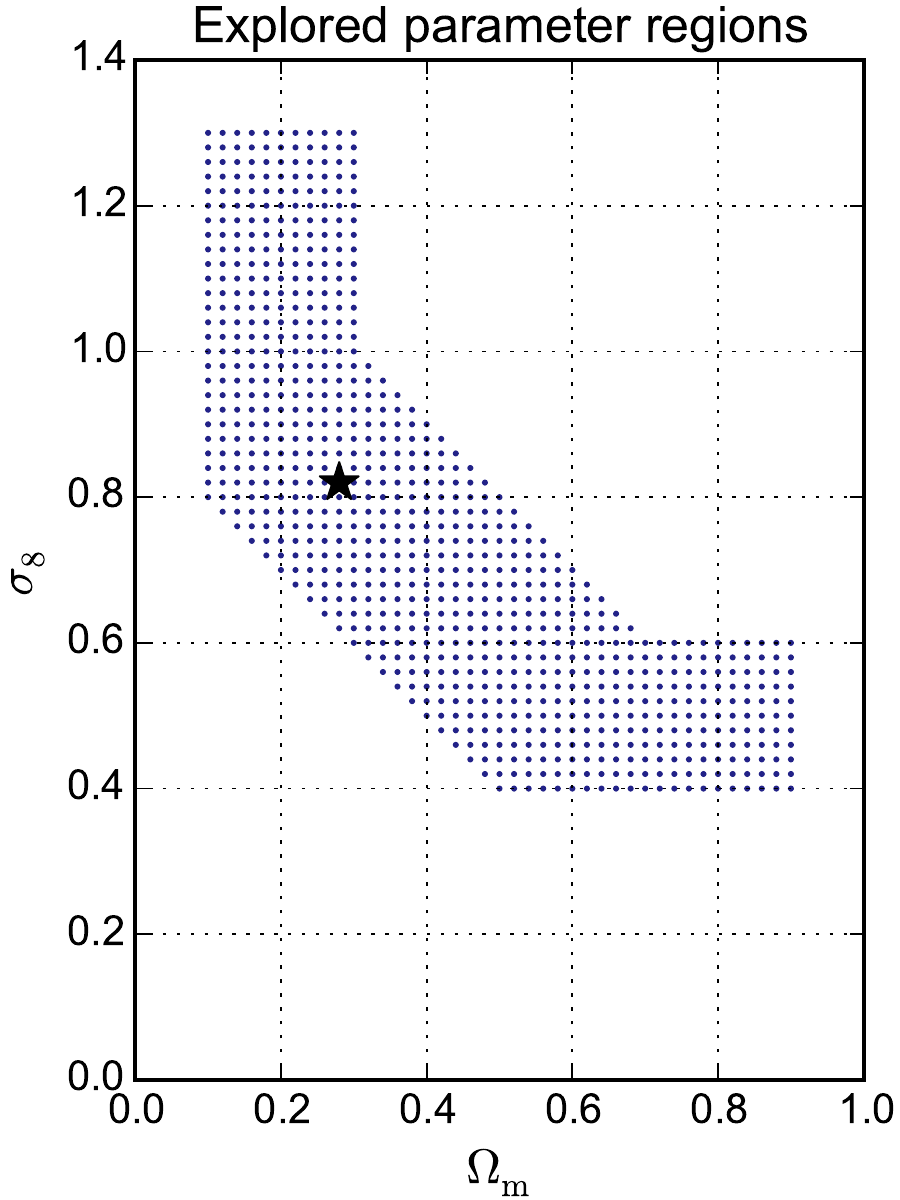}
	\caption{Distribution of evaluated parameter points on the $\OmegaM$-$\sigEig$ plane. This figure can be considered as a slice of points with the same $\wZero$. There are in total 46 slices of 816 points.}
	\label{fig:filtering:Explored_regions_2}
\end{figure}

\begin{figure}[tb]
	\centering
	\includegraphics[width=\textwidth]{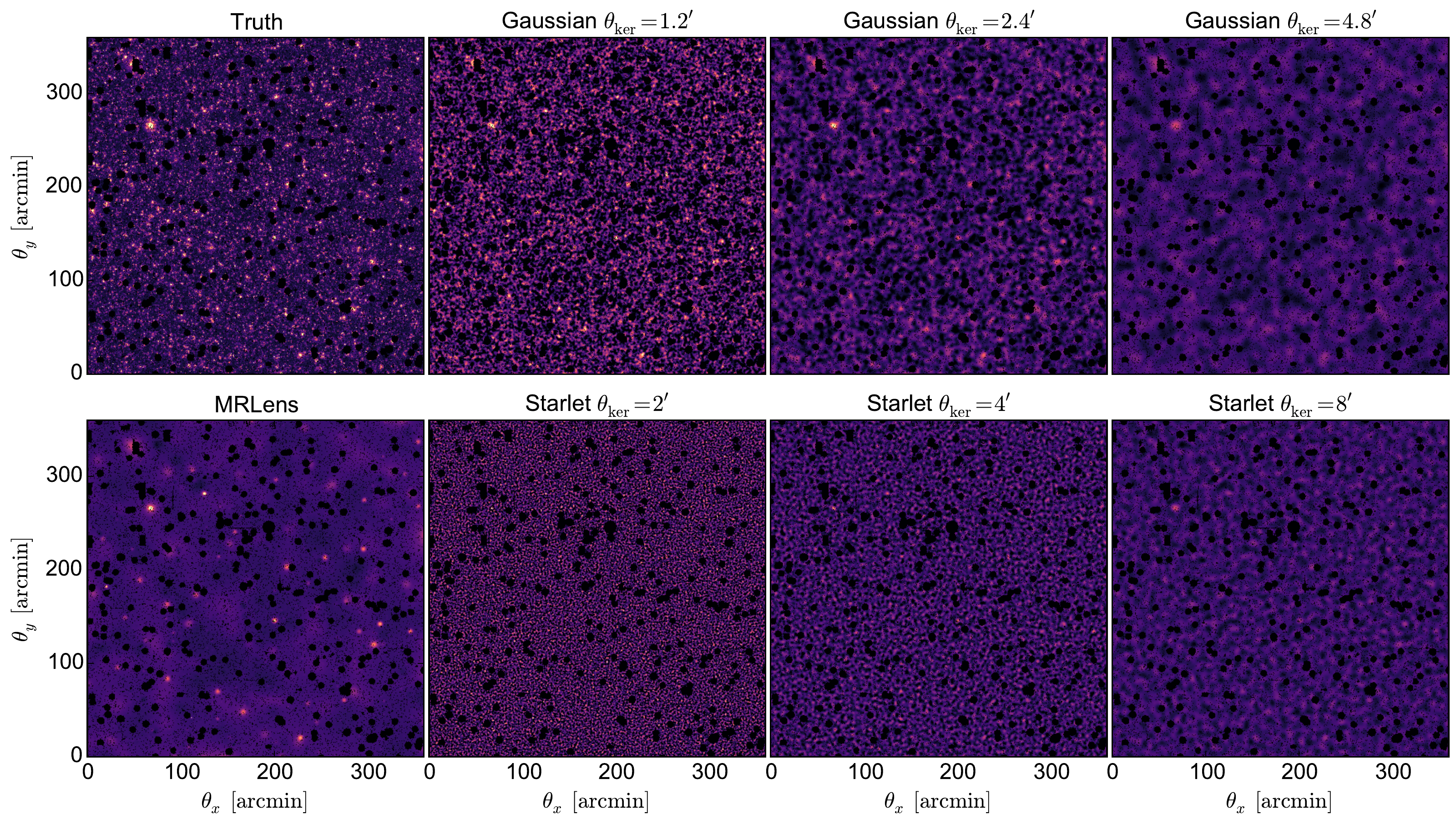}
	\caption{Maps taken from one of the simulations. The truth map is made by calculating $\kappa_\proj$ without noise. The panels of the rest are different filtering techniques applied on the map obtained from a \acro{KS} inversion after calculating $\epsilon\upp{\rmo}=(\epsilon\src+g_\proj)/(1+g_\proj^*\epsilon\src)$. The black areas are masks. The unit of kernel sizes is arcmin.}
	\label{fig:filtering:allFilters}
\end{figure}

In the likelihood analysis, the copula likelihood (Eq. \ref{for:constraint:L_vc}) is evaluated on a grid. The range of $\wZero$ is [-1.8, 0], with $\Delta\wZero=0.04$. Concerning $\OmegaM$ and $\sigEig$, only some particular values are chosen for evaluation in order to reduce the computing costs. This results in 816 points in the $\OmegaM$-$\sigEig$ plane, as displayed in \fig{fig:filtering:Explored_regions_2}, and the total number of parameter sets is 37536. For each parameter set, $N=400$ realizations of our model are performed. Each realization produces data vectors for the Gaussian kernel, the starlet kernel, and \MRLens, so that the comparisons between cases are based on the same stochasticity. The aperture mass is not included here because of the time consuming convolution of the unbinned shear catalogue with the filter $Q$. A map example is displayed in \fig{fig:filtering:allFilters} for the three cases and the input simulated $\kappa$ field.

\begin{table}[t]
	\centering
	\begin{tabular}{clcc}
		\hline\hline
		Filter          & $\theta_{\ker}$ [arcmin] or $\alpha$ & Number of bins  & $d$\\
		\hline
		Gaussian        & $\theta_{\ker}=$ 1.2, 2.4, 4.8       & 9 $\nu$ bins    & 27\\
		Starlet         & $\theta_{\ker}=$ 2, 4, 8             & 9 $\nu$ bins    & 27\\
		$M_\ap$ $\tanh$ & $\theta_{\ker}=$ 2.125, 4.25, 8.5    & 9 $\nu$ bins    & 27\\
		\MRLens         & $\alpha=0.05$                        & 6 $\kappa$ bins & 6\\
		\hline\hline
	\end{tabular}
	\caption{Definition of the data vector $\bx$ for \acro{PMC} \acro{ABC} runs. The 9 bins of $\nu$ are [1, 1.5, 2, $\ldots$, 4, 4.5, 5, $+\infty$[, and the 6 bins of $\kappa$ are [0.02, 0.03, 0.04, 0.06, 0.10, 0.16, $+\infty$[. The symbol $d$ is the total dimension of $\bx$, and $\alpha$ stands for the input value of \acro{FDR} for \MRLens.}
	\label{tab:filtering:x_mod_ABC}
\end{table}

In the \acro{ABC} analysis, all four filters are used. For the three first linear cases, the data vector $\bx$ is composed of three scales. Each scale has the same nine \acro{S/N} bins as in the likelihood analysis, which results in 27 bins in total. For \MRLens, $\bx$ is a 6-bin $\kappa$ histogram, also identical to the likelihood analysis (\tab{tab:filtering:x_mod_ABC}).

Concerning the \acro{ABC} parameters, we use 1500 particles in the \acro{PMC} process. The iteration stops when the success ratio of accept-reject processes falls below 1\%. Finally, we test two distances. Between the sampled data vector $\bx$ and the observed one, $\bx^\obs$, we consider a simplified distance $D_1$ and a fully correlated one $D_2$, which are respectively defined as
\begin{align}
	D_1\left(\bx, \vect{y}\right) &\equiv \sqrt{\sum_i\frac{\left(x_i-y_i\right)^2}{C_{ii}}}, \label{for:filtering:D_1}\\
	D_2\left(\bx, \vect{y}\right) &\equiv \sqrt{\left(\bx-\vect{y}\right)^T \bC\inv\left(\bx-\vect{y}\right)}, \label{for:filtering:D_2}
\end{align}
where $C_{ii}$ and $\bC\inv$ are now independent from cosmology, estimated under $(\OmegaM, \sigEig, \wZero)=(0.28, 0.82, -0.96)$.

\section{Results}

We again use the uncertainty on $\Sigma_8$ and the \acro{FoM} on the $\OmegaM$-$\sigEig$ plane to qualify constraints. Here, $\Sigma_8$ is defined differently:\index{Constraint indicator}
\begin{align}\label{for:filtering:CSE}
	\Sigma_8 \equiv \left(\frac{\OmegaM+\beta}{1-\alpha}\right)^{1-\alpha} \left(\frac{\sigma_8}{\alpha}\right)^{\alpha}.
\end{align}
The motivation for this definition is to measure the ``contour width'' independently from $\alpha$. With the common definition $\Sigma_8 \equiv \sigma_8(\OmegaM/\text{pivot})^\alpha$, the uncertainty on $\Sigma_8$ under different $\alpha$ does not correspond to the same measure. The 1-$\sigma$ error bar on $\Sigma_8$, $\DCSE$, is obtained using the same method as in \chap{sect:constraint}.

\subsection{With the copula likelihood}

First, we test the maximum information that Gaussian kernels can extract. \tab{tab:filtering:FoM_likelihood} shows the \acro{FoM} from the marginalized likelihood. We can see that adding $\theta_{\ker}=$ 2.4 and 4.8 arcmin to the filter with 1.2 arcmin has no significant effect on constraints. The constraints from the smallest filter are the most dominant ones among all.

\begin{figure}[tb]
	\centering
	\includegraphics[width=0.49\textwidth]{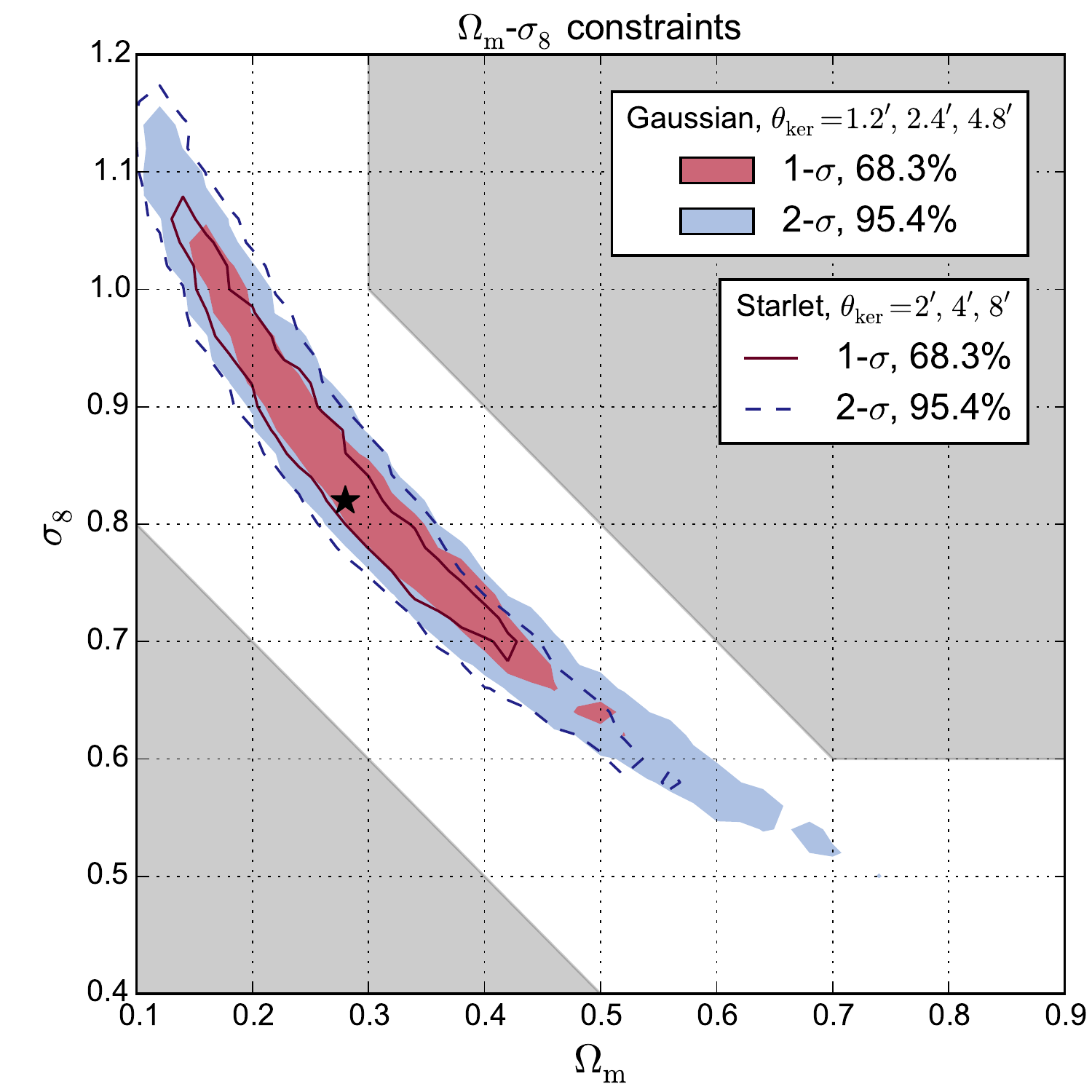}\hfill
	\includegraphics[width=0.49\textwidth]{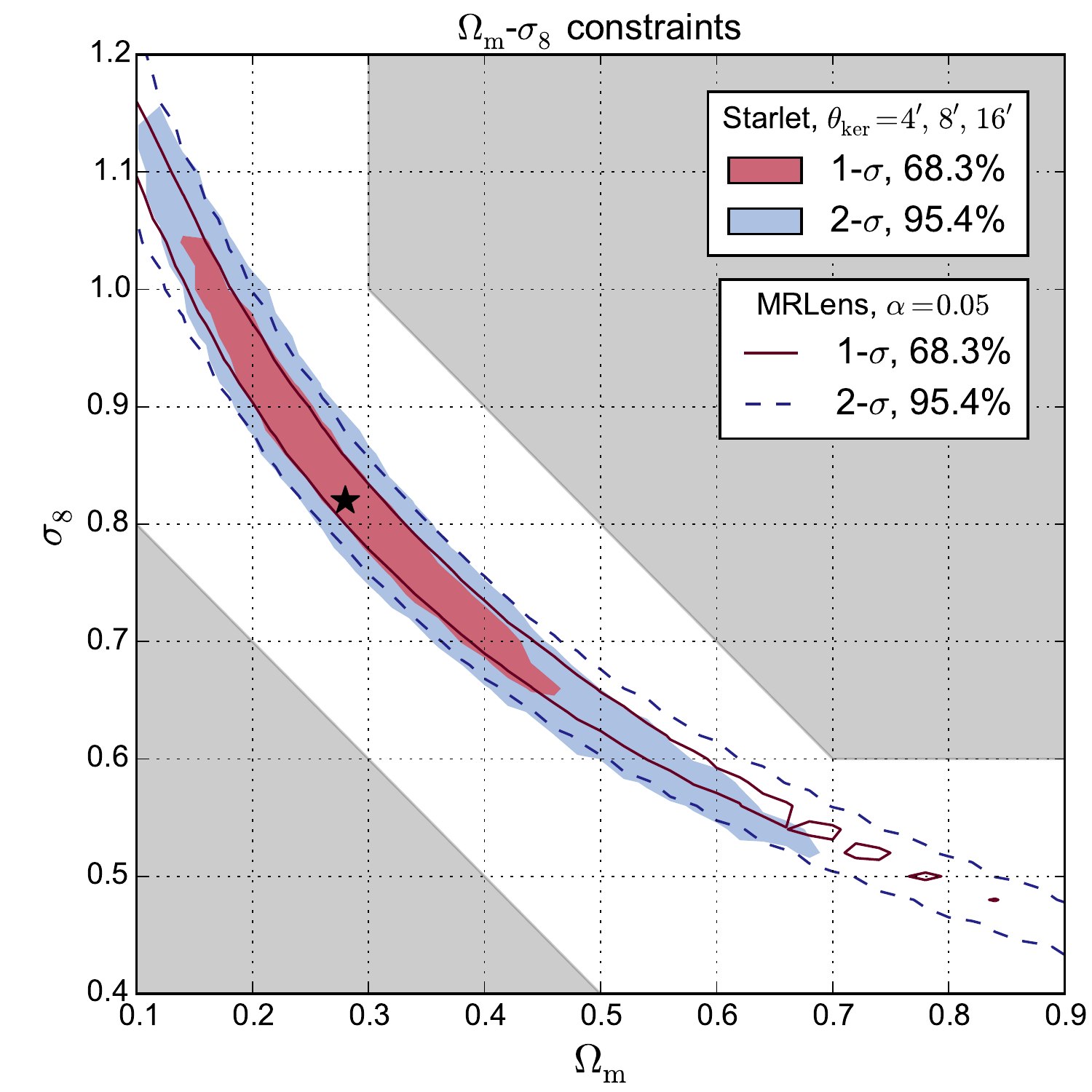}
	\caption{$\OmegaM$-$\sigEig$ constraints from four different cases. Left panel: the Gaussian case (colored regions) and the starlet case with three corresponding scales based on the noise level (solid and dashed contours, $\theta_{\ker}=$ 2, 4, and 8 arcmin). Right panel: the starlet case with three corresponding scales based on number counts (colored regions, $\theta_{\ker}=$ 4, 8, and 16 arcmin) and the \MRLens\ case (solid and dashed contours). The Gaussian and count-based starlet cases yield almost identical constraints. Between four cases, the best result is given by the noise-based starlet case. Black stars represent the input cosmology. Grey zones are excluded in this analysis.}
	\label{fig:filtering:contour_comp_fixedAxis2_star1}
\end{figure}

Next, we use all three Gaussian scales as the reference for the comparisons with the starlet function. As mentioned in \sect{sect:filtering:methodology:comparisons}, for the Gaussian filter scales of 1.2, 2.4, and 4.8 arcmin, we chose scales for the starlet based on two criteria: for an equal noise level, these are 2, 4, and 8 arcmin, and for equal number counts the corresponding scales are 4, 8, and 16 arcmin. The results are shown in \fig{fig:filtering:contour_comp_fixedAxis2_star1}. For the equal-number-count criterion, we see that if each scale gives approximately the same number of peaks, the $\OmegaM$-$\sigEig$ constraints obtained from the Gaussian and the starlet are similar (colored regions in the left and right panels). However, the starlet kernel leads to tighter constraints than the Gaussian when we match the same noise levels (lines and colored regions in the left panel). This results suggests that compensated kernels could be more powerful to extract cosmological information than non-compensated filters.

\begin{figure}[tb]
	\centering
	\includegraphics[width=0.49\textwidth]{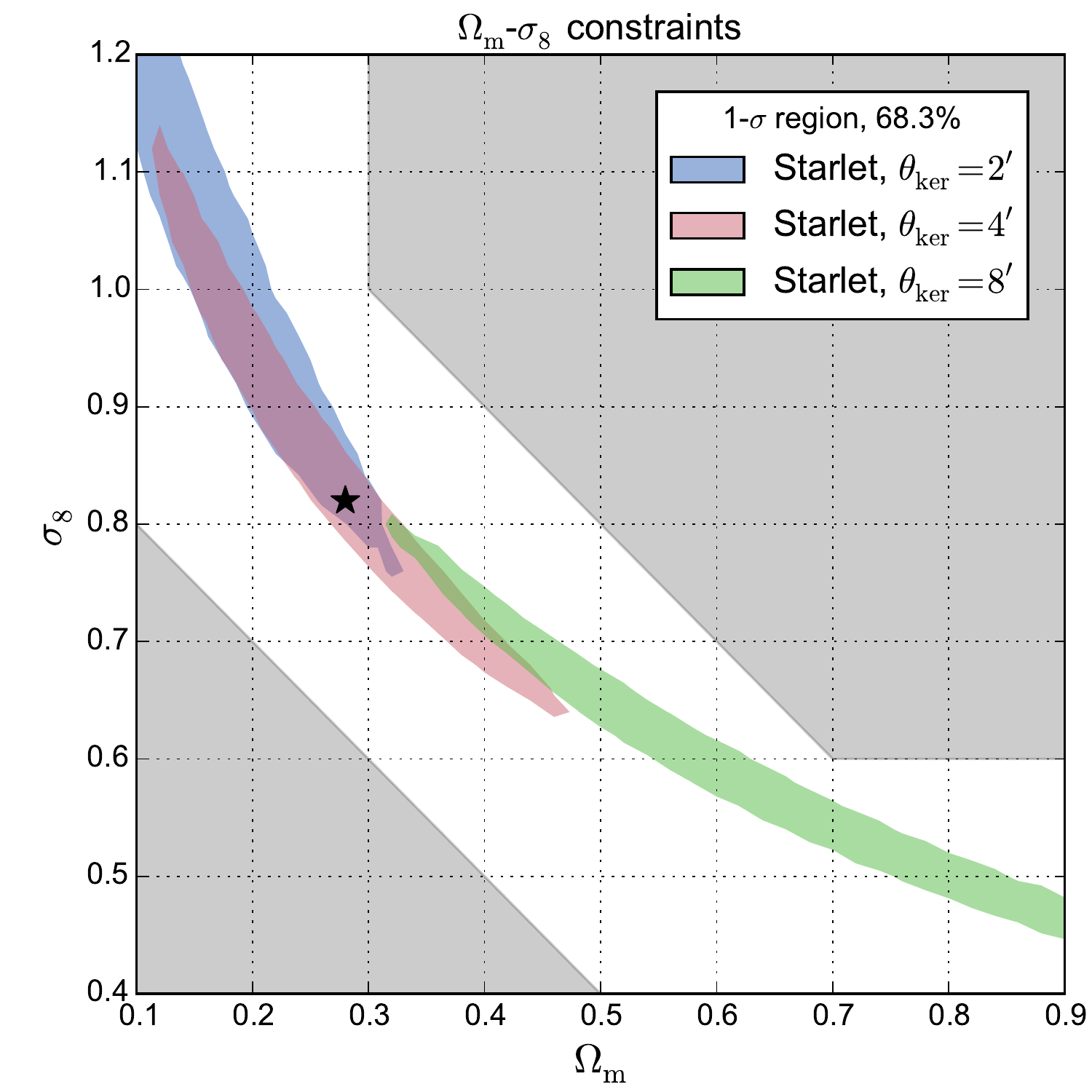}\hfill
	\includegraphics[width=0.49\textwidth]{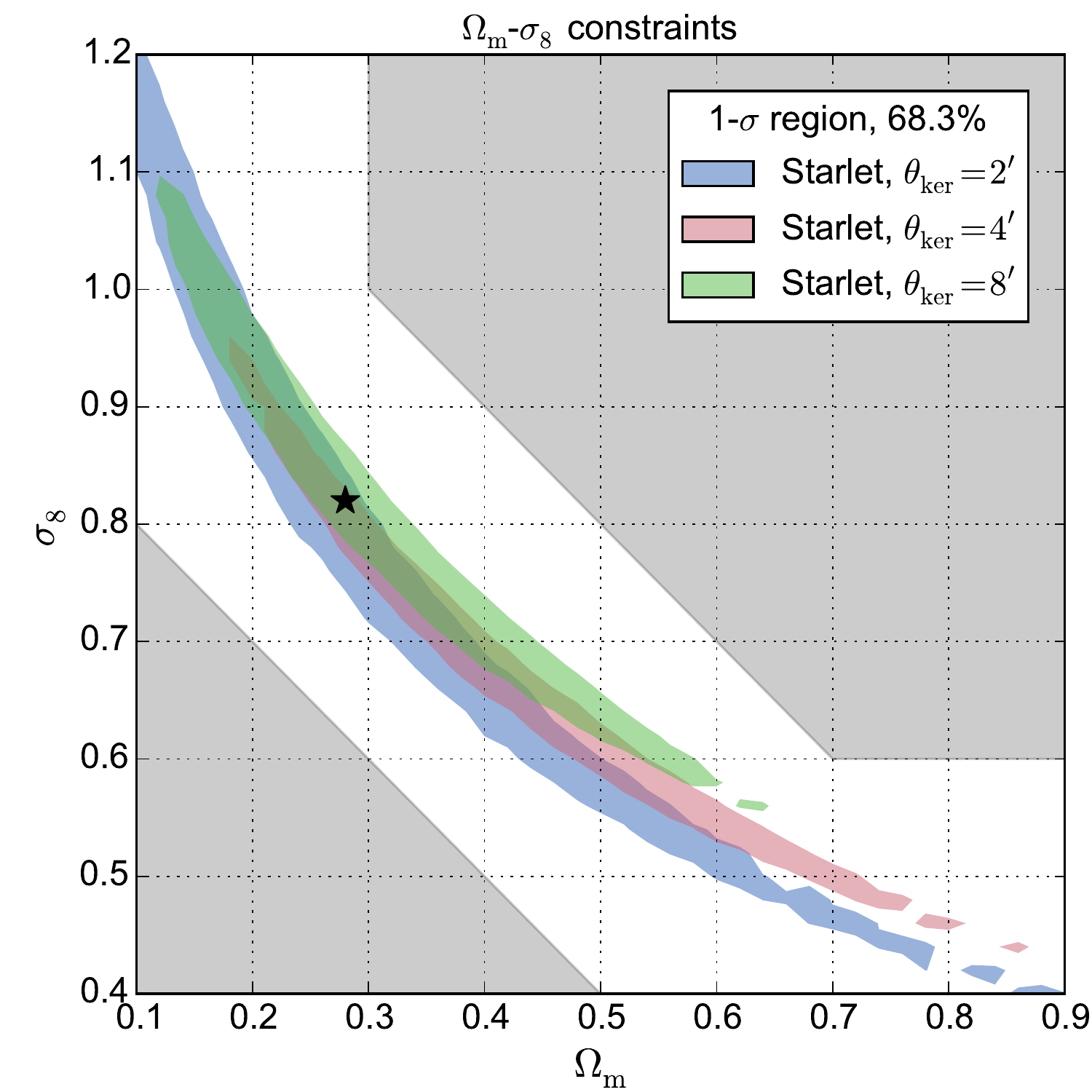}
	\caption{$\OmegaM$-$\sigEig$ 1-$\sigma$ region from individual scales of the starlet kernel. The plotted scales are 2 (blue), 4 (red), and 8 arcmin (green). The left panel is the result from the observation vector used in the analysis; the right panel represents the constraints from another realization of the observation. We see that the behaviors between different scales is more likely due to stochasticity. Grey zones are excluded in this analysis.}
	\label{fig:filtering:contour_comp_fixedAxis2_star3}
\end{figure}

We also draw constraints from individual scales of the starlet filter in (\fig{fig:filtering:contour_comp_fixedAxis2_star3}, left panel). It shows a very different behavior and seems to suggest that different scales could be sensitive to different cosmologies. However, this is actually a stochastic effect. We verify this statement by redoing the constraints with other observation vectors. It turns out that the scale-dependent tendency disappears (\fig{fig:filtering:contour_comp_fixedAxis2_star3}, right panel). Nevertheless, when different cosmologies are preferred by different scales, the effect is less pronounced for the Gaussian filter. This is likely due to the fact that the starlet is a compensated filter, which is a band-pass function in the Fourier space. Since different filtering scales could be sensitive to different mass ranges of the mass function, band-pass filters weaken the correlations between scales (\fig{fig:filtering:corr_skip2}) and separate better the multiscale information. The stochasticity of the observation vector suggests that the simulated field of view is rather small. While this should not affect the filter comparison nor contour sizes, actual cosmological constraints seem to require substantially larger data sets.

The right panel of \fig{fig:filtering:contour_comp_fixedAxis2_star1} shows the constraints from nonlinear filtering using \MRLens\ (solid and dashed lines). We observe that \MRLens\ conserves a strong degeneracy between $\OmegaM$ and $\sigEig$. The reasons for this result are various. First, large and small scales tend to be sensitive to different halo masses which could help break the degeneracy. Using the combined strategy loses this advantage. Second, a strict \acro{FDR} has been chosen. This rules out most of the spurious peaks, but also a lot of the signal. Third, as mentioned before, it is inappropriate to define signal-to-noise ratio when the filter is not linear. As a consequence, it is hardly possible to find bins for $\kappa$ peaks which are equivalent to $\nu$ bins in linear filtering. This is supported by \fig{fig:filtering:allFilters}, where we observe less peaks in the \MRLens\ map than in the other maps. Last, because of a low number of peaks, the binwidths need to be enlarged to contain larger number counts and to get closer to a Gaussian distribution, and large binwidths also weaken the signal.

A possible solution for exploring the \MRLens\ technique is to enhance the \acro{FDR} and to redesign the binning. By increasing the number of peaks, thinner bins would be allowed. Another solution to better account for rare events in the current configuration is to use the Poisson likelihood. Finally, one could adopt the separated strategy, i.e. turning back to the methodology used by \citet{Pires_etal_2009a} that consists in estimating the peak abundance in the different scales before final reconstruction. In this study, the comparison between ``linear and nonlinear techniques'' is basically the one between the ``separated and combined strategies''.

\begin{table}[tb]
	\centering
	\begin{tabular}{clcc}
		\hline\hline
		Filter   & $\theta_{\ker}$ [arcmin] or $\alpha$  & $\DCSE$ & \acro{FoM}\\
		\hline
		Gaussian & $\theta_{\ker}=$ 1.2                 & 0.045   & 19.1\\
		Gaussian & $\theta_{\ker}=$ 1.2, 2.4, 4.8       & 0.046   & 20.7\\
		Starlet  & $\theta_{\ker}=$ 2, 4, 8             & 0.046   & 23.4\\
		Starlet  & $\theta_{\ker}=$ 4, 8, 16            & 0.044   & 21.2\\
		Starlet  & $\theta_{\ker}=$ 2, 4, 8, 12, 16     & 0.045   & 24.8\\
		\MRLens  & $\alpha=$ 0.05                       & 0.046   & 16.2\\
		\hline\hline
	\end{tabular}
	\caption{Quality indicators for $\OmegaM$-$\sigEig$ constraints with likelihood. All cases figured below use number counts on $g$ peaks. The quantity $\DCSE$ stands for the width of the contour, while the \acro{FoM} is related to the area. In our study, combining five scales of starlet yield the best result in terms of \acro{FoM}.}
	\label{tab:filtering:FoM_likelihood}
\end{table}

\tab{tab:filtering:FoM_likelihood} measures numerical qualities for constraints with different filtering techniques. It indicates that the width of contours does not vary significantly. The tightest constraint that we obtain is derived from a compensated filter.

\begin{figure}[tb]
	\centering
	\includegraphics[width=0.49\textwidth]{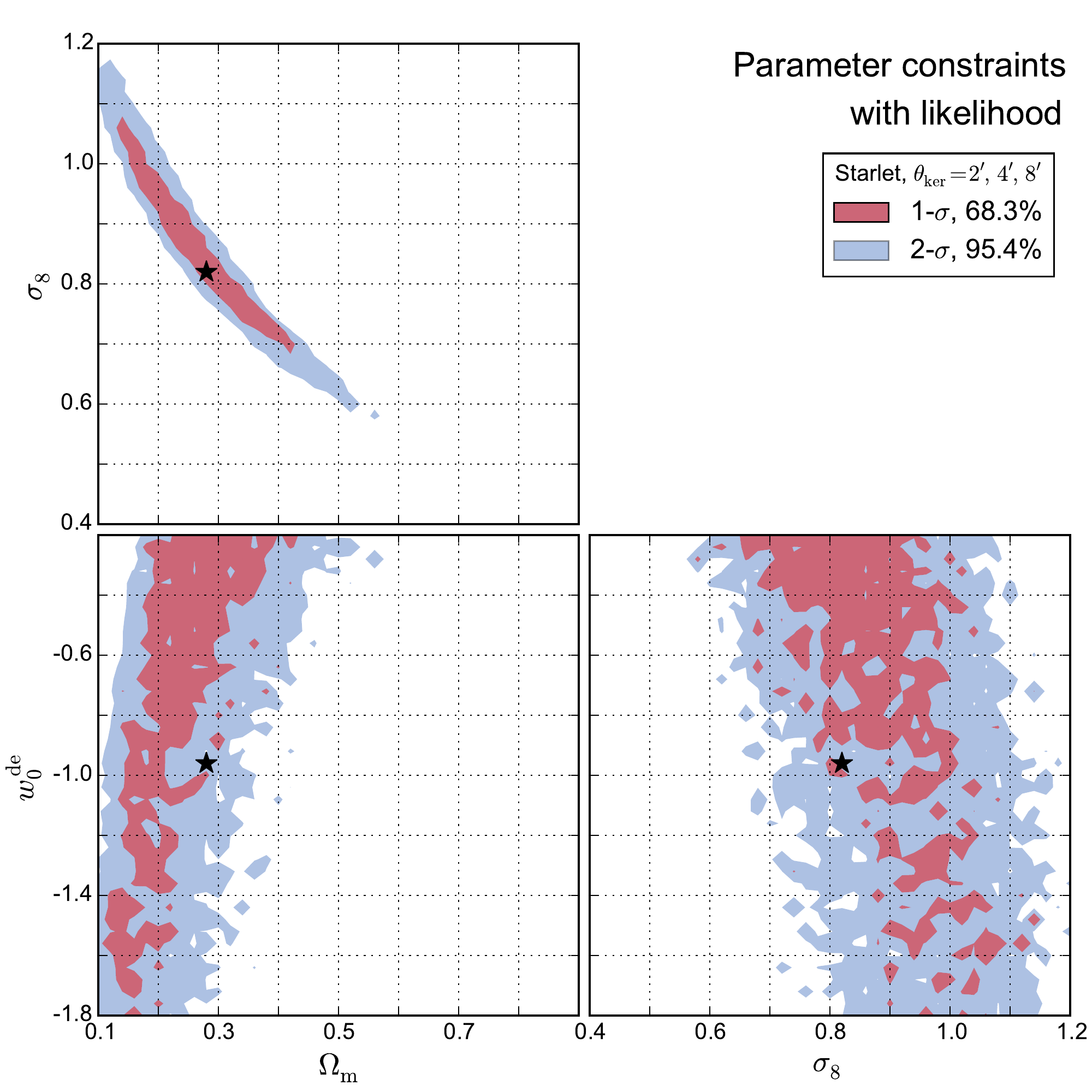}\hfill
	\includegraphics[width=0.49\textwidth]{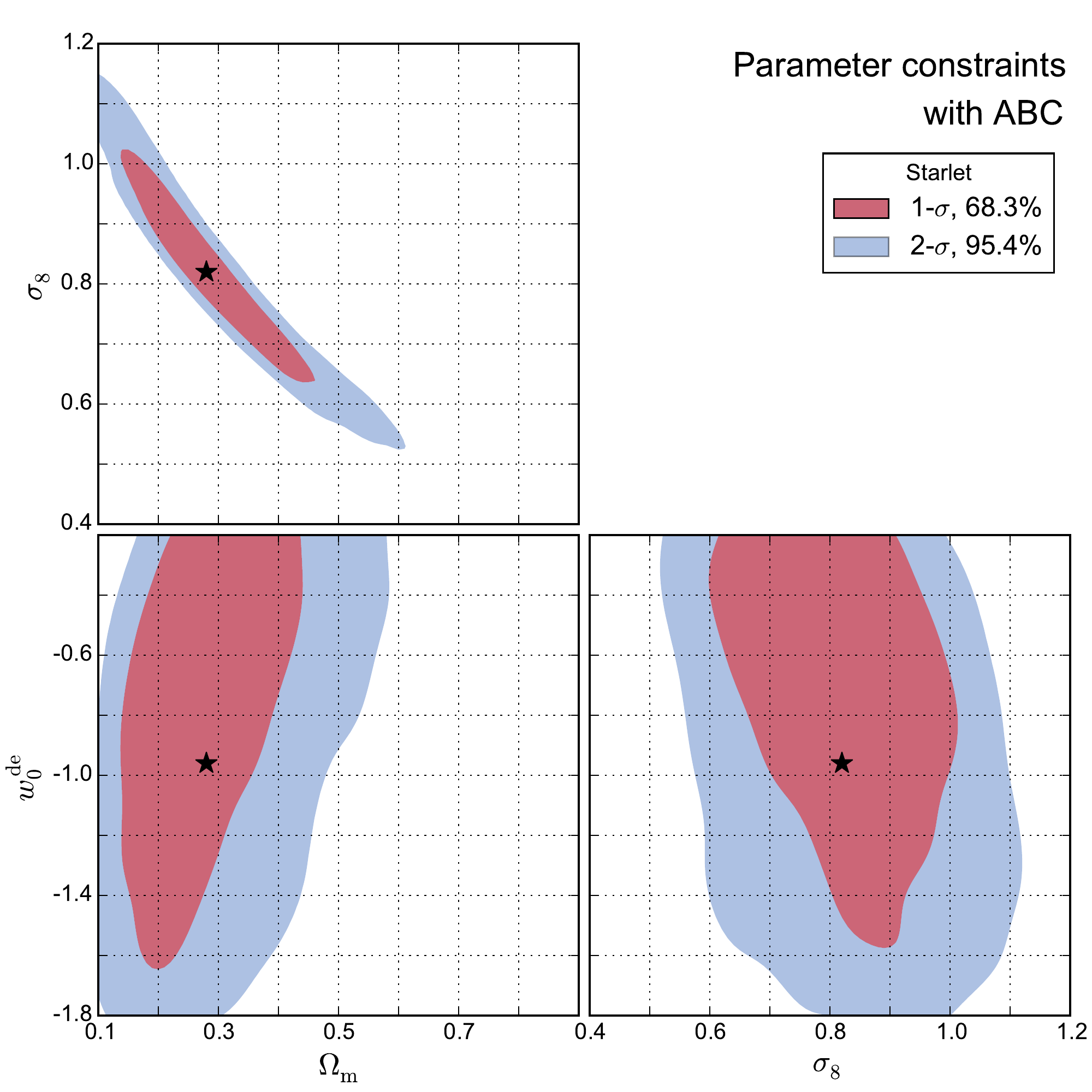}
	\caption{$\OmegaM$-$\sigEig$-$\wZero$ constraints using starlet with three scales. Left panel: constraints with the likelihood. Fluctuations on both lower panels are due to usage of the copula likelihood. Right panel: constraints with \acro{ABC} using the distance $D_2$. Each panel represents the marginalization over one of the three parameters. Black stars are the input cosmology. As far as $\wZero$ is concerned, the constraints are weak, but the degeneracies are clear.}
	\label{fig:filtering:contour_stair_star}
\end{figure}

Regarding results for $\wZero$, we show a representative case of starlet with $\theta_{\ker}=$ 2, 4, and 8 arcmin. The left panel of \fig{fig:filtering:contour_stair_star} presents the marginalized constraints of each doublet of parameters that we study. Those containing $\wZero$ are noisy because of the usage of the copula likelihood. We may see that the current configuration of our model does not allow to impose constraints on $\wZero$. To measure this parameter, it could be useful to perform a tomography analysis to separate information of different stages of the late-time Universe. Nevertheless, our results successfully highlight the degeneracies of $\wZero$ with two other parameters. We fit the posterior density with:
\begin{align}
	I_1 &= \OmegaM - a_1 \wZero,\\ 
	I_2 &= \sigEig + a_2 \wZero. 
\end{align}
We obtain for the slopes $a_1=0.108$ and $a_2=0.128$ for the left panel of \fig{fig:filtering:contour_stair_star}. The results for the other filter functions are similar.

\subsection{With PMC ABC}

We perform parameter constraints using the \acro{PMC} \acro{ABC} algorithm for our four cases. In the right panel of \fig{fig:filtering:contour_stair_star}, we show the results derived from the starlet case using the fully correlated distance $D_2$. The contours are marginalized posteriors for all three pairs of parameters. They show the same degeneracy as we have found with the likelihood. We measure $a_1$ and $a_2$ from the \acro{ABC} posteriors and obtain $a_1=0.083$ and $a_2=0.084$.

\begin{table}[tb]
	\centering
	\begin{tabular}{cccc}
		\hline\hline
		Filter       & Constraints & $\DCSE$ & \acro{FoM}\\
		\hline
		Gaussian     & Likelihood  & 0.046   & 20.7\\
		Gaussian     & ABC, $D_1$  & 0.043   & 16.3\\
		Gaussian     & ABC, $D_2$  & 0.059   & 11.7\\
		Starlet      & Likelihood  & 0.054   & 23.4\\
		Starlet      & ABC, $D_1$  & 0.050   & 15.5\\
		Starlet      & ABC, $D_2$  & 0.054   & 15.7\\
		$M_\ap$ tanh & ABC, $D_1$  & 0.037   & 19.4\\
		$M_\ap$ tanh & ABC, $D_2$  & 0.043   & 15.5\\
		\MRLens      & Likelihood  & 0.046   & 16.2\\
		\MRLens      & ABC, $D_1$  & 0.045   & 11.5\\
		\MRLens      & ABC, $D_2$  & 0.045   & 12.5\\
		\hline\hline
	\end{tabular}
	\caption{Quality indicators for $\OmegaM$-$\sigEig$ constraints with \acro{PMC} \acro{ABC}. The quantity $\DCSE$ stands for the width of the contour, while the \acro{FoM} is related to the area. \acro{ABC} is used with two different distances $D_1$ and $D_2$ respectively given by Eqs. \eqref{for:filtering:D_1} and \eqref{for:filtering:D_2}. Here, we also put values from likelihood constraints using the same scales in this table for comparison. The kernel sizes for linear methods are defined in \tab{tab:filtering:x_mod_ABC}.}
	\label{tab:filtering:FoM_ABC}
\end{table}

\begin{figure}[tb]
	\centering
	\includegraphics[width=0.9\textwidth]{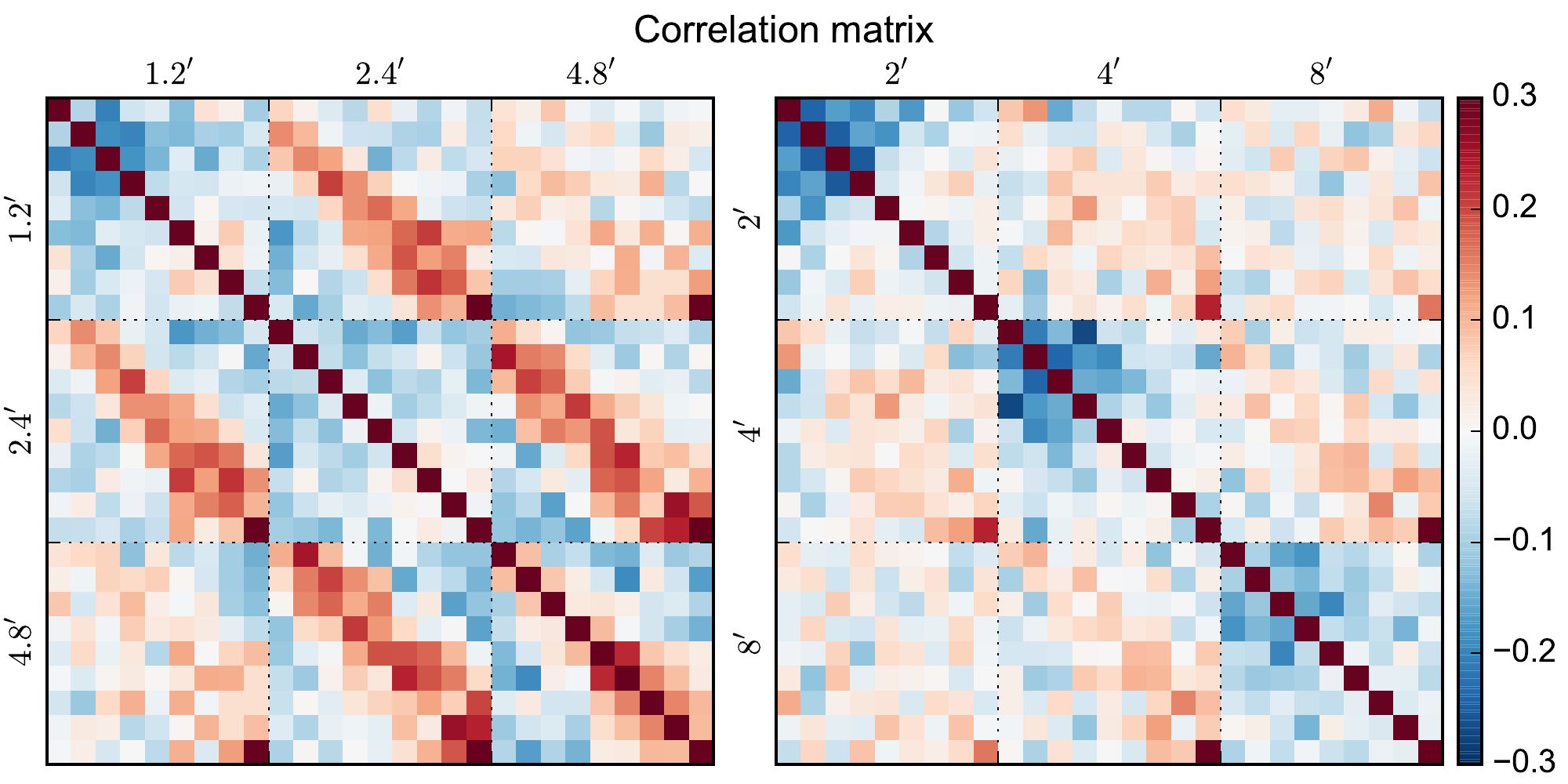}
	\caption{Correlation coefficient matrices under the input cosmology. Left panel: the Gaussian case with $\theta_{\ker}=1.2$, 2.4, and 4.8 arcmin. Right panel: the starlet case with $\theta_{\ker}=2$, 4, and 8 arcmin. Each of the 3$\times$3 blocks corresponds to the correlations between two filter scales. With each block, the \acro{S/N} bins are [1, 1.5, 2, $\ldots$, 5, +$\infty$[. The data vector by starlet is less correlated.}
	\label{fig:filtering:corr_skip2}
\end{figure}

Using the same starlet filters, we compare two distances used for \acro{PMC} \acro{ABC} runs. When $D_1$ is used with the starlet, i.e. data are treated as if uncorrelated, we find that the contour sizes do not change (see \tab{tab:filtering:FoM_ABC}) compared to $D_2$. For the Gaussian case, however, constraints from $D_1$ are tighter than those from $D_2$. This phenomenon is due to the off-diagonal elements of the covariance matrix. For non-compensated filters, the cross-correlations between bins are much stronger, as shown in \fig{fig:filtering:corr_skip2}. If these cross-correlations are ignored, the repeated peak counts in different bins are not properly accounted for. This overestimates the additional sensitivity to massive structures, and therefore produces overly tight constraints. As shown in \fig{fig:filtering:corr_skip2}, in the Gaussian case, adjacent filter scales show a 20--30\% correlation. The blurring of the off-diagonal stripes indicate a leakage to neighboring \acro{S/N} bins due to noise, and the fact that clusters produce \acro{WL} peaks with different \acro{S/N} for different scales. On the contrary, in the case of the starlet, except for the highest \acro{S/N} bin there are negligible correlations between different scales.

\begin{figure}[tb]
	\centering
	\includegraphics[width=0.55\textwidth]{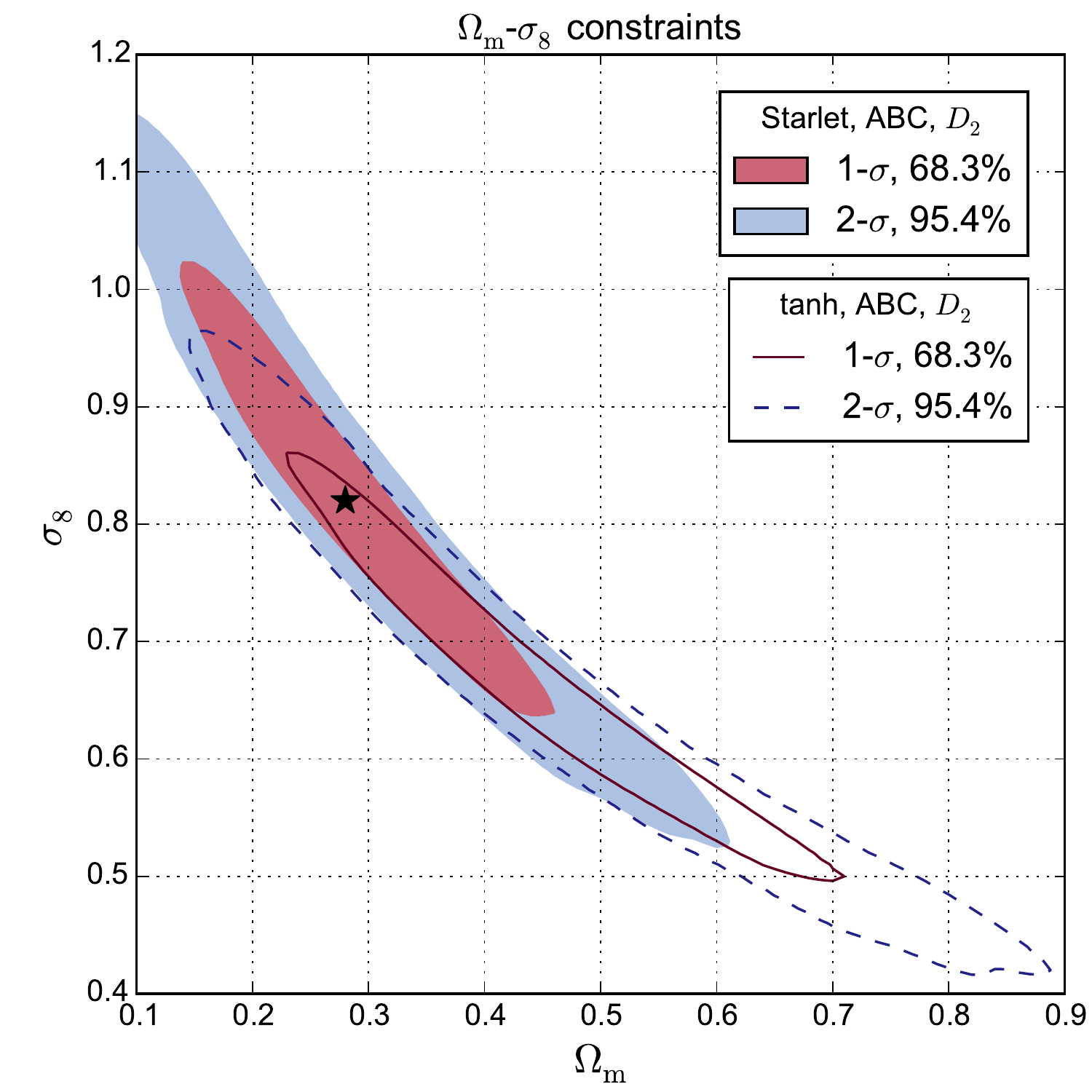}
	\caption{\acro{ABC} $\OmegaM$-$\sigEig$ constraints from the starlet and the aperture mass. The distance $D_2$ is used in both cases. The black star is the input cosmology. The difference between two cases is that another observation data vector is created for the aperture mass and the direct comparison is not valid anymore.}
	\label{fig:filtering:contour_ABC_star_vs_tanh}
\end{figure}

\tab{tab:filtering:FoM_ABC} shows the \acro{ABC} constraints from both the aperture mass and the starlet. We find that the \acro{FoM} are close. However, in \fig{fig:filtering:contour_ABC_star_vs_tanh}, we see that the contours from the aperture mass is shifted toward high-$\OmegaM$ regions. The explanation for this shift is once again the stochasticity. We simulated another observation data vector for $M_\ap$, and the maximum-likelihood point for different methods do not coincide.

From \tab{tab:filtering:FoM_ABC}, one can see that the difference between \MRLens\ and linear filters using \acro{ABC} is similar to using the likelihood. This suggests once again that the combined strategy leads to less tight constraints than the separated strategy. Note that we also try to adjust $\alpha$ and run \acro{PMC} \acro{ABC}. However, without modifying the $\kappa$ bin choice, the resulting constraints do not differ substantially from $\alpha=0.05$.

\begin{figure}[tb]
	\centering
	\includegraphics[width=0.49\textwidth]{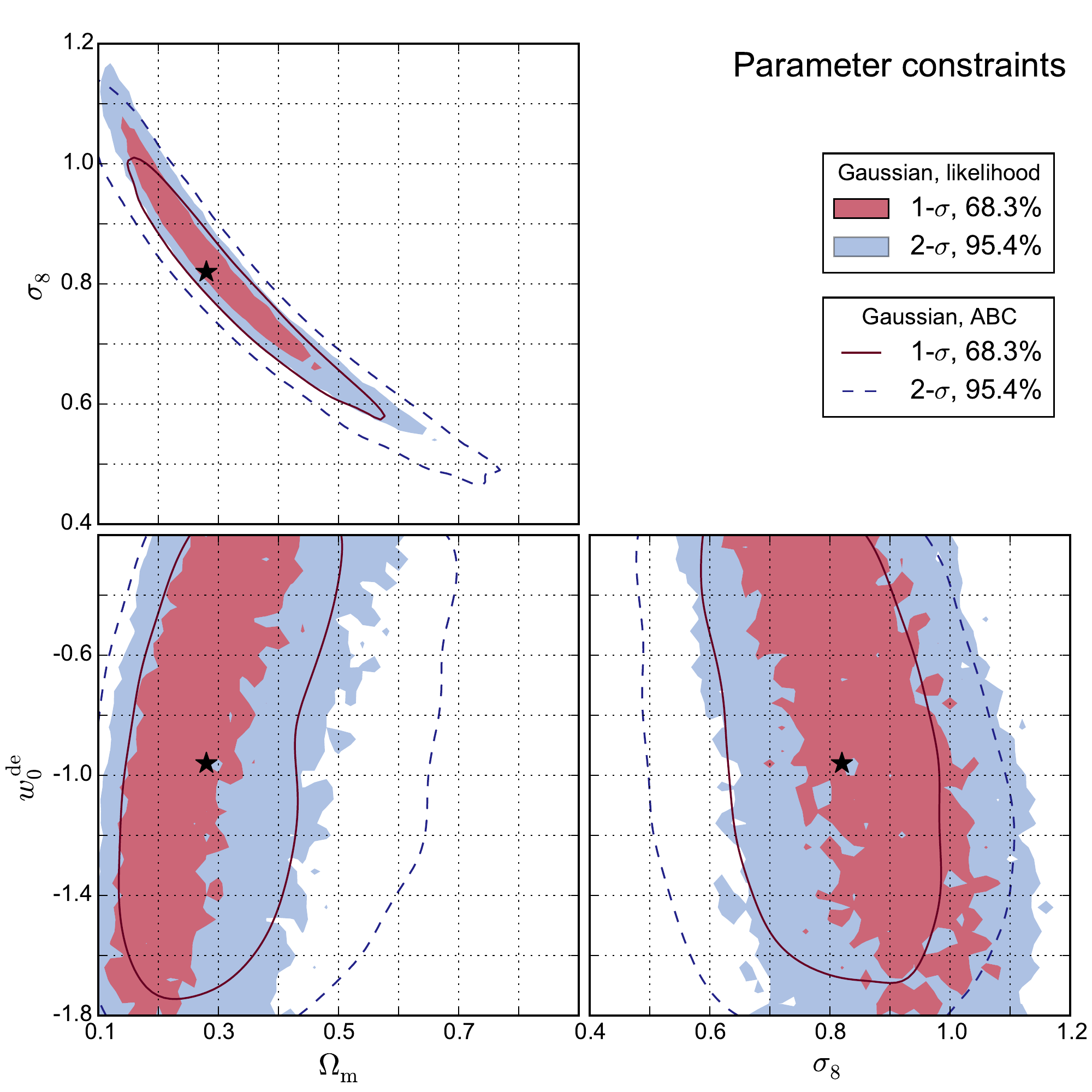}\hfill
	\includegraphics[width=0.49\textwidth]{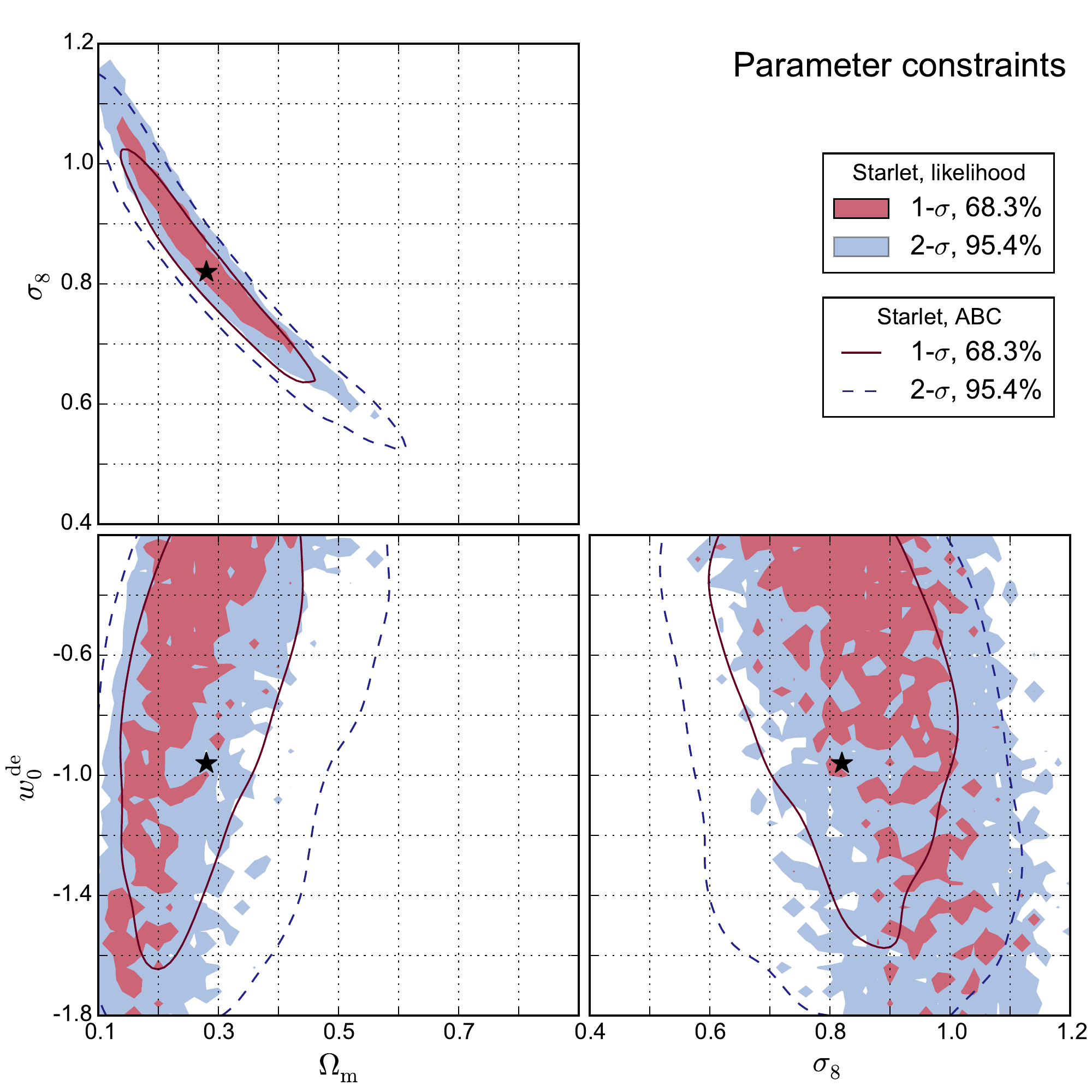}
	\caption{Comparison of $\OmegaM$-$\sigEig$-$\wZero$ constraints between likelihood and \acro{ABC}. Left panel: constraints with Gaussian smoothing. Right panel: constraints with starlet filtering. Although \acro{ABC} tolerates higher $\OmegaM$ and lower $\sigEig$ in both cases, two methods agree with each other.}
	\label{fig:filtering:contour_stair_comp_gauss}
\end{figure}

Finally, we show the likelihood and \acro{ABC} constraint contours for the Gaussian and starlet cases in \fig{fig:filtering:contour_stair_comp_gauss}. It turns out that \acro{ABC} contours are systematically larger in the high-$\OmegaM$, low-$\sigEig$ region. This phenomenon was not observed previously. We speculate that by including a third parameter $\wZero$ the contour becomes less precise, and \acro{ABC} might be more sensitive to this effect. Note also that \acro{KDE} is a biased estimator of posteriors. It smooths the posterior and makes contours broader. Nevertheless, the \acro{ABC} and likelihood constraints agree with each other. To be free from the bias, a possible alternative is to map the samples to a Gaussian distribution via some nonlinear mapping techniques \citep{Schuhmann_etal_2016}.

\subsubsection{Summary}

In this chapter, a new standard of comparisons between various filtering techniques has been presented. Comparing directly the constraint contours instead of purity and completeness of the cluster detection leads to a more direct measurement of cosmological information extraction.

The study shows a preference of compensated filters rather than non-compensated ones. Also, the separated strategy outperforms the combined strategy in terms of multiscale information extraction.

Constraints from the likelihood seem to be tighter than from \acro{ABC}. Since in \sect{sect:constraint:nonParam} we found that the copula likelihood closely approximates the true one, \acro{ABC} probably overestimate the true parameter errors.

Without tomography, $\wZero$ could not be constrained tightly. However, we observe that it is degenerate with $\OmegaM$ and $\sigEig$.

Our model has been improved to be adapted to more realistic observational conditions. We will now present a data application in the next chapter.

\clearpage
\thispagestyle{empty}
\cleardoublepage


\chapter{Data applications}
\label{sect:data}
\fancyhead[LE]{\sf \nouppercase{\leftmark}}
\fancyhead[RO]{\sf \nouppercase{\rightmark}}

\subsubsection{Overview}

In this chapter, we apply our analysis to three data sets. The cosmological constraints are performed using our peak-count model and \acro{ABC}. We will show preliminary results, possible improvements, and the perspectives of this work at the end.

\section{Data descriptions}

Three data sets are processed: the full data from the Canada-France-Hawaii Telescope Lensing Survey (\acro{CFHTLenS}), the data releases 1 and 2 from the Kilo-Degree Survey (\acro{KiDS} \acro{DR1/2}), and the science verification data from the Dark Energy Survey (\acro{DES} \acro{SV}).

\subsubsection{CFHTLenS}

The \acro{CFHTLenS} data \citep{Heymans_etal_2012, Erben_etal_2013} are composed of four separated fields W1, W2, W3, and W4. The total area is 154 deg$^2$. The total unmasked area is about 126 deg$^2$. Shape measurement was done with \textsc{Lensfit}, which is a model-fitting algorithm \citep{Miller_etal_2013}. Galaxies are selected with redshift, measurement weight, and mask flags, respectively as $0.2 <$ \texttt{Z\_B} $< 1.3$, $\mathtt{weight} > 0$, and $\mathtt{MASK} < 2$ \footnote{All science analyses of the \acro{CFHTLenS} team are performed with $\mathtt{MASK} < 2$. These objects can be safely used for most scientific purposes. \citep{Erben_etal_2013}}. This results in 6.1 million galaxies. The raw density is 13.53 arcmin$\invSq$.

\begin{figure}[!t]
	\centering
	\includegraphics[width=0.49\textwidth]{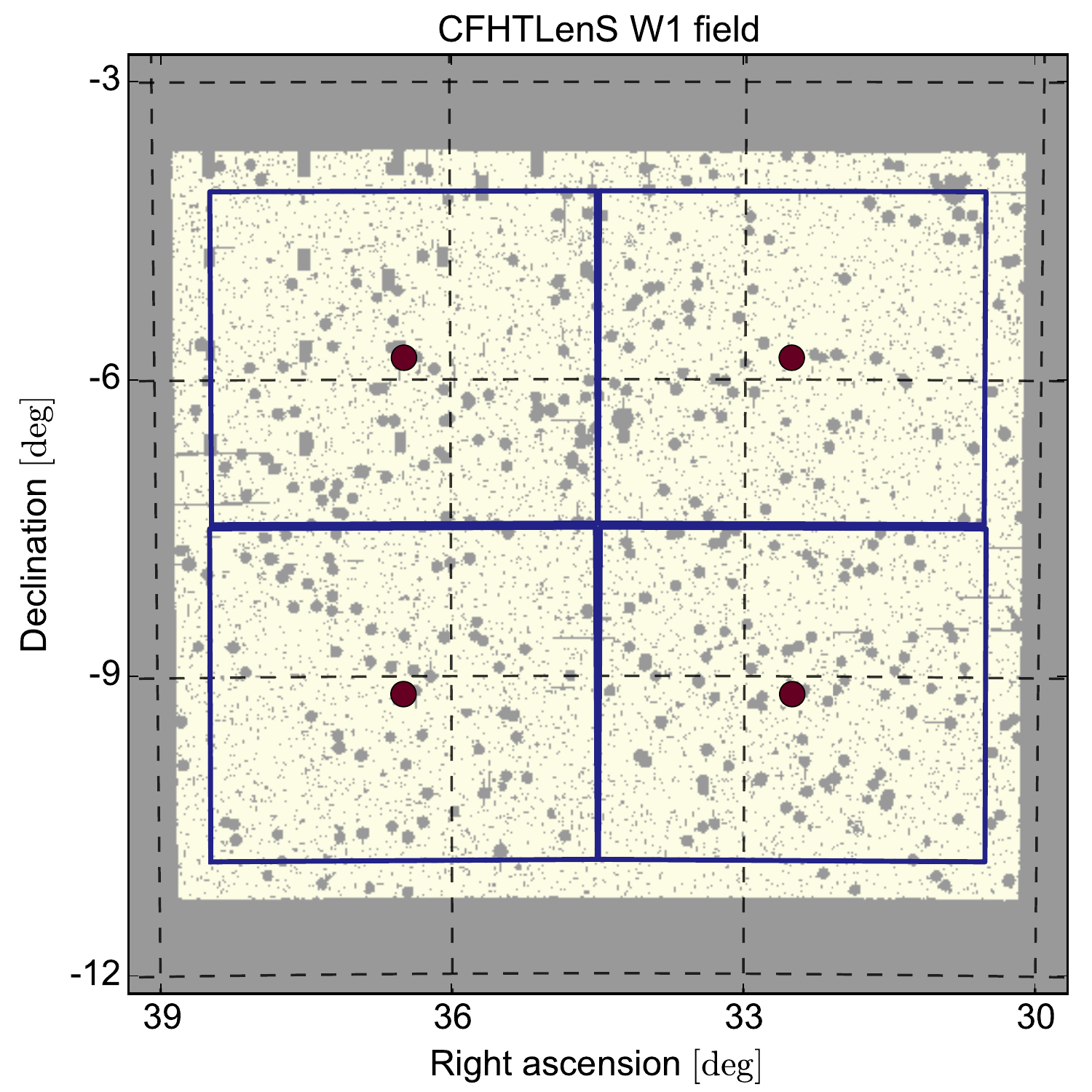}\hfill
	\includegraphics[width=0.49\textwidth]{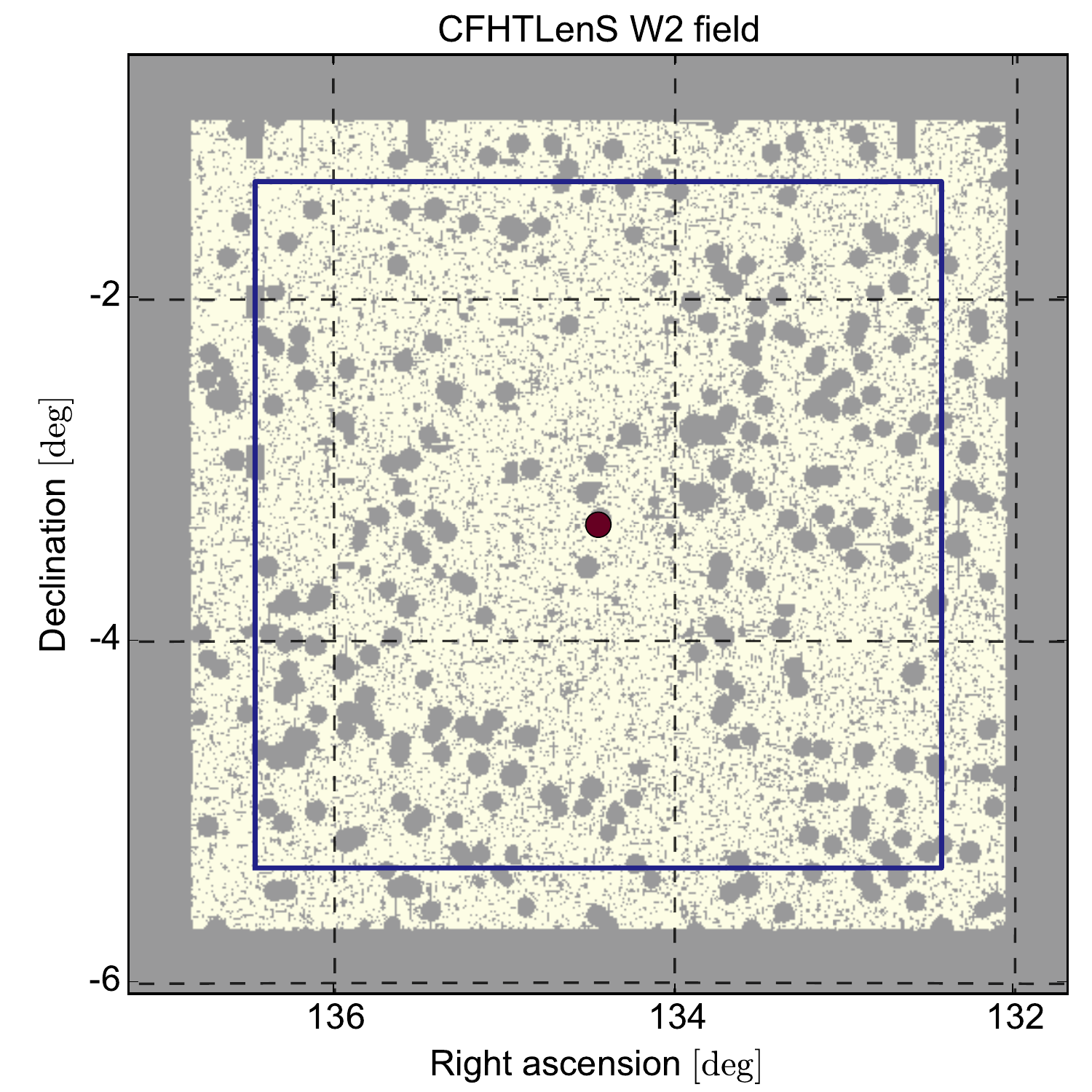}\\
	\includegraphics[width=0.49\textwidth]{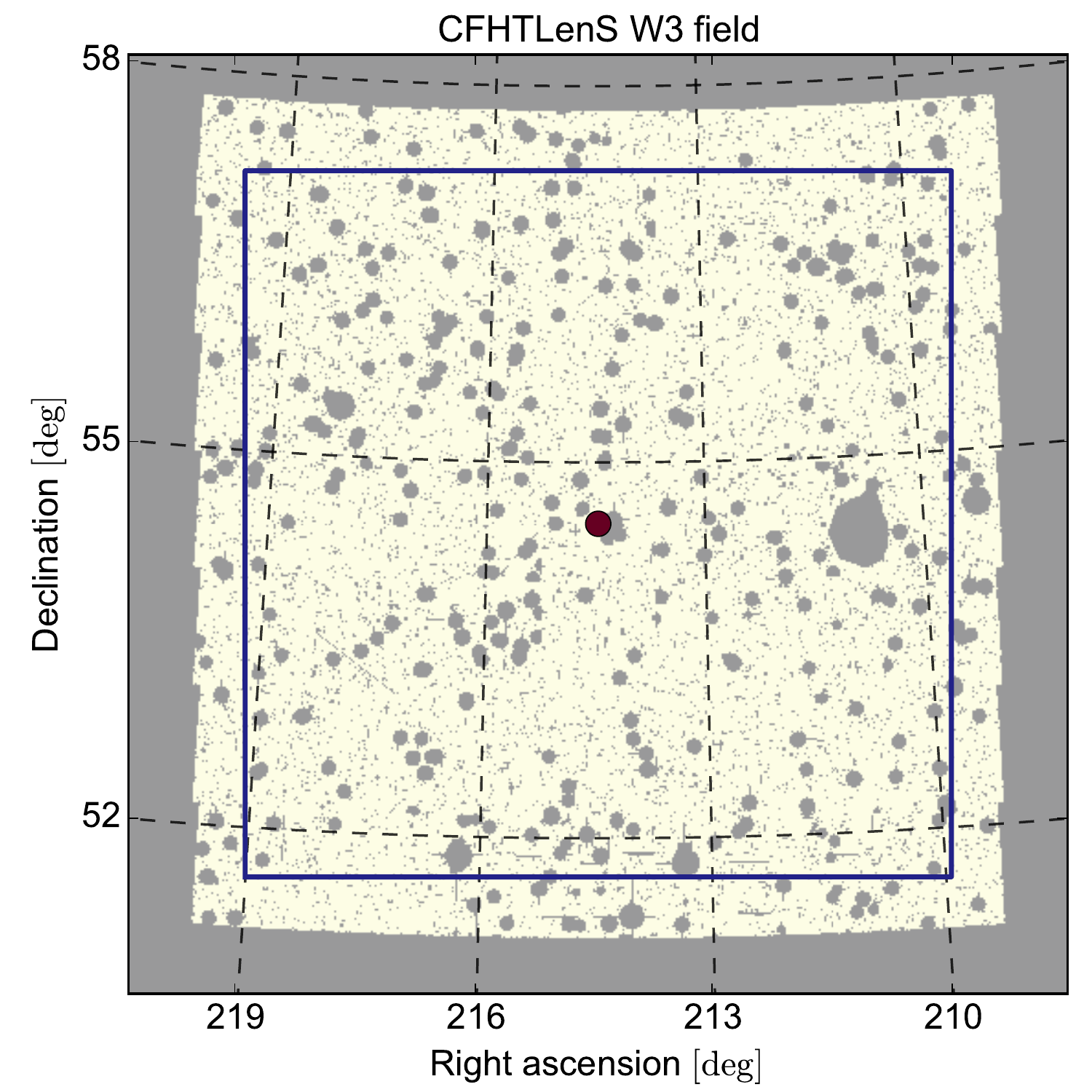}\hfill
	\includegraphics[width=0.49\textwidth]{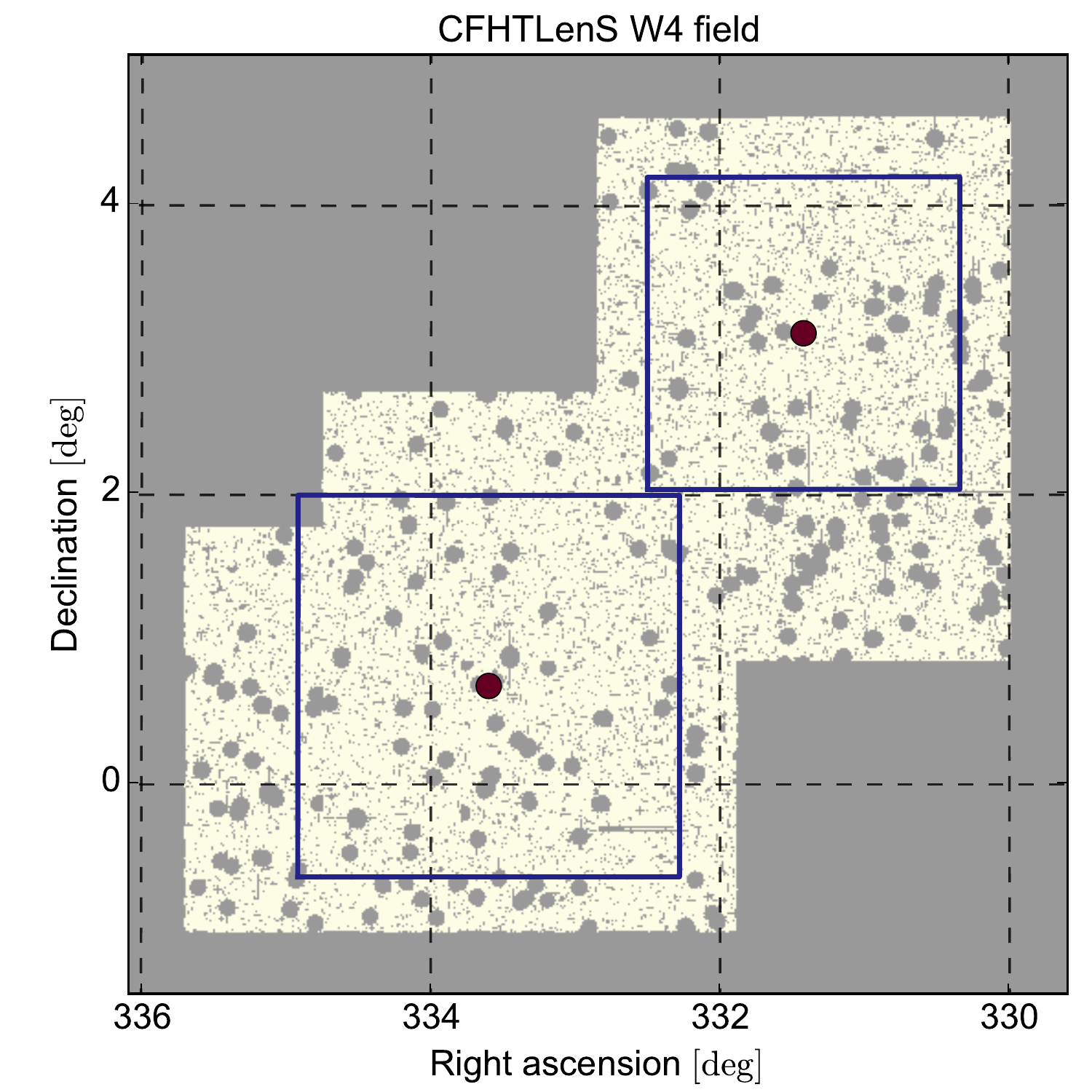}
	\caption{Illustrations of all eight patches (blue frames) from \acro{CFHTLenS}. The four panels represent respectively the W1, W2, W3, and W4 fields. Red points are the center of patches. Unmasked areas are showed in yellow.}
	\label{fig:data:CFHTLenS}
\end{figure}

To process peak counts, the W1 field, which is relatively large, is cut into four patches. Also, the W4 field which has a irregular shape is separated into two patches. Thus, the whole data set is represented by eight patches of different sizes (\fig{fig:data:CFHTLenS}). For each patch, border effects are carefully removed. This means that some galaxies near the edges are sacrificed and only peaks in the inner area are counted. We define it as the \textit{effective area}\index{Effective area} in the same way as in previous chapters, which is 112 deg$^2$ in this case.

\subsubsection{KiDS DR1/2}

\acro{KiDS} \acro{DR1/DR2} \citep{deJong_etal_2015} contains 109 tiles of 1 deg$^2$ all near the equator. Concerning the lensing catalogue, \acro{KiDS} uses the same algorithm as \acro{CFHTLenS}. We select sources following the criteria suggested by \citet{Kuijken_etal_2015}: \texttt{MAN\_MASK} $= 0$, $0.005 <$ \texttt{Z\_B} $< 1.2$, $\mathtt{weight} > 0$, and $\mathtt{SNratio} > 0$. In addition to these, due to calibration limits, we include also \texttt{c1\_best} $> -50$ and \texttt{c2\_best} $> -50$ to cut out values of -99 \citep{Kuijken_etal_2015}. The final selection includes 2.4 million galaxies distributed on an unmasked area of 75 deg$^2$. The raw density is 8.87 arcmin$\invSq$.

\begin{figure}[!t]
	\centering
	\includegraphics[width=0.49\textwidth]{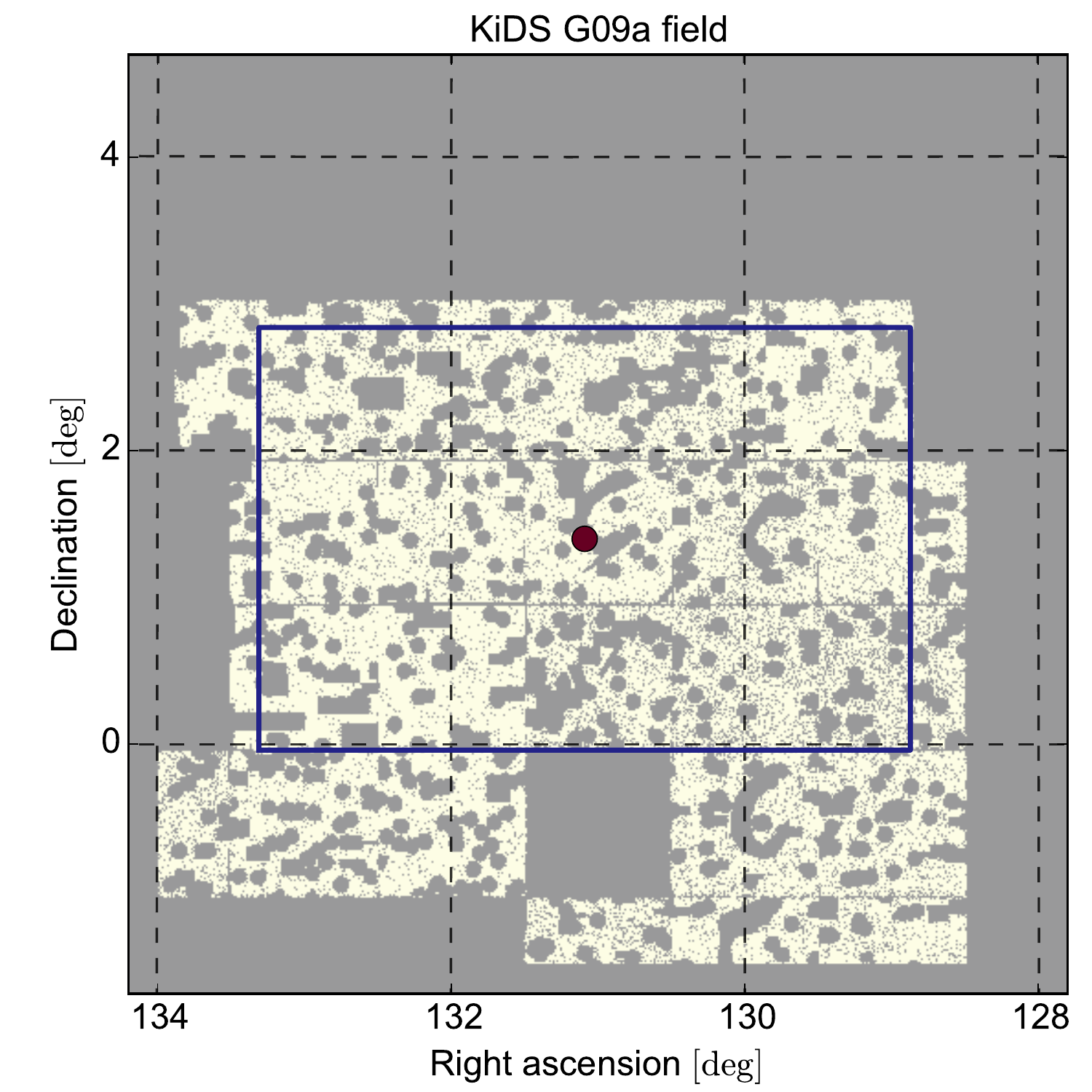}\hfill
	\includegraphics[width=0.49\textwidth]{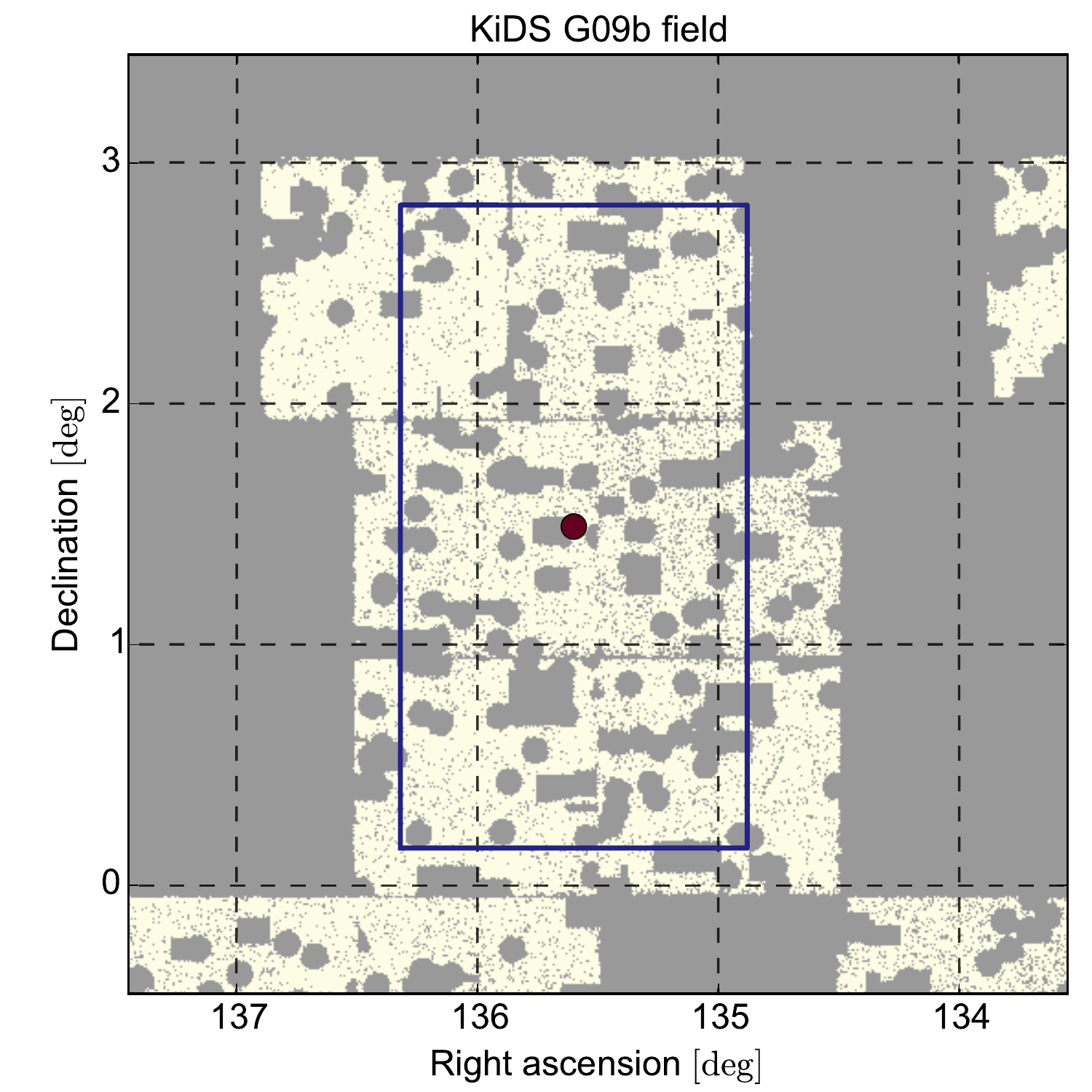}\\
	\includegraphics[width=0.49\textwidth]{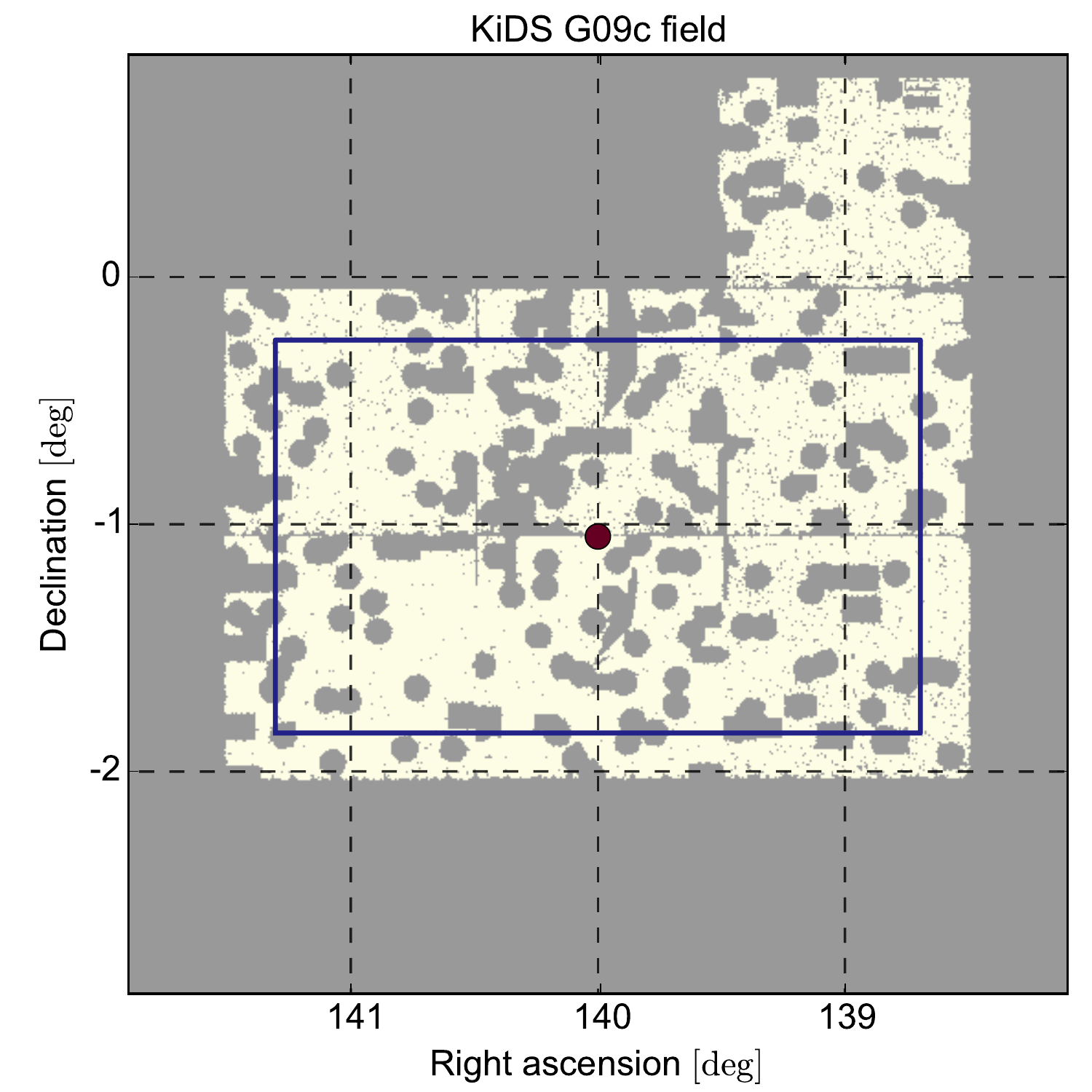}\hfill
	\includegraphics[width=0.49\textwidth]{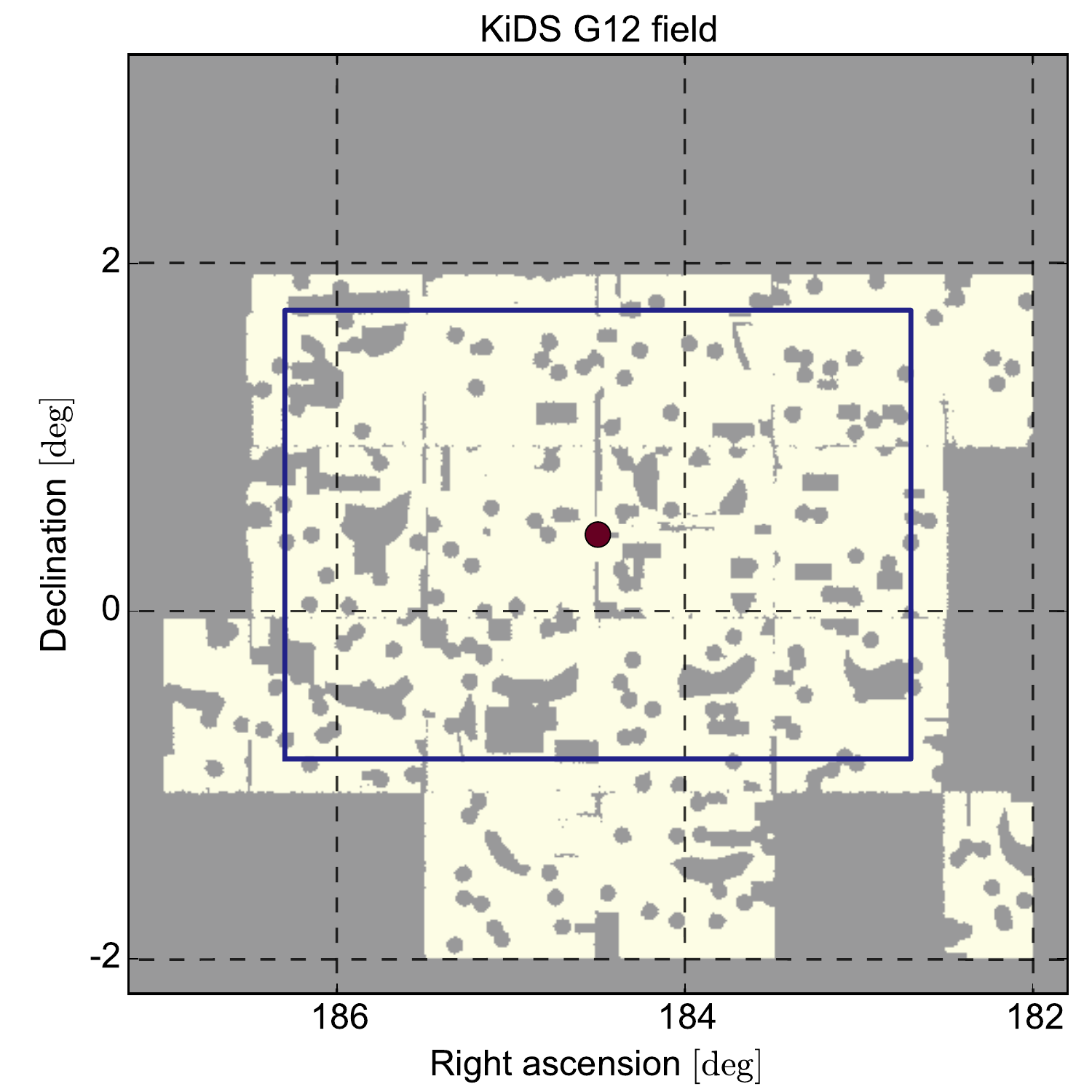}
	\caption{Illustrations of all four patches (blue frames) from \acro{KiDS DR1/2}. Three of them are taken from the G09 field. The rest is taken from the G12 field. Red points are the center of patches. Unmasked areas are showed in yellow.}
	\label{fig:data:KiDS}
\end{figure}

Regarding the fact that (1) the \acro{KiDS} data are more contaminated by masks and (2) some 1-deg$^2$ tiles are missing from a contiguous region, we only retain a small part of the whole data set. This is organized as four patches with a total area of 41 deg$^2$ (\fig{fig:data:KiDS}). After cutting out edges, the effective area is 30 deg$^2$.

\subsubsection{DES SV}

The \acro{DES} \acro{SV} data \citep{DarkEnergySurveyCollaboration_etal_2016, Jarvis_etal_2016} contains a large field named SPT-E and some other small fields. Here we only focus on the SPT-E field, which covers an unmasked area of 138 deg$^2$. We use the \textsc{ngmix} catalogue provided by the collaboration. The flag \texttt{NGMIX\_FLAG} $= 0$, which is a combination of several flags, validates that a source has a good \textsc{ngmix} measurement \footnote{See \url{http://des.ncsa.illinois.edu/}.}. An additional criterion is the redshift information, with $0.3 <$ \texttt{MEAN\_PHOTOZ} $< 1.3$, suggested by \citet{Kacprzak_etal_2016}. The total number of galaxies is 3.3 millions, with raw density 6.63 arcmin$^2$.

\begin{figure}[tb]
	\centering
	\includegraphics[width=0.65\textwidth]{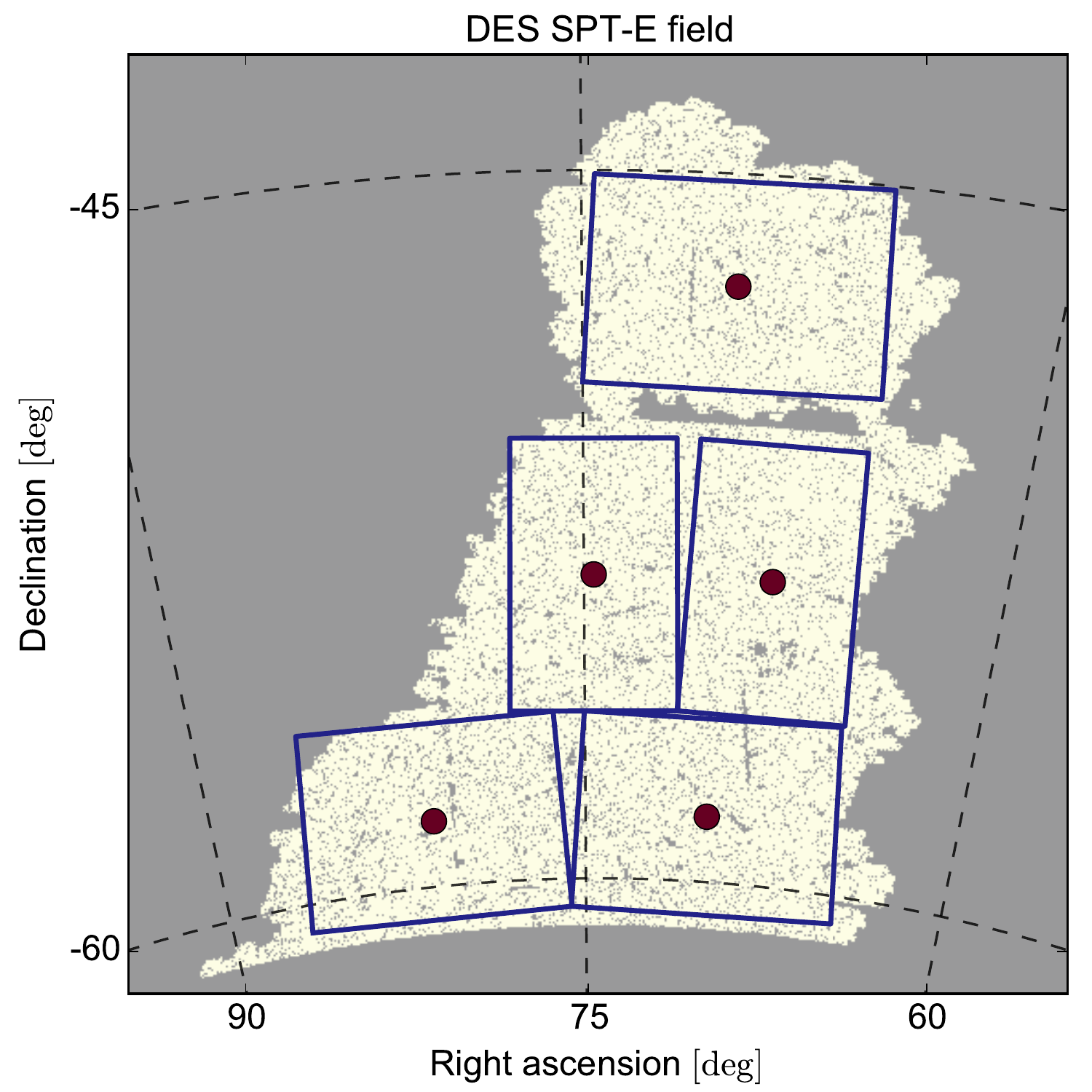}
	\caption{Illustrations of all five patches (blue frames) from \acro{DES SV}. Red points are the center of patches. Unmasked areas are showed in yellow.}
	\label{fig:data:DES}
\end{figure}

In this study, the SPT-E field is divided into five patches (\fig{fig:data:DES}). The resulting effective area is 115 deg$^2$.

\section{Methodology}

\begin{table}[tb]
	\centering
	\begin{tabular}{lcccc}
		\hline\hline
		Parameter                              & Symbol            & \acro{CFHTLenS} & \acro{KiDS DR1/2} & \acro{DES SV}\\
		\hline
		Source distribution parameter          & $z_\minn$         & 0.2             & 0.005             & 0.3\\
		Source distribution parameter          & $z_\maxx$         & 1.3             & 1.2               & 1.3\\
		Source distribution parameter          & $\alpha_\gala$    & 0.474           & 1.255             & 2.958\\
		Source distribution parameter          & $\beta_\gala$     & 5.990           & 0.698             & 1.048\\
		Source distribution parameter          & $z_0$             & 1.157           & 0.128             & 0.217\\
		Galaxy number density [arcmin$\invSq$] & $n_\gala$         & 13.53           & 8.87              & 6.63\\
		Ellipticity dispersion                 & $\sigma_\epsilon$ & 0.449           & 0.516             & 0.400\\
		Pixel size [arcmin]                    & $\theta_\pix$     & 0.72            & 0.9               & 1.05\\
		Number of filtering scales             & -                 & 4               & 3                 & 3\\
		\hline\hline
	\end{tabular}
	\caption{List of parameter values adopted in the study of this chapter.}
	\label{tab:data:parameters}
\end{table}

The analysis is processed with the same method as described in \sect{sect:filtering:methodology:pipeline}, with $M_\minn = 2\dixx{12}\ \Msol$. Each survey is modelled separately with the corresponding parameters such as $n_\gala$, $\sigma_\epsilon$, $\theta_\pix$, filtering scales, and the source redshift distribution (see \tab{tab:data:parameters}). At the end, three peak histograms are joined. The details are described in the following paragraphs.

\subsubsection{Source redshift distribution}

The redshift distribution of the sources is assumed to be a general gamma distribution:
\begin{align}
	p(z) \propto \left(\frac{z}{z_0}\right)^{\alpha_\gala}\exp\left(-\left(\frac{z}{z_0}\right)^{\beta_\gala}\right),
\end{align}
where the normalization is computed for the interval [$z_\minn$, $z_\maxx$]. For each survey, $z_\minn$ and $z_\maxx$ are given by the suggested flags. Then, we fit $\alpha_\gala$, $\beta_\gala$, and $z_0$ to the redshift distribution of the catalogues.

\subsubsection{Ellipticity dispersion and galaxy density}

The noise from the catalogue has two origins: intrinsic shape dispersion and shape measurement errors. Unfortunately, it is impossible to separate these two components from observations. For this reason, cosmologists define the effective number density, such that when this quantity is combined with the assumed intrinsic dispersion, the total variance will match the one in the catalogue. Therefore, the effective density is always smaller than the raw density. The less the galaxies, the higher the Poisson noise; and this Poisson part will account for the measurement errors neglected in the effective formalism.

However, this is actually a redundant step. First, the true shape dispersion is not Gaussian, neither are the measurement errors. Separating both components in the Gaussian formalism seems very unrealistic. Second, the intrinsic dispersion is never known. Although \citet{Jarvis_etal_2016} claim that it is possible via a fine-tuned combination of weights, the results obtained by this technique are not consistent between different surveys. Finally, we never need the value of shape dispersion alone in the model. Modelling the combination of both components as a Gaussian noise is much simpler than considering them with separate models and assuming further approximations.

As a result, we adopt a different approach to determine the density and the dispersion. The idea is simple: given a catalogue with its total variance, we look for the Gaussian dispersion which corresponds to the raw density, so that the total variance matches. This leads to two advantages. First, only one assumption is needed: the Gaussianity of the combination of the intrinsic dispersion and the shape measurement. Second, the number density that will be used in our model is larger. This reduces the bias generated by the \acro{KS} inversion when the distribution of sources is too irregular. 

More explicitly, the total variance is defined as the empirical weighted variance taken the multiplicative bias into account:
\begin{align}
	\sigma_\tot^2 = \frac{\sum_i w_i^2 \left(\epsilon_{i,1}^{(\rmc)2} + \epsilon_{i,2}^{(\rmc)2}\right)}{\sum_i w_i^2(1+m_i)^2},
\end{align}
where $\epsilon\upp{\rmc}_{i,j} \equiv \epsilon\upp{\rmo}_{i,j} - c_{i,j}$ is the $j$-th component of the observed ellipticity $\epsilon_i\upp{\rmo}$ of the $i$-th galaxy corrected with the additive bias $c_i$, $w_i$ is the weight, $m_i$ is the multiplicative bias, and the sum runs over all galaxies in the catalogue. The Gaussian dispersion $\sigma_\epsilon$ corresponding to the raw density satisfies
\begin{align}
	\sigma_\tot^2 = \frac{\sigma_\epsilon^2}{2N_\gala},
\end{align}
where $N_\gala$ is the total number of sources. In this case, the derived $\sigma_\epsilon$ can have a relatively large number. In practice, we truncate at $|\epsilon|=\pm 1$ to avoid unphysical values. The violation of the Gaussian variance by the truncation is ignored.

\subsubsection{Pixel size and filter scales}

The pixel size $\theta_\pix$ is determined based on $n_\gala$. The goal is to avoid having pixels with a low number of galaxies. We expect $\theta_\pix$ to satisfy
\begin{align}
	 P_{\bar{N}}(N=0) + P_{\bar{N}}(N=1) + P_{\bar{N}}(N=2) \leq 0.03,
\end{align}
where $\bar{N}=n_\gala\theta_\pix^2$ is the expected number of galaxies in a pixel, and $P_\lambda(N)$ is the probability associated at number $N$ for the Poisson distribution with parameter $\lambda$. This means that we expect the probability of having less than three galaxies in a pixel to be smaller than 3\%. The values in \tab{tab:data:parameters} satisfy this criterion.

\begin{table}[tb]
	\centering
	\begin{tabular}{ll}
		\hline\hline
		Data set                               & \acro{CFHTLenS}\\
		Filter scales $\theta_{\ker}$ [arcmin] & 1.44, 2.88, 5.76, 11.52\\
		\acro{S/N} bins                        & [2.5, 2.75, 3.0, 3.25, 3.5, 3.75, 4.0, 4.25, 4.75, $+\infty$[\\
		\hline
		Data set                               & \acro{KiDS DR1/2}\\
		Filter scales $\theta_{\ker}$ [arcmin] & 1.8, 3.6, 7.2\\
		\acro{S/N} bins                        & [2.5, 2.75, 3.0, 3.25, 3.5, $+\infty$[\\
		\hline
		Data set                               & \acro{DES SV}\\
		Filter scales $\theta_{\ker}$ [arcmin] & 2.1, 4.2, 8.4\\
		\acro{S/N} bins                        & [2.5, 2.75, 3.0, 3.25, 3.5, 3.75, 4.0, 4.5, $+\infty$[\\
		\hline\hline
	\end{tabular}
	\caption{List of scales and bins used in the study of this chapter. The dimension of the data vector is $4\times9+3\times5+3\times8=75$.}
	\label{tab:data:scales}
\end{table}

Concerning the filter, we only use the starlet filtering in this study. The filter scales are chosen to be some even multiples of the pixel size. The reason is that if $\theta_{\ker}=2N\theta_\pix$ where $N=1,2,\ldots$, then the starlet is strictly compensated in its discrete sum. These values are some natural choices for wavelet functions. We pick four scales for \acro{CFHTLenS} and three scales for the others. This results in 1.44, 2.88, 5.76, and 11.52 arcmin for \acro{CFHTLenS}; 1.8, 3.6, and 7.2 arcmin for \acro{KiDS}; and 2.1, 4.2, and 8.4 arcmin for \acro{DES}.

\subsubsection{Data vector}

\begin{figure}[tb]
	\centering
	\includegraphics[width=0.65\textwidth]{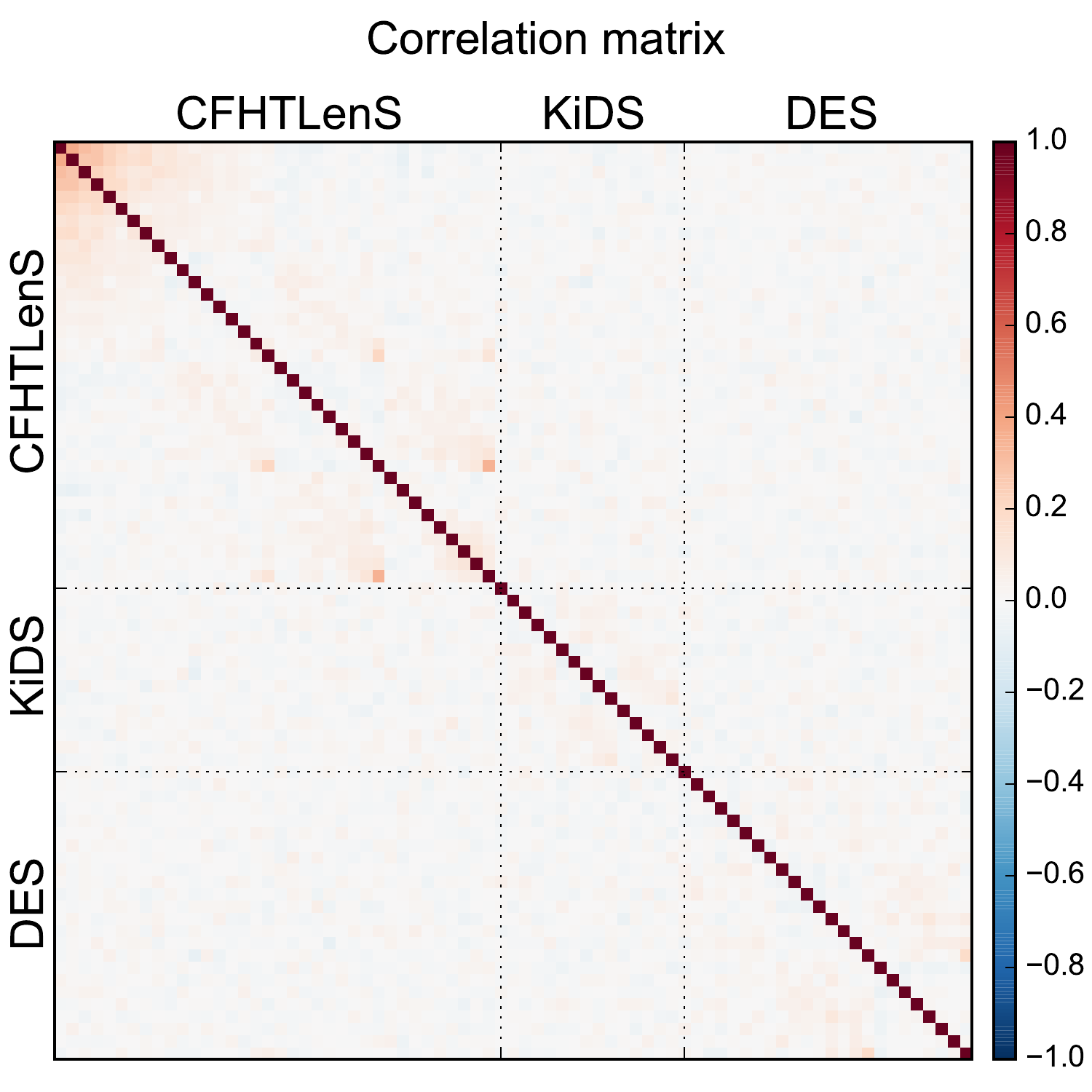}
	\caption{Correlation matrix from the data vector defined in this study. The components are very weakly correlated with each other.}
	\label{fig:data:covariance_CFHTLenS_KiDS_DES}
\end{figure}

The data vector is defined as the concatenation of peak histograms of all scales from three surveys. The binwidth is defined slightly differently depending on the survey (see \tab{tab:data:scales}). The reason for this is to have enough number counts in each bin ($\gtrsim 8$) such that the Poisson distribution can be approximated by a Gaussian one. This results in a data vector of dimension 75. The correlation matrix of this data vector is shown in \fig{fig:data:covariance_CFHTLenS_KiDS_DES}.

\subsubsection{Parameter sampling}

We only perform the \acro{ABC} analysis in this chapter. Two runs have been carried out. The first focuses on three free parameters, $\OmegaM$, $\sigEig$, and $\wZero$. The second one includes two additional parameters, $c_0$ and $\beta$, which define the halo \acro{$M$-$c$} relation (Eq. \ref{for:structure:M_c_relation}).

Concerning the configuration of \acro{ABC}, we take a flat prior in the studied parameter space. The number of particles is 2400. The shutoff parameter is set to 1\%. The distance is the same as \for{for:filtering:D_2}, accounting for the full correlation.

\section{Preliminary results}

\begin{figure}[!t]
	\centering
	\includegraphics[width=\textwidth]{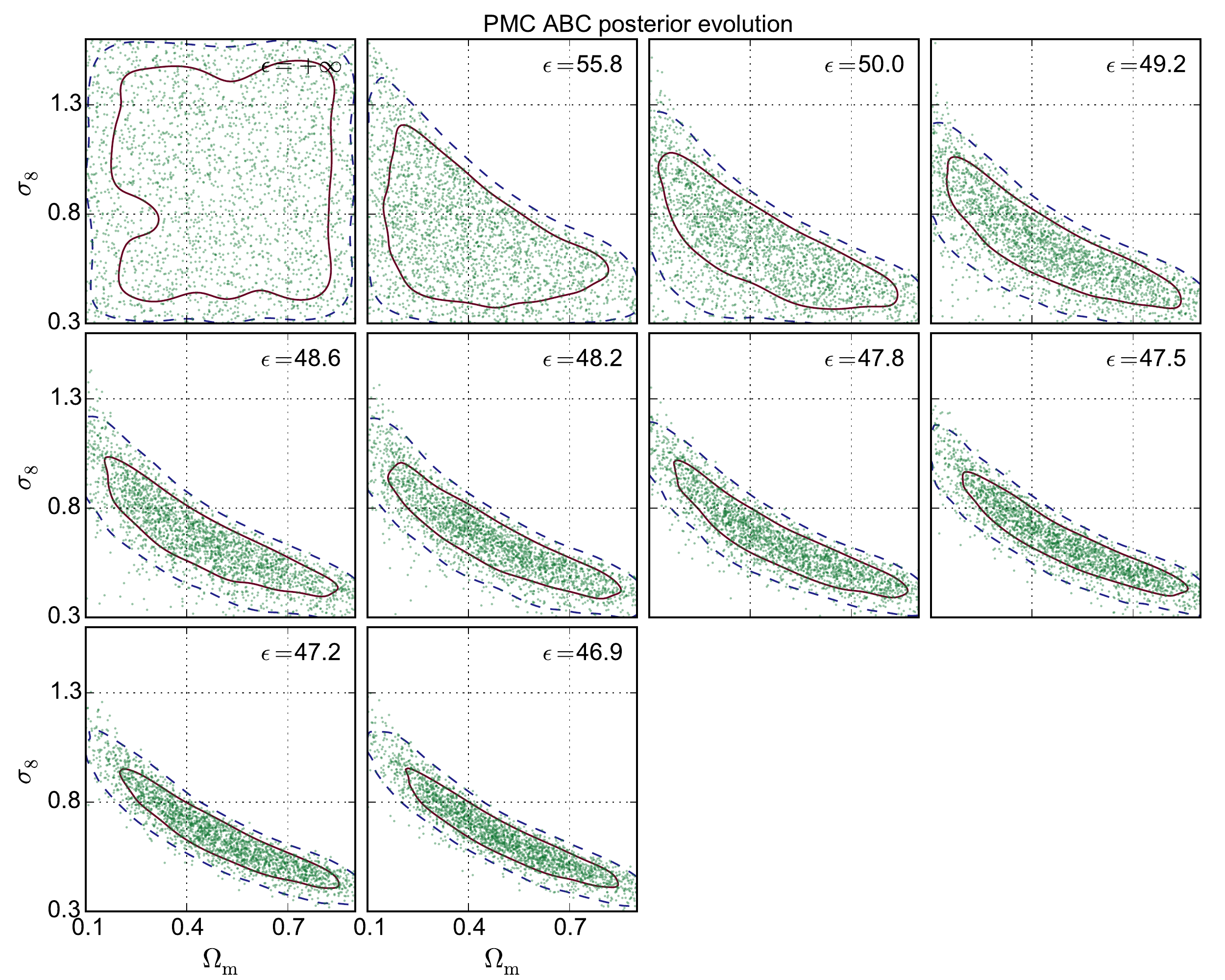}
	\caption{Preliminary result of $\OmegaM$-$\sigEig$ constraints with \acro{CFHTLenS}-\acro{KiDS}-\acro{DES} joint data sets. This figure shows the evolution of the \acro{PMC} \acro{ABC}. The blue solid and the red dashed lines represent respective 1- and 2-$\sigma$ contours. Green points are accepted \acro{ABC} particles.}
	\label{fig:data:evolution_OM_s8_run2}
\end{figure}

\begin{figure}[!t]
	\centering
	\includegraphics[width=0.6\textwidth]{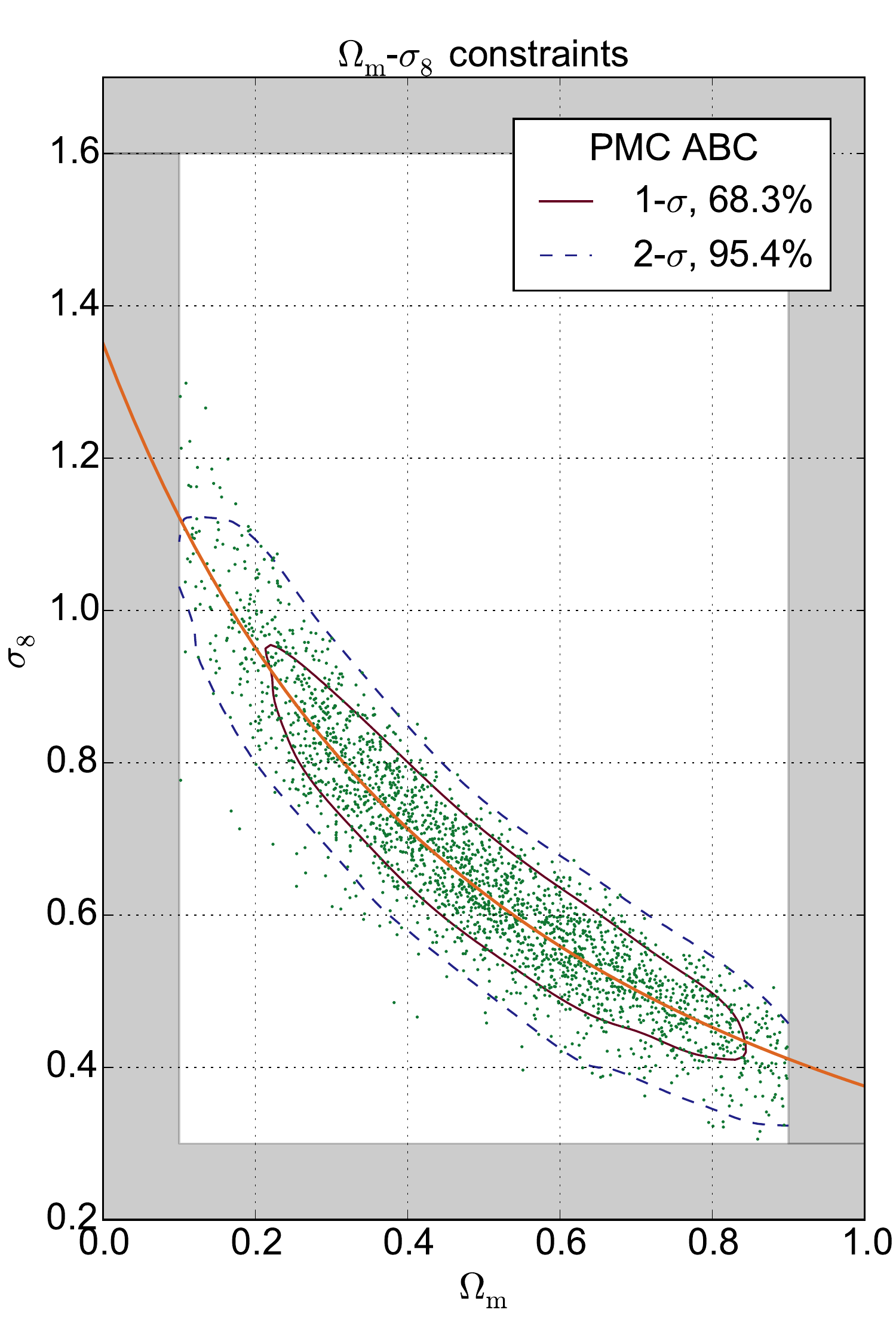}
	\caption{Preliminary result of $\OmegaM$-$\sigEig$ constraints with \acro{CFHTLenS}-\acro{KiDS}-\acro{DES} joint data sets. This figure shows the posterior from the final iteration. The blue solid and the red dashed lines represent respective 1- and 2-$\sigma$ contours. Green points are accepted \acro{ABC} particles. The orange curve is the best power-law fit.}
	\label{fig:data:contour_run2_t-1_fit}
\end{figure}

\figFull{fig:data:evolution_OM_s8_run2} shows the primary results that we obtain from the joint data constraints. In this figure, we only focus on $\OmegaM$ and $\sigEig$ from the run with three free parameters. As $\epsilon$ decreases, the contour size reduces. The constraints from the last iteration are shown in \fig{fig:data:contour_run2_t-1_fit}. The orange curve shows the best power-law fit (i.e. best $\Sigma_8$ fit, Eq. \ref{for:filtering:CSE}) between $\OmegaM$ and $\sigEig$. It indicates a very good agreement with the Planck cosmology ($\OmegaM=0.308\pm0.012$, $\sigEig=0.8149\pm0.0093$). The contour is consistent with constraints from other lensing surveys using two-point statistics (\acro{CFHTLenS}: \citealt{Kilbinger_etal_2013}, \citealt{Joudaki_etal_2016}; \acro{KiDS}: \citealt{Hildebrandt_etal_2016}; \acro{DES}: \citealt{TheDarkEnergySurveyCollaboration_etal_2015}) or peaks (\acro{CFHTLenS}: \citetalias{Liu_etal_2015} \citeyear{Liu_etal_2015}; \acro{CFHT} Stripe 82: \citetalias{Liu_etal_2015a} \citeyear{Liu_etal_2015a}; \acro{DES}: \citealt{Kacprzak_etal_2016}). However, since our results are only preliminary, the comparisons are not shown. Here, we only give the characteristics of the preliminary constraints: the error on the orange curve is $\DCSE=0.13$, obtained by the same method as \chap{sect:filtering}, and the \acro{FoM} is 5.2.

In this preliminary result, the constraints seem to be larger than expected. If we compare the \acro{FoM} from data with those from \tab{tab:filtering:FoM_ABC}, then the difference is about a factor of 3. Since the settings for the results from \tab{tab:filtering:FoM_ABC} have been designed to be very similar to the \acro{CFHTLenS} data set, we believe that they can be compared directly. The reason of these looser contours is believed to be the selection of \acro{S/N} bins and the bias from our model. A quick comparison (not plotted) with the \acro{Mice} simulations (5200 deg$^2$) indicates that the peak function given by our model is tilted from the one derived from $N$-body simulations. Typically, our model overestimates high peaks and underestimates low peaks. If this is the case, then changing the input cosmology of our model, which leads to a global change in all bins of the peak functions, could result in similar differences with regard to the observation, therefore degrades the constraint power. In this scenario, the bad accuracy reduces the precision, and this is susceptible to be the case of our preliminary result. For the future, the improvement will be to compare closely our model with the \acro{Mice} simulations to quantify the bias, and to adopt only bins with weak bias for constraints. With this refinement, we expect that the final results yield a similar \acro{FoM} and $\DCSE$ to the values from \tab{tab:filtering:FoM_ABC}.

\begin{figure}[tb]
	\centering
	\includegraphics[width=0.7\textwidth]{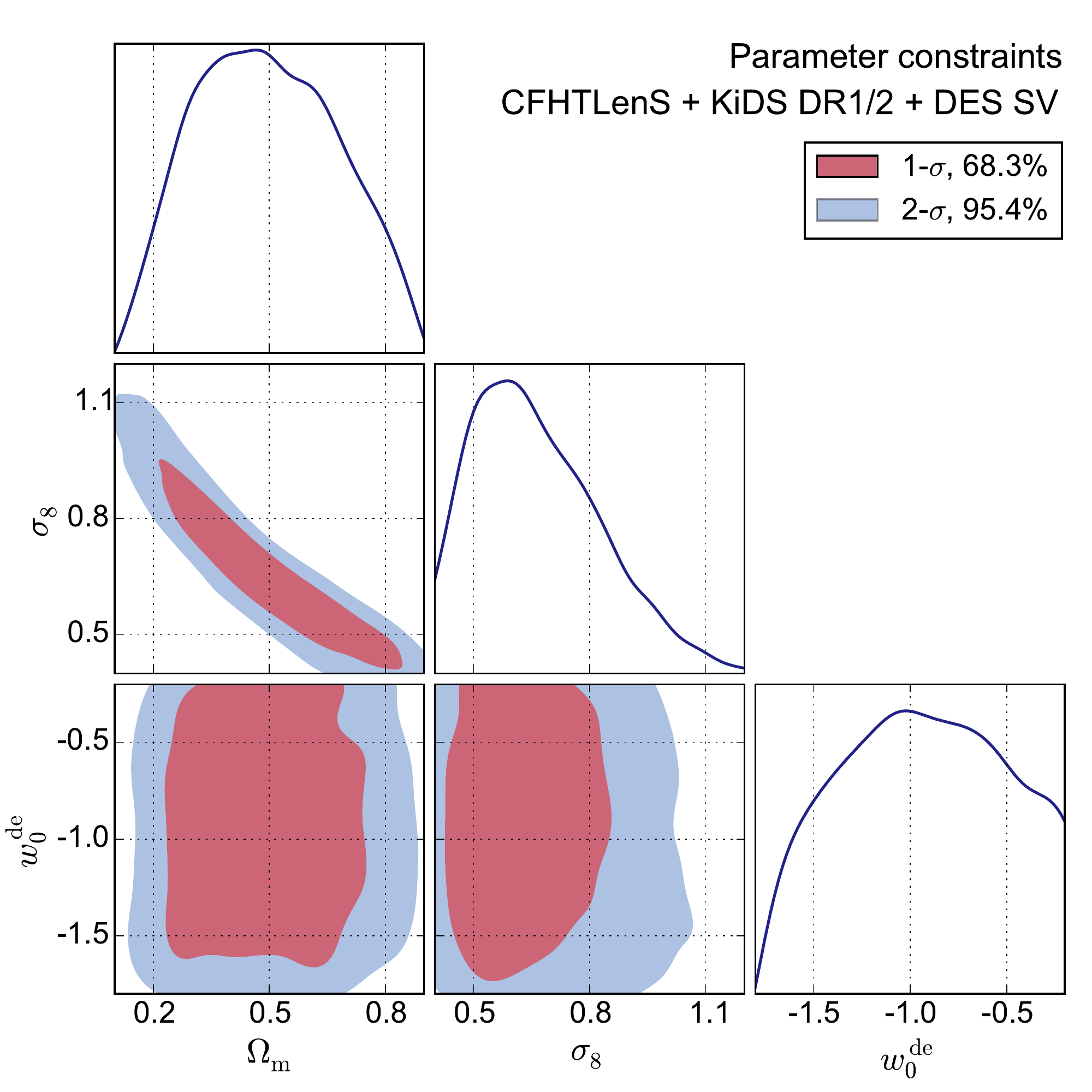}
	\caption{Preliminary result of $\OmegaM$-$\sigEig$-$\wZero$ constraints with \acro{CFHTLenS}-\acro{KiDS}-\acro{DES} joint data sets.}
	\label{fig:data:contour_stair_f3}
\end{figure}

\begin{figure}[tb]
	\centering
	\includegraphics[width=0.85\textwidth]{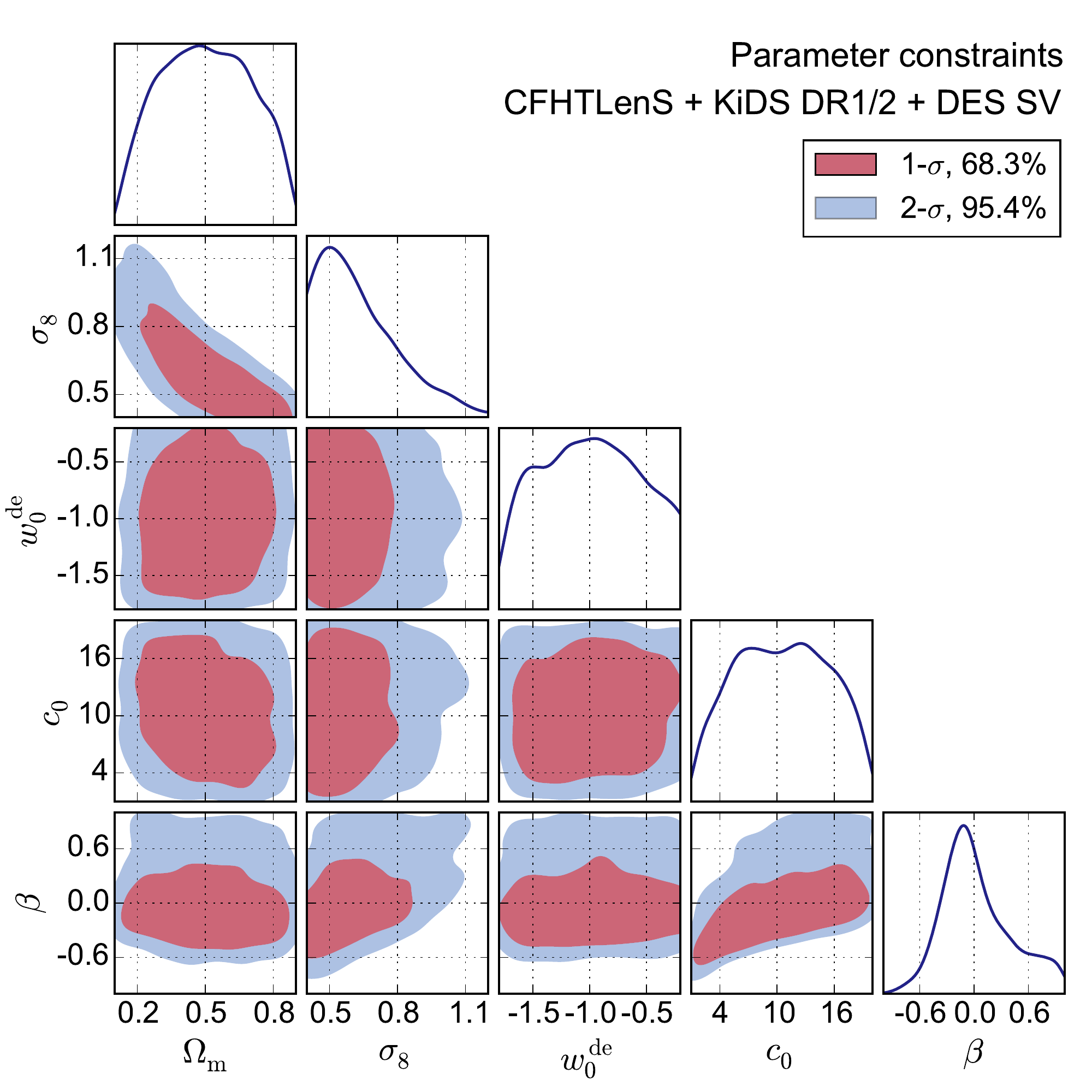}
	\caption{Preliminary result of $\OmegaM$-$\sigEig$-$\wZero$-$c_0$-$\beta$ constraints with \acro{CFHTLenS}-\acro{KiDS}-\acro{DES} joint data sets.}
	\label{fig:data:contour_stair_f5}
\end{figure}

\figFull{fig:data:contour_stair_f3} shows all marginalized constraints from the run with three free parameters. The degeneracy between $\wZero$ and other parameters is less significant because of \acro{ABC}. Kernel density estimation smooths the contours. In \fig{fig:data:contour_stair_f5}, we show the constraints with five free parameters. We notice that the constraint on $c_0$ is very broad. However, concerning $\beta$, the values below $-0.6$ are already been excluded by the preliminary study. By including two additional parameters, the constraints on $\OmegaM$ and $\sigEig$ become larger, which is expected.

\section{Perspectives}

In spite of the three-year investment, several questions concerning peak-count modelling still remained unanswered. Below is a list non-exhausted of ideas and perspectives that future studies could investigate.
\vspace*{-2.5ex}

\paragraph{Halo modelling} All along the thesis, we adopt the spherical \acro{NFW} modelling for halo density profiles. However, $N$-body simulations indicate well the asphericity of dark-matter halos. Without reproducing their true shapes, triaxial halos or substructures can be a more realistic option for lens modelling. Meanwhile, baryonic effects can also modify the profiles. These different possible improvements concerning halo geometry are worth taking a closer look. Next, in \fig{fig:modelling:M_c_relation}, we have already seen that the \acro{$M$-$c$} relation of halos suffers from a huge scatter. Therefore, taking this intrinsic scatter into account can lead to a more accurate modelling than a deterministic relation. Finally, halo correlation is another subject to investigate in detail. We have seen that neglecting correlation leads to a decrease of peak counts which is compensated by the replacement of $N$-body halo profiles with their \acro{NFW} proxies. Why do these two effects vary in opposite ways? Do they really compensate? These are some questions to investigate for the future.
\vspace*{-2.5ex}

\paragraph{Source modelling} The improvement for source modelling includes at least three aspects: \acro{photo-$z$}, galaxy clustering, and intrinsic alignment. The actual photometric estimation of galaxy redshifts is scattered and can be biased. For real surveys, we need to correct the bias for peak counts, and to take the propagation of the scatter into account. Meanwhile, instead of using randomized sources as in this work, adopting the exact source catalogue would reduce the stochasticity. Using the survey source catalogue will also account for galaxy clustering, which is a potential source of bias for peak counts \citep{Kacprzak_etal_2016}. Finally, intrinsic alignment plays an important role in weak lensing analyses. Understanding its impacts on weak-lensing peaks will be crucial for future large and high-quality surveys.
\vspace*{-2.5ex}

\paragraph{Filtering} To improve our knowledge about $\wZero$, tomography filters will be required. This allows us to extract information from different cosmic eras. Concerning mask effects, inpainting could be interesting as a tool to enhance the quantity of information. Even if the inpainted information is not included in statistics, the technique will still help to regularize mask effects and reduce the systematics.
\vspace*{-2.5ex}

\paragraph{Larger data sets} Ongoing lensing surveys such as \acro{KiDS}, \acro{DES}, \acro{HSC}, and \acro{J-PAS}, and upcoming surveys such as Euclid, \acro{LSST}, \acro{WFIRST} all aim for providing larger data sets than \acro{CFHTLenS}, considered as the current state of the art. Information from deeper and wider fields will help us refine cosmological constraints with weak-lensing peak counts.
\vspace*{-2.5ex}

\paragraph{Beyond $\vect{\Lambda}$CDM} Weak-lensing peak counts are sensitive to cosmological models beyond $\LCDM$ which change the mass function. Probing modified gravity is a future task of weak-lensing peaks. Constraints on parameters of modified gravity have recently been studied using \acro{CFHTLenS} data \citep{Liu_etal_2016}. Similarly, peak counts can also be used for studying the effective number of neutrinos and the neutrino mass. The shape of the mass function changes as different models are considered.
\vspace*{-2.5ex}

\paragraph{Beyond peak counts} Beyond the abundance, peaks could provide other information potentially interesting for extracting cosmological information. For example, \citet{Marian_etal_2013} estimated the Fisher information derived from tangential shear profiles of peaks and peak-peak correlations. These are some observables to explore for future studies.

\subsubsection{Summary}

We provide constraints on $\OmegaM$, $\sigEig$, $\wZero$, $c_0$, and $\beta$ with data sets from \acro{CFHTLenS}, \acro{KiDS DR1/2}, and \acro{DES SV}. This is done with our fast stochastic forward model and \acro{ABC}.

We also indicate that the result can be improved by excluding bins with large number-count bias, which will be done in the near future.

Finally, possible improvements for peak-count modelling are listed. We look forward to larger statistics provided by future surveys for improving our knowledge about the Universe.

\clearpage
\thispagestyle{empty}
\cleardoublepage


\chapter{Conclusion}
\label{sect:conclusion}
\fancyhead[LE]{\sf \nouppercase{\leftmark}}
\fancyhead[RO]{\sf \nouppercase{\rightmark}}

\hfill\textit{\large Science may be described as the art of systematic over-simplification.}

\hfill--- Karl Popper
\vskip4ex

This thesis studies cosmology with weak-lensing peak counts. In my work, a new model has been proposed, peak statistics has been analyzed, and cosmological constraints have been obtained.

The new model adopts a stochastic forward approach. This strategy permits to include real-world effects and guarantees flexibility. In addition, it is time-saving compared to the $N$-body process. These advantages make our model the ideal choice for modelling peak counts.

Starting with a mass function, our model simulates uncorrelated halos with \acro{NFW} profiles. It computes the projected mass from this simulation, and adds noise to generate galaxy shape catalogues. Then it applies a mass-mapping technique to obtain a convergence map, and counts and regroups peaks by their \acro{S/N}. The model is implemented by the public code \acro{Camelus}.

The model is based on two hypotheses. First, diffuse and unbound matter has little impact; second, halo correlation contributes little to peak counts. This halo approach, corresponding to the one-halo term, leads to a very good approximation to peaks from $N$-body simulations. We examined the model in detail and discovered two compensated effects: the replacement of irregular halo profiles by the NFW ones enhances peak counts, and decorrelation decreases numbers.

In this study, the Gaussian likelihood assumption has been tested in different ways. On the one hand, the cosmological dependence of the covariance can not be neglected. Assuming a constant covariance leads to a loss of constraint power. On the other hand, the constraints from the Gaussian likelihood agree well with the results from generalized techniques: the copula likelihood and non-parametric methods such as the true likelihood and the $p$-value. The conclusion is that the Gaussian likelihood is a good approximation to weak-lensing peak counts.

Beyond the likelihood analysis, approximate Bayesian computation (\acro{ABC}) provides an alternative for parameter constraints. It is especially competitive for problems with complex likelihood shapes. It has been shown that \acro{ABC} yields a consistent result compared with the likelihood in our studied framework, and can reduce the computation time by two orders of magnitude.

This study has also provided a framework in which filtering techniques can be fairly compared. Regarding the complexity of how cosmology is encrypted in the lensing observables, it is very difficult to determine directly an optimal filter which minimizes the contour sizes from peak counts. However, ranking filters by their performance is possible. This methodology has been followed for filter comparisons. We concluded that compensated filters are more competitive than the usual Gaussian functions and that the \textit{separated strategy} is well more recommended than the \textit{combined strategy} when dealing with multiscale information.

This peak-count analysis is applied to three survey data jointly: \acro{CFHTLenS}, \acro{KiDS DR1/2}, and \acro{DES SV}. So far, preliminary results agree well with Planck, also with two-point-statistic results from \acro{CFHTLenS}, \acro{KiDS}, and \acro{DES}. The quality of the constraints is characterized by $\DCSE=0.13$ and \acro{FoM} $=5.2$.

Various improvements for the new model are possible. How to properly model the halo correlations? How to deal with the intrinsic scatter of the halo concentration parameters? Does baryonic physics influence peak counts? How do peaks response to intrinsic alignment of the galaxies? What additional information can we extract from a tomography filter? What are the numbers of peaks in a massive neutrino cosmology and in a modified gravity model? These are the questions worth investigating for the future.

Conceptually, two ideas have been promoted and put forward by this thesis. The first one is the notion of \textit{fast stochastic forward modelling}, which provides an approach to predict non-Gaussian observables and to account for complex survey effects; the other one is \textit{approximate Bayesian computation}, which is a promising likelihood-free parameter inference tool. As the need for exploring higher-order statistics increases, it is expected that these two concepts rise for future cosmological studies.

\clearpage
\thispagestyle{empty}
\cleardoublepage

\chapter*{Abbreviations}
\addcontentsline{toc}{chapter}{Abbreviations} 
\fancyhead[LE]{\sf \nouppercase{Abbreviations}}
\fancyhead[RO]{}

\newcommand{\abbrev}[1]{{\sffamily #1}}
\newcommand{\abbrevCode}[1]{{\sffamily \scshape #1}}

\begin{longtable}{r@{\hspace{1em}}l}
	\abbrev{1D}                    & one-dimensional\\
	\abbrev{2D}                    & two-dimensional\\
	\abbrev{2PCF}                  & two-point-correlation function\\
	\abbrev{3D}                    & three-dimensional\\
	\abbrev{3PCF}                  & three-point-correlation function\\
	\abbrev{ABC}                   & approximate Bayesian computation\\
	\abbrev{AGN}                   & active galactic nucleus\\
	\abbrev{BAO}                   & baryon acoustic oscillations\\
	\abbrevCode{Calclens}          & Curved-sky grAvitational Lensing for Cosmological LightconE simulatioNS\\
	\abbrevCode{Camelus}           & Counts of Amplified Mass Elevations from Lensing with Ultrafast Simulation\\
	\abbrev{CCD}                   & charge-coupled device\\
	\abbrev{CDC}                   & cosmology-dependent covariance\\
	\abbrev{CDF}                   & cumulative distribution function\\
	\abbrev{CDM}                   & cold dark matter\\
	\abbrev{CFHT}                  & Canada-France-Hawaii Telescope\\
	\abbrev{CFHTLenS}              & Canada-France-Hawaii Telescope Lensing Survey\\
	\abbrev{CFHTLS}                & Canada-France-Hawaii Telescope Legacy Survey\\
	\abbrev{CMB}                   & cosmic microwave background\\
	\abbrev{COSEBI}                & complete orthogonal sets of $E$-/$B$-mode integral\\
	\abbrev{COSMOS}                & Cosmological Evolution Survey\\ 
	\abbrev{CPU}                   & central processing unit\\
	\abbrev{cst}                   & constant\\
	\abbrev{DES}                   & Dark Energy Survey\\
	\abbrev{DES SV}                & Dark Energy Survey science verification\\
	\abbrev{DIM-ACAV}              & Domain d'Int\'er\^et Majeur en Astrophysique et Conditions d'Apparition de la Vie\\
	\abbrev{DLS}                   & Deep Lens Survey\\
	\abbrev{DM}                    & dark matter\\
	\abbrevCode{FASTLens}          & FAst STatistics for weak Lensing\\
	\abbrev{FFT}                   & fast Fourier transform\\
	\abbrev{FLRW}                  & Friedmann-Lemaître-Robertson-Walker\\
	\abbrev{FOF}                   & friends of friends\\
	\abbrev{FoM}                   & figure of merit\\
	\abbrev{FORS}                  & FOcal Reducer Spectrograph\\
	\abbrev{FSF}                   & fast stochastic forward\\
	\abbrev{FSL}                   & Fan-Shan-Liu\\
	\abbrev{GaBoDS}                & Garching-Bonn Deep Survey\\
	\abbrevCode{Gadget}            & GAlaxies with Dark matter and Gas intEracT\\
	\abbrevCode{Glimpse}           & Gravitational Lensing Inversion and MaPping with Sparse Estimators\\
	\abbrev{GPU}                   & graphics processing unit\\
	\abbrev{HEALPix}               & Hierarchical Equal Area isoLatitude Pixelization\\
	\abbrev{HOD}                   & halo occupation distribution\\
	\abbrev{HSC}                   & Subaru Hyper Suprime-Cam\\
	\abbrev{HST}                   & Hubble Space Telescope\\
	\abbrev{IA}                    & intrinsic alignment\\
	\abbrev{iid}                   & independent and identically distributed\\
	\abbrev{J-PAS}                 & Javalambre Physics of the Accelerating Universe Astrophysical Survey\\
	\abbrev{KDE}                   & kernel density estimation\\
	\abbrev{KiDS}                  & Kilo-Degree Survey\\
	\abbrev{KiDS DR1/2}            & Kilo-Degree Survey data releases 1 and 2\\
	\abbrev{KS}                    & Kaiser-Squires\\
	\abbrev{LPT}                   & Lagrangian perturbation theory\\
	\abbrev{LSS}                   & large-scale structure\\
	\abbrev{LSST}                  & Large Synoptic Survey Telescope\\
	\abbrev{$\mathsf{M}$-$\mathsf{c}$} & mass-concentration\\
	\abbrev{MCMC}                  & Markov Chain Monte Carlo\\
	\abbrev{MF}                    & Minkowski functional\\
	\abbrev{Mice}                  & Marenostrum Institut de Ciencias de l'Espai\\
	\abbrev{MPI}                   & message passing interface\\
	\abbrevCode{MRLens}            & Multi-Resolution methods for gravitational LENSing\\
	\abbrev{NFW}                   & Navarro-Frenk-White\\
	\abbrevCode{Nicaea}            & NumerIcal Cosmology And lEnsing cAlculations\\
	\abbrev{PDF}                   & probability density function\\
	\abbrev{photo-$\mathsf{z}$}    & photometric redshift\\
	\abbrevCode{Pinocchio}         & PINpointing Orbit-Crossing Collapsed HIerarchical Objects\\
	\abbrev{PMC}                   & population Monte Carlo\\
	\abbrev{PNCG}                  & Programme National de Cosmologie et Galaxies\\
	\abbrev{PSF}                   & point spread function\\
	\abbrevCode{Rockstar}          & Robust Overdensity Calculator using $K$-Space Topologically Adaptive Refinement\\
	\abbrev{SCDM}                  & standard cold dark matter\\
	\abbrev{SDSS}                  & Sloan Digital Sky Survey\\
	\abbrev{SIS}                   & singular isothermal sphere\\
	\abbrev{SMC}                   & sequential Monte Carlo\\
	\abbrev{S/N}                   & signal-to-noise ratio\\
	\abbrev{SNIa}                  & supernova of type Ia\\
	\abbrev{SO}                    & spherical overdensity\\
	\abbrev{SZ}                    & Sunyaev–Zel'dovich\\
	\abbrev{VLT}                   & Very Large Telescope\\
	\abbrev{WFIRST}                & Wide Field Infrared Survey Telescope\\
	\abbrev{WL}                    & weak lensing\\
	\abbrev{WMAP}                  & Wilkinson Microwave Anisotropy Probe\\
	\abbrev{$\mathsf{\Lambda}$CDM} & cosmological constant and cold dark matter
\end{longtable}

\clearpage
\thispagestyle{empty}
\cleardoublepage


\chapter*{Notation}
\addcontentsline{toc}{chapter}{Notation} 
\fancyhead[LE]{\sf \nouppercase{Notation}}
\fancyhead[RO]{}

\small
\begin{longtable}{r@{\hspace{1em}}l}
	boldface            & vector or matrix\\
	lower case          & physical coordinates and radii\\
	upper case          & comoving coordinates and radii, except for $w$, $f_K(w)$\\
	$\dot{\mbox{}}$\hspace*{0.3em} & time derivative\\
	$\widehat{\mbox{}}$\hspace*{0.3em} & estimator\\
	$\widetilde{\mbox{}}$\hspace*{0.3em} & Fourier transform\\
	$^*$                & complex conjugate\\
	$\ast$              & convolution operator\\
	$\nabla$            & gradient operator\\
	$\nabla_\perp$      & transverse derivative operator\\
	$\mathbbm{1}$       & indicator function\\
	$a$                 & scale factor\\
	$A$                 & amplitude of the power spectrum\\
	$A_\pix$            & pixel area\\
	$A_\epsilon$        & probability of passing the one-sample test\\
	$\mathcal{A}$       & distortion matrix\\
	$B$                 & brightness (flux density)\\
	$\rmc$              & light speed\\
	$c$                 & additive correction for shape measurement\\
	$c$                 & copula density\\
	$c$                 & halo concentration\\
	$c_0$               & amplitude parameter for the halo $M$-$c$ relation\\
	$C$                 & copula\\
	$C$                 & two-point-correlation function\\
	$\bC$               & covariance matrix\\
	$d$                 & dimension of a data vector\\
	$D$                 & differential operator\\
	$D$                 & distance function\\
	$D_+$               & growth factor\\
	$D_\rmA$            & angular diameter distance\\
	$D_\rmH$            & Hubble distance at the current time\\
	$\DL$               & angular diameter distance between the lens and the observer\\
	$\DLs$              & angular diameter distance between the lens and the source\\
	$D_\rmL$            & luminosity distance\\
	$D_\rmp$            & proper distance\\
	$\DS$               & angular diameter distance between the source and the observer\\
	$e_N$               & change of variables used in \chap{sect:modelling}\\
	$E$                 & energy\\
	$E$                 & ratio of Hubble parameters\\
	$f$                 & dimension of a parameter vector\\
	$f$                 & filling factor\\
	$f$                 & mass function, multiplicity function\\
	$f$                 & quantity for \acro{NFW} profiles\\
	$f_K$               & comoving transverse distance\\
	$f_\mathrm{NL}$     & amplitude of the primordial non-Gaussianity\\
	$F$                 & cumulative distribution function\\
	$g$                 & lens efficiency\\
	$g$                 & lensing reduced shear\\
	$\vect{g}$          & gravitational field\\
	$g_{\mu\nu}$        & metric\\
	$\rmG$              & gravitational constant\\
	$G$                 & dimensionless projected mass\\
	$G$                 & Green function\\
	$G_{\mu\nu}$        & Einstein tensor\\
	$h$                 & dimensionless Hubble parameter at the current time\\
	$H$                 & Hubble parameter\\
	$\vect{H}$          & bandwidth matrix for kernel density estimation\\
	$H_0$               & Hubble parameter at the current time\\
	$I$                 & integral\\
	$I$                 & surface brightness (brightness density)\\
	$k$                 & norm of the vector $\vect{k}$\\
	$\vect{k}$          & wave vector in \acro{3D} Fourier space\\
	$K$                 & smoothed noiseless convergence\\
	$K$                 & spacetime curvature\\
	$K_N$               & smoothed noisy convergence\\
	$\vect{\ell}$       & wave vector in \acro{2D} Fourier space\\
	$L$                 & linear differential operator\\
	$L$                 & log-likelihood function\\
	$L$                 & luminosity\\
	$\mathcal{L}$       & likelihood function\\
	$\ln$               & natural logarithm, to base e\\
	$\log$              & common logarithm, to base 10\\
	$m$                 & multiplicative correction for shape measurement\\
	$M$                 & mass\\
	$M_\star$           & pivot mass\\
	$\Msol$             & solar mass\\
	$M_\ap$             & aperture mass\\
	$M_\maxx$           & upper limit of mass sampling\\
	$M_\minn$           & lower limit of mass sampling\\
	$n$                 & mass function, halo number density\\
	$n$                 & noise field\\
	$n_\gala$           & galaxy number density\\
	$n_\peak$           & peak function, \acro{PDF} of having a peak with a specific \acro{S/N}\\
	$n_\rms$            & scalar index, power law index of the scalar field power spectrum\\
	$N$                 & number of realizations\\
	$N$                 & smoothed noise field\\
	$N_\gala$           & number of galaxies\\
	$\mathcal{N}(\vect{\mu}, \bC)$ & normal law with the mean $\vect{\mu}$ and the covariance $\bC$\\
	$p$                 & $p$-value\\
	$p$                 & pressure\\
	$p$                 & source redshift distribution\\
	$p_i$               & $p$-value corresponding to the $i$-$\sigma$ significance\\
	$P$                 & distribution in the data space\\
	$P$                 & power spectrum\\
	$P(\cdot|\bpi)$     & probability density of a stochastic model under a parameter set $\bpi$\\
	$\mathcal{P}$       & distribution in the parameter space\\
	$\mathcal{P}(\cdot)$ & prior distribution\\
	$\mathcal{P}(\cdot|\bx^\obs)$ & posterior distribution\\
	$P_\kappa$          & \acro{2D} convergence power spectrum\\
	$Q$                 & aperture mass filter in shear space\\
	$Q$                 & number of particles\\
	$Q$                 & second-order moment tensor\\
	$r$                 & success ratio\\
	$r_\rms$            & scale radius\\
	$r_\mathrm{stop}$   & shutoff parameter\\
	$r_\vir$            & physical virial radius\\
	$\vect{r}$          & physical coordinates\\
	$R$                 & Ricci scalar\\
	$R_\vir$            & comoving quantity corresponding to the physical virial radius\\
	$R_{\mu\nu}$        & Ricci tensor\\
	$\vect{R}$          & comoving coordinates\\
	$s$                 & summary statistic\\
	$S$                 & source function\\
	$S$                 & surface\\
	$t$                 & cosmic time\\
	$T$                 & number of iterations\\
	$T_{\mu\nu}$        & stress-energy tensor\\
	$\mathcal{T}$       & optical tidal matrix\\
	$u$                 & percentile\\
	$U$                 & aperture mass filter in convergence space\\
	$\vect{u}$          & velocity\\
	$\vect{v}$          & peculiar velocity\\
	$V$                 & volume\\
	$w$                 & comoving radial distance\\
	$w$                 & equation of state\\
	$\wZero$            & equation of state of dark energy\\
	$W$                 & filter function\\
	$W_{\bC}$           & Gaussian smoothing kernel with covariance $\bC$\\
	$W_R$               & filter function with characteristic size $R$\\
	$x_N$               & change of variables used in \chap{sect:modelling}\\
	$\bx$               & data vector, vector of observables\\
	$\bx^\model$        & data vector from the model prediction\\
	$\bx^\obs$          & data vector from the observation\\
	$z$                 & complex number\\
	$z$                 & redshift\\
	$z_\rms$            & source redshift\\
	$\alpha$            & \acro{FDR} parameter for \MRLens\\
	$\alpha$            & inner slope parameter for \acro{NFW} profiles\\
	$\alpha$            & slope parameter for brightness power law\\
	$\alpha$            & slope parameter for $\Sigma_8$\\
	$\alpha_\gala$      & source distribution parameter\\
	$\beta$             & dragging value for $\Sigma_8$\\
	$\beta$             & mass power law index for the halo mass-concentration relation\\
	$\beta_\gala$       & source distribution parameter\\
	$\bbeta$            & unlensed angular position\\
	$\gamma$            & lensing shear\\
	$\gamma_\ast$       & quantity used in \chap{sect:modelling}\\
	$\gamma_+$          & tangential shear\\
	$\gamma_\times$     & cross shear\\
	$\delta$            & density contrast\\
	$\delta$            & difference between summary statistics\\
	$\delta$            & Dirac delta function\\
	$\delta$            & Kronecker delta\\
	$\delta_\rmc$       & linear threshold for density contrast\\
	$\Delta$            & overdensity\\
	$\Delta^2$          & dimensionless power spectrum\\
	$\epsilon$          & observed ellipticity\\
	$\epsilon$          & tolerance level\\
	$\epsilon\src$      & intrinsic ellipticity\\
	$\theta$            & change of variables used in \chap{sect:structure}\\
	$\theta$            & distance in the angular space, norm of the vector $\btheta$\\
	$\btheta$           & \acro{2D} angular position coordinates\\
	$\theta_\ast$       & quantity used in \chap{sect:modelling}\\
	$\theta_1$          & angular coordinate component, aligned with right ascension\\
	$\theta_2$          & angular coordinate component, aligned with declination\\
	$\theta_\rmG$       & Gaussian kernel size\\
	$\theta_{\ker}$     & kernel size\\
	$\theta_N$          & change of variables used in \chap{sect:modelling}\\
	$\theta_\pix$       & pixel width\\
	$\theta_\rms$       & angular scale radius\\
	$\theta_\vir$       & angular virial radius\\
	$\Theta$            & Heaviside step function\\
	$\kappa$            & lensing convergence\\
	$\Delta\kappa$      & difference between the true convergence and $\kappa_\halo$\\
	$\kappa_0$          & free constant from the \acro{KS} inversion\\
	$\kappa_1$          & correction for $\kappa_\proj$\\
	$\kappa_\halo$      & lensing convergence from the projected mass of a halo\\
	$\kappa_\proj$      & lensing convergence from the projected mass along a line of sight\\
	$\kappa_n$          & convergence with noise\\
	$\lambda$           & affine parameter\\
	$\lambda$           & eigenvalue\\
	$\lambda$           & mass proportion within a specified range\\
	$\lambda$           & mass-sheet degeneracy\\
	$\lambda$           & threshold for filling factor\\
	$\lambda$           & wavelength\\
	$\Lambda$           & cosmological constant\\
	$\mu$               & lensing magnification\\
	$\nu$               & \acro{S/N} of peaks\\
	$\bxi$              & physical separation vector between two geodesics\\
	$\bXi$              & comoving separation vector between two geodesics\\
	$\bpi$              & parameter vector\\
	$\bpi^\inp$         & input parameter vector\\
	$\rho$              & matter density\\
	$\bar{\rho}$        & background matter density\\
	$\bar{\rho}_0$      & background matter density at the current time\\
	$\rho_\crit$        & critical density at the current time\\
	$\rho_\rms$         & central mass density\\
	$\sigma$            & density dispersion\\
	$\sigEig$           & matter fluctuation under a spherical smoothing with radius 8 $\Mpc/h$\\
	$\sigma_i$          & $i$-th order moment of a smoothed noise field\\
	$\sigma_\noise^2$   & variance of a smoothed noise field, squared zero-order moment\\
	$\sigma_\pix^2$     & variance of noise in a pixel\\
	$\sigma_\epsilon^2$ & variance of the intrinsic ellipticity dispersion, sum of both components\\
	$\Sigma$            & surface mass density\\
	$\Sigma_8$          & tilted matter fluctuation defined by \for{for:constraint:Sigma_8} or \for{for:filtering:CSE}\\
	$\DCSE$             & width of the banana-shaped contour on the $\OmegaM$-$\sigEig$ plane\\
	$\Sigma_\crit$      & critical surface mass density\\
	$\tau$              & conformal time\\
	$\phi$              & probability density of a normal distribution\\
	$\phi$              & reduced Newtonian potential\\
	$\varphi$           & rotation angle\\
	$\Phi$              & Newtonian potential\\
	$\chi_f^2$          & probability density of the chi-squared distribution with $f$ degrees of freedom\\
	$\psi$              & lensing potential\\
	$\omega$            & weight, weight function\\
	$\Omega$            & solid angle\\
	$\OmegaB$           & dimensionless baryon density at the current time\\
	$\Omega_K$          & dimensionless density corresponding to the space-time curvature\\
	$\OmegaM$           & dimensionless matter density at the current time\\
	$\Omega_\rmr$       & dimensionless radiation density at the current time\\
	$\OmegaL$           & dimensionless density corresponding to the cosmological constant for a \acro{$\LCDM$} model
\end{longtable}
\normalsize

\clearpage
\thispagestyle{empty}
\cleardoublepage


\fancyhead[LE]{\sf \nouppercase{Bibliography}}
\fancyhead[RO]{}

\ifnumequal{\entreeBiblio}{0}{}{
	\selectlanguage{english}
	\settocbibname{\nomBiblio}

	\ifnumequal{\entreeBiblio}{1}{ 
		\bibliography{BibList_WLPC,BibList_others,BibList_literature}
	}{}

	\ifnumequal{\entreeBiblio}{2}{ 
		
	}{}

	\clearpage
	\thispagestyle{empty}
	\cleardoublepage
}

\fancyhead[LE]{\sf \nouppercase{Index}}
\fancyhead[RO]{}
\printindex
\clearpage
\thispagestyle{empty}
\cleardoublepage

\end{document}